\documentclass[a4paper,11pt]{article}
\usepackage[dvipsnames]{xcolor}
\usepackage[utf8]{inputenc}
\usepackage[bbgreekl]{mathbbol}
\usepackage{geometry}

\usepackage{xcolor}
\usepackage{amsmath, array, amssymb, amsfonts,amsthm}
\usepackage{upgreek}
\usepackage{xfrac}
\usepackage[inline]{enumitem}
\setlist{nolistsep}
\usepackage[french, english]{babel}
\usepackage[all]{xy}
\usepackage{txfonts}  
\usepackage{sectsty}
\usepackage{booktabs}
\usepackage{caption}
\usepackage{dsfont}
\usepackage{mathtools}
\usepackage{slashed}
\usepackage[makeroom]{cancel}
\usepackage[hidelinks]{hyperref}
\usepackage{textcomp}
\usepackage{multicol}
\usepackage[title]{appendix}
\usepackage[square,numbers,sort,compress,semicolon,merge]{natbib}
\let\cite\citep 
\usepackage{hyperref}
\hypersetup{colorlinks=true, urlcolor=blue, citecolor=blue, linktoc=page}

\usepackage{tikz}
\usepackage{tikz-cd}
\usetikzlibrary{cd}
\tikzcdset{
arrow style=tikz,
diagrams={>={Straight Barb[scale=0.8]}}
}

\DeclareMathAlphabet{\mathpzc}{OT1}{pzc}{m}{it}

\usepackage{mathrsfs}
\usepackage{scrextend}

\makeatletter
\renewcommand*\env@matrix[1][\arraystretch]{%
  \edef\arraystretch{#1}%
  \hskip -\arraycolsep
  \let\@ifnextchar\new@ifnextchar
  \array{*\c@MaxMatrixCols c}}
\makeatother

\newcommand{\defeq}{\vcentcolon=}
\newcommand{\rdefeq}{=\vcentcolon}

\newcommand\M{\mathcal{M}}

\newcommand\RR{\mathbb{R}}
\newcommand\CC{\mathbb{C}}
\newcommand\C{\mathcal{C}}

\newcommand\id{\textit{id}}
\newcommand\T{\mathcal{T}}

\renewcommand\H{\mathcal{H}}
\newcommand\A{\mathcal{A}}

\renewcommand\S{\mathcal{S}}

\newcommand\U{\mathcal{U}}
\newcommand\SO{\mathcal{SO}}

\renewcommand\O{\mathcal{O}}

\newcommand\D{\mathcal{D}}
\newcommand\W{\mathcal{W}}
\newcommand\vphi{\varphi}

\newcommand\sS{\mathsf{S}}

\renewcommand\epsilon{\varepsilon}

\newcommand\rarrow{\rightarrow}

\newcommand\aut{\mathfrak{aut}}
\newcommand\diff{\mathfrak{diff}}

\newcommand\LieG{\mathfrak{g}}

\renewcommand\t{\tilde}

\renewcommand\b{\bar }
\newcommand\w{\wedge}
\renewcommand\d{\partial}
\newcommand\s{\sigma}
\newcommand\bs{\boldsymbol}
\newcommand \omegaf{\bs\omega_{\text{\tiny$0$}}}

\renewcommand\-{^{-1}}
\newcommand\Ad{\text{Ad}}

\renewcommand\id{\text{id}}

\def\munderline#1#2{\color{#1}\underline{{\color{black}#2}}\color{black}}
\newcommand{\bline}[2][4]{\underline{#2\mkern-#1mu}\mkern#1mu }

\makeatletter

\newcommand{\Rmnum}[1]{\expandafter\@slowromancap\romannumeral #1@}
\makeatother

\makeatletter
\newcommand{\leqnomode}{\tagsleft@true\let\veqno\@@leqno}
\newcommand{\reqnomode}{\tagsleft@false\let\veqno\@@eqno}
\makeatother

\DeclareMathOperator{\Diff}{Diff}
\DeclareMathOperator{\Aut}{Aut}
\DeclareMathOperator{\Der}{Der}

\DeclareMathOperator{\im}{Im}

\newtheorem{thm}{Theorem}

\newtheorem{prop}[thm]{Proposition}
\theoremstyle{definition}

\makeatother

\begin{document}


\title{ 
The dressing field method for diffeomorphisms: \\
a relational framework
 }
 \author{J. T. \textsc{François} \textsc{André}}
\date{}

\maketitle
\begin{center}
\vskip -0.8cm
\noindent
Department of Mathematics \& Statistics, Masaryk University -- MUNI.\\
Kotlářská 267/2, Veveří, Brno, Czech Republic.\\[1.5mm]
 
Department of Philosophy -- University of Graz. \\
 Heinrichstraße 26/5, 8010 Graz, Austria.\\[1.5mm]
 
Department of Physics, Mons University -- UMONS.\\
 Service \emph{Physics of the Universe, Fields \& Gravitation}.\\
20 Place du Parc, 7000 Mons, Belgium.\\[1.5mm]

{\small Email: francois@math.muni.cz, \  Orcid: 0000-0002-3808-6343}\\[1.5mm]

\end{center}
%
\vspace{-2mm}

\begin{abstract}
The dressing field method is a tool to reduce gauge symmetries. Here we extend it to cover the case of diffeomorphisms. The resulting framework is a  systematic scheme to produce $\Diff(M)$-invariant objects, which has a natural relational interpretation. 

Its precise formulation relies on a clear understanding of the bundle geometry of field space. By detailing it, among other things we stress the geometric nature of field-independent and field-dependent diffeomorphisms, and highlight that the heuristic ``extended bracket" for field-dependent vector fields often featuring in the covariant phase space literature can be understood as arising from  the Frölicher-Nijenhuis bracket. 
Furthermore, by articulating this bundle geometry with the covariant phase space approach, we  give a streamlined account of the elementary objects of the (pre)symplectic structure of a $\Diff(M)$-theory: Noether charges and their bracket, as induced by the standard prescription for the presymplectic potential and 2-form. We  give conceptually transparent expressions allowing to read the integrability conditions and the circumstances under which the bracket of charge is Lie, and the resulting Poisson algebras of charges are central extensions of the Lie algebras of field-independent ($\diff(M)$) and field-dependent vector fields.

We show that, applying the dressing field method, one obtains a $\Diff(M)$-invariant and manifestly relational formulation of a general relativistic field theory. 
Relying on results just mentioned, we easily derive the ``dressed" (relational) presymplectic structure of the theory. 
This reproduces or extends results from the gravitational edge modes and gravitational dressings literature. 
In addition to simplified technical derivations, the conceptual clarity of the framework supplies several insights and allows us to dispel misconceptions. 
\end{abstract}


\vspace{-3mm}

\tableofcontents

\bigskip

 
\section{Introduction}  
\label{Introduction}  

The dressing field method (DFM) is an  algorithm to build basic forms on a bundle, i.e. to obtain gauge-invariants in gauge theories. It was first introduced systematically in \cite{GaugeInvCompFields, Francois2014, Attard_et_al2017}, and its implications regarding the philosophy of gauge theories  has been first expounded in \cite{Francois2018}. 
Its applications range from the construction of twistors and tractors  in conformal Cartan geometry \cite{Attard-Francois2016_II, Attard-Francois2016_I}, to reformulation of electroweak physics dispensing with the notion of  spontaneous symmetry breaking (SSB) \cite{Berghofer-et-al2023}. 
More recently, it was shown in \cite{Francois2021, Francois-et-al2021} that it is the  geometric underpinning of the notion of \emph{edge modes} introduced in recent years in the study of the presymplectic structure of gauge theories over bounded regions  \cite{DonnellyFreidel2016, Geiller2017, Speranza2018, Geiller2018, Geiller2019, Freidel-et-al2020-1, Freidel-et-al2020-2, Freidel-et-al2020-3, Teh-et-al2021} -- which was first hinted at in \cite{Teh-et-al2020}. 

  The DFM has been first developed  for theories with internal gauge symmetries, i.e. Yang-Mills theories and gauge theories of gravity formulated via Cartan geometry. The interested reader can find self-contained expositions in sections 3 and 4.3.1 of \cite{Francois2021}, section 2.3 of \cite{Francois-et-al2021}, or chapter 5 of \cite{Berghofer-et-al2023}. See also \cite{Riello-Schiavina2023b, Riello-Schiavina2023a} for another mathematical development and 
  applications.
   Here, we aim to provide a detailed exposition of its natural extension to theories with $\Diff(M)$-symmetry, and to show how the resulting framework encompasses and unifies a diversity of notions in old and new literature on gravity (to be named  below). 
 The proper mathematical formulation of the DFM requires some familiarity with the geometry of the field space $\Phi$ of $\Diff(M)$-theories.
 The~paper is thus organised as follows. 
\medskip

In section \ref{Geometry of field space}, in a systematic manner, we  lay the elementary notions pertaining to the geometry of $\Phi$ as an infinite dimensional  fiber \mbox{bundle} with structure group $\Diff(M)$. 
First, we will be interested in the bundle structure of $\Phi$: the smooth action of $\Diff(M)$,  
the definition of its group of vertical diffeomorphisms (otherwise known as ``transformations under \emph{field-dependent} gauge parameters"), giving in particular the geometric definition of gauge transformations.
 Then, we consider the differential structure: tangent bundle and remarkable vector fields,  the de Rham complex, remarkable spaces of differential forms, and the graded algebra of their derivations.
 
 This will  be the occasion to show that the heuristic bracket for field-dependent gauge parameters (vector fields) introduced in \cite{Bergmann-Komar1972} and \cite{Salisbury-Sundermeyer1983}, then again in \cite{Barnich-Troessaert2009} -- and often used in covariant phase space literature, see e.g. \cite{Gomes-et-al2018, Freidel-et-al2021, Freidel-et-al2021bis, Chandrasekaran-et-al2022} -- is  but a special case of the Frölicher-Nijenhuis bracket for vector-valued differential forms on $\Phi$. The field space Lie derivative along such field-dependent vector fields is simply the Nijenhuis-Lie derivative \cite{Kolar-Michor-Slovak}. Of course, both these facts are true when an internal gauge group is taken as structure group of $\Phi$ instead of $\Diff(M)$. 

Once these elementary notions properly introduced, we will mention two natural notions of connections  $\Phi$ can be endowed with: Ehresmann connections, and a generalisation known as ``twisted connections" \cite{Francois2019_II}.  \mbox{The latter are} directly relevant 
to understanding the geometry behind classical and/or quantum $\Diff(M)$-anomalies. 

Finally, we will give an account of integration on $M$ as a natural pairing operation on a space we call the \emph{associated bundle of regions}: a fiber bundle associated to field space via the defining representation of the structure group $\Diff(M)$, i.e. the field of open sets $U\subset M$. This allows us to show how field-dependent diffeomorphisms act (non-trivially) on integrals.
\medskip

Section \ref{The dressing field method} details the dressing field method (DFM) for diffeomorphisms. 
It is a systematic procedure to produce \emph{basic} forms on $\Phi$, invariant under vertical diffeomorphisms,  provided one can identify or build a \emph{dressing field}. A~notion we define precisely. 
We consider the question of residual symmetries, arising in case of partial symmetry reduction, or of possible ambiguities in the choice of dressing field. We show that in the latter case, the space of dressed fields has its own bundle structure, mirroring that of $\Phi$. 
The notion of dressed integrals is then defined, and we discuss its relevance to physical applications (to be detailed later), especially regarding the variational principle. 

The conceptual implications of the DFM are discussed, and contact with the literature is made.
It is  argued that the framework developed, technically implementing Einstein's point coincidence argument to field theory \cite{Norton1993, Stachel2014, Giovanelli2021}, has a natural relational interpretation. 

We then aim to showcase a natural application of the DFM to the covariant presymplectic structure of $\Diff(M)$-theories, to highlight how it reproduces results 
pertaining to this literature. 
%

\medskip

In section \ref{Covariant phase space methods},  we thus give our account of covariant phase space methods for $\Diff(M)$-theories, making explicit use of the bundle geometry of $\Phi$ exposed in \ref{Geometry of field space}. Given a choice of Lagrangian, we  define Noether currents and charges for both field-independent and field-dependent vector fields, and show that, equipped with the bracket induced by the presymplectic 2-form, their Poisson algebras are  respectively central extensions of the Lie algebras of the structure group of field space, $\diff(M)$, and  of the group of vertical diffeomorphisms of field space. We give (hopefully) conceptually transparent and concrete expressions for all relevant objects.

Then we move to deriving geometrically the vertical transformations 
of key objects: Lagrangian, field equations, presymplectic potential and 2-form.
This will show how and why $\Diff(M)$-theories enjoy a much larger ``covariance group" under field-dependent diffeomorphisms, as hinted at in \cite{Bergmann-Komar1972}. 
  This is used to derive naturally the previously mentioned Noether charges for field-dependent parameters together with their bracket. 
  
 Our  goal here is twofold. Firstly, we aim to be both synthetic and systematic in our presentation of elementary covariant phase space notions: 
  by streamlining their technical derivation and improving the conceptual clarity, we hope to lower the cost of entry for newcomers (PhDs and postdocs most notably, whose time is precious). 
 Secondly, as stated above, we will rely on the derived results  to showcase the DFM.
 \medskip

This we do in a final short section \ref{Relational formulation}, where  the DFM is applied to easily write down the dressed version of a $\Diff(M)$-theory together with their \emph{basic} presymplectic structure.  
This  will make manifest that the DFM is the unifying  geometric framework underpinning various notions featuring in the literature on gravity: Notably in the past decade that of  \emph{gravitational edge modes} as introduced in  \cite{DonnellyFreidel2016} 
 and  developed e.g. in \cite{Speranza2018, Freidel-et-al2020-1, Freidel-et-al2020-2, Freidel-et-al2020-3, Chandrasekaran_Speranza2021, Speranza2022, Chandrasekaran-et-al2022}, 
 and of \emph{gravitational dressings} as proposed in \cite{Giddings-Donnelly2016, Giddings-Donnelly2016-bis, Giddings-Kinsella2018, Giddings2019, Giddings-Weinberg2020} -- see also  \cite{Kabel-Wieland2022}  -- as well as the more recent idea of   ``dynamical reference frames" as expounded in \cite{carrozza-et-al2022, Hoehn-et-al2022}, or yet that of ``embedding maps/fields" as advocated in \cite{Speranza2019, Ciambelli-et-al2022,  Speranza2022, Ciambelli2023}. 
  
 But a more fundamental fact, we will argue again, is  that the DFM makes manifest the \emph{relational} character of general relativistic physics:
 By systematically implementing a notion of \emph{relational observables} \cite{Tamborino2012}, or $\Diff(M)$-\emph{Dirac variables},  
it allows to reformulate a general relativistic field theory  in a $\Diff(M)$-invariant and \emph{manifestly relational} way. 
This provides some conceptual insights. In particular it  dispels the often encountered notion that ``boundaries breaks $\Diff(M)$-invariance" (a statement equivalent to the famous hole argument). It will also suggest a resolution as to why general relativistic theories could be successfully experimentally tested before solving the issue of their observables. The technical relational formulation provided by the DFM encompassed various notions of ``scalar coordinatisation" proposed for GR, e.g. \cite{Rovelli1991, Brown-Kuchar1995, Rovelli2002b, Pons-et-al2009, Pons-et-al2009b}. 
\medskip

We conclude by  taking stock of our main results and hint at future applications and developments. Technical appendices complete the main text.

\section{Geometry of field space}  
\label{Geometry of field space}  

In this section we give a sense of the bundle geometry of field space as an infinite dimensional manifold. 
We aim for a correct conceptual description rather than for a perfectly rigorous account,  we thus refer to the the relevant literature, e.g. \cite{Frolicher-Kriegl1988, Kriegl-Michor1997}, to back the soundness of extending  standard notions defined in the finite dimensional context to the infinite dimensional setting.  

\subsection{Field space as a principal bundle}  
\label{Field space as a principal bundle}  


As we are interested in $\Diff(M)$-theories, our field space is  $\Phi=\Gamma(\T)$ where $\T$ denotes collectively any kind of tensor or pseudo-tensor bundles. Generically these are vector and affine bundles associated to the frame bundle $LM$ of  $M$, and,  if we are interested in  gauge field theories based on a principal bundle $P(M,  H)$ with structure group $H$, it may include the affine bundle of (Ehresmann or Cartan) connections $\C$ and/or any vector bundles $E$ associated to $P$ via  representations of $H$. If we have  furthermore  a $S\!O$-reduction $OM$ of $LM$, we may consider also  spinor bundles $\sS$ associated to $OM$ via spin representations, and a fortiori the bundles $\sS \otimes E$ whose sections would describe gauge-charged spinors. A set of sections of these bundles will be noted collectively as $\phi \in \Phi$, i.e. a single point in field space.

 The group $\Diff(M)$  has a natural  action by pullback on sections of  those bundles:
  Given $\phi \in \Phi$ and $\psi \in \Diff(M)$ we have,  $\phi^\psi := \psi^*\phi$ -- where the left-hand side is a notation for the pullback on the right-hand side. We will use it when convenient. 
 Due to the well-known relation $(f\circ g)^*=g^* \circ f^*$ for any two smooth maps $M \xrightarrow{g}N \xrightarrow{f} Q$, this action is a right-action.\footnote{This is clear on forms, a.k.a. contravariant tensors. On vector fields and more general covariant tensors, the pullback action is $\psi^* \defeq {\psi^{-1}}_*$, the pushforward by the inverse diffeomorphism. So, on general (mixed) tensors, $\psi^*$ is indeed a well-defined right action of $\Diff(M)$. } 
 We therefore note,
 \begin{equation}
  \begin{aligned}
  \label{Right-action-on-Phi}
  \Phi \times\Diff(M) &\rarrow \Phi,  \\
               (\phi, \psi) &\mapsto R_\psi \phi:= \psi^*\phi,
\end{aligned}
 \end{equation}              
with indeed, for another  $\psi' \in \Diff(M)$: $R_{\psi'} R_\psi \phi:= {\psi'}^*\psi^*\phi= (\psi \circ \psi')^*\phi=:R_{\psi \circ \psi'} \phi$. The field space is fibered by the action of $\Diff(M)$, the fiber through a point $\phi$ being its orbit $\mathcal O(\phi)$ under diffeomorphisms. The set of orbits, or moduli space, we denote $\Phi/\Diff(M) =: \M$. Under adequate restrictions of either of $\Phi$ or $\Diff(M)$, this means that the field space $\Phi$ can be understood as an infinite dimensional principal fiber bundle over $\M$, its \emph{base} space, with structure group $\Diff(M)$.\footnote{We refer to \cite{Kriegl-Michor1997} for rigorous results on how to describe a fibered space with the non-compact structure group $\Diff(M)$.}\footnote{In particular, this requires that all points $\phi$ have trivial stability groups, meaning that we a priori reject Killing symmetries of $\phi$. 
} We note, 
 \begin{equation}
  \begin{aligned}
  \Phi &\xrightarrow{\pi} \M,  \\
   \phi &\mapsto \pi(\phi)=: [\phi]
\end{aligned}
 \end{equation}     
The projection $\pi$ of course satisfies the usual relation $\pi \circ R_\psi= \pi$. 
The fiber over a point $[\phi] \in \M$ can be noted $\pi\-([\phi])=\mathcal O(\phi)$, and it is diffeomorphic to the structure group $\Diff(M)$ as a manifold.

As a fiber bundle, $\Phi$ is  locally trivialisable. Meaning, for  any open subset $\mathcal U \subset \M$ there is a morphism of $\Diff(M)$-spaces $t: \Phi_{|\U}  \mapsto \U \times \Diff(M)$, $\phi \mapsto (\pi(\phi), \uppsi(\phi))$, called a \emph{trivialisation}. In others words, it is a local coordinate system on $\Phi$, and $\uppsi$ is the fiber coordinate of $\phi$. It is s.t. $t(\phi^{\psi}) = R_\psi \, t(\phi)= t(\phi)^\psi = (\pi(\phi), \uppsi(\phi)^\psi)$. A~trivialisation is equivalent to the datum of a \emph{local section}, $\bs \s: \U \rarrow \Phi_{|\U}$, $[\phi] \mapsto \bs \s([\phi])\rdefeq\phi_0$, s.t. $\pi \circ \bs\s =\id^{}_\U$. One considers that $t(\phi_0)=([\phi], \id)$, so that any other point $\phi$ of the fiber over $[\phi]$ can be reached via the action of the structure group: $\phi=R_\uppsi \bs\s([\phi])=\bs\s([\phi])^\uppsi$. 
Unless $\Phi$ is trivial, there are no global section $\bs \s: \M \rarrow \Phi$. A choice of local section $\bs\s$ is a \emph{gauge fixing} -- a choice of representative in each gauge orbit over $\U$ -- and the previous statement simply expresses the inexistence of a globally valid gauge fixing procedure, i.e. the Gribov-Singer obstruction \cite{Gribov, Singer1978}.

Two choices of local sections $\bs\s'$ and $\bs\s$ are related via $\bs\s'= R_{\bs g} \bs\s$, where $\bs g: \U \rarrow \Diff(M)$, $[\phi] \mapsto \bs g([\phi])$, belongs to the set of \emph{transition functions}  of $\Phi$.\footnote{The latter constitute \emph{passive} gauge transformations, or \emph{gluings}, on  $\M$. We will be concerned throughout this paper with \emph{active} gauge transformations  associated to  vertical automorphisms on $\Phi$ (and more generally with vertical diffeomorphisms of $\Phi$). Conceptually, these two are in the same relationship as coordinate changes, a.k.a. passive diffeomorphisms,  and $\Diff(M)$, a.k.a active diffeomorphisms.}
We stress that this means that a choice of gauge-fixing/local section \emph{cannot be invariant} under a further action of the structure group $\Diff(M)$.

\subsubsection{Natural transformation groups} 
\label{Natural transformation groups}  

As an infinite dimensional manifold $\Phi$ has a diffeomorphism group $\bs{\Diff}(\Phi)$, but \emph{as a principal bundle} its maximal transformation group is its group of automorphisms $\bs{\Aut}(\Phi) := \big\{ \Xi \in \bs{\Diff}(\Phi) \, |\, \Xi \circ R_\psi = R_\psi \circ \Xi  \big\}$. Its elements  preserve the fibration structure, and thus project naturally as elements of $\bs{\Diff}(\M)$. 

The subgroup of \emph{vertical} diffeomorphisms $\bs{\Diff}_v(\Phi) \defeq \left\{ \Xi \in \bs{\Diff}(\Phi)\, |\, \pi \circ \Xi = \pi \right\}$  induce the identity transformation on $\M$: 
These are moving along fibers, therefore  to $\Xi \in \bs{\Diff}_v(\Phi)$ corresponds a unique $\bs\psi: \Phi \rarrow \Diff(M)$ s.t. $\Xi(\phi)=R_{\bs\psi(\phi)} \phi := [\bs\psi(\phi)]^* \phi$. In other words  $\bs{\Diff}_v(\Phi) \simeq C^\infty\big(\Phi, \Diff(M)\big)$. It should be noticed that the map composition law for $\bs{\Diff}_v(\Phi)$ gives rise to a peculiar composition operation for $C^\infty\big(\Phi, \Diff(M)\big)$. 
Indeed, for $\Xi, \Xi' \in \bs{\Diff}_v(\Phi)$ to which correspond $\bs\psi, \bs\psi' \in C^\infty\big(\Phi, \Diff(M)\big)$, one has: 
$
\Xi' \bs \circ\,\! \Xi \,(\phi) = R_{\bs\psi' \left( \Xi(\phi) \right)}\, \Xi(\phi) = R_{\bs\psi' \left( \Xi(\phi) \right)}\, R_{\bs\psi(\phi)}\phi = R_{\bs\psi(\phi) \circ \bs\psi' \left( \Xi(\phi) \right) } \phi.
$
Thus,  
\begin{align}
\label{twisted-comp-law}
 \Xi' \bs \circ\,\! \Xi \in  \bs{\Diff}_v(\Phi) \ \ \text{ corresponds to }\ \ \bs\psi \circ \big( \bs\psi' \bs \circ R_{\bs \psi}\big) \in C^\infty\big(\Phi, \Diff(M)\big). 
\end{align}
Notice we distinguish the composition law $\bs \circ$ of maps on $\Phi$, from the composition law $\circ$ of maps on $M$. 

This is an example of the composition law of the infinite dimensional group of bisections of a Lie groupoid \cite{Mackenzie2005, Schmeding-Wockel2015, Maujouy2022}. As a mater of fact, what we have been discussing above can be reframed in the groupoid framework as  follows: 
One defines the generalised action groupoid $\bs\Gamma  \rightrightarrows \Phi$ with $\bs \Gamma = \Phi \rtimes \Big(C^\infty\big(\Phi, \Diff(M)\big) \simeq \bs\Diff_v(\Phi)\Big)$, and with source and target maps 
$s: \bs\Gamma \rarrow \Phi$, $(\phi, \bs\psi \simeq \Xi) \mapsto \phi$,  and 
$t: \bs\Gamma \rarrow \Phi$, $(\phi, \bs\psi \simeq \Xi) \mapsto \Xi(\phi)=\bs\psi(\phi)^*\phi$.
The associative composition law: $g \circ f \in \bs\Gamma$ for $f,  g\in \bs\Gamma$ is defined whenever $t(f)=s(g)$.
 It generalises the action groupoid $\b{\bs\Gamma} =  \Phi \rtimes \Diff(M) \rightrightarrows \Phi$ associated with the right action of $\Diff(M)$ on $\Phi$. The group of bisections $\mathcal B(\bs\Gamma)$ of $\bs\Gamma$ is the set of sections of  $s$ --  maps $\sigma : \Phi \rarrow \bs\Gamma$, $\phi \mapsto \s(\phi)=(\phi, \bs\psi \sim \Xi)$  s.t. $s\circ \s =\id_\Phi$ -- such that $t \circ \s: \Phi\rarrow \Phi$ is invertible: thus $t \circ \s \in \bs\Diff(\Phi)$. We have indeed, for  $\s \in \mathcal B(\bs\Gamma)$, that $t \circ \s=\Xi \, (=\bs\psi^*) \in \bs\Diff_v(\Phi)\simeq C^\infty\big(\Phi, \Diff(M) \big)$. 
The group law of bisections is defined as 
$\big(\s_2 \star \s_1\big)(\phi) \defeq \s_2\big( (t \circ \s_1 (\phi) \big) \circ \s_1(\phi)$. 
After composition with the target map $t$ on the left, this reproduces precisely the peculiar group law \eqref{twisted-comp-law}. 
Then, slightly abusing the terminology, we may refer to $C^\infty\big(\Phi, \Diff(M) \big)$ as the group of bisections of the generalised action groupoid $\bs\Gamma$ associated to $\Phi$. 
In the rest of this paper though, we will simply, and accurately, refer to it as the group of generating maps of $\bs\Diff_v(\Phi)$.

The latter group  generalises the better known, and more often discussed, subgroup of \emph{vertical automorphisms} $ \bs{\Aut}_v(\Phi) := \big\{ \Xi \in \bs{\Aut}(\Phi) \, |\, \pi~\circ~\Xi~=~\pi  \big\}=\bs\Diff_v(\Phi)  \cap \bs\Aut(\Phi)$, isomorphic to the \emph{gauge group} 
 \begin{align}
 \label{GaugeGroup}
 \bs{\Diff}(M) := \big\{ \bs\psi : \Phi \rarrow   \Diff(M) \, |\,  \bs\psi(\phi^\psi) =\psi^{-1} \circ \bs\psi(\phi) \circ \psi   \big\} 
 \end{align}
via $\Xi(\phi)=R_{\bs\psi(\phi)}\, \phi$ still.\footnote{It is indeed easy to check that the equivariance condition on $\bs\psi \in \bs{\Diff}(M)$ flows from the definition of an automorphism: On the one hand  $\Xi \circ R_\psi (\phi):= \Xi(\phi^\psi):= R_{\bs\psi(\phi^\psi)} \phi^\psi := [\bs\psi(\phi^\psi)]^*\phi^\psi = [\bs\psi(\phi^\psi)]^*\psi^*\phi =  [\psi \circ  \bs\psi(\phi^\psi)]^*\phi$. On the other hand $R_\psi \circ \Xi (\phi):= R_\psi ( [\bs\psi(\phi)]^* \psi ) = \psi^*  [\bs\psi(\phi)]^* \phi = [\bs\psi(\phi) \circ \psi]^*\phi $. The equality of both expression imposes: $\bs\psi(\phi^\psi) =\psi^{-1} \circ \bs\psi(\phi) \circ \psi$, or $R^\star_\psi \bs\psi = \psi^{-1} \circ \bs\psi \circ \psi$. This condition is thus necessary if a map $\bs\psi$ is to induce a vertical automorphism of $\Phi$. }
The defining equivariance of elements $\bs\psi$ of $\bs\Diff(M)$ implies that to $ \Xi' \bs \circ\,\! \Xi \in  \bs{\Aut}_v(\Phi)$ \mbox{corresponds} $\bs\psi' \circ  \bs\psi \in\bs\Diff(M)$: i.e. the composition operation $\bs\circ$ in $ \bs{\Aut}_v(\Phi)$ translates to  the usual (pointwise) composition operation $\circ$ of the group $\bs\Diff(M)$ -- as expected and familiar in the Yang-Mills case. 

We have that $\bs{\Aut}(\Phi)$ is in the normaliser of  $\bs{\Diff}_v(\Phi)$, and as a group is a subgroup of its normaliser we get:  $N_{\bs\Diff(\Phi)}\big(\bs\Diff_v(\Phi) \big) = \bs\Diff_v(\Phi) \cup \bs\Aut(\Phi)$. As a special case, we have $N_{\bs\Diff(\Phi)}\big(\bs\Aut_v(\Phi) \big) = \bs\Aut(\Phi)$, i.e.  $ \bs\Aut_v(\Phi) \triangleleft  \bs\Aut(\Phi)$, which gives the short exact sequence (SES) of groups characteristic of a principal bundle:
 \begin{align}
 \label{SESgroup}
\bs{\Diff}(M) \simeq \bs{\Aut}_v(\Phi)  \rarrow   \bs{\Aut}(\Phi) \rarrow \bs{\Diff}(\M),
 \end{align}
where the image of each arrow is in the kernel of the next. We remark that, as physical degrees of freedom are gauge invariant, they should be in the moduli space $\M$. Therefore, if anything is deserving of the name of ``\emph{physical} symmetries", it is the right-most group $ \bs{\Diff}(\M)$.  

In the physics literature, especially on the covariant phase space approach  to gravity theories, one often encounters the notion of ``\emph{field-dependent}" gauge transformations, or diffeomorphisms. 
We submit that the group $\bs\Diff_v(\Phi) \simeq C^\infty\big(\Phi, \Diff(M) \big)$ is the correct mathematical embodiment of this notion. We nonetheless stress that stricto sensu, geometric (active) gauge transformations on $\Phi$ are defined by the action of the subgroup $\bs{\Aut}_v(\Phi) \simeq \bs{\Diff}(M)$. Still, the linearised action of $\bs\Diff_v(\Phi)$ gives rise to the so-called ``extended bracket" for infinitesimal  field-dependent gauge transformations, or diffeomorphisms, as introduced by Bergmann \& Komar \cite{Bergmann-Komar1972} and Salisbury \& Sundermeyer \cite{Salisbury-Sundermeyer1983}, then again by Barnich \& Troessaert (BT) in  \cite{Barnich-Troessaert2009} -- we will further indicate that this bracket is just  the Frölicher-Nijenhuis (FN) bracket of vector-valued differential forms  \cite{Kolar-Michor-Slovak}. Obviously, the structure group $\Diff(M)$ supplies the notion of  ``\emph{field-independent}" gauge transformations or diffeomorphisms.

The linearization of \eqref{SESgroup} is the Atiyah Lie algebroid of the principal bundle $\Phi$. To understand it, and more generally the differential structure of $\Phi$, let us first remind a general result that we will use often. 
\medskip

Consider the manifolds $M$, $N$ and their tangent bundles $TM$, $TN$, together with a diffeomorphism $\psi: M \rarrow N$ and the flow $\upphi_\tau:N \rarrow N$ of a vector field $X \in \Gamma(TN)$, s.t. $X_{|\upphi_0}=\tfrac{d}{d\tau} \upphi_\tau |_{\tau=0} \in T_{\upphi_0}N$. One defines the flow 
 \begin{align}
 \label{Psi-conjugation}
\vphi_\tau:= \psi\- \circ \upphi_\tau \circ \psi : M \rarrow M
\end{align}
of a vector field $Y \in \Gamma(TM)$  related to $X$ as: 
 \begin{align}
 \label{Psi-relatedness}
Y:=  \tfrac{d}{d\tau} \left( \psi\- \circ \upphi_\tau \circ \psi \right)\big|_{\tau=0}= (\psi\-)_*\, X \circ \psi.
 \end{align}
Indeed as maps we have $M \xrightarrow{\psi} N \xrightarrow{X} TN \xrightarrow{(\psi\-)_*} TM$. Their composition give the above vector field $Y: M \xrightarrow{} TM $.
We say that $Y$ and $X$ are $\psi$-\emph{related}. It is a standard result that $\psi$-relatedness is a morphism of Lie algebras, that is: $[Y, Y']=  (\psi\-)_* [X, X'] \circ \psi$. 
We  will therefore  remember that $X\in \Gamma(TN)$ and $Y\in \Gamma(TM)$ are $\psi$-related \eqref{Psi-relatedness} when their flows are $\psi$-\emph{conjugated} \eqref{Psi-conjugation}.

The following lemma can be considered a generalisation of the above: Suppose $\phi$ is some tensor field on $M$, and $\mathfrak L_X \phi$ its Lie derivative along $X \in \Gamma(TM)$ with flow $\varphi_\tau$. Then, for  $\psi \in \Diff(M)$ we have, 
\begin{align}
\label{lemma1}
\psi^*\big( \mathfrak L_X \phi \big) &=  \psi^* \tfrac{d}{d\tau} \varphi^*_\tau \phi \big|_{\tau=0} =\tfrac{d}{d\tau} (\varphi_\tau \circ \psi )^* \phi \big|_{\tau=0} 
						     =\tfrac{d}{d\tau} (\varphi_\tau \circ \psi )^* \ (\psi\-)^*\psi^* \phi \big|_{\tau=0}, \notag \\
						     &=\tfrac{d}{d\tau} ( \psi\- \circ \varphi_\tau \circ \psi )^* \ \psi^* \phi \big|_{\tau=0}, \notag\\
						     & =: \mathfrak L_{[(\psi\-)_*X \circ \psi]} \, (\psi^* \phi). 
\end{align}

\subsection{Differential structure} 
\label{Differential structure}  

As a manifold, $\Phi$ has a tangent bundle $T\Phi$, a cotangent bundle $T^\star\Phi$, or more generally a space of forms $\Omega^\bullet(\Phi)$. 
We~will consider both structures in turn. It will be useful to distinguish easily the pushforward and pullback on $M$ and $\Phi$; we thus reserve $*$ to denote these notions on $M$, and use $\star$ for their counterparts on $\Phi$.  

\subsubsection{Tangent bundle and subbundles} 
\label{Tangent bundle and subbundles}  

Sections of the tangent bundle  $\mathfrak X : \Phi \rarrow T\Phi$ are vector fields on $\Phi$, we note $\mathfrak X \in \Gamma(T\Phi)$. They form a Lie algebra under the bracket of vector field $[\ , \, ]: \Gamma(T\Phi) \times \Gamma(T\Phi) \rarrow \Gamma(T\Phi)$. 
We may write a vector field at $\phi \in \Phi$ as $\mathfrak X_{|\phi} =\tfrac{d}{d\tau}  \Psi_\tau(\phi) \, \big|_{\tau=0}$, with  $\Psi_\tau \in \bs\Diff(\Phi)$ its flow s.t.  $\Psi_{\tau=0}(\phi)=\phi$.
Considered as derivations of the algebra of functions $C^\infty(\Phi)$ we would write, in analogy with the finite dimensional case: $\mathfrak X = \mathfrak X(\phi) \tfrac{\delta}{\delta \phi} $, where $\tfrac{\delta}{\delta \phi}$ is of course the functional differentiation w.r.t. $\phi$, and $\mathfrak X(\phi)$ are the functional components. 
 
 The pushforward by the projection is the map $\pi_\star: T_\phi\Phi \rarrow T_{\pi(\phi)}\M=T_{[\phi]}\M$. 
The pushforward by the right action of $\psi \in \Diff(M)$ is the map $R_{\psi\star}: T_\phi\Phi \rarrow T_{\psi^*\phi}\Phi$.  
 In general $R_{\psi\star} \mathfrak X_{|\phi} \neq \mathfrak X_{|\psi^*\phi}$, meaning that a generic vector field ``rotates" as it is pushed vertically along fibers.  For this reason, in general $\pi_\star \mathfrak X$ is not a well-defined vector field on the base $\M$: Indeed, at $[\phi]\in \M$  the vector  obtained would vary depending on where on the fiber over $[\phi]$ the projection is taken.

There is a remarkable Lie subalgebra of $\Gamma(T\Phi)$, the \emph{right-invariant} vector fields:
 \begin{align}
 \label{Inv-vector-fields}
\Gamma_{\text{\!\tiny{inv}}}(T\Phi):=\left\{\mathfrak X \in \Gamma(T\Phi)\, |\, R_{\psi\star} \mathfrak X_{|\phi}=  \mathfrak X_{|\psi^*\phi}  \right\}.
 \end{align}
 Those do not rotate as they are pushed vertically. Consequently, they project as well-defined vector fields on $\M$: 
Indeed, for $\mathfrak X \in \Gamma_{\text{\!\tiny{inv}}}(T\Phi)$, we have $\pi_\star \mathfrak X_{|\psi^*\phi}=\pi_\star R_{\psi\star} \mathfrak X_{|\phi} = (\pi \circ R_\psi)_\star \mathfrak X_{|\phi} = \pi_\star \mathfrak X_{|\phi} =: \mathfrak Y_{|[\phi]}  \in T_{[\phi]}\M$. The result is  independent of  where on the fiber over $[\phi]$ the projection is taken, so $\pi_\star \mathfrak X =: \mathfrak Y \in \Gamma(T\M)$ is a well defined vector field. Furthermore, the defining property of invariant vector fields implies that their flow is an automorphism of $\Phi$: Indeed we have,
 \begin{align}
R_{\psi\star} \mathfrak X_{|\phi}=\tfrac{d}{d\tau} R_\psi \Psi_\tau (\phi)\, \big|_{\tau=0} \quad \text{and} \quad  \mathfrak X_{|R_\psi \phi}=\tfrac{d}{d\tau}  \Psi_\tau (R_\psi \phi)\, \big|_{\tau=0}.
 \end{align}
 The equality of both implies $R_\psi \circ \Psi_\tau = \Psi_\tau \circ R_\psi$. Therefore, the Lie subalgebra of right-invariant vector fields is the Lie algebra of the group $\bs\Aut(\Phi)$:
  \begin{align}
  \label{LieAlg-Aut}
  \bs\aut(\Phi)=\big(\Gamma_{\text{\!\tiny{inv}}}(T\Phi); [\ ,\,] \big).
  \end{align}

The tangent bundle $T\Phi$ has a canonical subbundle, the \emph{vertical tangent bundle} defined as  $V\Phi := \ker \pi_\star$. In other words, vertical vector fields are elements of $\Gamma(V\Phi):=\left\{  \mathfrak X \in \Gamma(T\Phi)\, |\, \pi_\star \mathfrak X=0 \right\}$. Since $V\Phi$ is a subbundle,  $\Gamma(V\Phi)$ is a Lie subalgebra of $\Gamma(T\Phi)$, even an ideal. Indeed,  since $\pi_\star: \Gamma(T\Phi) \rarrow \Gamma(T\M)$ is a Lie algebra morphism, we have, for $\mathfrak X \in \Gamma(V\Phi)$ and $\mathfrak Y \in \Gamma(T\Phi)$: $\pi_\star[\mathfrak X, \mathfrak Y]=[\pi_\star \mathfrak X, \pi_\star \mathfrak Y ]=[0, \pi_\star \mathfrak Y]=0$, i.e. $[\mathfrak X, \mathfrak Y] \in \Gamma(V\Phi)$. 

 Vertical vector fields can be understood as arising from the linearization of the right action of $\Diff(M)$.
Its characterization requires that we describe the Lie algebra of $\Diff(M)$. It is well know that as a vector space it it given by $\Gamma(TM)$, the vector fields on $M$, but as a Lie algebra its Lie bracket is \emph{minus} the bracket of vector fields: 
  \begin{align}
\mathfrak{diff}(M)=\big(\Gamma(TM); -[\ , \, ]\big). 
  \end{align}
We may use the notation $[X,Y]_{\text{{\tiny $\mathfrak{diff}$}}}:=-[X , Y]_{\text{{\tiny $\Gamma(TM)$}}}$ when useful.

Let us have a look at the types of vertical vector fields induced by the respective actions of $\diff(M)$, $\bs\diff(M)$ and $\bs\diff_v(\Phi)$. 
\smallskip

 A \emph{fundamental} vertical vector field at $\phi \in \Phi$ generated by  $X=\tfrac{d}{d\tau} \psi_\tau \big|_{\tau=0} \in \mathfrak{diff}(M)$ with flow $\psi_\tau \in \Diff(M)$ is: 
 \begin{align}
 \label{Fund-vect-field}
 X^v_{|\phi} := \tfrac{d}{d\tau} R_{\psi_\tau} \phi\, \big|_{\tau=0} = \tfrac{d}{d\tau}  \psi^*_\tau \phi \, \big|_{\tau=0} =: \mathfrak L _X \phi, 
 \end{align}
where $ \mathfrak L _X \phi$ is the spacetime Lie derivative of the field $\phi$ along the vector field $X \in T(M)$. The spacetime Lie derivative is also given by the Cartan formula $ \mathfrak L _X=[\iota_x, d]=\iota_Xd + d\iota_X$, with $d$ the de Rham exterior derivative on $M$ -- It is derivation of degree $0$ of the algebra $\Omega^\bullet(M)$ of forms on $M$, since $\iota_X$ is of degree $-1$ and $d$ is of degree $1$. Naturally, $X^v_{|\phi}$ is tangent to the $\Diff(M)$-orbit $\mathcal O[\phi]$ at $\phi \in \Phi$, hence the name. As expected, it satisfies $\pi_\star X^v\equiv 0$, since 
$\pi_\star X^v_{|\phi} =\tfrac{d}{d\tau} \pi \circ R_{\psi_\tau} \phi\, \big|_{\tau=0} = \tfrac{d}{d\tau} \pi(\phi)\, \big|_{\tau=0}$. 
As is standard, one shows (see Appendix \ref{Lie algebra (anti)-isomorphisms}) that the map $|^v :\diff(M) \rarrow \Gamma(V\Phi)$, $X \mapsto X^v$, is a Lie algebra morphism: i.e. $([X,Y]_{\text{{\tiny $\diff$}}})^v=(-[X, Y]_{\text{{\tiny $\Gamma(TM)$}}})^v=[X^v, Y^v]$.

The pushforward by the right-action of a fundamental vertical vector field is:
 \begin{align}
 \label{Pushforward-fund-vect}
 R_{\psi\star} X^v_{|\phi} :=&\, \tfrac{d}{d\tau} R_\psi \circ R_{\psi_\tau} \phi \,  \big|_{\tau=0} 
                                         =  \tfrac{d}{d\tau} R_{\psi_\tau \circ \psi} \phi \, \big|_{\tau=0} =  \tfrac{d}{d\tau} R_{\psi_\tau \circ \psi} \, R_{\psi^{-1} \circ \psi} \, \phi \, \big|_{\tau=0},  \notag\\
                                         =&\,  \tfrac{d}{d\tau} R_{(\psi\- \circ \psi_\tau \circ \psi)}  \, R_{\psi} \, \phi \, \big|_{\tau=0} =   \tfrac{d}{d\tau} R_{(\psi\- \circ \psi_\tau \circ \psi)}  \, \psi^*\phi \, \big|_{\tau=0} ,  \notag\\
                                         =&\!: \left(  (\psi\-)_* \, X \circ \psi \right)^v_{|\psi^*\phi}. 
 \end{align}
Where we use the relation between $\psi$-relatedness of vector fields and $\psi$-conjugation of their flows.\footnote{This can be understood as the incarnation of the lemma \eqref{lemma1} on field space.} A fundamental vector field generated by the action of the (Lie algebra of the) structure group is thus not right-invariant. 

But the fundamental vector fields induced by $\bs{\diff}(M)$, the Lie algebra of the gauge group $\bs\Diff(M)$, are. For $\bs\psi_\tau \in \bs\Diff(M)$, we have of course $\bs X =\tfrac{d}{d\tau}\, \bs\psi_\tau\, \big|_{\tau=0} \bs{\diff}(M)$. Now, given the definition  \eqref{GaugeGroup} of the gauge group, the transformation property of whose elements can be written $R^\star_\psi \bs\psi = \psi\- \circ \bs\psi \circ \psi $ -- and reminding the link between $\psi$-relatedness of vector fields and $\psi$-conjugation of their flow --  the former is,
 \begin{align}
 \label{LieAlg-GaugeGroup}
 \bs{\diff}(M):=\left\{\, \bs X: \Phi \rarrow \Gamma(TM)\  |\ R^\star_\psi \bs X = (\psi\-)_*\, \bs X \circ \psi \,  \right\}.
 \end{align}
 This transformation property can also be written as: $\bs X(\phi^\psi)= \bs X(\psi^*\phi) =  (\psi\-)_*\, \bs X(\phi) \circ \psi $. 
We may remark that the infinitesimal version is given by the Lie derivative on $\Phi$ along the corresponding fundamental vector field:
 \begin{align}
 \label{inf-equiv-gauge-Lie-alg}
\bs L_{X^v} \bs X = X^v(\bs X) = \tfrac{d}{d\tau} \, R^\star_{\psi_\tau} \bs X \, \big|_{\tau=0} = \tfrac{d}{d\tau} \, (\psi\-_\tau)_*\, \bs X \circ \psi_\tau  \, \big|_{\tau=0} 
				=: \mathfrak L_X \bs X =[X, \bs X]_{\text{{\tiny $\Gamma(TM)$}}} =[\bs X, X]_{\text{{\tiny $\diff$}}}. 
 \end{align}
 Which is natural, and as one would expect. Now a fundamental vector field generated by $\bs X \in \bs{\diff}(M)$ is,
  \begin{align}
 \bs X^v_{|\phi} := \tfrac{d}{d\tau} R_{\bs\psi_\tau(\phi)} \phi\, \big|_{\tau=0} = \tfrac{d}{d\tau}  (\bs\psi_\tau(\phi))^* \phi \, \big|_{\tau=0} =: \mathfrak L _{\bs X} \phi.
 \end{align}
 Its pushforward by the right-action of the structure group is:
  \begin{align}
 R_{\psi\star} \bs X^v_{|\phi} :=&\, \tfrac{d}{d\tau} R_\psi \circ R_{\bs\psi_\tau(\phi)} \phi \,  \big|_{\tau=0} 
                                                =  \tfrac{d}{d\tau} R_{(\psi\- \circ \bs\psi_\tau(\phi) \circ \psi)}  \, R_{\psi} \, \phi \, \big|_{\tau=0} =   \tfrac{d}{d\tau} R_{\bs\psi_\tau(\psi^*\phi)}  \, \psi^*\phi \, \big|_{\tau=0}
                                         \rdefeq \bs X^v_{|\psi^*\phi}. 
 \end{align}
 These are right-invariant, as advertised. 
Furthermore, one shows  (see Appendix \ref{Lie algebra (anti)-isomorphisms}) that the map $|^v : \bs{\diff}(M) \rarrow \Gamma_{\text{\!\tiny{inv}}}(V\Phi)$, $\bs X \mapsto \bs X^v$, is a Lie algebra \emph{anti}-morphism: i.e. $([\bs X, \bs Y]_{\text{{\tiny $\diff$}}})^v=(-[\bs X, \bs Y]_{\text{{\tiny $\Gamma(TM)$}}})^v=-[\bs X^v, \bs Y^v]$.
 Therefore, since the Lie subalgebra of right-invariant vertical vector fields is the Lie algebra of the group $\bs\Aut_v(\Phi)$, we have 
  \begin{align}
  \label{Gauge-Lie-alg-morph}
 \bs{\diff}(M)  \simeq \bs\aut_v(\Phi)=\big(\Gamma_{\text{\!\tiny{inv}}}(V\Phi); [\ ,\,] \big).
  \end{align}
We can thus write the  infinitesimal version of \eqref{SESgroup}, i.e. the SES describing the Atiyah Lie algebroid of the  bundle $\Phi$:
  \begin{align}
 \label{Atiyah-Algebroid}
0\rarrow \bs{\diff}(M)  \simeq \bs\aut_v(\Phi)  \xrightarrow{|^v}   \bs{\mathfrak{aut}}(\Phi) \xrightarrow{\pi_\star} \bs{\diff}(\M) \rarrow 0,
  \end{align}
where again, the ``\emph{physical} symmetries" are in the right-most Lie algebra $ \bs{\diff}(\M)$.\footnote{A splitting of this SES, i.e. the datum of a map $ \bs{\mathfrak{aut}}(\Phi) \rarrow \bs{\diff}(M)$ -- or equivalently of a map $\bs{\diff}(\M) \rarrow  \bs{\mathfrak{aut}}(\Phi)$ -- which would allow to decompose a (right-invariant) vector field on $\Phi$ as a sum of a gauge element and a vector field on $\M$, is supplied by a choice of Ehresmann connection $1$-form on $\Phi$. See later. }

Finally, consider the Lie algebra of the group of vertical automorphisms $\bs\Diff_v(\Phi)$:
\begin{align}
\label{Lie-Alg_diff_v}
\bs\diff_v(\Phi)\defeq \left\{ \bs X_{|\phi}^v =\tfrac{d}{d\tau} \bs\Xi_\tau(\phi) \big|_{\tau=0}=\tfrac{d}{d\tau} R_{\bs\psi_\tau(\phi)} \phi \big|_{\tau=0} \in \Gamma(V\Phi) \right\},
\end{align}
 where  $\bs\Xi_\tau \in \bs\Diff_v(\Phi)$, and $\bs\psi_\tau \in  C^\infty\big(\Phi, \Diff(M) \big)$ is the flow of $ \bs X: \Phi \rarrow \Gamma(TM)\simeq \diff(M)$. 
 Therefore, we have that $\bs\diff_v(\Phi) \simeq C^\infty\big(\Phi, \diff(M) \big)$. 
The pushforward of $\bs X^v$ by the action of the structure group $\Diff(M)$ is the same as for a fundamental vector field \eqref{Pushforward-fund-vect},
 \begin{align}
 \label{Pushforward-fund-vect_bis}
 R_{\psi\star} \bs X^v_{|\phi} = \left(  (\psi\-)_* \, \bs X \circ \psi \right)^v_{|\psi^*\phi}. 
 \end{align}
 Now, here the map $|^v : C^\infty\big(\Phi, \diff(M) \big) \rarrow \Gamma(V\Phi)$ is also a Lie algebra morphism, yet the bracket on $C^\infty\big(\Phi, \diff(M) \big)$ is not simply the bracket in $\diff(M)$ but an ``extended" bracket that take into account the $\Phi$-dependence of its elements. 
 Indeed, one shows that for $\bs X^v, \bs Y^v \in \bs\diff_v(\Phi)$: 
 \begin{align}
 \label{extended-bracket1}
\big[ \bs X^v, \bs Y^v\big] &= -[  \bs X,  \bs Y]_{\text{{\tiny $\Gamma(TM)$}}}^v +  [ \bs X^v(\bs Y)]^v -  [\bs Y^v(\bs X)]^v, \notag\\
			    &= \left\{  [\bs X, \bs Y]^{\phantom{-}}_{\text{\tiny{$\diff(M)$}}} +  \bs X^v(\bs Y) - \bs Y^v(\bs X)  \right\}^v \rdefeq \{\bs X, \bs Y\}^v.  
 \end{align}
 The result is proven in the short  note \cite{Francois2023-b}  (for the finite dimensional case). 
 The bracket on $C^\infty\big(\Phi, \diff(M) \big) $ is thus,
 \begin{align}
 \label{extended-bracket2}
 \{\bs X, \bs Y\} :=  [\bs X, \bs Y]^{\phantom{-}}_{\text{\tiny{$\diff(M)$}}} +  \bs X^v(\bs Y) - \bs Y^v(\bs X).
 \end{align}
 From this follows naturally that,
  \begin{align}
 [\bs L_{\bs{ X}^v}, \bs L_{ \bs{ Y}^v }]  = \bs L_{[ \bs{ X}^v,  \bs{ Y}^v]}= \bs L_{\{ \bs{ X},  \bs{ Y}\}^v }.     \label{Lie-deriv-ext-bracket}
 \end{align}
 We may observe that the result \eqref{extended-bracket1} reproduces the results of Appendix \ref{Lie algebra (anti)-isomorphisms}: 
 i.e. on the one hand if $\bs X, \bs Y  \rarrow  X,Y \in \diff(M)$ the second and third term of the extended bracket vanish and $[X^v, Y^v]= (-[  X,  Y]_{\text{{\tiny $\Gamma(TM)$}}})^v = ( [ X,  Y]_{\text{\tiny{$\diff(M)$}}})^v$, 
 and on the other hand if  if $\bs X, \bs Y  \rarrow \bs X,\bs Y \in \bs\diff(M)$ we may use \eqref{inf-equiv-gauge-Lie-alg} to obtain the second and third term of the extended bracket so that  $[\bs X^v,\bs Y^v]= ([ \bs X, \bs Y]_{\text{{\tiny $\Gamma(TM)$}}})^v = ( -[\bs X,\bs  Y]_{\text{\tiny{$\diff(M)$}}})^v$.
 Naturally, the relation \eqref{extended-bracket1} is the infinitesimal counterpart of \eqref{twisted-comp-law}, with \eqref{extended-bracket2}  reflecting the peculiar composition law in $C^\infty\big(\Phi, \Diff(M) \big)$.

Echoing our discussion of the action groupoid $\bs\Gamma$, let us briefly recall the Lie algebroid picture of the above \cite{Mackenzie2005, Crainic-Fernandez2011}. Associated to the action of $\diff(M)$ on $\Phi$, i.e. to the Lie algebra morphism $\alpha = |^v: \diff(M) \rarrow \Gamma(V\Phi) \subset \Gamma(T\Phi)$, is the action (or transformation) Lie algebroid $\bs A=\Phi \rtimes \diff(M) \rarrow \Phi$ with anchor $\rho : \bs A \rarrow V\Phi \subset T\Phi$, $(\phi, X) \mapsto \alpha(X)_{|\phi}=X^v_{|\phi}$, inducing the Lie algebra morphism $\t \rho: \Gamma(\bs A) \rarrow \Gamma(V\Phi) \subset \Gamma(T\Phi)$. 
The space of sections $\Gamma(\bs A)=\{ \Phi \rarrow \bs A, \phi \mapsto \big(\phi, \bs X(\phi)\big) \}$ is naturally identified with the space $C^\infty\big(\Phi , \diff(M) \big)$,\footnote{Sections of $\Gamma(\bs A)$ are just the graphs of elements of $C^\infty\big(\Phi , \diff(M) \big)$.} so~that $\t \rho=\alpha=|^v$.  
The~Lie~algebroid bracket $[\ \,,\  ]_{\text{\tiny{$\Gamma(\bs A)$}}}$ is 
uniquely determined by the Leibniz condition $[\bs X, f \bs Y]_{\text{\tiny{$\Gamma(\bs A)$}}}=f[\bs X, \bs Y]_{\text{\tiny{$\Gamma(\bs A)$}}} + \t \rho(\bs X) f \cdot \bs Y$, for $f\in C^\infty(\Phi)$,
and the requirement that  $[\ \,,\ ]_{\text{\tiny{$\Gamma(\bs A)$}}} = [\ \, , \  ]_{\text{\tiny{$\diff(M)$}}}$ on constant sections.
It is found to be: $[\bs X, \bs Y]_{\text{\tiny{$\Gamma(\bs A)$}}} = [\bs X, \bs Y  ]_{\text{\tiny{$\diff(M)$}}} + \t \rho (\bs X) \bs Y - \t\rho(\bs Y) \bs X$.
The action Lie algebroid bracket indeed reproduces \eqref{extended-bracket2} above -- which was obtained via a computation which amounts to showing explicitly  
that $\t\rho=|^v$ is indeed a Lie algebra morphism. 

Quite intuitively, the action Lie algebroid $\bs A$ is the limit of the action Lie groupoid $\bs\Gamma$ when the target and source maps are infinitesimally close. Hence $C^\infty\big(\Phi, \diff(M) \big)\simeq \Gamma(A)$ is the Lie algebra of the group (of bisections) $C^\infty\big(\Phi, \Diff(M)\big)$, as we've shown  explicitly above. 
 \medskip
 
 To the best of my knowledge, the bracket \eqref{extended-bracket2} has been first introduced in physics by 
 Bergmann \& Komar \cite{Bergmann-Komar1972} (for $\phi=g_{\mu\nu}$) and Salisbury \& Sundermeyer \cite{Salisbury-Sundermeyer1983} in their investigations of the largest possible  symmetry group of General Relativity -- see eq.(3.1)-(3.2) in \cite{Bergmann-Komar1972} and eq.(2.1) in  \cite{Salisbury-Sundermeyer1983}. It was latter reintroduced by Barnich \& Troessaert in their study  \cite{Barnich-Troessaert2009} of asymptotic symmetries of gravity in the flat limit at null infinity. This Bergmann-Komar-Salisbury-Sundermeyer (BKSS) bracket 
 $ \{ \bs{ X},  \bs{ Y}\}_{\text{\tiny{BKSS}}} =- \{ \bs{ X},  \bs{ Y}\}$
 appears in eq.(8) in \cite{Barnich-Troessaert2009} and more recently e.g. in eq.(3.3) in \cite{Freidel-et-al2021}, eq.(2.12) in \cite{Freidel-et-al2021bis}, eq.(2.21) in \cite{Gomes-et-al2018}, eq.(2.3) in \cite{Chandrasekaran-et-al2022}, eq.(1.1) in \cite{Speranza2022}.  
 As this references show, this bracket is commonly used in the literature concerned with the covariant symplectic structure of gravity and with the analysis of its asymptotic symmetries (BMS and extensions).
 This bracket was first interpreted as a bracket of action Lie algebroid  in \cite{Barnich2010} (eq. (28) -- or (3.11) of the preprint). 
 In addition of showing here explicitly how  \eqref{extended-bracket2}  arises simply from standard bundle geometry,  we show in the next section below that it is also an instance in degree $0$ of the Frölicher-Nijenhuis bracket of vector valued forms. 
 \medskip

 Finally, let us state an elementary yet important result (proven in Appendix  \ref{Pushforward by a vertical diffeomorphism}): The pushforward by a vertical diffeomorphism $\Xi \in \bs\Diff_v(\Phi)$, to which corresponds $\bs \psi \in C^\infty\big(\Phi, \Diff(M) \big)$,
  is  a map $\Xi_\star: T_\phi\Phi \rarrow T_{\Xi(\phi)}\Phi=T_{\bs\psi^*\phi}\Phi$. Applied on a generic $\mathfrak X \in \Gamma(T\Phi)$ it results in, 
 \begin{align}
 \label{pushforward-X}
 \Xi_\star \mathfrak X_{|\phi} &= R_{\bs\psi(\phi) \star} \mathfrak X_{|\phi} + \left\{   \bs\psi(\phi)\-_* \bs d \bs\psi_{|\phi}(\mathfrak X_{|\phi})   \right\}^v_{|\Xi(\phi)},  \notag\\
 					    &=R_{\bs\psi(\phi) \star} \left( \mathfrak X_{|\phi} +  \left\{  \bs d \bs\psi_{|\phi}(\mathfrak X_{|\phi})  \circ \bs\psi(\phi)\-  \right\}^v_{|\phi} \right). 
 \end{align}
The proof is independent from the equivariance of $\Xi \sim \bs\psi$, thus holds the same for  $\Xi \in \bs\Aut_v(\Phi)\sim\bs \psi \in \bs\Diff(M)$. 
This~relation can be used to obtain the formula for repeated pushforwards:  e.g. to obtain the result for $(\Xi' \bs \circ \Xi)_\star \mathfrak X_{|\phi}$, per \eqref{twisted-comp-law},  one only needs to substitute $\bs\psi \rarrow \bs\psi \circ \big( \bs\psi' \bs \circ R_{\bs \psi}\big)$. In case  $\Xi, \Xi' \in \bs{\Aut}_v(\Phi)$, one substitutes  $\bs\psi \rarrow \bs\psi'  \circ  \bs\psi$.
This~is~key to the geometric definition of general vertical transformations and gauge transformations on field space.

\subsubsection{Differential forms and their derivations} 
\label{Differential forms and their derivations}  

As a manifold, $\Phi$ has a space of forms $\Omega^\bullet(\Phi)$, together with the graded Lie algebra of its derivations 
$\Der_\bullet \big(\Omega^\bullet(\Phi) \big)=\bigoplus_k \Der_k \big(\Omega^\bullet(\Phi) \big)$  whose graded brackets is $[D_k, D_l]=D_k \circ D_l - (-)^{kl} D_l \circ D_k$, with $D_i \in \Der_i\big(\Omega^\bullet(\Phi) \big)$.  

 The de Rham complex of $\Phi$ is $\big( \Omega^\bullet(\Phi); \bs d  \big)$ with $\bs d \in \Der_1$ the de Rham (exterior) derivative, which is nilpotent -- $\bs d ^2 =0=\sfrac{1}{2}[\bs d, \bs d]$ -- and defined via the Koszul formula.  Given the exterior product $\w$ defined as usual on scalar-valued forms, we have that  
 $\big( \Omega^\bullet(\Phi, \mathbb K), \w, \bs d \big)$ is a differential graded algebra.\footnote{The exterior product can also be defined on the space $\Omega^\bullet(\Phi, \sf A)$ of variational differential forms with values in an algebra $(\sf A, \cdot)$, using the product in $\sf A$ instead of the product in the field $\mathbb K$. So  $\big( \Omega^\bullet(\Phi, \sf A), \w, \bs d \big)$ is a again a differential graded algebra. On the other hand, an exterior product cannot be defined on $\Omega^\bullet(\Phi, \bs V)$ where $\bs V$ is merely a vector space.}

One may define the subset of vector-field valued differential forms $\Omega^\bullet(\Phi, T\Phi)=\Omega^\bullet(\Phi) \otimes T\Phi$. Then, the subalgebra of 
 ``\emph{algebraic}" derivations is defined as $D_{|\Omega^0(\Phi)}=0$, they have the form $\iota_{\bs K} \in \Der_{k-1}$ for $\bs K \in \Omega^{k}(\Phi, T\Phi)$, with $\iota$ the inner product. 
 For $  \omega  \otimes  \mathfrak X \in \Omega^\bullet(\Phi, T\Phi)$ we have : $\iota_{\bs K}( \omega  \otimes  \mathfrak X)  := \iota_{\bs K} \omega\otimes  \mathfrak X = \bs\omega \circ \bs K \otimes  \mathfrak X$.
 On $\Omega^\bullet(\Phi, T\Phi)$,  the \emph{Nijenhuis-Richardson bracket} (or \emph{algebraic} bracket) is defined by: 
\begin{align}
\label{NR-bracket} 
 [\bs K, \bs L]_{\text{\tiny{NR}}}:=\iota_{\bs K} \bs L -(-)^{(k-1)(l-1)} \, \iota_{\bs L} \bs K. 
 \end{align}
 It generalises the inner contraction of a form on a vector field and makes the map, 
  \begin{align}
  \iota: \Omega^\bullet(\Phi, T\Phi) &\rarrow \Der_\bullet \big(\Omega^\bullet(\Phi) \big) \\ 
  					\bs K &\mapsto  \iota_{\bs K}         \notag 
  \end{align}
a graded Lie algebra morphism; 
\begin{align}
\label{NR-bracket-id}
[\iota_{\bs K}, \iota_{\bs L}]=\iota_{[\bs K, \bs L]_{\text{\tiny{NR}}}}. 
\end{align}
The \emph{Nijenhuis-Lie derivative} is the map,  
\begin{equation}  
\label{NL-derivative}
  \begin{aligned}    
  \bs L:=[\iota, \bs d] : \Omega^\bullet(\Phi, T\Phi) &\rarrow \Der_\bullet \big(\Omega^\bullet(\Phi) \big) \notag\\
  					\bs K &\mapsto  \bs L_{\bs K}:= \iota_{\bs K} \bs d - (-)^{k-1} \bs d  \iota_{\bs K}      
  \end{aligned}
  \end{equation}     
We have $\bs L_{\bs K} \in \Der_k$ for $\bs K \in \Omega^{k}(\Phi, T\Phi)$.  It generalises the Lie derivative along vector fields, $\bs L_\mathfrak{X} \in \Der_0$. 
It is such that $[\bs L_{\bs K}, \bs d]=0$, and it is a  morphism of graded Lie algebras: 
\begin{align}
\label{NL-deriv-morph}
[\bs L_{\bs K}, \bs L_{\bs J}]=\bs L_{[\bs K, \bs J]_{\text{\tiny{FN}}}},
\end{align}
 where $[\bs K, \bs J]_{\text{\tiny{FN}}}$ is the \emph{Frölicher-Nijenhuis bracket}.  Explicitely, for $\bs K = K \otimes \mathfrak X \in \Omega^k(\Phi, T\Phi)$ and   $\bs J = J \otimes \mathfrak Y \in \Omega^l(\Phi, T\Phi)$, it is: 
  \begin{align}
  \label{FN-bracket}
  [\bs K, \bs J]_{\text{\tiny{FN}}}= K \w J \otimes [\mathfrak X, \mathfrak Y]  + K \w \bs L_\mathfrak{X} J \otimes \mathfrak Y
  														  -  \bs L_\mathfrak{Y} K \w J \otimes \mathfrak X
														  + (-)^k \big(  dK \w \iota_\mathfrak{X} J \otimes \mathfrak Y
														                       + \iota_\mathfrak{Y} K \w dJ \otimes \mathfrak X   \big).
  \end{align}
We further have the relations: 
\begin{equation}
  \begin{aligned}
  \label{Relations-L-iota}
[\bs L_{\bs K}, \iota_{\bs J}] &= \iota^{}_{ [\bs K, \bs J]_{\text{\tiny{FN}}} } - (-)^{k(l-1)} \bs L_{(\iota_{\bs K} \bs J)},  \\
[\iota_{\bs J}, \bs L_{\bs K}] &= \bs L_{(\iota_{\bs K} \bs J)} +(-)^k \, \iota^{}_{ [\bs J, \bs K]_{\text{\tiny{FN}}} }. 
  \end{aligned}
  \end{equation}
  We refer to \cite{Kolar-Michor-Slovak} (Chap II, section 8) for a systematic presentation of these notions in the finite dimensional case, and for proofs of the above relations.
  \medskip
  
The FN bracket \eqref{FN-bracket} reproduces as a special case the  bracket \eqref{extended-bracket1}: 
Indeed, specialising the above formulae in degree $0$, for  $\bs f = f \otimes \mathfrak X$ and $\bs g = g \otimes \mathfrak Y \in \Omega^0(\Phi, T\Phi)$ we have: 
\begin{equation}
  \begin{aligned}
  \label{NR-bracket-special}
  [\bs f, \bs{dg}]_{\text{\tiny{NR}}} &= \iota^{}_{\bs f } \bs{dg} -(-)^0  \cancel{\iota^{}_{\bs{dg}} \bs f} = \big(  f \w \iota_\mathfrak{X} \bs d g\big)\otimes \mathfrak Y = \big(  f \w \bs L_\mathfrak{X}  g\big)\otimes \mathfrak Y,   \\
   [\bs{df}, \bs{g}]_{\text{\tiny{NR}}} &= \cancel{  \iota^{}_{\bs{df}} \bs{g} }  -(-)^0  \iota^{}_{\bs{g}} \bs{d f} = -  [\bs{g}, \bs{df}]_{\text{\tiny{NR}}} =-  \big(  g \w \bs L_\mathfrak{Y}  f\big)\otimes \mathfrak X.
  \end{aligned}
  \end{equation}
 So that, 
    \begin{align}
    \label{FN-bracket-special}
     [\bs f, \bs g]_{\text{\tiny{FN}}} &= f \w g \otimes [\mathfrak X, \mathfrak Y]  + f \w \bs L_\mathfrak{X} g \otimes \mathfrak Y
  														  -  \bs L_\mathfrak{Y} f \w g \otimes \mathfrak X, \notag \\
						&= f \w g \otimes [\mathfrak X, \mathfrak Y] +  [\bs f, \bs{dg}]_{\text{\tiny{NR}}}  -  [\bs{g}, \bs{df}]_{\text{\tiny{NR}}},
    \end{align}
 and 
   \begin{align}
   \label{FN-NL-der}
   [\bs L_{\bs f}, \iota_{\bs g}] &= \iota^{}_{ [\bs f, \bs g]_{\text{\tiny{FN}}} }. 
    \end{align}      

Now, the map  $|^v : C^\infty\big(\Phi, \diff(M)\big) \rarrow  \Gamma(V\Phi)$,  $\bs{ X} \mapsto \bs{  X}^v$, allows to think of $\bs{ X}^v \in \diff_v(\Phi)$ as a (vertical) vector-valued 0-form on $\Phi$, i.e. $ \bs{ X}^v \in \Omega^0(\Phi, V\Phi) \subset \Omega^\bullet(\Phi, T\Phi)$. 
Therefore, the Nihenhuis-Richardson and Frölicher-Nijenhuis brackets naturally apply: in particular, specialising further eq.\eqref{NR-bracket-special} we have, for $\bs{ X}^v, \bs{  Y}^v \in \Omega^0(\Phi, V\Phi)$, 
\begin{equation}
\label{NR-bracket-field-dep-diff}
\begin{split}
 [\bs{ X}^v, \bs d \bs{ Y}^v ]_{\text{\tiny{NR}}} &= \{ \iota_{\bs{ X}^v} \bs{d}\bs{ Y} \}^v = \{ \bs{ X}^v(\bs{ Y})\}^v, \\
  [\bs d \bs{ X}^v,  \bs{ Y}^v ]_{\text{\tiny{NR}}} &= -[\bs{ Y}^v, \bs d \bs{ X}^v ]_{\text{\tiny{NR}}} = -\{ \iota_{\bs{ Y}^v} \bs{d}\bs{ X} \}^v = - \{ \bs{ Y}^v(\bs{ X})\}^v, 
  \end{split}
\end{equation}
 so that the FN bracket for 0-forms \eqref{FN-bracket-special} specialises to:
 \begin{align}
 \label{FN-bracket-field-dep-diff=BT-bracket}
   [\bs{ X}^v,  \bs{ Y}^v ]_{\text{\tiny{FN}}} &=\big( [\bs{ X}, \bs{ Y}]_{\text{{\tiny $\diff$}}}\big)^v + [\bs{ X}^v, \bs d\bs{ Y}^v]_{\text{\tiny{NR}}}  - [\bs{ Y}^v, \bs d\bs{ X}^v]_{\text{\tiny{NR}}}, \notag \\
    &=  \big( - [\bs{ X}, \bs{ Y}]_{\text{{\tiny $\Gamma(TM)$}}}  +  \bs{ X}^v(\bs{ Y}) -   \bs{ Y}^v(\bs{ X}) \,\big)^v   = \{ \bs{ X},  \bs{ Y}\}^v. 
 \end{align}
 Then of course we have the following special cases of identities eq.\eqref{NR-bracket-id}, \eqref{Relations-L-iota}, and \eqref{NL-deriv-morph} among derivations in $\Der^\bullet$: 
 \begin{align}
 [\iota_{\bs{ X}^v}, \iota_{\{\bs{d Y}\}^v}] &= \iota^{}_{ [\bs{ X}^v, \bs d \bs{ Y}^v ]_{\text{\tiny{NR}}} } = \iota^{}_{ \{ \iota_{\bs{ X}^v} \bs{d}\bs{ Y} \}^v },
 \label{identity-iota}  \\
  [\bs L_{\bs{ X}^v}, \iota_{ \bs{ Y}^v }] &= \iota^{}_{[\bs{ X}^v,  \bs{ Y}^v ]_{\text{\tiny{FN}}}}, \label{identity-L-iota} \\
 [\bs L_{\bs{ X}^v}, \bs L_{ \bs{ Y}^v }] &= \bs L_{[\bs{ X}^v,  \bs{ Y}^v ]_{\text{\tiny{FN}}}} 
 										    = \bs L_{\{ \bs{ X},  \bs{ Y}\}^v }.     \label{NL-derr-ext-bracket}
 \end{align}
 As one should expect, \eqref{NL-derr-ext-bracket} reproduces \eqref{Lie-deriv-ext-bracket}.
 These identities were derived heuristically in the appendices of various works, sometimes at quite a computational cost.
 For example, eq.\eqref{NL-derr-ext-bracket}/\eqref{Lie-deriv-ext-bracket} reproduces e.g. eq.(3.1) in \cite{Freidel-et-al2021}, eq.(2.13) in \cite{Freidel-et-al2021bis}. 
  We show here that they flow naturally from well-established general geometric structures.\footnote{See indeed appendix A of \cite{Chandrasekaran-et-al2022}, which hints at a similar take. }
 
 \medskip

\paragraph{Remarkable forms}  The structure group $\Diff(M)$ of $\Phi$ acts on a form $\bs \alpha \in \Omega^\bullet(\Phi)$ by pullback, $R^\star_\psi \alpha$, defining its \emph{equivariance}. 
The action by pullback of   $\bs\Diff_v(\Phi)\simeq C^\infty\big( \Phi, \Diff(M)\big)$, i.e. $\bs\alpha^{\bs\psi}\defeq \Xi^\star \bs\alpha$, defines  \emph{general vertical transformations},
while the action by pullback of $\bs\Aut_v(\Phi)\simeq \bs{\Diff}(M)$ similarly defines  \emph{gauge transformations}. 

Let us express such a generic form at $\phi\in \Phi$ as,
\begin{align}
\bs\alpha_{|\phi} = \alpha\big(\! \w^\bullet \!\bs d\phi_{|\phi}; \phi \big),
\end{align}
where $\bs d\phi \in \Omega^1(\Phi)$ is the basis 1-form on $\Phi$ (the infinite dimensional analogue of $dx^{\,\mu}$ on a manifold $M$), and $\alpha(\ ;\ )$ is the functional expression of $\bs\alpha$, alternating multilinear in the first arguments, and whose dependence in the second argument $\phi$ is a priori arbitrary -- but often in physics it will be polynomial. Then, its equivariance and general vertical transformation are: 
\begin{align}
R^\star_\psi \bs\alpha_{|\phi^\psi} &= \alpha\big(\! \w^\bullet \! R^\star_\psi \bs d\phi_{|\phi^\psi};\,  R_\psi \phi \big) = \alpha\big(\! \w^\bullet \! R^\star_\psi \bs d\phi_{|\phi^\psi}; \, \phi^\psi \big), 
\quad  \text{for } \ \psi \in \Diff(M), \\
\bs\alpha^{\bs\psi}_{|\phi} \defeq \Xi^\star \bs\alpha^{}_{|\Xi(\phi)} &= \alpha\big(\! \w^\bullet \! \Xi^\star \bs d\phi^{}_{|\Xi(\phi)};\  \Xi( \phi) \big) = \alpha\big(\! \w^\bullet \! \Xi^\star \bs d\phi_{|\phi^{\bs\psi}};\,  \phi^{\bs\psi} \big),
\quad \text{for } \  \Xi \in \bs \Diff_v(\Phi) \sim \bs\psi \in C^\infty\big( \Phi, \Diff(M)\big).    \label{GT-general}
\end{align}
The infinitesimal equivariance and general vertical transformations are given by the (Nijenhuis-) Lie derivative along the elements of $\Gamma(V\Phi)$ generated  respectively by $\diff(M)$ and $C^\infty\big( \Phi, \diff(M)\big)$ :
\begin{align}
\bs L_{X^v}\bs\alpha = \tfrac{d}{d\tau}  R^\star_{\psi_\tau} \bs\alpha \big|_{\tau=0}  \quad  \text{ with } \ X \in \diff(M),  
 \qquad \quad  \bs L_{\bs X^v}\bs\alpha = \tfrac{d}{d\tau}  \Xi_\tau^\star \bs\alpha \big|_{\tau=0}\quad  \text{ with } \ \bs X \in C^\infty\big( \Phi, \diff(M)\big),  \label{GT-inf-general}
\end{align}
Concrete results may be computed explicitly. But for some types of forms, there are shortcut for structural reasons. In that regard, and as part of the elementary bundle geometry of field space, we may describe forms of special interest. 
\medskip

First, \emph{equivariant} forms are those whose equivariance is well-behaved in some sense. \emph{Standard} equivariant forms are valued in representations $(\rho, \bs V)$ of the structure group $\Diff(M)$ and s.t.: 
\begin{align}
\Omega_\text{eq}^\bullet(\Phi, \rho) \defeq \left\{ \, \bs\alpha \in \Omega^\bullet(\Phi, \bs V)\,|\, R^\star_\psi\bs\alpha_{|\phi^\psi}=\rho(\psi)\-\bs\alpha_{|\phi}\, \right\}. 
 \end{align} 
The infinitesimal version of the equivariance property is $\bs L_{X^v}\bs\alpha=-\rho_*(X) \bs\alpha$.

The latter is a subspace of \emph{twisted} equivariant forms \cite{Francois2019_II}: their equivariance is controlled by a 1-cocycle for the action of $\Diff(M)$ on $\Phi$, 
 i.e. a map:
 \begin{equation}
\begin{aligned}
\label{cocycle}
C: \Phi \times \Diff(M) &\rarrow G, \quad \text{$G$ some Lie group (possibly infinite dimensional).}   \\
	(\phi, \psi) &\mapsto C(\phi; \psi) \qquad \text{s.t.} \quad C(\phi; \psi'\circ \psi) =C(\phi; \psi') \cdot C(\phi^{\psi'}; \psi). 
 \end{aligned} 
  \end{equation}
 Manifestly,   $\phi$-independent 1-cocycles are just group morphisms, thus typical 1-cocycles generalise   representations. 
 From the 1-cocycle property \eqref{cocycle} follows that $C(\phi; \id_M)=\id_G=C(\phi^\psi; \id_M)$, thus that $C(\phi; \psi)\-\!= C(\phi^\psi; \psi\-)$. 
\mbox{If $\bs V$ is} a $G$-space, twisted equivariant forms are defined as, 
 \begin{align}
 \label{C-eq-form}
 \Omega^\bullet_\text{eq}(\Phi, C) \defeq \left\{ \bs\alpha \in \Omega^\bullet(\Phi, \bs V)\, | \, R^\star_\psi \bs\alpha_{|\phi^\psi} = C(\phi  ; \psi)\- \bs\alpha_{|\psi}\right\}.
 \end{align} 
 The 1-cocycle relation \eqref{cocycle} ensures compatibility with the right action: $R_{\psi'}^\star R^\star_{\psi} = R^\star_{\psi' \circ\,\psi}$. The infinitesimal equivariance  is $\bs L_{X^v}\bs\alpha=-a(X; \phi) \bs\alpha$, where $a(X, \phi):= \tfrac{d}{d\tau}\, C(\phi, \psi_\tau) |_{\tau=0}$ is a 1-cocycle for the action of ${\diff}(M)$~on~$\Phi$:
  \begin{equation}
 \begin{aligned}
 \label{inf-cocycle}
a: \Phi \times {\diff}(M) &\rarrow \mathfrak g, \quad \text{$\mathfrak g$ the Lie algebra of $G$.}   \\
	(\phi, X) &\mapsto a(X; \phi) \qquad \text{s.t.} \quad X^v\! \cdot a(Y; \phi)- Y^v\!\cdot a(X; \phi)+ \big[a(X; \phi)\,,\ a(Y; \phi)\big]_{\mathfrak g}=a([X,Y]_{\text{{\tiny $\diff$}}}; \phi). 
 \end{aligned} 
   \end{equation}
 The infinitesimal 1-cocycle relation \eqref{inf-cocycle} ensures compatibility with the right action: $[\bs L_{X^v}, \bs L_{X^v}]=\bs L_{[X^v, Y^v]}=\bs L_{([X, Y]_{\text{{\tiny $\diff$}}})^v}$. The reader familiar with gauge anomalies may discern that it is a non-Abelian generalisation of the Wess-Zumino consistency condition for a $\Diff(M)$ (Einstein) anomaly $a(X; \phi)$ --  reproduced for $G$ Abelian. 
 
 The subspace of \emph{invariant} forms are those whose  equivariance is trivial: $\Omega_\text{inv}^\bullet(\Phi)=\left\{ \, \bs\alpha \in \Omega^\bullet(\Phi)\,|\, R^\star_\psi \bs\alpha=\bs\alpha \,\right\}$. Infinitesimally we have $\bs L_{X^v}\bs\alpha=0$. 
 
  \medskip
The space of of \emph{horizontal} forms is $\Omega_\text{hor}^\bullet(\Phi)=\left\{ \, \bs\alpha \in \Omega^\bullet(\Phi)\,|\, \iota_{X^v}\bs\alpha=0 \right\}$. Now, a form which is both equivariant and horizontal is said \emph{tensorial}. We have thus standard tensorial forms  
 \begin{align}
 \label{tens-forms}
\Omega_\text{tens}^\bullet(\Phi, \rho)\defeq \left\{ \, \bs\alpha \in \Omega^\bullet(\Phi, \bs V)\,|\, R^\star_\psi\bs\alpha=\rho(\psi)\-\bs\alpha,\, \text{ \& }\ \iota_{X^v}\bs\alpha=0\, \right\}. 
\end{align} 
And similarly, we have the generalisation: the space of \emph{twisted} tensorial forms, 
 \begin{align}
 \label{twisted-tens-forms}
\Omega_\text{tens}^\bullet(\Phi, C)\defeq \left\{ \, \bs\alpha \in \Omega^\bullet(\Phi, \bs V)\,|\, R^\star_\psi\bs\alpha=C(\phi; \psi)\-\bs\alpha,\, \text{ \& }\ \iota_{X^v}\bs\alpha=0\, \right\}. 
\end{align} 
In either case, we have obviously $\Omega_\text{tens}^0(\Phi)=\Omega_\text{eq}^0(\Phi)$. 

One observes that any field theory, given by a Lagrangian $L : \Phi \rarrow  \Omega^n(M)$, $\phi \mapsto L(\phi)$, with $n=$dim$M$, or an action $S : \Phi \rarrow  \RR$, $\phi \mapsto S(\phi)$, is actually a twisted tensorial 0-form. 
 Indeed, let us define  $Z\in \Omega^0(\Phi, \CC^*)$,  i.e. $Z: \Phi \rarrow  \CC^*$, by $ \phi \mapsto Z(\phi) \defeq \exp{i S(\phi)}$. The action of $\Diff(M)$ is  $R^\star_\psi L = \psi^* L$, one may then define the object $C: \Phi \times \Diff(M) \rarrow U(1)$, by $C(\phi; \psi)\defeq \exp{-i \int c(\phi; \psi)}$, where $c(\ \,; \psi) \defeq R^\star_\psi L - L$. 
 This implies that $R^\star_\psi Z = C(\ \,; \psi)\- Z$. 
 Furthermore,  it is easily checked that $c(\phi; \psi'\circ \psi) = c(\phi; \psi') + c(\phi^{\psi'};\psi)$, i.e. $C(\phi; \psi)$ satisfies \eqref{cocycle},  meaning that it is a $U(1)$-valued $\Diff(M)$-1-cocycle. 
 Therefore $Z \in \Omega_\text{tens}^0(\Phi, C)$. 
 
 The infinitesimal equivariance is thus $\bs L_{X^v} Z = - a(X; \phi)\, Z$, with 
 $- a(X; \ \, )=\tfrac{d}{d\tau} C( \ \,; \psi_\tau)\-\big|_{\tau=0} = \tfrac{d}{d\tau}  i \int c( \ \,; \psi_\tau)\big|_{\tau=0}$$ = i\int \tfrac{d}{d\tau} R^\star_{\psi_\tau} L - L\, \big|_{\tau=0}
 = i\int \tfrac{d}{d\tau} \psi_\tau^* L \, \big|_{\tau=0} = i \int \mathfrak L_X L \rdefeq i \int \alpha (X; \ \,)$. In other words, $a (X; \phi)$ is the classical $\Diff(M)$-anomaly.  In case $Z$ is instead the path integral of $S$, the quantity $a (X; \phi)$ is the quantum $\Diff(M)$-anomaly, or Einstein anomaly \cite{Bertlmann, Bonora2023}. In both cases, since Lie$U(1)=i\RR$ is abelian, \eqref{inf-cocycle} 
 is $X^v\! \cdot a(Y; \phi)- Y^v\!\cdot a(X; \phi) =a([X,Y]_{\text{{\tiny $\diff$}}}; \phi)$, i.e. the Wess-Zumino consistency condition for the anomaly. 

Remark that this view of the non-trivial equivariance of $Z= \exp{i S}$ relies on a convention for the action of $\Diff(M)$ on integrals (over $M$) that is not the one we use (and argue for) in this paper.\footnote{This is not the case for the path integral $Z(\phi)=\int \bs d \phi  \exp{\tfrac{i}{\hbar} S(\phi)}$ defining quantum theories. Even with our convention, giving an invariant action $S$, the twisted equivariance of $Z$ comes from the non-trivial transformation of the measure $\bs d \phi$: the 1-cocycle $C(\phi; \psi)\defeq \exp{-i \int c(\phi; \psi)}$ is then the integrated quantum $\Diff(M)$-anomaly. See e.g. chapter 12 of \cite{Bertlmann}. \label{foontote-Q}} 
We elaborate on this issue in section \ref{Associated bundles}, in connection to a wider discussion of associated bundles (whose space of sections are the 0-degree of the spaces of tensorial forms  \eqref{tens-forms}-\eqref{twisted-tens-forms}) and of the fundamental representation of $\Diff(M)$. 
\medskip

Let us highlight the fact that the de Rham derivative $\bs d$ does not preserve the space of tensorial forms (one loses the property of horizontality).\footnote{This is easy to check: starting on a 1-form $\bs\alpha$, using the Koszul formula for $\bs d$ we have that for any $\mathfrak X, \mathfrak Y \in \Gamma(T\Phi)$: $\bs{d\alpha}(\mathfrak X, \mathfrak Y)= (\bs L_\mathfrak{X} \bs \alpha)(\mathfrak Y) - \mathfrak Y \cdot \bs \alpha(\mathfrak X)$. So, for $\mathfrak X = X^v$, $\bs{d\alpha}(X^v, \mathfrak Y)= (\bs L_{X^v} \bs \alpha)(\mathfrak Y) - \mathfrak Y \cdot \bs \alpha( X^v)$. If $\bs\alpha$ is horizontal, $\bs{d\alpha}(X^v, \mathfrak Y)= (\bs L_{X^v} \bs \alpha)(\mathfrak Y)$ and is non-zero unless $\bs\alpha$ is also invariant. See next.}
This is one reason one needs to introduce an adequate notion of \emph{connection} on $\Phi$ so as to define a \emph{covariant derivative} on the space of tensorial forms. For standard tensorial forms, one needs an Ehresmann connection 1-form, while for twisted tensorial forms one needs a generalisation called \emph{twisted} connection  \cite{Francois2019_II}. See section \ref{Connections on field space} below.  
\medskip

Finally, forms that  are both  invariant and horizontal are called  \emph{basic}: 
 \begin{align}
\Omega_\text{basic}^\bullet(\Phi)\defeq \left\{ \, \bs\alpha \in \Omega^\bullet(\Phi)\,|\, R^\star_\psi\bs\alpha=\bs\alpha\, \text{ \& }\ \iota_{X^v}\bs\alpha=0\, \right\}. 
\end{align} 
This space \emph{is} preserved by $\bs d$,\footnote{See previous footnoote. In other words, $\bs d$  is a covariant derivative for basic forms!} which means that 
$\big( \Omega^\bullet_\text{basic}(\Phi), \bs d \big)$ is a subcomplex of the de Rham complex of $\Phi$: the \emph{basic subcomplex}.  
Alternatively, basic forms can be defined as  Im$(\pi^\star)$ (hence the name), that is: 
 \begin{align}
\Omega_\text{basic}^\bullet(\Phi)\defeq\left\{ \, \bs\alpha \in \Omega^\bullet(\Phi)\,|\, \exists\, \bs\beta \in \Omega^\bullet(\M) \text{ s.t. } \bs\alpha=\pi^\star\bs\beta \,  \right\}. 
\end{align} 
Since $[\bs d, \pi^\star]=0$, the cohomology of the basic subcomplex is isomorphic  to the cohomology  $\big(\Omega^\bullet(\M), \bs d \big)$ of the moduli space: The cohomology of $\big( \Omega^\bullet_\text{basic}(\Phi), \bs d \big)$ is  known as the \emph{equivariant cohomology} of $\Phi$.

Remark that the ``form" analogue of $\Gamma_{\text{\!\tiny{inv}}}(T\Phi)$ --  projecting to well defined vector fields  $\Gamma(TM)$ -- is not $\Omega^\bullet_\text{inv}(\Phi)$, 
but $\Omega_\text{basic}^\bullet(\Phi)$: Only the latter, per the second definition, projects to well-defined forms in $\Omega^\bullet(\M)$.
It will be a main endeavor of this paper to  show how, given some form $\bs\alpha \in \Omega^\bullet(\Phi)$, to formally built its basic counterpart $\bs\alpha^b \in \Omega_\text{basic}^\bullet(\Phi)$. This will be achieved via the dressing field method, section \ref{The dressing field method}.

\subsection{General vertical transformations, and gauge transformations} 
\label{General vertical transformations, and gauge transformations}  

As  already stated above, the general vertical transformation a form $\bs\alpha \in \Omega^\bullet(\Phi)$ is its  pullback by $\bs\Diff_v(\Phi)\simeq C^\infty\big( \Phi, \Diff(M)\big)$
: $\bs\alpha^{\bs\psi} \defeq \Xi^\star \bs\alpha$.
  The notation on the left-hand side is operationally defined by the right-hand side, and justified by the fact that the vertical transformation is expressible in term of the generating  element $\bs\psi \in C^\infty\big( \Phi, \Diff(M)\big)$ associated to $\Xi \in \bs\Diff_v(\Phi)$. 
  Performing two vertical transformations, using \eqref{twisted-comp-law}, one has 
  \begin{align}
  \label{2-Vert-Trsf}
  \big( \bs\alpha^{\bs\psi}\big)^{\bs\psi'} \defeq {\Xi'}^\star \Xi^\star \bs\alpha = \big( \Xi \bs \circ \Xi' \big)^\star \bs\alpha \rdefeq \bs\alpha^{ \bs\psi' \circ\, (\bs\psi \bs\circ R_{\bs\psi'}) }.
  \end{align} 
   In the case of gauge transformations defined by the action of $\bs\Aut_v(\Phi)\simeq \bs{\Diff}(M)$, since the defining equivariance of gauge group elements is $R^\star_\psi \bs\psi = \bs\psi \bs\circ R_\psi =  \psi\- \circ \bs\psi \circ \psi$, the above expression simplifies to a more familiar form
     \begin{align}
  \label{2-Gauge-Trsf}
  \big( \bs\alpha^{\bs\psi}\big)^{\bs\psi'} = \bs\alpha^{ \bs\psi \,\circ\, \bs\psi' }.
  \end{align} 
  Naturally, infinitesimal general vertical transformations are given by the action of $\bs\diff_v(\Phi) \simeq C^\infty\big(\Phi, \diff(M) \big)$ via the Nijenhuis-Lie derivative: 
  \begin{align}
  \label{Inf-2-Vert-Trsf}
  \bs L_{\bs X^v} \bs\alpha = \left\{   \begin{matrix} \tfrac{d}{d\tau} \, \Xi_\tau^\star \bs\alpha\, \big|_{\tau=0} \\[3mm]
					                                       \hspace{-.2cm} [\iota_{\bs X^v}, \bs d]\, \bs \alpha \end{matrix} \right. \qquad \text{so} \qquad    
 [\bs L_{\bs{ X}^v}, \bs L_{ \bs{ Y}^v }] \, \bs\alpha = \bs L_{[\bs{ X}^v,  \bs{ Y}^v ]_{\text{\tiny{FN}}}} \, \bs\alpha = \bs L_{\{ \bs{ X},  \bs{ Y}\}^v }\, \bs\alpha.    
  \end{align}
 The second equation, using  \eqref{Lie-deriv-ext-bracket}/\eqref{NL-derr-ext-bracket}, is the infinitesimal reflection of \eqref{2-Vert-Trsf}. 
 For infinitesimal gauge transformations, given by the action of $\bs\aut_v(\Phi)\simeq \bs{\diff}(M)$, this simplifies to 
 $[\bs L_{\bs{ X}^v}, \bs L_{ \bs{ Y}^v }] \, \bs\alpha = \bs L_{( -[\bs X,\bs  Y]_{\text{\tiny{$\diff(M)$}}})^v}\, \bs\alpha = \bs L_{([ \bs X, \bs Y]_{\text{{\tiny $\Gamma(TM)$}}})^v}\, \bs\alpha$.

  To be more explicit, it is possible to  take a further step: For any $\mathfrak X,\mathfrak X', \ldots \in \Gamma(T\Phi)$, given \eqref{pushforward-X} and the duality between pullback and pushforward, one has that
\begin{equation}
\label{GT-geometric}
\begin{aligned}
\bs\alpha^{\bs\psi}_{|\phi} (\mathfrak X_{|\phi}, \ldots) =  \Xi^\star \bs\alpha_{|\Xi(\phi)} (\mathfrak X_{|\phi}, \ldots)= \bs\alpha^{}_{|\Xi(\phi)} (\Xi_\star \mathfrak X_{|\phi}, \ldots) &= \bs\alpha^{}_{|\phi^{\bs\psi(\phi)}}\left( R_{\bs\psi(\phi) \star} \left( \mathfrak X_{|\phi} +  \left\{  \bs d \bs\psi_{|\phi}(\mathfrak X_{|\phi})  \circ \bs\psi(\phi)\-  \right\}^v_{|\phi} \right), \ldots\right),  \\
&=R_{\bs\psi(\phi)}^\star\, \bs\alpha_{|\phi^{\bs\psi(\phi)}} \left(  \mathfrak X_{|\phi} +  \left\{  \bs d \bs\psi_{|\phi}(\mathfrak X_{|\phi})  \circ \bs\psi(\phi)\-  \right\}^v_{|\phi}, \ldots \right).
\end{aligned}
\end{equation}
Cleary therefore, the vertical transformation of a form is controlled by its equivariance and verticality properties. 
In particular, from \eqref{GT-geometric} follows that the vertical transformation of a \emph{tensorial} form is simply given by its equivariance: 
\begin{equation}
\label{GT-tensorial}
\begin{split}
&\text{For }\ \bs\alpha \in \Omega_\text{tens}^\bullet(\Phi, \rho), \quad \bs\alpha^{\bs\psi} = \rho(\bs\psi)\- \bs\alpha. \\
&\text{For }\ \bs\alpha \in \Omega_\text{tens}^\bullet(\Phi, C), \quad \bs\alpha^{\bs\psi} = C(\bs\psi)\- \bs\alpha.
\end{split}
\end{equation}
In the second line we introduced the shorter notation $[C(\bs\psi)](\phi) \defeq C\big(\phi; \bs\psi(\phi)\big)$, so the map $C(\bs\psi)$ has thus a double dependency on the point $\phi\in \Phi$.
Due to their horizontality, their vertical transformation is homogeneous, tensorial so to speak, hence their name. 
It should be remarked that $\bs\alpha^{\bs\psi}=\Xi^\star \bs\alpha  \notin \Omega_\text{tens}^\bullet(\Phi, \rho)$ unless $\bs\psi \in \bs\Diff(M) \sim \Xi \in \bs\Aut_v(\Phi)$ -- see \cite{Francois2023-b} -- making \emph{gauge transformations} special indeed. 

It is standard, and easy to show, that the infinitesimal versions of \eqref{GT-geometric} is, by definition \eqref{Inf-2-Vert-Trsf}, simply,
\begin{align}
\label{Inf-GT-general}
\bs L_{\bs X^v} \bs\alpha = \tfrac{d}{d\tau} \, R_{\bs\psi_\tau}^\star \bs\alpha\, \big|_{\tau=0} + \iota_{\{\bs{dX}\}^v} \bs\alpha.
\end{align}
where of course $\bs X=\tfrac{d}{d\tau}\bs\psi_\tau \big|_{\tau=0}\!$, and $\bs\psi=\bs\psi(\phi) \in \Diff(M)$ in the equivariance term is $\phi$-constant. 
Remark that $\{\bs{dX}\}^v$ can be considered an element of $\Omega^1(\Phi, V\Phi)$, so $\iota_{\{\bs{dX}\}^v}$ is an algebraic derivation (of degree 0) of the type discussed in section \ref{Differential forms and their derivations}. 
Again, it is clear that a vertical transformation depends on both the equivariance and the verticality properties of a form. 
For $\bs\alpha \in \Omega^\bullet_\text{eq}(\phi)$, depending if it is standard or twisted equivariant, \eqref{Inf-GT-general} specialises to:
\begin{align}
\label{Inf-GT-eq}
\bs L_{\bs X^v} \bs\alpha = \left\{   \begin{matrix}   -\rho_*(\bs X)\, \bs\alpha + \iota_{\{\bs{dX}\}^v} \bs\alpha, \\[3mm]
						                              -a(\bs X)\, \bs \alpha + \iota_{\{\bs{dX}\}^v} \bs\alpha,
					  \end{matrix} \right.
\end{align}
where we  introduce the notation $[a(\bs X)](\phi)\defeq a\big(\bs X(\phi); \phi\big)$ for the linearised 1-cocycle. 
And in particular for $\rho(X)\-=\psi^*$ the pullback representation, i.e. for  $\Omega^\bullet(M)$- (or tensor-) valued forms $\bs\alpha$, \eqref{Inf-GT-general} gives naturally: 
\begin{align}
\label{Inf-GT-diff-rep}
\bs L_{\bs X^v} \bs\alpha = \mathfrak L_{\bs X} \bs\alpha + \iota_{\{\bs{dX}\}^v} \bs\alpha.
\end{align}
We remark that the above formula features in an indirect way in (and clarify the meaning of) the so-called ``anomaly operator" appearing in recent covariant phase space literature: 
 $\Delta_{\bs X}\defeq \bs L_{\bs X^v} - \mathfrak L_{\bs X} - \iota_{\{\bs{dX}\}^v}$  in our notations
-- compare e.g. to  \cite{Hopfmuller-Freidel2018} eq.(2.39), \cite{Chandrasekaran_Speranza2021} eq.(2.1), \cite{Freidel-et-al2021} eq.(2.13),  \cite{Freidel-et-al2021bis} eq.(2.3), or \cite{Speziale-et-al2023} eq.(2.2).
Of course, this operator can be non-zero on $\Phi$ only in theories that are \mbox{fundamentally} ``non-general-relativisitic" in admitting background non-dynamical structures or fields \mbox{``breaking"} $\Diff(M)$-covariance (see section \ref{The dressing field method} for further elaboration on this comment).

The infinitesimal versions of \eqref{GT-tensorial}, for tensorial forms, are then obviously: 
\begin{align}
\label{GT-inf-tensorial}
\bs L_{\bs X^v} \bs \alpha = -\rho_*(\bs X) \bs \alpha, \quad\text{or } \quad \bs L_{\bs X^v} \bs \alpha = -a(\bs X) \bs \alpha, 
\end{align}
From applying  \eqref{Lie-deriv-ext-bracket}/\eqref{NL-derr-ext-bracket}, the commutativity property  of the Nijenhuis-Lie derivative, on a twisted tensorial form $\bs \alpha$: 
$[ \bs L_{\bs X^v}, \bs L_{\bs Y^v}] \, \bs \alpha=\bs L_{[\bs X^v, \bs Y^v]_{\text{{\tiny FN}}}}\,  \bs \alpha$ $= \bs L_{\{\bs X, \bs Y\}^v}\,  \bs \alpha$, we get the following relation for the infinitesimal 1-cocycle, 
\begin{equation}
\label{BB}
\begin{aligned}
& \bs X^v\big( a(\bs Y; \phi) \big) -  \bs Y^v\big( a(\bs X; \phi) \big)    - a(\{\bs X, \bs Y\}; \phi)  + [a(\bs X; \phi), a(\bs Y; \phi)]_{\text{{\tiny $\LieG$}}} =0, \\
&\bs X^v\big( a(\munderline{red}{\bs Y}; \phi) \big) -  \bs Y^v\big( a( \munderline{red}{\bs X}; \phi) \big)    - a([\bs X, \bs Y]_{\text{{\tiny $\diff(M)$}}}; \phi)  + [a(\bs X; \phi), a(\bs Y; \phi)]_{\text{{\tiny $\LieG$}}}=0.  
\end{aligned}
\end{equation}
One obtains the second line from the first using  the FN bracket \eqref{FN-bracket-field-dep-diff=BT-bracket}/\eqref{extended-bracket1}: The notation $\munderline{red}{\bs Y},\munderline{red}{\bs X}$  is meant to indicate that the element $\bs Y, \bs X$ are considered $\phi$-independent in the formula, so that $\bs X^v, \bs Y^v$ pass through it. Therefore,  \eqref{BB} simply reproduces the defining  infinitesimal 1-cocycle property  \eqref{inf-cocycle} even for $\phi$-dependent $\diff(M)$-parameters. 
 
 As  examples, we may consider the case of elements of the gauge group $\bs\Diff(M)$ and its Lie algebra $\bs\diff(M)$. These are 0-forms, so trivially horizontal, and their equivariance are specified by definition: they are thus tensorial 0-forms. For $\bs\eta \in \bs\Diff(M)$ whose equivariance is given by $\psi$-conjugation, see \eqref{GaugeGroup}, and for $\bs Y \in \bs\diff(M)$ whose equivariance if given by $\psi$-relatedness, see \eqref{LieAlg-GaugeGroup}, we have thus: 
 \begin{align}
 \bs \eta^{\bs \psi} = \bs\psi \- \circ \bs \eta \circ \bs \psi, \qquad \text{and} \quad \bs Y^{\bs\psi}= (\bs\psi\-)_* \bs Y\circ \bs\psi 
 \end{align}
 The infinitesimal gauge transformation of  $\bs Y$ is thus:  $\bs L_{\bs X^v} \bs Y = \mathfrak L_{\bs X} \bs Y = [\bs Y, \bs X]_{\text{{\tiny $\diff(M)$}}}$ -- in coherence with its  infinitesimal equivariance given by \eqref{inf-equiv-gauge-Lie-alg}. 
 
And as a special case of the above, or given their definition, basic forms are strictly gauge invariant: 
\begin{align}
\label{GT-basic}
\text{For }\ \bs\alpha \in \Omega_\text{basic}^\bullet(\Phi): \quad \bs\alpha^{\bs\psi} = \bs\alpha, \quad \text{so}\quad  L_{\bs X^v} \bs \alpha=0. 
\end{align}
 That is another way to  detect a variational form who would induce (or come from) a corresponding form on the moduli space $\M$, where presumably physical degree of freedom (d.o.f.) live. 
 \medskip
 
 An example worth looking at closely is that of the basis 1-form $\bs d\phi  \in \Omega^1(\Phi)$, 
 given that its vertical transformation $\bs d\phi^{\bs\psi}\defeq \Xi^\star \bs d \phi$ intervene in the general formulae \eqref{GT-general}  for the vertical transformation of a generic form.
 To compute it geometrically via \eqref{GT-geometric}  we must know the equivariance and verticality properties of $\bs d\phi$. These are given by definition: 
 \begin{align}
 \label{basis-1-form}
R^\star_\psi \bs d\phi \defeq \psi^* \bs d \phi, \qquad \text{and} \qquad \bs \iota_{X^v}\bs d\phi \defeq \mathfrak L_X \phi. 
 \end{align}
 That is, the equivariance is controlled by the pullback representation, while the verticality just reproduces the infinitesimal (field-independent) diffeo/gauge transformation of the field $\phi$.
 It is therefore almost immediate that its vertical  transformation is, 
 \begin{align}
  \label{GT-basis-1-form}
  \bs d\phi^{\bs\psi}   \defeq \Xi^\star \bs d \phi = \bs\psi^*   \big(  \bs d\phi   +   \mathfrak L_{ [\bs d \bs\psi  \circ \bs\psi\-]} \phi  \big).   
 \end{align}
 It is maybe pedagogically useful to see it unfold in more detail: 
  \begin{align*}
\bs d\phi^{\bs\psi}_{|\phi} (\mathfrak X_{|\phi} ) \defeq \Xi^\star \bs d \phi^{}_{|\phi}  (\mathfrak X_{|\phi} ) = \bs d \phi^{}_{|\Xi(\phi)}  (\Xi_\star \mathfrak X_{|\phi} )
&
=\bs d\phi^{}_{|\phi^{\bs\psi(\phi)}}\left( R_{\bs\psi(\phi) \star} \big( \mathfrak X_{|\phi} +  \left\{  \bs d \bs\psi_{|\phi}(\mathfrak X_{|\phi})  \circ \bs\psi(\phi)\-  \right\}^v_{|\phi} \big) \right), \notag \\
                    &=R_{\bs\psi(\phi)}^\star \bs d\phi_{|\phi^{\bs\psi(\phi)}} \left(  \mathfrak X_{|\phi} +  \left\{  \bs d \bs\psi_{|\phi}(\mathfrak X_{|\phi})  \circ \bs\psi(\phi)\-  \right\}^v_{|\phi}  \right), \notag\\
                    &=\bs\psi(\phi)^*  \bs d\phi^{}_{|\phi} \left(  \mathfrak X_{|\phi} +  \left\{  \bs d \bs\psi_{|\phi}(\mathfrak X_{|\phi})  \circ \bs\psi(\phi)\-  \right\}^v_{|\phi}  \right), \notag\\
                    &= \bs\psi(\phi)^*  \bs d\phi^{}_{|\phi} ( \mathfrak X_{|\phi})  +   \bs\psi(\phi)^* \mathfrak L_{ [\bs d \bs\psi_{|\phi}(\mathfrak X_{|\phi})  \circ \bs\psi(\phi)\-]} \phi, \notag\\
                    &=\left(  \bs\psi(\phi)^*   \big(  \bs d\phi^{}_{|\phi}    +   \mathfrak L_{ [\bs d \bs\psi_{|\phi}  \circ \bs\psi(\phi)\-]} \phi  \big) \right)  (\mathfrak X_{|\phi} ),  
 \end{align*}
 This reproduces a result now standard in the literature on the covariant phase space of gravity: See e.g. eq.(3.5) and (3.6) in  \cite{DonnellyFreidel2016} where the relation is proven in Appendix B via  more  heuristic algebraic computations. 
 Notice the gain in efficiency stemming from the geometric approach. 
 By \eqref{Inf-GT-diff-rep}, the infinitesimal version is 
  \begin{align}
  \label{GT-inf-basis-1-form}
 \bs L_{\bs X^v} \bs d\phi    
					 =  \mathfrak L_{\bs X} \bs d \phi +   \mathfrak L_{\bs d \bs X} \phi.
 \end{align}
 The same could be found via $ \bs L_{\bs X^v} \bs d\phi = \bs d (\iota_{\bs X^v} \bs d\phi) = \bs d \big( \mathfrak L_{\bs X} \phi \big)$.

\subsection{Connections on field space}  
\label{Connections on field space}  

As we've observed above, the exterior derivative $\bs d$ does not preserve $\Omega^\bullet_{\text{tens}}(\Phi)$  of standard/twisted tensorial forms. To build a first order linear differential operator that does, the covariant derivative, one needs to endow  $\Phi$ with  an adequate notion of connection 1-form.

\subsubsection{Ehresmann connections} 
\label{Ehresmann connections} 

A Ehresmann connection 1-form $\bs \omega \in \Omega^1_{\text{eq}}\big(\Phi, \diff(M) \big)$ is defined by:
\begin{equation}
\label{Variational-connection}
\begin{aligned}
\bs\omega_{|\phi} \big( X^v_{|\phi}\big)&=X, \quad \text{for } X\in \diff(M), \\
R^\star_\psi \bs\omega_{|\phi^\psi} &= \psi\-_* \, \bs\omega_{|\phi}  \circ\, \psi. 
\end{aligned}
\end{equation}
Infinitesimally, the equivariance of the connection under $\diff(M)$ is,
\begin{align}
\label{inf-equiv-connection}
\bs L_{X^v} \bs\omega = \tfrac{d}{d\tau}\, R^\star_{\psi_\tau} \bs\omega\, \big|_{\tau=0} = \tfrac{d}{d\tau}\,  {\psi_\tau\-}_* \, \bs\omega  \circ\, \psi_\tau \,\big|_{\tau=0} 
																   = [X, \bs \omega]_{\text{{\tiny $\Gamma(TM)$}}} = [\bs\omega, X]_{\text{{\tiny $\diff(M)$}}}.
\end{align}
The space of connection $\C$ is an affine space modelled on the vector space $\Omega^1_{\text{tens}}\big(\Phi, \diff(M)\big)$: Indeed, it is clear that for $\bs\omega, \bs\omega' \in \C$, we have that $\bs\beta\defeq \bs \omega' - \bs\omega \in \Omega^1_{\text{tens}}\big(\Phi, \diff(M)\big)$. Or, given $\bs\omega \in \C$ and $\bs\beta \in \Omega^1_{\text{tens}}\big(\Phi, \diff(M)\big)$, we have that $\bs\omega'=\bs\omega + \bs \beta \in \C$. This means that in general one cannot add connections.\footnote{Only affine combinations like $\bs\omega_\tau \defeq \tau \bs\omega' + (1-\tau) \bs\omega = \bs\omega + \tau \bs\beta$,  $\tau \in [0, 1],$ are possible. Then the midpoint $\bs\omega_{\tau=\sfrac{1}{2}}=\tfrac{1}{2}(\bs\omega+\bs\omega')$ is an averaged sum of two connections.}

A connection allows to define the horizontal subbundle $H\Phi \defeq \ker \bs\omega$ complementary to the vertical subbundle, $T\Phi=V\Phi \oplus H\Phi$. The horizontal projection is thus the map $|^h : T\Phi \rarrow H\Phi$, $\mathfrak X \mapsto \mathfrak X^h \defeq \mathfrak X - [\bs\omega(\mathfrak X)]^v$, as clearly $\bs\omega (\mathfrak X^h)=0$.

The covariant derivative is thus defined as $\bs D\defeq \bs d \bs \circ |^h :  \Omega^\bullet_\text{eq} \big(\Phi, \rho) \rarrow   \Omega^{\bullet+1}_\text{tens} \big(\Phi, \rho)$. On tensorial forms is has the familiar expression, $\bs D :  \Omega^\bullet_\text{tens} \big(\Phi, \rho) \rarrow   \Omega^{\bullet+1}_\text{tens} \big(\Phi, \rho)$, $\bs\alpha \mapsto \bs{ D\alpha}=\bs{d\alpha} +\rho_*(\bs\omega) \bs\alpha$. Which is the first order linear operator we were looking for. 

The curvature 2-form is defined as $\bs\Omega \defeq \bs{d \omega} \bs \circ |^h$, from which follows that $\bs \Omega \in \Omega^2_\text{tens} \big(\Phi, \diff(M)\big)$. As is well known, it is also given by Cartan structure equation, 
\begin{align}
\label{Curvature-Cartan-eq}
\bs\Omega = \bs{d \omega} +\tfrac{1}{2}[\bs \omega, \bs \omega]_{\text{{\tiny $\diff(M)$}}}.
\end{align}
From this it is easy to see that it satisfies Bianchi identity, $\bs{D\Omega}=\bs{d\Omega}+[\bs\omega, \bs \Omega]_{\text{{\tiny $\diff(M)$}}}=0$. On tensorial forms we have $\bs D \circ \bs D =\rho_*(\bs \Omega)$. We may remark that the FN bracket  \eqref{FN-bracket-field-dep-diff=BT-bracket}/\eqref{extended-bracket1} plays a key role in the last step of proving \eqref{Curvature-Cartan-eq}, which requires to show that both sides of the equality vanish on $ \bs X^v, \bs Y^v \in \bs\diff_v(\Phi)$ with $\bs X, \bs Y \in C^\infty\big(\Phi, \diff(M)\big)$: It is the case of the left-hand side by definition -- since $(\bs X^v)^h\equiv0$ -- while for the right-hand side  one has, 
\begin{align}
\bs{d \omega}\big(\bs X^v, \bs Y^v \big) +[\bs \omega(\bs X^v), \bs \omega(\bs Y^v)]_{\text{{\tiny $\diff(M)$}}}
&=\bs X^v\big(\omega( \bs Y^v)\big) - \bs Y^v\big(\omega(\bs X^v)\big) - \bs \omega\big([\bs X^v, \bs Y^v]\big)    + [\bs X, \bs Y]_{\text{{\tiny $\diff(M)$}}}  \notag\\
&= \bs X^v\big( \bs Y\big) - \bs Y^v\big(\bs X\big)    - \bs \omega\big(\{\bs X, \bs Y\}^v\big)  + [\bs X, \bs Y]_{\text{{\tiny $\diff(M)$}}}  \notag\\
&=   [\bs X, \bs Y]_{\text{{\tiny $\diff(M)$}}} +  \bs X^v\big( \bs Y\big) - \bs Y^v\big(\bs X\big)   - \{\bs X, \bs Y\} \equiv 0.  \notag
\end{align}
The Koszul formula for $\bs d$ was used in the first line, and  \eqref{extended-bracket1} in the last to conclude.
\medskip

Given the defining equivariance and verticality properties \eqref{Variational-connection} of a connection, it is easy, using \eqref{pushforward-X}/\eqref{GT-geometric}, to see that its general vertical transformation under $\bs\Diff_v(\Phi)\simeq C^\infty\big( \Phi, \Diff(M)\big)$ is
\begin{align}
\label{Vert-trsf-connection}
\bs\omega^{\bs\psi }\defeq \Xi^\star \bs\omega =\bs\psi\-_* \, \bs\omega  \circ \bs\psi + \bs\psi\-_*\bs{d\psi}. 
\end{align}
Naturally, the formula is the same for its  gauge transformation under $\bs\Aut_v(\Phi)\simeq \bs\Diff(M)$: The difference arises upon repeated transformations of each type, as we stressed in section \ref{General vertical transformations, and gauge transformations}, see \eqref{2-Vert-Trsf}--\eqref{2-Gauge-Trsf}. 
Here again, it should be remarked that $\bs\omega^{\bs\psi}=\Xi^\star \bs\omega  \notin \C$ unless $\bs\psi \in \bs\Diff(M) \sim \Xi \in \bs\Aut_v(\Phi)$ -- see \cite{Francois2023-b} -- highlithing again the special status of \emph{gauge transformations}. 
Infinitesimally, by \eqref{Inf-2-Vert-Trsf}, transformations of a connection under $\bs\diff_v(\Phi)\simeq C^\infty\big( \Phi, \diff(M)\big)$  are given by the Nijenhuis-Lie derivative, 
\begin{align}
\label{Inf-Vert-trsf-connection}
\bs L_{\bs X^v} \bs\omega =\bs{dX}+ [\bs\omega, \bs X]_{\text{{\tiny $\diff(M)$}}}. 
\end{align}
Infinitesimal gauge transformations, under $\bs\aut_v(\Phi)\simeq \bs\diff(M)$, are given by the same relation, but can be written $\bs L_{\bs X^v} \bs\omega =\bs{DX}$ as $\bs X \in \bs\diff(M)$ is a tensorial 0-form. 
In the same way, finite and infinitesimal general vertical transformations of the curvature are given, as special cases of \eqref{GT-tensorial} and \eqref{GT-inf-tensorial}, by:
\begin{align}
\label{GT-curvature}
\bs\Omega^{\bs\psi }\defeq \Xi^\star \bs\Omega =\bs\psi\-_* \, \bs\Omega  \circ \bs\psi, \qquad \text{so} \qquad \bs L_{\bs X^v} \bs\Omega =[\bs\Omega, \bs X]_{\text{{\tiny $\diff(M)$}}}. 
\end{align}
The same hold for its gauge transformations, with the same caveat  expressed above.
\medskip

The above, eq. \eqref{Vert-trsf-connection}, allows to write a little lemma that we will occasionally use: For $\bs\alpha, \bs{D\alpha} \in \Omega^\bullet_\text{tens}(\Phi, \rho)$, 
we have on the one hand $\bs d\, \Xi^\star \bs\alpha = \bs d\big( \rho(\bs \psi)\- \bs\alpha \big)$. On the other hand, by $\Xi^\star \bs {D\alpha} =  \rho(\bs \psi)\- \bs{D\alpha}$, we have that 
\begin{align*}
\Xi^\star \bs {d\alpha} &= \rho(\bs \psi)\- \bs{D\alpha} - \Xi^\star \big(\rho_*(\bs\omega)\bs\alpha \big) 
				   =  \rho(\bs \psi)\- \bs{d\alpha} +  \rho(\bs \psi)\-\,  \rho_*(\bs\omega) \bs\alpha  -  \rho_*(\bs\omega^{\bs \psi})\bs\alpha^{\bs \psi}
				   = \rho(\bs \psi)\- \bs{d\alpha}  -  \rho_*(\bs \psi\-_* \bs d\bs\psi) \rho(\bs\psi)\-\bs\alpha,\\
				 &= \rho(\bs \psi)\- \left(  \bs{d\alpha} -  \rho_*( \bs d\bs\psi \circ \bs\psi\-) \bs\alpha \right). 
\end{align*}
By naturality of the pullback $[\Xi^\star, \bs d]=0$, we  obtain the identity:
\begin{align}
\label{Formula-general-rep}
\bs d\big( \rho(\bs \psi)\- \bs\alpha \big) = \rho(\bs \psi)\- \left(  \bs{d\alpha} -  \rho_*( \bs d\bs\psi \circ \bs\psi\-)\, \bs\alpha \right)
\end{align}
In particular, for the pullback representation, $\rho(\psi)\-\!= \psi^*$ and $-\rho_*(X) =\mathfrak L_X$, this is:
  \begin{align}
\label{Formula-pullback-rep}
\bs d\big( \bs \psi^* \bs\alpha \big) = \bs \psi^* \left(  \bs{d\alpha} + \mathfrak L_{\bs d\bs\psi \circ \bs\psi\-} \bs\alpha \right). 
\end{align}
The latter special case of the lemma appears in the covariant phase space literature, e.g. in \cite{DonnellyFreidel2016} and \cite{Speranza2018}. 
Despite the superficial similarity, it ought not to be confused with with \eqref{GT-basis-1-form} as the two results are distinct geometric statements.
\smallskip

\noindent{\bf Horizontalisation of forms.}
The definition of the curvature showcases the operation dual to the horizontal projection for vector fields: 
the horizontalisation of forms,  
\begin{align}
\label{Horiz-forms}
{\color{gray}|}^h : \Omega^\bullet(\Phi) \rarrow \Omega^\bullet_{\text{hor}}(\Phi), \quad \bs\alpha \mapsto \bs\alpha^h\defeq \bs\alpha \bs\circ |^h .
\end{align} 
Of course, ${\color{gray}|}^h : \Omega^\bullet_{\text{eq}}(\Phi) \rarrow \Omega^\bullet_{\text{tens}}(\Phi)$ and ${\color{gray}|}^h : \Omega^\bullet_{\text{inv}}(\Phi) \rarrow \Omega^\bullet_{\text{basic}}(\Phi)$. So, a connection can be used to extract the \emph{basic} version of some forms, invariant ones. 

In recent years, this has been used in the covariant phase space literature to offer a possible solution of the boundary problem arising in the study of the (pre)symplectic structure of a gauge field theory over a bounded region: The technical core of the problem being that the invariant presymplectic potential and 2-forms lacks horizontality due to boundary terms contributions to their verticality properties, thus fail to be invariant under $\bs\Diff_v(\Phi)$ or $\bs\Aut_v(\Phi)$. An issue we will spell out in more detail in section \ref{Covariant phase space methods}. 
The proposal, developed by Gomes and Riello, was to use a connection to horizontalize the presymplectic potential -- see \cite{Gomes-Riello2017, Gomes-Riello2018, Gomes-et-al2018, Gomes-Riello2021}, and  \cite{Gomes2019,Gomes2019-bis, Gomes2021} for a more philosophical discussion of the proposal. 

For a 1-form $\bs\alpha \in  \Omega^1(\Phi)$ we have the explicit expression in term of the connection 
\begin{align}
\label{Horiz-1-form}
\bs\alpha^h \defeq \bs\alpha -\iota_{[\bs\omega(\  )]^v} \bs\alpha \quad \in  \Omega^\bullet_{\text{hor}}(\Phi).
\end{align}
In particular for the basis 1-form $\bs d \phi \in  \Omega^\bullet_{\text{eq}}(\Phi)$, with properties \eqref{basis-1-form}:
\begin{align}
\label{Horiz-basis-1-form}
\bs{d}\phi^h \defeq \bs d\phi -\iota_{[\bs\omega(\  )]^v} \bs d \phi = \bs d\phi -  \mathfrak L_{\bs\omega} \phi \quad \in  \Omega^\bullet_{\text{tens}}(\Phi).
\end{align}
We remark that applying the covariant derivative to it, as a tensorial form, one finds that $\bs D(\bs d\phi^h)=-\mathfrak L_{\bs\Omega}\phi$.
So, from \eqref{Horiz-basis-1-form} we have that the horizontalisation map \eqref{Horiz-forms}  is also written explicitly as,
\begin{align}
\label{Horiz-forms-explicit}
\bs\alpha =  \alpha\big(\! \w^\bullet \!\bs d\phi; \phi \big) \quad \mapsto \quad \bs\alpha^h = \alpha\big(\! \w^\bullet \!\bs d\phi^h; \phi \big).
\end{align}
Then, \eqref{Horiz-1-form}  is also written as,
\begin{align}
\label{Horiz-forms-explicit}
\bs\alpha^h = \bs\alpha - \alpha\big(\mathfrak L_{\bs\omega} \phi; \phi \big) \quad \in  \Omega^\bullet_{\text{hor}}(\Phi).
\end{align}
This relation is key to the horizontalisation of the presymplectic potential $\bs\theta$ via a connection, i.e. to the the extraction of a \emph{basic counterpart} $\bs\theta^{\ \!\!b}$ of $\bs\theta \in  \Omega^\bullet_{\text{inv}}(\Phi)$. 
For an invariant 1-form $\bs\alpha \in  \Omega^1_{\text{inv}}(\Phi)$, we have $\bs\alpha^h\rdefeq \bs\alpha^b \ \in  \Omega^1_{\text{basic}}(\Phi)$ and one shows that the following relation between the covariant derivative and the horizontalised form holds, 
\begin{align}
\label{Formula-D-h}
\bs D\bs \alpha = \bs d\bs\alpha^b + \iota_{[\bs\Omega]^v} \bs \alpha=   \bs d\bs\alpha^b + \alpha(\mathfrak L_{\bs\Omega} \phi; \phi). 
\end{align}
Therefore for a flat connection $\mathring{\bs\omega}$, $\mathring{\bs D}\bs \alpha = \bs d\bs\alpha^b $.
The  formula \eqref{Formula-D-h} generalises the special case applying to the presymplectic potential $\bs\alpha^b=\bs\theta^b$ discussed by Gomes \& Riello (in the YM case) in corollary 3.2 and section 3.4 of \cite{Gomes-Riello2021},  also at the end of section 3.1. in \cite{Riello2021}.
\medskip

\noindent{\bf Non-uniqueness.}
One may highlight that since a priori the choice of connection is not unique (nor canonical),\footnote{Unless there is a natural choice, as would be e.g. the case if a connection was derived from a natural metric on $\Phi$: This is what happens in the Yang-Mills (YM) case, where $\Phi$ is the space of connection $\A$ of a $H$-bundle $P$, on which a bundle metric induces the Singer-DeWitt connection. This is the approach followed in \cite{Gomes-Riello2017, Gomes-Riello2018, Gomes-et-al2018, Gomes-Riello2021}. See section 2.2 of \cite{Francois-et-al2021} for a discussion. } nor is the horizontalisation map \eqref{Horiz-forms}. 
 For any $\bs \beta \in \Omega^1_\text{tens}\big(\Phi, \diff(M)\big)$, both $\bs\omega$  and $\bs\omega' = \bs\omega +\bs\beta$ are valid choices to implement \eqref{Horiz-forms}. Given $\bs\alpha \in \Omega^\bullet(\Phi)$,  the corresponding horizontal forms obtained are $\bs\alpha^h_{\bs\omega}$ and $\bs\alpha^h_{\bs\omega'}$. 
 
 One may ask how they are related. 
 The answer is easily found for 1-forms:
 Consider the affine path $\bs\omega_\tau \defeq \bs\omega + \tau \bs \beta$, with $\tau \in [0,1]$, in $\C$.  Given $\bs\alpha \in \Omega^1(\Phi)$, define 
 \begin{align}
 \label{Ambiguity-hor}
 \bs \alpha^h_{\bs\omega_\tau} &= \bs\alpha - \iota_{[\bs\omega_\tau]^v} \bs\alpha \quad \in \Omega^1_\text{hor}(\Phi).   \notag  \\[1.5mm]
\text{So, } \quad  \textstyle\int_0^1d\tau\ \tfrac{d}{d\tau}  \bs \alpha^h_{\bs\omega_\tau}& = \bs\alpha^h_{\bs\omega'} - \bs\alpha^h_{\bs\omega} = - \iota_{[\bs\beta]^v} \bs\alpha, \notag \\[1mm]
& \hookrightarrow \quad  \bs\alpha^h_{\bs\omega'} =  \bs\alpha^h_{\bs\omega} - \iota_{[\bs\beta]^v} \bs\alpha \quad  \in \Omega^1_\text{hor}(\Phi).
 \end{align} 
 In particular, for the basis 1-form $\bs d \phi \in \Omega^1_\text{eq}(\Phi)$ we get
  \begin{align} 
  \label{Ambiguity-hor-dphi}
 \bs d\phi^h_{\bs\omega'} =  \bs d\phi ^h_{\bs\omega} - \iota_{[\bs\beta]^v} \bs d \phi = \bs d\phi ^h_{\bs\omega} - \mathfrak L_{\bs\beta} \phi \quad \in \Omega^1_\text{tens}(\Phi).
\end{align} 
From this, one can work out the precise relation between $\bs\alpha^h_{\bs\omega'}=\alpha\big(\!\w^\bullet\! \bs d\phi^h_{\bs\omega'}\, ;\,  \phi \big)$ and $\bs\alpha^h_{\bs\omega}=\alpha\big(\!\w^\bullet\! \bs d\phi^h_{\bs\omega}\, ;\,  \phi \big)$.
 In the case  $\bs\alpha = \alpha\big(\bs d \phi; \phi \big)\in \Omega^1_\text{inv}(\Phi)$, we thus have
 \begin{align} 
 \label{Ambiguity-basic-1-form}
 \bs\alpha^b_{\bs\omega'} &=  \bs\alpha^b_{\bs\omega} - \iota_{[\bs\beta]^v} \bs\alpha \quad  \in \Omega^1_\text{basic}(\Phi), \notag \\
 					&=\bs\alpha^b_{\bs\omega} - \alpha\big( \mathfrak L_{\bs\beta} \phi; \phi \big). 
\end{align} 
And, since on basic forms $\bs D=\bs d$, we have that
 \begin{align} 
 \label{Ambiguity-basic-2-form}
\bs d \bs\alpha^b_{\bs\omega'} = \bs d\bs\alpha^b_{\bs\omega} - \bs d \alpha\big( \mathfrak L_{\bs\beta} \phi; \phi \big)   \quad  \in \Omega^2_\text{basic}(\Phi).  	
\end{align} 
This result generalises the special case $\bs\alpha^b=\bs\theta^b$ (for YM theories) discussed in section 6.7 in \cite{Gomes-Riello2021} (compare also to eq.(83) of \cite{Riello2021}), where it is interpreted as a ``gluing relation" between the basic presymplectic structures stemming from different choices of connections made by different observers on adjacent  regions.
\medskip

After this digression connecting to recent literature on covariant phase space, let us come back to our initial question and discuss next the connections adapted to twisted tensorial forms.

\subsubsection{Twisted connections} 
\label{Twisted connections} 

We remind that twisted equivariant/tensorial forms $\bs\alpha$ have values in a $G$-space $\bs V$, $G$ a (possibly infinite dimensional) Lie group, 
and  their equivariance is  given by a 1-cocycle \eqref{cocycle}
 for the action of ${\Diff}(M)$ on $\Phi$,
$C: \Phi \times \Diff(M) \rarrow G$,   $(\phi, \psi) \mapsto C(\phi; \psi)$, such that
\begin{align}
\label{twist-equiv}
R^\star_\psi \bs\alpha = C(\phi; \psi)\- \bs\alpha, \qquad \text{with} \quad C(\phi; \psi'\circ \psi) =C(\phi; \psi') \cdot C(\phi^{\psi'}; \psi).
\end{align}
Their infinitesimal equivariance is then given by $\bs L_{X^v} \bs\alpha = -a(X, \phi) \bs \alpha$, with 
 $a(X, \phi):= \tfrac{d}{d\tau}\, C(\phi, \psi_\tau) |_{\tau=0}$ a 1-cocycle \eqref{inf-cocycle}  for the action of ${\diff}(M)$ on $\Phi$. 
 As we remarked above, in section \ref{Differential forms and their derivations},  Lagrangians/actions are twisted 0-forms: A~Lagrangian being clearly always $\diff(M)$-anomalous. 

In particular, if the target group of the 1-cocycle is a (subgroup of the) diffeomorphism group, $G=\Diff(M)$, and $\bs V$ is a space of tensors of $M$, \eqref{twist-equiv} specialises to
\begin{align}
\label{twist-equiv-diff}
R^\star_\psi \bs\alpha = C(\phi; \psi)^* \bs\alpha, \qquad \text{with} \quad C(\phi; \psi'\circ \psi) =C(\phi; \psi') \circ C(\phi^{\psi'}; \psi).
\end{align}
These, we may observe, is the type of mathematical objects that underly the ``relational observables"  introduced in \cite{Hoehn-et-al2022}: 
Indeed, eq.(3.183)-(3.185) in section 3.4.4.  displays the fact that, up to a left/right action convention, these ``relational observables" are twisted equivariant/tensorial 0-forms.\footnote{We have a right action convention, so that e.g. if $\bs V$ are purely vectors the equivariance \eqref{twist-equiv-diff} is  $R^\star_\psi \bs\alpha = [C(\phi; \psi)^{-1}]_*\, \bs\alpha$, to be compared to the equivariance eq.(3.183) in \cite{Hoehn-et-al2022}. Compare also the cocycle property \eqref{twist-equiv} to their eq.(3.185).} 
Then, eq.(3.190) is an instance of $\bs\Diff_v(\Phi)\simeq C^\infty \big(\Phi, \Diff(M)\big)$ transformation \eqref{GT-tensorial} of twisted tensorial forms. 
\medskip

A \emph{twisted} connection 1-form $\bs \varpi \in \Omega^1_{\text{eq}}\big(\Phi, \LieG \big)$ is defined by:
\begin{equation}
\label{Variational-twisted-connection}
\begin{aligned}
\bs\varpi_{|\phi} \big( X^v_{|\phi}\big)&=a(X, \phi) \ \in \LieG, \quad \text{for } X\in \diff(M), \\
R^\star_\psi \bs\varpi_{|\phi^\psi} &= \Ad\big( C(\phi; \psi)\big)\- \, \bs\varpi_{|\phi} + C(\phi; \psi)\-\bs d C(\ \, ; \psi)_{|\phi}. 
\end{aligned}
\end{equation}
In the special case $G=\Diff(M)$, the equivariance of $\bs \varpi \in \Omega^1_{\text{eq}}\big(\Phi, \diff(M) \big)$  is 
\begin{align}
\label{Special-eq-twisted}
R^\star_\psi \bs\varpi_{|\phi^\psi} &= [C(\phi; \psi)\big)\-]_* \, \bs\varpi_{|\phi} \circ  C(\phi; \psi) +  [C(\phi; \psi)\big)\-]_*  \bs d C(\ \, ; \psi)_{|\phi},
\end{align}
to be compared to that \eqref{Variational-connection} of an Ehresmann connection.
The infinitesimal equivariance under $\diff(M)$ is
\begin{align}
\label{inf-equiv-twisted-connection}
\bs L_{X^v} \bs\varpi = \tfrac{d}{d\tau}\, R^\star_{\psi_\tau} \bs\varpi\, \big|_{\tau=0} = \bs d a(X; \ ) + [\bs\varpi, a(X;\ )]_{\text{{\tiny $\LieG$}}} 
\end{align}
Or, in the case  $G=\Diff(M)$, $ \bs L_{X^v} \bs\varpi= \bs d a(X; \ ) + [\bs\varpi, a(X;\ )]_{\text{{\tiny $\diff(M)$}}} $. 

The space of twisted connection $\b\C$ is an affine space modelled on the vector space $\Omega^1_{\text{tens}}\big(\Phi, \LieG\big)$: Clearly, for $\bs\omega, \bs\omega' \in \C$, we have that $\bs\beta\defeq \bs \omega' - \bs\omega \in \Omega^1_{\text{tens}}\big(\Phi, \LieG\big)$. Or, given $\bs\omega \in \b\C$ and $\bs\beta \in \Omega^1_{\text{tens}}\big(\Phi, \LieG\big)$, we have that $\bs\omega'=\bs\omega + \bs \beta \in \b\C$. This means that in general one cannot add connections, only affine paths are possible. 

A \emph{twisted covariant derivative} is  defined as  $\b{\bs D} :  \Omega^\bullet_\text{eq} \big(\Phi, C) \rarrow   \Omega^{\bullet+1}_\text{tens} \big(\Phi, C)$, $\bs\alpha \mapsto \b{\bs D} \bs\alpha \defeq \bs{d\alpha} +\rho_*(\bs\varpi) \bs\alpha$. This is the first order linear operator adapted to twisted equivariant/tensorial forms. 

As a relevant example, consider $Z=\exp{iS} \in \Omega^0_\text{eq} \big(\Phi, C)$ as defined earlier to illustrate \eqref{twisted-tens-forms}: we claim that $\b{\bs D} Z \in \Omega^1_\text{eq} \big(\Phi, C)$. 
Indeed, in this case $\bs \varpi \in \Omega^1_{\text{eq}}\big(\Phi, i\RR \big)$, so \eqref{Variational-twisted-connection} gives 
$\bs\varpi_{|\phi} \big( X^v_{|\phi}\big)=a(X, \phi) = - i \int \alpha(X; \phi)$ -- the classical/quantum $\Diff(M)$-anomaly -- and 
$R^\star_\psi \bs\varpi_{|\phi^\psi} =  \bs\varpi_{|\phi} - i  \bs d c(\ \, ; \psi)_{|\phi}$. From this is easily verified that $R^\star_\psi \b{\bs D}Z = C(\phi; \psi)\-\! \b{\bs D}Z$ and $\b{\bs D}Z (X^v)= 0$. 
We may remark that, in this case, $ \b{\bs D}Z = (i\bs d S + \bs\varpi) Z$. Now since $Z$ and $ \b{\bs D}Z$ are both tensorial, it means that 
\begin{align}
\label{Generalised-WZ-trick}
i\bs d S + \bs\varpi \ \in \Omega^1_\text{basic} \big(\Phi). 
\end{align}
This is a kind of generalisation of the Wess-Zumino trick to build counterterms and ``improve" and action by restoring its gauge invariance: the original trick is recovered for flat $\bs\varpi_{\text{\tiny$0$}}$, equivalent to a dressing field, as we show in section \ref{Building basic forms via dressing} below. 
Conversely, the WZ trick is seen to come from a special case of twisted covariant derivative. 

Contrary to an Ehresmann connection, a twisted connection does not to defined an equivariant horizontal distribution on $\Phi$. Yet, its curvature 2-form can be defined, via Cartan structure equation
\begin{align}
\label{Twisted-curvature}
\b{\bs\Omega} \defeq \bs{d \varpi} +\tfrac{1}{2}[\bs \varpi, \bs \varpi]_{\text{{\tiny $\LieG$}}} \ \  \in  \Omega^2_\text{tens} \big(\Phi, \LieG \big)
\end{align}
It satisfies  Bianchi identity, $\b{\bs D} \b{\bs\Omega}=\bs d\b{\bs\Omega}+[\bs\varpi, \b{\bs \Omega}]_{\text{{\tiny $LieG$}}}=0$. And it is still true that $\bs D \circ \bs D =\rho_*(\bs \Omega)$.
We may observe that the horizontality of $ \b{\bs\Omega}$ encode interesting information. For $X^v, Y^v \in \Gamma(V\Phi)$ with $X, Y \in \diff(M)$, we have
\begin{align}
 \b{\bs\Omega}(X^v, Y^v)&= X^v\big(\omega(  Y^v)\big) -  Y^v\big(\omega( X^v)\big) - \bs \omega\big([ X^v,  Y^v]\big)   +[\bs \omega( X^v), \bs \omega( Y^v)]_{\text{{\tiny $\LieG$}}} \notag\\
0&=  X^v\big( a(Y; \phi) \big) -  Y^v\big( a(X; \phi) \big)    - a([X, Y]_{\text{{\tiny $\diff(M)$}}}; \phi)  + [a(X; \phi), a(Y; \phi)]_{\text{{\tiny $\LieG$}}},  \label{C}
\end{align}
which reproduces the infinitesimal 1-cocycle property  \eqref{inf-cocycle}  generalizing the Wess-Zumino consistency condition on gauge/$\Diff(M)$ anomalies.

Given the defining equivariance and verticality properties \eqref{Variational-twisted-connection} of a twisted connection, one shows using \eqref{pushforward-X}/\eqref{GT-geometric},  that its general vertical transformation under $\bs\Diff_v(\Phi)\simeq C^\infty\big( \Phi, \Diff(M)\big)$ is
\begin{align}
\label{Vert-trsf-twisted-connection}
\bs\varpi^{\bs\psi }\defeq \Xi^\star \bs\varpi =\Ad\big( C(\bs\psi)\- \big) \, \bs\varpi  + C(\bs\psi)\-\bs{d} C(\bs\psi), 
\end{align}
where we remind the short-hand notation $[C(\bs\psi)](\phi) \defeq C\big(\phi; \bs\psi(\phi)\big)$. In case $G=\Diff(M)$ the above specialises to: $\bs\varpi^{\bs\psi }=C(\bs\psi)\-_* \, \bs\varpi  \circ C(\bs\psi) + C(\bs\psi)\-_*\bs{d} C(\bs\psi)$.
Gauge transformations under $\bs\Aut_v(\Phi)\simeq \bs\Diff(M)$ are given by the same formula: The difference being noticed upon repeated transformations of each type, as  stressed by \eqref{2-Vert-Trsf}--\eqref{2-Gauge-Trsf} in section \ref{General vertical transformations, and gauge transformations}. 
Infinitesimally, transformations of a twisted connection under $\bs\diff_v(\Phi)\simeq C^\infty\big( \Phi, \diff(M)\big)$  are given by the Nijenhuis-Lie derivative, 
\begin{align}
\label{Inf-Vert-trsf-twisted-connection}
\bs L_{\bs X^v} \bs\varpi =\bs{d} a(\bs X)+ [\bs\varpi, a(\bs X)]_{\text{{\tiny $\LieG$}}}. 
\end{align}
where we remind that $[a(\bs X)](\phi)\defeq a\big(\bs X(\phi); \phi\big)$. 
Infinitesimal gauge transformations, under $\bs\aut_v(\Phi)\simeq \bs\diff(M)$, are given by the same relation. 
The difference being seen upon iteration, as reflected by  the commutation property  \eqref{Lie-deriv-ext-bracket}/\eqref{NL-derr-ext-bracket}  of the Nijenhuis-Lie derivative along $\bs\diff_v(\Phi) \simeq C^\infty \big(\Phi, \diff(M)\big)$, $[ \bs L_{\bs X^v}, \bs L_{\bs Y^v}]=\bs L_{[\bs X^v, \bs Y^v]_{\text{{\tiny $FN$}}}} = \bs L_{\{\bs X, \bs Y\}^v}$, which, in the case $\bs\aut_v(\Phi)\simeq \bs\diff(M)$, simplifies $[\bs L_{\bs{ X}^v}, \bs L_{ \bs{ Y}^v }]  = \bs L_{( -[\bs X,\bs  Y]_{\text{\tiny{$\diff(M)$}}})^v} = \bs L_{([ \bs X, \bs Y]_{\text{{\tiny $\Gamma(TM)$}}})^v}$ -- as we observed in section \ref{General vertical transformations, and gauge transformations}. 

Finite and infinitesimal general vertical transformations of the curvature are given by, 
\begin{align}
\label{GT-twisted-curvature}
\b{\bs\Omega}^{\bs\psi }\defeq \Xi^\star \b{\bs\Omega} =\Ad\big( C(\bs\psi)\- \big) \, \, \b{\bs\Omega}, \qquad \text{so} \qquad \bs L_{\bs X^v}\b{\bs\Omega} =[\b{\bs\Omega}, a(\bs X)]_{\text{{\tiny $\LieG$}}}. 
\end{align}
or, $\b{\bs\Omega}^{\bs\psi }\defeq \Xi^\star \b{\bs\Omega} =\bs\psi\-_* \, \b{\bs\Omega}  \circ \bs\psi$ and $\bs L_{\bs X^v}\b{\bs\Omega} =[\b{\bs\Omega}, \bs X]_{\text{{\tiny $\diff(M)$}}}$ in the case $G=\Diff(M)$. 
These are  special cases of \eqref{GT-tensorial} and \eqref{GT-inf-tensorial}. The same hold for its gauge transformations, with the usual caveat.

For $\bs X^v, \bs Y^v \in \bs\diff_v(\Phi)$ with $\bs X, \bs Y \in C^\infty \big(\Phi, \diff(M)\big)$, we have
\begin{equation}
\label{B}
\begin{aligned}
 \b{\bs\Omega}(\bs X^v, \bs Y^v)&= \bs X^v\big(\omega(  \bs Y^v)\big) - \bs Y^v\big(\omega(\bs X^v)\big) - \bs \omega\big([ \bs X^v,  \bs Y^v]\big)   +[\bs \omega( \bs X^v), \bs \omega( \bs Y^v)]_{\text{{\tiny $\LieG$}}} \\
0&=  \bs X^v\big( a(\bs Y; \phi) \big) -  \bs Y^v\big( a(\bs X; \phi) \big)    - a(\{\bs X, \bs Y\}; \phi)  + [a(\bs X; \phi), a(\bs Y; \phi)]_{\text{{\tiny $\LieG$}}},  \\
0&=\bs X^v\big( a(\munderline{red}{\bs Y}; \phi) \big) -  \bs Y^v\big( a( \munderline{red}{\bs X}; \phi) \big)    - a([\bs X, \bs Y]_{\text{{\tiny $\diff(M)$}}}; \phi)  + [a(\bs X; \phi), a(\bs Y; \phi)]_{\text{{\tiny $\LieG$}}}, 
\end{aligned}
\end{equation}
where the FN bracket \eqref{FN-bracket-field-dep-diff=BT-bracket}/\eqref{extended-bracket1} is used. 
This reproduces \eqref{BB}, where the notation of the last line was first used.
As the relation holds for  $\bs X^v, \bs Y^v \in \bs\aut_v(\Phi)$,  the horizontality properties \eqref{C}-\eqref{B} of the twisted curvature therefore encode the same generalisation of the Wess-Zumino consistency condition. 

\subsection{Associated bundles, fondamental representation of $\Diff(M)$, and integration on $M$}
\label{Associated bundles}  

Starting from a principal bundle, one of the most natural construction one can perform is that of its associated bundles. The standard theory specify how a class of them is built from representations of the structure group, while a natural extension of this classic construction involve the use of 1-cocycles for the action of the structure group instead of representations \cite{Francois2019_II}. Let us consider them in turn in our case, where the principal bundle is $\Phi$ with structure group $\Diff(M)$.

Given a representation space $(\rho,  \bs V)$ of $\Diff(M)$, consider the direct product space $\Phi \times \bs V$, with the two natural projections: $\pi_\Phi : \Phi \times \bs V \rarrow \Phi$ and  $\pi_{\bs V} : \Phi \times \bs V \rarrow \bs V$.
One defines a right action of $\Diff(M)$ on  $\Phi \times \bs V$ by: 
\begin{equation}
\label{Diff-action-PxV}
\begin{aligned}
\big(\Phi \times \bs V \big) \times \Diff(M) &\rarrow \Phi \times \bs V, \\
\big((\phi,  v), \psi \big) & \mapsto  \big(\psi^*\phi,\, \rho(\psi)\-  v\big)=\big(R_\psi \phi,\, \rho(\psi)\-  v\big) \rdefeq \b R_\psi (\phi,  v). 
\end{aligned}
\end{equation}
 The  bundle $\bs E$ associated to $\Phi$ via the representation $\rho$ is then defined as the quotient of the product space $\Phi \times \bs V$ by this right action:  
 \begin{equation}
\begin{aligned}
\bs E = \Phi \times_\rho \bs V \defeq   \Phi \times  \bs V /\!\sim
\end{aligned}
\end{equation}
where $(\phi',  v') \sim (\phi,  v)$ when $\exists\, \psi \in \Diff(M)$ s.t. $(\phi',  v')=\b R_\psi (\phi,  v)$. 
We may write $\b \pi_{\bs E} :  \Phi \times \bs V \rarrow \bs E$. 
A point in $\bs E$ is thus an equivalence class $e=[\phi,  v]$. The projection of $\bs E \xrightarrow{\pi_{\bs E}} \M$ is defined by $\pi_{\bs E}([\phi,  v]) \defeq \pi(\phi)=[\phi]$. 
It is a standard result of bundle theory that sections of $\bs E$ are in 1:1 correspondence with $V$-valued $\rho$-equivariant functions on $\Phi$: 
 \begin{equation}
 \label{iso-section-equiv-fct}
\begin{aligned}
\Gamma(\bs E)\defeq \big\{ \bs s: \M \rarrow \bs E\big\}  \   \
\simeq \ \  \Omega^0_\text{eq}\big(\Phi, \rho \big)\defeq \big\{ \bs\vphi: \Phi \rarrow V\, |\,  R^\star_\psi \bs\vphi =\rho(\psi)\- \bs\vphi  \big\} . 
\end{aligned}
\end{equation}
The isomorphism being $\bs s([\phi]) = [\phi, \bs \vphi(\phi)]$.  
Naturally, equivariant functions are tensorial 0-forms, and like all equivariant forms their vertical transformations (gauge transformations) is given by \eqref{GT-tensorial}-\eqref{GT-inf-tensorial}: 
\begin{align}
\bs\vphi^{\bs\psi} = \rho(\bs\psi)\- \bs\vphi, \quad \text{ and } \quad \bs L_{\bs X^v} \bs\vphi = -\rho_*(\bs X) \bs\vphi. 
\end{align} 
for $\bs\psi \in C^\infty \big(\Phi, \Diff(M)\big) \simeq \bs\Diff_v(\Phi)$ and $\bs X \in C^\infty \big(\Phi, \diff(M)\big) \simeq \bs\diff_v(\Phi)$.
A principal connection as discussed in section \ref{Ehresmann connections} is needed for their covariant differentiation. 

The right action of $\diff(M)$ on $\Phi \times \bs V$ defines the vertical subbundle $V(\Phi \times \bs V)\simeq V\Phi \oplus V\bs V \subset T(\Phi \times \bs V)\simeq T\Phi \oplus T\bs V$. By linearisation of \eqref{Diff-action-PxV}, it is: 
\begin{equation}
\label{diff-action-PxV}
\begin{alignedat}{2}
\big(\Phi \times \bs V \big) \times \diff(M) &\rarrow V(\Phi \times \bs V) \simeq  V\Phi  \oplus V\bs V, && \\
                               \big((\phi,  v), X \big) & \mapsto   \  X^v_{|(\phi, v)}\defeq  \tfrac{d}{d\tau} \b R_{\psi_\tau}(\phi,  v)\,  \big|_{\tau=0} &&= \tfrac{d}{d\tau} \big(\psi_\tau^*\phi,\, \rho(\psi_\tau)\-  v\big) \,  \big|_{\tau=0} = \big(\mathfrak L_X \phi, v \big) \oplus   \big( \phi, -\rho_*(X)v \big),  \\
				                               &\phantom{\mapsto   \  X^v_{|(\phi, v)}\defeq} && =   \left((\phi, v); \big( \mathfrak L_X \phi,-\rho_*(X)v \big) \right) \\	
				                               	&\phantom{\mapsto   \  X^v_{|(\phi, v)}\defeq}  && =	\big(\phi;  \mathfrak L_X \phi \big) \oplus \big(v; -\rho_*(X)v \big)
								 											\rdefeq X^v_{|\phi} + X^v_{|v}. 
\end{alignedat}
\end{equation}
Therefore, the  tangent bundle of $\bs E$ is defined as  $T\bs E\defeq T(\Phi \times \bs V)/V(\Phi \times \bs V)$
\medskip

The whole construction carries through if one replaces the representation  $\rho$ by a 1-cocycle 
$\ C\!: \Phi \times \Diff(M) \rarrow G$, as defined by \eqref{cocycle} in section \ref{Differential forms and their derivations}.
If $\bs V$ is a $G$-space, one may define a right action of $\Diff(M)$ on $ \Phi \times  \bs V$ via:
$\big(\Phi \times \bs V \big) \times \Diff(M) \rarrow \Phi \times \bs V$, 
$\big((\phi,  v), \psi \big)  \mapsto  \b R_\psi (\phi,  v)=\big(\psi^*\phi,\, C(\phi; \psi)\-  v\big)$. 
 The \emph{twisted} bundle $\widetilde{\bs E} \rarrow \M$ associated to $\Phi$ via the 1-cocycle $C$ is then defined as:
$\widetilde{\bs E} = \Phi \times_C \bs V \defeq   \Phi \times  \bs V /\!\sim\,$,
with $\big(\psi^*\phi,\, C(\phi; \psi)\-  v\big) \sim (\phi,  v)$. 
As above we have the isomorphism, 
 \begin{equation}
\begin{aligned}
\Gamma\big(\widetilde{\bs E}\big)\defeq \big\{ \t{\bs s}: \M \rarrow \widetilde{\bs E}\big\}  \   \
\simeq \ \  \Omega^0_\text{eq}\big(\Phi,  C \big)\defeq \big\{ \t{\bs\vphi}: \Phi \rarrow V\, |\,  R^\star_\psi \t{\bs\vphi} = C(\phi; \psi)\- \t{\bs\vphi}  \big\},
\end{aligned}
\end{equation}
where the latter is the space of \emph{twisted} equivariant function on $\Phi$. 
As we highlighted in section \ref{Differential forms and their derivations}, neglecting the action of $\Diff(M)$ on the integration domain, the action functional $S$ of a theory defines a twisted equivariant function $Z=\exp(iS)$ on $\Phi$ -- i.e. a section of a twisted associated ($\CC$-line) bundle.  See \cite{Francois2019_II, Francois2021, Francois-et-al2021}. 
In this paper, we adopt the more natural stance taking into account the natural action of $\Diff(M)$ on integration domains,\footnote{Under which quantum functionals $Z=\int \bs d\phi\, \exp(\tfrac{i}{\hbar}S)$ remains twisted equivariant, i.e. sections of a twisted $\CC$-bundle. See footnote \ref{foontote-Q}.} see below.
The vertical transformation of such  twisted equivariant functions is given by  \eqref{GT-tensorial}-\eqref{GT-inf-tensorial}. 
A twisted connection as discussed in section \ref{Twisted connections} is needed for their covariant differentiation. 

\medskip

\subsubsection{Associated bundle of regions} 
\label{Associated bundle of regions} 

There is a particular bundle, $\bs E= \b {\bs U}(M)$, canonically associated to $\Phi$, as it is built via the \emph{defining representation} of its structure group $\Diff(M)$:\footnote{In that respect it is the conceptual equivalent of e.g. $TM$ for the frame bundle $LM$ of $M$. The tangent bundle can indeed be seen as associated to $LM$ via the defining representation $\RR^n$ of its structure group $GL(n)$: $TM = LM \times_{GL(n)} \RR^n$.} 
the field (or $\sigma$-algebra) of open sets of $M$, $\bs V=\bs U(M)\defeq \big\{ U \subset M\, | \, U \text{ open set} \big\}$.\footnote{Purists will notice that this highlights the fact that $\Diff(M)$ is a (Lie) pseudo-group rather than a (Lie) group. See e.g. \cite{Kobayashi1972}. Indeed, one may aim to generalise the whole formalism presented here to pseudo-groups, or even groupoids.}
The~right action of $\Diff(M)$ on the product space $ \Phi \times \bs U(M) $ is,
 \begin{equation}
 \label{right-action-diff}
\begin{aligned}
\big( \Phi \times \bs U(M) \big) \times \Diff(M) &\rarrow  \Phi \times \bs U(M), \\
\big( (\phi, U), \psi \big) &\mapsto \b R_\psi (\phi,  U) \defeq \big( \psi^*\phi, \psi\-(U) \big). 
\end{aligned}
\end{equation}
Notice the right action on $\bs U(M)$, $R_\psi U=\psi\-(U)$ (preimage convention), instead of the familiar defining left action  $L_\psi U=\psi(U)$ (direct image convention), is not accidental here but required by the bundle geometry framework (where the right action convention is now settled standard practice). 
The associated bundle is thus:
 \begin{equation}
 \label{Assoc-bundle-regions}
\begin{aligned}
\b{\bs U}(M) = \Phi \times_\text{{\tiny $\Diff(M)$}} \bs U(M) \defeq   \Phi \times  \bs U(M) /\!\sim.
\end{aligned}
\end{equation}
Call it the ``\emph{associated bundle of regions}" of $M$. 
Its space of sections $\Gamma\big(\b{\bs U}(M) \big)\defeq \big\{ \b{\bs s}: \M \rarrow \b{\bs U}(M) \big\}$ is isomorphic to
 \begin{equation}
 \label{equiv-regions}
\begin{aligned}
\Omega^0_\text{eq}\big(\Phi,  U(M)  \big)\defeq \big\{ \bs U: \Phi \rarrow \bs U(M) \, |\,  R^\star_\psi {\bs U} = \psi\-({\bs U})  \big\}.
\end{aligned}
\end{equation}
We may interpret   $\phi \rarrow \bs U(\phi)$ as being ``field-dependent" open sets of $M$, or yet, regions of $M$ defined in a ``$\phi$-\emph{relative}" -- and $\Diff(M)$-equivariant -- way.
Their  transformations under $\bs\Diff_v(\Phi) \simeq C^\infty \big(\Phi, \Diff(M)\big)$ and $\bs\diff_v(\Phi) \simeq C^\infty \big(\Phi, \diff(M)\big)$ are respectively, by \eqref{GT-tensorial}-\eqref{GT-inf-tensorial}: 
\begin{align}
\bs U^{\bs \psi} = \bs\psi\-(\bs U), \quad \text{ and } \quad \bs L_{\bs X^v} \bs U = -\bs X(\bs U). 
\end{align} 
 An example of such equivariant functions is provided by a familiar operation: integration. 
 To show this, in the next section, we frame integration as a special case of a natural construction over the product space $\Phi \times \bs U(M)$. 
 
 \subsubsection{Integration map} 
\label{Integration map} 

To obtain an associated bundle, we defined above a right action of $\Diff(M)$ on $\Phi \times \bs U(M)$: $\b R_\psi (\phi,  v)\defeq \big(\psi^*\phi,\, \rho(\psi)\-  v\big)$. 
Let us now also define a corresponding action of $C^\infty\big(\Phi, \Diff(M)\big) \simeq \bs\Diff_v(\Phi)$:
\begin{equation}
\label{Diff_v-action-PxV}
\begin{aligned}
\big(\Phi \times \bs V \big) \times C^\infty\big(\Phi, \Diff(M)\big) &\rarrow \Phi \times \bs V, \\
\big((\phi,  v), \bs\psi \big) & \mapsto  \big(\bs\psi^*\phi,\, \rho(\bs\psi)\-  v\big) =  \big(\Xi(\phi),\, \rho(\bs\psi)\-  v\big) \rdefeq \b \Xi (\phi,  v). 
\end{aligned}
\end{equation}
It is easily checked that in particular, for $\bs\psi \in \bs\Diff(M)$, we have $\b \Xi \bs\circ \b R_\psi = \b R_\psi \bs\circ \b \Xi$. 
The action of $\C^\infty\big(\Phi, \diff(M)\big)\simeq \bs\diff_v(\Phi)$ may be seen as the linearisation of the above, 
$\big(\Phi \times \bs V \big) \times C^\infty\big(\Phi, \diff(M)\big) \rarrow V(\Phi \times \bs V)\simeq V\Phi \oplus V\bs V \subset T(P\times \bs V)$, 
which is as \eqref{diff-action-PxV} under the replacement $X \rarrow \bs X$.

The induced actions of $\Diff(M)$ and $C^\infty\big(\Phi, \Diff(M)\big)\simeq \bs\Diff_v(\Phi)$ on the space $ \Omega^\bullet(\Phi) \times \bs V$ are 
\begin{equation}
\begin{aligned}
\big( \Omega^\bullet(\Phi) \times \bs V \big) \times \Diff(M) &\rarrow \Omega^\bullet(\Phi)\times \bs V, \\
\big((\bs\alpha,  v), \psi \big) & \mapsto  \big(R^\star_\psi \bs\alpha,\, \rho(\psi)\-  v\big)  \rdefeq \t R_\psi (\bs\alpha,  v). 
\end{aligned}
\end{equation}
and 
\begin{equation}
\label{Diff_v-action-PxV}
\begin{aligned}
\big(\Omega^\bullet(\Phi) \times \bs V \big) \times C^\infty\big(\Phi, \Diff(M)\big) &\rarrow \Omega^\bullet(\Phi) \times \bs V, \\
\big((\bs\alpha,  v), \bs\psi \big) & \mapsto  \big(\Xi^\star \bs\alpha ,\, \rho(\bs\psi)\-  v\big)  \rdefeq \t \Xi (\bs\alpha,  v). 
\end{aligned}
\end{equation}
The induced actions of $\diff(M)$ and $C^\infty\big(\Phi, \diff(M)\big)\simeq \bs\diff_v(\Phi)$ are the linearisations:
\begin{equation}
\label{linear-versions}
\begin{aligned}
\big((\bs\alpha,  v), X \big) & \mapsto  \tfrac{d}{d\tau}\, \t R_{\psi_\tau} (\bs\alpha,  v) \,\big|_{\tau=0} = \big(\bs L_{X^v} \bs\alpha,\,  v\big) \oplus \big( \bs\alpha,\, -\rho_*(X)  v\big), \\
\big((\bs\alpha,  v), \bs X \big) & \mapsto  \tfrac{d}{d\tau}\, \t \Xi_\tau (\bs\alpha,  v)\, \big|_{\tau=0} = \big(\bs L_{\bs X^v} \bs\alpha,\,  v\big) \oplus \big( \bs\alpha,\, -\rho_*(\bs X)  v\big)  
\end{aligned}
\end{equation}
We further observe that given a representation $(\t \rho, \bs W)$ of $\Diff(M)$:
\begin{equation}
\begin{aligned}
&\text{if }\ \bs\alpha \in \Omega^\bullet_{\text{eq}}(\Phi, \bs W)  \  \text{ then } \   
 \t R_\psi (\bs\alpha,  v) =  \big(R^\star_\psi \bs\alpha,\, \rho(\psi)\-  v\big)=  \big( \t\rho(\psi)\- \bs\alpha,\, \rho(\psi)\-  v\big), \\
&\text{if }\ \bs\alpha \in \Omega^\bullet_{\text{tens}}(\Phi, \bs W)  \  \text{ then } \   
 \t \Xi (\bs\alpha,  v) =  \big(\Xi^\star \bs\alpha,\, \rho(\bs \psi)\-  v\big)=  \big( \t\rho(\bs\psi)\- \bs\alpha,\, \rho(\bs\psi)\-  v\big),
\end{aligned}
\end{equation}
Linear versions are obviously read from \eqref{linear-versions}.
Remark that the exterior derivative $\bs d$ on $\Phi$ a priori extends to $\Phi \times \bs V$ as ${\bs d} \rarrow  \b{\bs d}=\bs d \times  \id$, which we may still write $\bs d$ for simplicity. Yet, following the action of  $C^\infty\big(\Phi, \diff(M)\big)\simeq \bs\diff_v(\Phi)$, it will also act on the second factor $\rho(\bs\psi)\-  v$ due to the $\phi$-dependence of $\bs\psi$. 

Consider $(\b \rho, \bs V^*)$ a representation of $\Diff(M)$ \emph{dual} to $(\rho, \bs V)$ w.r.t. a non-degenerate $\Diff(M)$-invariant \emph{pairing} 
\begin{equation}
\begin{aligned}
\langle\ , \ \rangle: \bs V^* \times \bs V &\rarrow \RR, \\
                                                       (w, v) &\mapsto \langle w , v \rangle, \quad \text{s.t.} \quad \langle\, \b \rho(\psi) w ,   \rho(\psi)v \rangle =  \langle w , v \rangle. 
\end{aligned}
\end{equation}
Invariance of the pairing means that under $\diff(M)$, for which we have the induced representation $\b \rho_*$ and $\rho_*$, the following identity holds:
\begin{align}
\label{infinitesimal-equiv-pairing}
\langle\, \b \rho_*(X) w ,   v \rangle + \langle\,  w ,   \rho_*(X)v \rangle =0
\end{align}
For $\bs\alpha \in \Omega^\bullet(\Phi, \bs V^*)$, let us then define the operation $\mathcal I$ 
on $\Omega^\bullet(\Phi, \bs V^*) \times \bs V$ by,
\begin{equation}
\begin{aligned}
\mathcal I: \Omega^\bullet(\Phi, \bs V^*)  \times \bs V &\rarrow \Omega^\bullet(\Phi),\\
					(\bs\alpha, v) &\mapsto \mathcal I(\bs\alpha, v) \defeq  \langle \bs\alpha , v \rangle. 
\end{aligned}
\end{equation}
Which can also be seen as an object on $\Phi \times \bs V$ by
\begin{equation}
\begin{aligned}
\mathcal I(\bs\alpha, \ ): \Phi \times \bs V &\rarrow \Lambda^\bullet(\Phi),\\
					(\phi, v) &\mapsto \mathcal I(\bs\alpha_{|\phi}, v) \defeq  \langle \bs\alpha_{|\phi} , v \rangle. 
\end{aligned}
\end{equation}
Naturally, we thus have:
\begin{align}
\label{der-eval-integr-obj}
\bs d \mathcal I(\bs\alpha, \ ) = \mathcal I(\bs{d\alpha}, \ ), \quad \text{ and } \quad \iota_{\mathfrak X}  \mathcal I(\bs\alpha, \ ) = \mathcal I( \iota_{\mathfrak X} \bs\alpha, \ ) \quad \text{ for } \mathfrak X \in \Gamma(T\Phi).
\end{align}
The induced actions of $\Diff(M)$ and $C^\infty\big(\Phi, \Diff(M) \big) \simeq \bs\Diff_v(\Phi)$ on such an object are:
\begin{equation}
\begin{aligned}
 &\t R^\star_\psi \mathcal I(\bs\alpha,\   )_{|(\psi^*\phi,\  \rho(\psi)\- v)} \defeq  \langle\ , \ \rangle \circ \t R_\psi (\bs \alpha, v) = \langle R^\star_\psi \bs\alpha_{|\psi^*\phi},\, \rho(\psi)\-  v\rangle, \\
& \t \Xi^\star \mathcal I(\bs\omega,\   )_{|(\Xi (\phi),\  \rho(\bs\psi)\- v)} \defeq  \langle\ , \ \rangle \circ \t \Xi (\bs \alpha, v) = \langle \Xi^\star \bs\alpha_{|\Xi(\phi)},\, \rho(\bs\psi)\-  v\rangle. 
 \end{aligned}
\end{equation}
The induced actions of $\diff(M)$ and $C^\infty\big(\Phi, \diff(M)\big)\simeq \bs\diff_v(\Phi)$ are thus (with simplified but obvious notation):
\begin{equation}
\label{linear-versions-pairing}
\begin{aligned}
&\tfrac{d}{d\tau}\, \t R_{\psi_\tau}^\star \mathcal I (\bs\alpha,  v) \,\big|_{\tau=0} = \langle \bs L_{X^v} \bs\alpha,\,  v \rangle 
																			   + \langle \bs\alpha,\, -\rho_*(X)  v \rangle, \\
&\tfrac{d}{d\tau}\, \t \Xi_\tau^\star \mathcal I  (\bs\alpha,  v)\, \big|_{\tau=0} = \langle \bs L_{\bs X^v} \bs\alpha,\,  v \rangle 
																			 + \langle \bs\alpha,\, -\rho_*(\bs X)  v \rangle.  
\end{aligned}
\end{equation}

We then observe that, if $\bs\alpha \in \Omega^\bullet_{\text{eq}}(\Phi, \bs V^*)$:
\begin{equation}
\label{Int-eq}
\begin{aligned}
 \t R^\star_\psi \mathcal I(\bs\alpha,\   )_{|(\psi^*\phi,\  \rho(\psi)\- v)} \defeq 
 &\, \langle R^\star_\psi \bs\alpha_{|\psi^*\phi},\, \rho(\psi)\-  v\rangle, \\
 =&\,  \langle \b\rho(\psi)\- \bs\alpha_{|\phi},\, \rho(\psi)\-  v\rangle = \langle  \bs\alpha_{|\phi},  v\rangle \rdefeq \mathcal I(\bs\alpha,\ )_{|(\phi, v)}. 
 \end{aligned}
\end{equation}
From which follows, by \eqref{linear-versions-pairing}, the identity:
\begin{equation}
\label{General-cont-eq1}
\begin{aligned}
 \langle \bs L_{X^v} \bs\alpha,\,  v \rangle + \langle \bs\alpha,\, -\rho_*(X)  v \rangle &= 0, \qquad X \in \diff(M).\\
  \langle  -\b \rho_*(X)\, \bs\alpha,\,  v \rangle + \langle \bs\alpha,\, -\rho_*(X)  v \rangle &= 0
\end{aligned}
\end{equation}
If $\bs\alpha \in \Omega^\bullet_{\text{tens}}(\Phi, \bs V^*)$:
 \begin{equation}
  \label{Int-tens}
\begin{aligned}
 \t \Xi^\star \mathcal I(\bs\alpha,\   )_{|(\Xi (\phi),\  \rho(\psi)\- v)} \defeq  
 &\, \langle \Xi^\star \bs\alpha_{|\Xi(\phi)},\, \rho(\bs\psi)\-  v\rangle, \\
 =&\,  \langle \b\rho(\bs\psi)\- \bs\alpha_{|\phi},\, \rho(\bs\psi)\-  v\rangle = \langle  \bs\alpha_{|\phi},  v\rangle \rdefeq \mathcal I(\bs\alpha,\ )_{|(\phi, v)}. 
 \end{aligned}
\end{equation}
From which follows, by \eqref{linear-versions-pairing}, the identity:
\begin{equation}
\label{General-cont-eq2}
\begin{aligned}
 \langle \bs L_{\bs X^v} \bs\alpha,\,  v \rangle + \langle \bs\alpha,\, -\rho_*(\bs X)  v \rangle &= 0, \qquad \bs X \in C^\infty\big(\Phi, \diff(M)\big).\\
  \langle  -\b \rho_*(\bs X)\, \alpha,\,  v \rangle + \langle \bs\alpha,\, -\rho_*(\bs X)  v \rangle &= 0
\end{aligned}
\end{equation}
This last case, $\bs\alpha$ tensorial,  ensures that $\mathcal I(\bs\alpha,\ )$ is ``basic" on $\Phi\times \bs V$, thus well-defined as an object on $\bs E=\Phi \times \bs V/\sim$. 
Being constant along a $\Diff(M)$-orbit in $\Phi \times \bs V$,  $\mathcal I(\bs \alpha,\ )$ then allows to define 
$\bs\vphi_{\mathcal I(\bs \alpha)}\! \in \Omega^0_\text{eq}(\Phi, \rho)$ 
 via the prescription: 
 \begin{equation}
 \label{Induced-equiv-fct}
\begin{aligned}
\bs\vphi_{\mathcal I(\bs \alpha)}(\phi)&\defeq \pi_{\bs V}(\phi,  v)_{|\mathcal I(\bs\alpha_{|\phi}, v) =cst} \equiv v, \\
\bs\vphi_{\mathcal I(\bs \alpha)}(\psi^*\phi)& \defeq \pi_{\bs V}(\psi^*\phi, \rho(\psi)\- v)_{|\mathcal I(\bs\alpha_{|\phi}, v) =cst} \equiv \rho(\psi)\- v.
 \end{aligned}
\end{equation}
Which in turn is equivalent to a section $\bs s_{\mathcal I(\bs \alpha)} :\M \rarrow \bs E$, by \eqref{iso-section-equiv-fct}.

We may also observe that   \eqref{Int-tens} implies that for $\bs\alpha \in \Omega^\bullet_{\text{tens}}(\Phi, \bs V^*)$ one has 
$\bs d \, \t \Xi^\star \mathcal I(\bs\alpha,\   ) = \bs d \mathcal I(\bs\alpha,\   )= \mathcal I(\bs{d\alpha},\   )$. 
This is shown explicitly using Lemma \eqref{Formula-general-rep}: 
\begin{align}
\bs d \,\langle \Xi^\star \bs\alpha,\, \rho(\bs\psi)\-v \rangle  
					&= \langle  \bs d \,  \Xi^\star  \bs\alpha, \, \rho(\bs\psi)\-v \rangle 
								\ + \ \langle  \Xi^\star \bs\alpha, \, \bs d \rho(\bs\psi)\-v\rangle, \notag \\
					&= \langle  \bs d \, \b\rho(\bs\psi)\-  \bs\alpha, \, \rho(\bs\psi)\-v \rangle 
								\ + \ \langle \b\rho(\bs\psi)\-  \bs\alpha, \, \bs d \rho(\bs\psi)\-v\rangle, \notag \\
					&= \langle  \b\rho(\bs\psi)\- \big(  \bs d \bs \alpha -\b\rho_*(\bs{d\psi}\circ \bs\psi\-)\, \bs\alpha \big),\, \rho(\bs\psi)\-v\rangle 
								\ +\  \langle   \bs\alpha, \, \rho(\bs\psi)\bs d \rho(\bs\psi)\-v\rangle, \notag \\
					&= \langle    \bs d \bs \alpha -\b\rho_*(\bs{d\psi}\circ \bs\psi\-) \bs\alpha,\, v\rangle 
								\ +\  \langle   \bs\alpha, \, \rho_*(\bs\psi\bs d \bs\psi\-)\,v\rangle,  \notag\\
					&= \langle    \bs d \bs \alpha, v \rangle+  \underbrace{\langle -\b\rho_*(\bs{d\psi}\circ \bs\psi\-) \bs\alpha,\, v\rangle 
				+ \langle   \bs\alpha,  -\rho_*(\bs d \bs\psi\circ \bs\psi\-)\,v\rangle}_{=0 \,\text{ by \eqref{infinitesimal-equiv-pairing}} }, \notag\\[1mm]
	\text{i.e.} \quad \bs d \, \t \Xi^\star \mathcal I(\bs\alpha,\   ) &= \mathcal I(\bs{d\alpha},\   )= \bs d \mathcal I(\bs\alpha,\   ). \label{commute-d-vert-trsf}
\end{align}
Noticing that the first term on the right-hand side of the first line above is (stretching a bit the notation for simplicity)
\begin{align*}
\langle  \bs d \,  \Xi^\star  \bs\alpha, \, \rho(\bs\psi)\-v \rangle =  \langle  \Xi^\star \bs d \bs\alpha, \, \rho(\bs\psi)\-v \rangle \rdefeq  \t\Xi^\star \langle \bs{d\alpha}, v \rangle,
\end{align*}
the above computation also supplies the following identity:
\begin{equation}
\label{Vert-trsf-pairing-dalpha}
\begin{aligned}
 \t\Xi^\star \langle \bs{d\alpha}, v \rangle &= \langle \bs{d\alpha}, v \rangle + \langle \bs\alpha, -\rho_*(\bs{d\psi}\circ \bs\psi\-) v \rangle, \\
 							      &= \langle \bs{d\alpha}, v \rangle + \langle \b\rho_*(\bs{d\psi}\circ \bs\psi\-) \bs\alpha,  v \rangle.					
\end{aligned}
\end{equation}
\medskip

Let us specialise  the above construction, considering the ``fundamental" representation $\bs V = \bs U(M)$ on which  $\Diff(M)$ acts via the preimage convention, and the representation $\bs V^* = \Omega^\text{top}(U)$ of volume forms on $U \in \bs U(M)$ on which $\Diff(M)$ acts by pullback. These are dual under the invariant \emph{integration pairing}:
\begin{equation}
\begin{aligned}
\langle\ , \ \rangle: \bs \Omega^\text{top}(U) \times \bs U(M) &\rarrow \RR, \\[-2mm]
                                                       (\omega, U) &\mapsto \langle \omega , U \rangle \defeq \int_U \omega. 
\end{aligned}
\end{equation}
 The invariance property assumes the familiar form:
 \begin{align}
 \label{invariance-action}
\langle\, \psi^*\omega ,   \psi\-(U) \rangle =  \langle \omega , U \rangle \quad   \rarrow  \quad  \int_{\psi\-(U)} \psi^*\omega = \int_U \omega. 
 \end{align}
 This, as a special case of \eqref{infinitesimal-equiv-pairing} with $-\b \rho_*(X)=\mathfrak L_X$ and $-\rho_*(X)=-X$, gives the identity:
  \begin{align}
 \label{inf-invariance-action}
\langle\, \mathfrak L_X \omega ,   U \rangle + \langle\,  \omega ,   -X(U) \rangle =0  \quad
\rarrow \quad
  \int_U   \mathfrak L_X \omega  + \int_{-X(U)} \hspace{-2mm}\omega =0,
 \end{align}
 which amounts to a continuity equation for the action of $\diff(M)$. 
 The form is reminiscent of Stokes theorem, i.e. the duality between the DeRham derivative $d$ on $\Omega^\bullet(U)$ and the boundary operator $\d$ on $\bs U(M)$, adjoint operators w.r.t. to the integration pairing: 
 \begin{align}
 \langle\, d \omega ,   U \rangle = \langle\,  \omega ,   \d U \rangle 
 \rarrow \quad
  \int_U   d \omega  = \int_{\d U} \hspace{-2mm}\omega,
\end{align} 
degree and dimension being appropriately adjusted. 
 
Considering  $\bs \alpha \in \Omega^\bullet\big( \Phi,  \Omega^\text{top}(U) \big)$, the field-dependent volume forms, we define the integration map on $\Phi \times \bs U(M)$:
\begin{align}
\mathcal I(\bs\alpha_{|\phi}, U) = \langle \bs\alpha_{|\phi}, U \rangle \defeq \int_U \bs \alpha_{|\phi}.
\end{align}
We will sometimes use the simplified notation $\bs\alpha_U$ when more convenient. 
We have naturally $\bs d \mathcal I(\bs\alpha, U) = \mathcal I(\bs{d\alpha}, U)$, and $\iota_{\mathfrak X}  \mathcal I(\bs\alpha, U ) = \mathcal I( \iota_{\mathfrak X} \bs\alpha, U)$ for $\mathfrak X \in \Gamma(T\Phi)$ -- as a special case of \eqref{der-eval-integr-obj}. 
The induced actions of $\Diff(M)$ and $C^\infty\big(\Phi, \Diff(M) \big) \simeq \bs\Diff_v(\Phi)$ on integrals are:
\begin{align}
 &\t R^\star_\psi \mathcal I(\bs\alpha,\   )_{|\big(\psi^*\phi,\ \psi\-(U) \big)} \defeq  \langle R^\star_\psi \bs\alpha_{|\psi^*\phi},\, \psi\-(U) \rangle = \int_{ \psi\-(U)} R^\star_\psi \bs\alpha_{|\psi^*\phi}  \\
& \t \Xi^\star \mathcal I(\bs\alpha,\   )_{|\big(\Xi (\phi),\  \bs\psi\-(U) \big)} \defeq  \langle \Xi^\star \bs\alpha_{|\Xi(\phi)},\, \bs\psi\-(U)  \rangle = \int_{ \bs\psi\-(U)}  \Xi^\star \bs\alpha_{|\Xi(\phi)}. \label{Vert-trsf-int-generic}
 \end{align}
One may notice that, in the latter case, the  derivative $\bs d$ will now act also on the transformed region $\bs\psi\-(U)$ due to the $\phi$-dependence of $\bs\psi$. 
Using the  notation just mentioned, we may write the above as ${\bs\alpha_U}^\psi$ and ${\bs\alpha_U}^{\bs\psi}$ respectively. 
The induced actions of $\diff(M)$ and $C^\infty\big(\Phi, \diff(M)\big)\simeq \bs\diff_v(\Phi)$ on integrals are thus (with simplified notation):
\begin{equation}
\label{linear-versions-integral}
\begin{aligned}
&\tfrac{d}{d\tau}\, \t R_{\psi_\tau}^\star \mathcal I (\bs\alpha,  U) \,\big|_{\tau=0} = \langle \bs L_{X^v} \bs\alpha,\,  U \rangle 
																			   + \langle \bs\alpha,\, -X(U) \rangle
														           =\int_U  \bs L_{X^v} \bs\alpha + \int_{-X(U)} \bs\alpha, \\
&\tfrac{d}{d\tau}\, \t \Xi_\tau^\star \mathcal I  (\bs\alpha,  U)\, \big|_{\tau=0} = \langle \bs L_{\bs X^v} \bs\alpha,\,  U \rangle 
																			 + \langle \bs\alpha,\, -\bs X(U)   \rangle
															=\int_U  \bs L_{\bs X^v} \bs\alpha + \int_{-\bs X(U)} \bs\alpha.  
\end{aligned}
\end{equation}
When convenient, we may write the above as $\delta_X\ \!{\bs\alpha_U}$ and $\delta_{\bs X}\ \!{\bs\alpha_U}$ respectively. 

In case $\bs\alpha$ is  tensorial 
on $\Phi$, we have the special case of the results \eqref{Int-tens} 
above, so that  $\bs\alpha_U=\mathcal I(\bs\alpha_{|\phi}, U)$ is $C^\infty\big(\Phi, \Diff(M) \big)$-invariant:  ${\bs\alpha_U}^{\bs \psi} =\bs\alpha_U$. 
From which follows, as a special case of  \eqref{General-cont-eq2} and \eqref{linear-versions-integral}, that:
\begin{equation}
\label{Cont-eq-general-case}
\begin{aligned}
\delta_{\bs X} \bs\alpha_U =0 \quad \Rightarrow \quad \langle \bs L_{\bs X^v} \bs\alpha,\,  U \rangle + \langle \bs\alpha,\, -\bs X(U) \rangle &= 0, \qquad \bs X \in C^\infty\big(\Phi, \diff(M)\big).\\
  \langle  \mathfrak L_{\bs X} \alpha,\,  U \rangle + \langle \bs\alpha,\, -\bs X(U)   \rangle &= 0 \quad \rarrow \quad \int_U \mathfrak L_{\bs X} \alpha + \int_{-\bs X(U)} \bs\alpha =0.  
\end{aligned}
\end{equation}
This can be interpreted as a continuity equations or sort. For $\bs\alpha$ equivariant, we have $\Diff(M)$-invariance of its integral: ${\bs\alpha_U}^{\psi} =\bs\alpha_U$, and \eqref{Cont-eq-general-case} holds mutatis mutandis ($\bs X \rarrow X$). 
For $\bs\alpha$ tensorial, $\bs\alpha_U=\mathcal I(\bs\alpha, U)$ is thus well-defined on the bundle of regions $\b{\bs U}(M)=\Phi \times \bs U(M)/\sim$, and one may -- quite formally still -- define an equivariant $\bs U(M)$-valued function on $\Phi$ as in  \eqref{Induced-equiv-fct}.
It also mean that $\bs d({\bs\alpha_U}^{\bs\psi}) = \bs d \bs\alpha_U$, i.e. 
\begin{align}
\label{Invariance-tens-int}
\bs d \, \langle \bs\psi^* \bs\alpha, \bs\psi\-(U)\rangle =   \langle \bs d \bs\alpha, U\rangle = \bs d  \langle  \bs\alpha, U\rangle \quad \rarrow \quad
\bs d \int_{\bs\psi\-(U)}  \bs\psi^* \bs\alpha = \int_U \bs{d\alpha} = \bs d \int_U \bs \alpha.
\end{align}
Which is also proven in exactly the same way as  \eqref{commute-d-vert-trsf}, using the special form  \eqref{Formula-pullback-rep}  of the lemma \eqref{Formula-general-rep}  and concluding by the invariance property  \eqref{inf-invariance-action}. 
Also, specialising \eqref{Vert-trsf-pairing-dalpha} we get the identity: 
\begin{equation}
\label{Vert-trsf-int-dalpha}
\begin{aligned}
\bs (\bs{d\alpha}_U)^{\bs \psi}	 &= \bs{d\alpha}_U + \langle \mathfrak L_{\bs{d\psi}\circ \bs\psi\-} \bs\alpha,  U \rangle,  \\
 \langle  \bs{d\alpha}, U \rangle ^{\bs \psi} &= \langle\bs{d\alpha}, U \rangle + \langle \mathfrak L_{\bs{d\psi}\circ \bs\psi\-} \bs\alpha,  U \rangle. 
\end{aligned}
\end{equation}
These are  relevant to the variational principle.
\bigskip

Indeed, consider  a Lagrangian  $L \in \Omega^0_\text{tens}\big(\Phi, \Omega^\text{top}(U) \big)$, a tensorial 0-form on $\Phi$ and volume form~on~$U$. 
The corresponding action functional is the  invariant object on $\Phi \times \bs U(M)$: $S = \mathcal I(L, U) =\langle L, U \rangle= \int_U L $, 
or yet $S(\phi)=\mathcal I\big(L(\phi), U\big) = \langle L(\phi), U \rangle= \int_U L(\phi)$, and s.t. $S^{\bs\psi}=\langle L^{\bs\psi}, \bs\psi\-(U) \rangle=\langle \bs\psi^* L, \bs\psi\-(U) \rangle=\langle L, U \rangle = S$. 
The variational principle is expressed as $\bs d S = \bs d\mathcal I(L, U) = \mathcal I(\bs d L, U) = \int_U \bs d L \equiv0$.
Under the action of $C^\infty\big(\Phi, \Diff(M) \big)\simeq \bs\Diff_v(\Phi)$, we have on the one hand  by \eqref {Invariance-tens-int} that 
$\bs d (S^{\bs \psi})=\bs d S$, from which follows as a special case of \eqref{Cont-eq-general-case} that $\delta_{\bs X} S =0$, i.e. the relation $\langle \mathfrak L_{\bs X} L, U \rangle + \langle L , -\bs X(U) \rangle = 0$.
On the other hand, by \eqref{Vert-trsf-int-dalpha} above, we have the relation:
\begin{equation}
\label{Vert-trsf-dS}
\begin{aligned}
(\bs dS)^{\bs\psi} &= \bs d S + \langle  \mathfrak  L_{\bs d \bs\psi  \circ \bs\psi\-} L,  U  \rangle \\
			   &=\bs d S + \langle  d \iota_{\bs d \bs\psi  \circ \bs\psi\-} L,  U  \rangle
			   = \bs d S + \langle  \iota_{\bs d \bs\psi  \circ \bs\psi\-} L,  \d U  \rangle.
\end{aligned}
\end{equation}
The fact that $\bs d S$ and its transform differ by a boundary term implies that the variational principle remain well-defined under the action of $C^\infty\big(\Phi, \Diff(M) \big) \simeq \bs\Diff_v(\Phi)$, and that the space of solution $\S$ is stable under field-dependent transformations.\footnote{As will be shown more explicitly in section \ref{Vertical and gauge transformations}, when we obtain the transformation of the field equations \eqref{GT-E}.}
This, in other words, proves that $\Diff(M)$-covariant theories $L$ enjoy a larger $\bs\Diff_v(\Phi) \simeq C^\infty\big(\Phi, \Diff(M) \big)$-covariance -- a fact  pointed out in the particular case of GR by Bergmann and Komar in 1971 \cite{Bergmann-Komar1972}. 
\medskip

The equivariant $\bs U(M)$-valued function associated to $S$ is:
 \begin{equation}
\begin{aligned}
\bs U_S(\phi)&\defeq \pi_{\bs U(M)}(\phi,  U)_{|S=cst} = U, \\
\bs U_S(\psi^*\phi)& \defeq \pi_{\bs U(M)}\big(\psi^*\phi, \psi\-(U)\big)_{|S =cst} =\psi\-(U).
 \end{aligned}
\end{equation}
Its   $\bs\Diff_v(\Phi) \simeq C^\infty \big(\Phi, \Diff(M)\big)$ and $\bs\diff_v(\Phi) \simeq C^\infty \big(\Phi, \diff(M)\big)$ transformations are thus respectively, by \eqref{GT-tensorial}-\eqref{GT-inf-tensorial}: 
\begin{align}
\bs U_S^{\bs \psi} = \bs\psi\- ( \bs U_S), \quad \text{ and } \quad \bs L_{\bs X^v} \bs U_S = -\bs X ( \bs U_S). 
\end{align} 
 The function outputs the region on which the fields $\phi$, and volume form $\bs\alpha(\phi)=L(\phi)$, are defined and integrated over. 
So, it subverts the usual logic by making the region $U$ individuated by the fields it hosts. This is not surprising since the  viewpoint articulated here,``dual" to the usual one, reframes familiar geometric objects in a way that gives primacy to the bundle of fields $\Phi$, and sees regions as representation space associated to $\Phi$ and its transformation groups. 
This, we may say, is a way to define regions of $M$ in a ``$\phi$-\emph{relative}" and  $\Diff(M)$-\emph{equivariant} way. As we will see in section \ref{The dressing field method} next, the DFM will allow a step further, suggesting a way to define regions of \emph{spacetime} in a $\phi$-\emph{relational} and $\Diff(M)$-\emph{invariant} way -- in accordance with the key insight of general relativistic physics.



\section{The dressing field method}  
\label{The dressing field method}  

We now turn the main topic of this paper. 
The dressing field method (DFM) is a systematic algorithm to build basic forms on a bundle, i.e. to obtain gauge-invariants in gauge theories. 
It was developed primarily for  internal gauge groups, i.e. for Yang-Mills theories and gauge gravity theories formulated via Cartan geometry \cite{GaugeInvCompFields, Francois2014, Attard_et_al2017,  Attard-Francois2016_II, Attard-Francois2016_I, Francois2018, Berghofer-et-al2023}. 
Relatively complete and self-contained expositions can be found in \cite{Francois2021, Francois-et-al2021}.

%
  
  In this section we detail the natural extension of the DFM to theories with $\Diff(M)$ symmetry.
   As the  next sections will make manifest, it is the general unifying  geometric framework for the notions of \emph{gravitational edge modes} as introduced in  \cite{DonnellyFreidel2016} and further developed e.g. in \cite{Speranza2018, Chandrasekaran_Speranza2021, Speranza2022, Chandrasekaran-et-al2022}, of \emph{gravitational dressings} as proposed in \cite{Giddings-Donnelly2016, Giddings-Donnelly2016-bis, Giddings-Kinsella2018, Giddings2019, Giddings-Weinberg2020} -- see also  \cite{Kabel-Wieland2022}  -- as well as the more recent idea of 
  ``dynamical reference frames" as expounded in \cite{carrozza-et-al2022, Hoehn-et-al2022}, or yet that of ``embedding maps/fields" as advocated in \cite{Speranza2019, Ciambelli-et-al2022,  Speranza2022, Ciambelli2023}. 
  
 But a more fundamental fact, we will argue, is  that the DFM is a framework that formally makes manifest the \emph{relational} character of general relativistic physics:
 It does so by systematically implementing a notion of \emph{relational observables}, or \emph{Dirac variables}, for $\Diff(M)$-theories \cite{Tamborino2012}.

\subsection{Building basic forms via dressing}  
\label{Building basic forms via dressing}  

One may profitably compare what follows to the DFM for theories with internal gauge group as presented in \cite{Francois2021, Francois-et-al2021}. 
One defines the space of $\Diff(M)$-dressing fields as,
\begin{align}
\label{Dressing-fields-space}
\D r[N, M] \defeq \left\{\ u: N \rarrow M\  |\ u^\psi \defeq \psi\- \circ u, \ \text{ with }\  \psi \in \Diff(M) \right\},
\end{align}
where $N$ is a reference $n$-dim manifold. From this, the action of $\diff(M)$ in the dressing field $u$ must be $\delta_X u \defeq  -X \circ u$. Given a field $\phi \in \Phi$ in $M$, acted upon by $\Diff(M)$  as $R_\psi=\psi^* \phi$,  the dressing map is defined as: 
\begin{align}
\label{Dressed-field}
|^u : \Phi \ &\rarrow\ \Phi^u, \notag \\
\phi \ &\mapsto \ \phi^u\defeq  \ u^*\phi.
\end{align}
Clearly,  $\phi^u$, called the \emph{dressing} of $\phi$,  is $\Diff(M)$-invariant: $(u^*\phi)^\psi \defeq (u^\psi)^*(\phi^\psi) = (\psi\- \circ u )^*(\psi^*\phi)= u^*\phi$.  Which is unsurprising as the space $\Phi^u$ of dressed fields is a subset of fields living on $N$. 

More generally, one defines a \emph{field-dependent dressing field} as a smooth map,
\begin{align}
\label{Field-dep-dressing}
\bs u : \Phi \ &\rarrow\ \D r[N, M]  \notag \\
\phi \ &\mapsto \  \bs u(\phi), \qquad \text{ s.t. }\quad  R^\star_\psi \bs u = \psi\- \!\circ \bs u \quad\text{ i.e. } \quad \bs u(\phi^\psi)=\psi\-\! \circ \bs u(\phi). 
\end{align}
In other words, $\bs u$ is an equivariant 0-form on $\Phi$ -- $\D r[N, M]$ being a representation space for $\Diff(M)$. Its infinitesimal equivariance, i.e. $\diff(M)$-transformation, is then $\bs L_{X^v} \bs u = -X \circ \bs u$.
And its transformations under $\bs\Diff_v(\Phi)\simeq C^\infty \big(\Phi, \Diff(M) \big)$ and  $\bs\diff_v(\Phi)\simeq C^\infty \big(\Phi, \diff(M) \big)$ are  respectively:
\begin{align}
\label{Vert-trsf-dressing}
\bs {u^\psi}\defeq \Xi^\star \bs u = \bs\psi\- \circ \bs u, \qquad \text{ and }\qquad \bs L_{\bs X^v} \bs u= -\bs X \circ \bs u.
\end{align}

Such a $\phi$-dependent dressing field allows to define the map  
\begin{align}
\label{F-map}
F_{\bs u} : \Phi \ &\rarrow\ \M  \notag \\
\phi \ &\mapsto \  F_{\bs u}(\phi) \defeq \bs u(\phi)^* \phi \sim [\phi],  \qquad \text{ s.t. }\quad  F_{\bs u} \circ R_\psi = F_{\bs u}. 
\end{align}
Clearly, $F_{\bs u}(\phi^\psi) = \bs u(\phi^\psi)^*(\phi^\psi)=\big( \psi\-\! \circ \bs u(\phi) \big)^*\psi^*\phi= \bs u(\phi)^* \phi \rdefeq F_{\bs u}(\phi)$. This means  that the image of $F_{\bs u}$ is a ``coordinatisation" of $\M$ since each $\Diff(M)$-orbit $[\phi] \in \M$ of $\phi \in \Phi$ is represented by a tensor field $\bs u(\phi)^* \phi$ on $N$ (rather than on $M$). 
A very natural understanding of the dressed field $\phi^{\bs u} \defeq \bs u(\phi)^* \phi$, is that it is a \emph{relational variable}. Indeed, being $\Diff(M)$-invariant it represents the physical d.o.f. of a theory, but it also does so in a manifestly \emph{relational} way: the very expression $\bs u(\phi)^* \phi$ is explicitly a field-dependent coordinatisation of the physical d.o.f., i.e. a definition of physical d.o.f. with respect to each other. 
The dressed fields $\phi^{\bs u}$ are thus Dirac relational variables, or ``complete observables" in the terminology of \cite{Rovelli2002, Tamborino2012}. And as we are about to see, the DFM is then a framework allowing to systematically reformulate a theory in a manifestly $\Diff(M)$-invariant and relational way.

Notice that the map \eqref{F-map} is a realisation of the bundle projection, $F_{\bs u} \sim \pi$. 
Therefore, it becomes possible in principle to build basic forms on $\Phi$ since, as seen in section \ref{Differential forms and their derivations}, 
$\Omega^\bullet_{\text{Basic}}(\Phi) =$ Im$\,\pi^\star \simeq$ Im$\,F_{\bs u}^\star$.  
Given a form $\bs\alpha=\alpha\big(\!\w^\bullet\! \bs d\phi; \phi \big) \in \Omega^\bullet(\Phi)$, to build its basic counterpart we first consider its formal ``version" on the base space, $\b{\bs \alpha}=\alpha\big( \!\w^\bullet\! \bs d[\phi]; [\phi] \big) \in  \Omega^\bullet(\M)$, then simply define
\begin{align}
\label{Dressing-form}
\bs \alpha^{\bs u}\defeq F_{\bs u}^\star  \b{\bs \alpha} = \alpha \big(  \!\w^\bullet\! F_{\bs u}^\star \bs d[\phi]; F_{\bs u}(\phi)  \big) \quad \in \ \Omega^\bullet_{\text{Basic}}(\Phi).
\end{align}
This object, basic by definition, is called the \emph{dressing} of $\bs\alpha$. By construction it is invariant under $\bs\Diff_v(\Phi)\simeq C^\infty \big(\Phi, \Diff(M) \big)$ -- thus also gauge-invariant, under $\bs\Aut_v(\Phi)\simeq \bs\Diff(M)$: $ (\bs{\alpha^u})^{\bs \psi} = \bs{\alpha^u}$.
\medskip

To get a more operational expression for $\bs \alpha^{\bs u}$, one needs only to find the explicit results for $F_{\bs u}^\star \bs d[\phi]$ which is a basis for basic forms. 
This is done by first finding the general result of the pushforward ${F_{\bs u}}_\star : T_{\phi}\Phi \rarrow T_{F_{\bs u}(\phi)}\M$, $\mathfrak X_{|\phi} \mapsto {F_{\bs u}}_\star \mathfrak X_{|\phi}$. Indeed, given a generic $\mathfrak X \in \Gamma(T\Phi)$ with flow $\vphi_\tau : \Phi \rarrow \Phi$, s.t. $\mathfrak X_{|\phi}  = \tfrac{d}{d\tau} \vphi_\tau(\phi) \big|_{\tau =0} = \mathfrak X(\phi)\tfrac{\delta}{\delta \phi}$, one has $F_{\bs u}^\star \bs d[\phi]_{|F_{\bs u}(\phi)}\big(\mathfrak X_{|\phi} \big)= \bs d[\phi]_{|F_{\bs u}(\phi)} \big( {F_{\bs u}}_\star \mathfrak X_{|\phi} \big)$. So, 
\begin{align*}
 {F_{\bs u}}_\star \mathfrak X_{|\phi}  \defeq  {F_{\bs u}}_\star \tfrac{d}{d\tau}\ \vphi_\tau(\phi)\ \big|_{\tau =0}  =  \tfrac{d}{d\tau}\ {F_{\bs u}} \big( \vphi_\tau(\phi)   \big)\ \big|_{\tau =0}
 							&=  \tfrac{d}{d\tau} \ \bs u\big( \vphi_\tau(\phi)\big)^* \big( \vphi_\tau(\phi)   \big) \ \big|_{\tau =0}, \\
							&=  \tfrac{d}{d\tau} \bs u\big( \vphi_\tau(\phi)\big)^* \phi \ \big|_{\tau =0} +  \tfrac{d}{d\tau}\ \bs u(\phi)^* \big( \vphi_\tau(\phi)   \big)\  \big|_{\tau =0}
 \end{align*}
The first contribution is, inserting $\id_N =  \bs u(\phi)\- \!\circ \bs u(\phi)$,
\begin{align*}
 \tfrac{d}{d\tau}\ \bs u(\phi)^* {\bs u(\phi)\-}^{_{\text{\scriptsize	$*$} }}   \bs u\big( \vphi_\tau(\phi)\big)^* \phi \ \big|_{\tau =0} = 
 \bs u(\phi)^*  \tfrac{d}{d\tau}\ \big(    \bs u\big( \vphi_\tau(\phi)\big)   \circ   \bs u(\phi)\-   \big)^*\phi \ \big|_{\tau =0}.
\end{align*}
The term $\bs u\big( \vphi_\tau(\phi)\big)   \circ   \bs u(\phi)\-$ is a curve in $M$, so $ \tfrac{d}{d\tau}\   \bs u\big( \vphi_\tau(\phi)\big)   \circ   \bs u(\phi)\-   \ \big|_{\tau =0} = \bs{du}_{|\phi}(\mathfrak X_{|\phi}) \circ \bs u(\phi)\-  \ \in \Gamma(TM)$. Therefore, 
\begin{align}
\label{1st-term}
\tfrac{d}{d\tau} \bs u\big( \vphi_\tau(\phi)\big)^* \phi \ \big|_{\tau =0} = 
 \bs u(\phi)^*  \tfrac{d}{d\tau}\ \big(    \bs u\big( \vphi_\tau(\phi)\big)   \circ   \bs u(\phi)\-   \big)^*\phi \ \big|_{\tau =0} =   \bs u(\phi)^*  \mathfrak L_{\bs{du}_{|\phi}(\mathfrak X_{|\phi}) \circ \bs u(\phi)\-} \phi. 
\end{align}
The second contribution is 
$\tfrac{d}{d\tau}\ \bs u(\phi)^* \big( \vphi_\tau(\phi)   \big)\  \big|_{\tau =0} \rdefeq  \tfrac{d}{d\tau}\ F_{\bs u(\phi)} \big( \vphi_\tau(\phi)   \big)\  \big|_{\tau =0}
														 = F_{\bs u(\phi)\, \star} \mathfrak X_{|\phi}$, 
where $\bs u(\phi)$ is considered $\phi$-constant/ independent. Now, this is indeed a vector (field) on $\M$, so we can find its expression as a derivation by applying it to  $\bs g \in C^\infty(\M)$:
 \begin{align*}
\big[F_{\bs u(\phi)\, \star} \mathfrak X\,  \big( \bs g \big)\big]([\phi]) &=  \tfrac{d}{d\tau}\ \bs g \left( F_{\bs u(\phi)} \big( \vphi_\tau(\phi) \big) \right)\  \big|_{\tau =0}, \\
			  &=  \left( \tfrac{\delta}{\delta [\phi]} \bs g \right) \big(F_{\bs u(\phi)}(\phi)\big)    \ \underbrace{\tfrac{d}{d\tau}\, F_{\bs u(\phi)}\big( \vphi_\tau(\phi) \big) \, \big|_{\tau=0}}_{\big[\mathfrak X(F_{\bs u(\phi)})\big] (\phi)  }
			  =  \left( \tfrac{\delta}{\delta [\phi]} \bs g \right) \big(\underbrace{\bs u(\phi)^*\phi}_{\sim [\phi]} \big)    \  \underbrace{\big( \tfrac{\delta}{\delta \phi} F_{\bs u(\phi)}\big)(\phi)}_{\tfrac{\delta}{\delta \phi} \, \bs u(\phi)^* \phi\, =\, \bs u(\phi)^*}  \ \mathfrak X(\phi), \\
			  &= \left( \tfrac{\delta}{\delta [\phi]} \bs g \right) \big([\phi]\big) \ \bs u(\phi)^* \mathfrak X(\phi)
			   = \big[ \bs u(\phi)^* \mathfrak X(\phi)  \tfrac{\delta}{\delta [\phi]} (\bs g) \big]([\phi]).
\end{align*}
Then, gathering the two contributions, we have
\vspace{-2mm}
\begin{align}
\label{Pushforward-X-dress}
 {F_{\bs u}}_\star \mathfrak X_{|\phi} = \bs u(\phi)^* \left( \mathfrak X(\phi) +  \mathfrak L_{\bs{du}_{|\phi}(\mathfrak X_{|\phi}) \circ \bs u(\phi)\-} \phi  \right) \,  \frac{\delta}{\delta [\phi]}_{|F_{\bs u(\phi)(\phi)}}.\\[-8mm]
 ~ \notag
\end{align}
From this we deduce, for any $\mathfrak X \in \Gamma(T\Phi)$:
\begin{align*}
F_{\bs u}^\star \bs d[\phi]_{|F_{\bs u(\phi)(\phi)}}\, \big(\mathfrak X_{|\phi} \big) &= \bs d[\phi]_{|F_{\bs u(\phi)(\phi)}} \, \big( {F_{\bs u}}_\star \mathfrak X_{|\phi} \big), \\
						&= \bs d[\phi]_{|F_{\bs u(\phi)(\phi)}}\,  \left(  \bs u(\phi)^* \left( \mathfrak X(\phi) +  \mathfrak L_{\bs{du}_{|\phi}(\mathfrak X_{|\phi}) \circ \bs u(\phi)\-} \phi  \right) \,  \tfrac{\delta}{\delta [\phi]}_{|F_{\bs u(\phi)(\phi)}} \right), \\
						&=  \bs u(\phi)^* \left( \mathfrak X(\phi) +  \mathfrak L_{\bs{du}_{|\phi}(\mathfrak X_{|\phi}) \circ \bs u(\phi)\-} \phi  \right), \\
						&=  \bs u(\phi)^* \left( \bs d\phi_{|\phi}\, \big(\mathfrak X_{|\phi} \big) +  \mathfrak L_{\bs{du}_{|\phi}(\mathfrak X_{|\phi}) \circ \bs u(\phi)\-} \phi  \right), \\
						&=\left(   \bs u(\phi)^* \big( \bs d\phi_{|\phi}  +  \mathfrak L_{\bs{du}_{|\phi} \circ \bs u(\phi)\-} \phi  \big)   \right) (\mathfrak X_{|\phi}).
\end{align*}
We thus obtain the dressing of $\bs d\phi$, the basic basis 1-form:
\begin{align}
\label{Dressing-basis-1-form}
\bs d\phi^{\bs u}\defeq F_{\bs u}^\star \bs d [\phi] =  \bs u^* \big( \bs d\phi  +  \mathfrak L_{\bs{du} \circ \bs u\-} \phi \big) \quad \in \ \Omega^1_{\text{basic}}(\Phi), 
\end{align}
where $\bs u =\bs u(\phi)$. This can be compared e.g. to eq.(3.30) in \cite{DonnellyFreidel2016},  eq.(2.7) in \cite{Speranza2022}, or eq.(2.8) in \cite{Geiller2018}.
 The result \eqref{Dressing-basis-1-form} can then be inserted into \eqref{Dressing-form} to get the dressing of any form $\bs\alpha$:
\begin{align}
\label{Dressing-form-bis}
\bs \alpha^{\bs u} =  \alpha \big(  \!\w^\bullet\! \bs d\phi^{\bs u};\, \phi^{\bs u}  \big) \ \  \in \ \Omega^\bullet_{\text{Basic}}(\Phi),
\end{align}
which is basic by construction, thus invariant under $\bs\Diff_v(\Phi)\simeq C^\infty\big(\Phi, \Diff(M) \big)$ -- and under  $\bs\Aut_v(\Phi)\simeq \bs\Diff(M)$.

Given the formal similarity between $\Xi(\phi)=\bs\psi(\phi)^*\phi$ and $F_{\bs u}(\phi)=\bs u(\phi)^*\phi$ on the one hand, and between $\bs d\phi^{\bs \psi}$  \eqref{GT-basis-1-form} and $\bs d\phi^{\bs u}$ \eqref{Dressing-basis-1-form} on the other hand, which results into the formal similarity between $\bs\alpha^{\bs \psi}$  \eqref{GT-general} and $\bs\alpha^{\bs u}$ \eqref{Dressing-form}-\eqref{Dressing-form-bis}, we can spell out the following rule of thumb to obtain the dressing of any form $\bs\alpha$: 
First compute its  $\bs\Diff_v(\Phi)\simeq C^\infty\big(\Phi, \Diff(M) \big)$ transformation $\bs\alpha^{\bs \psi}$, then substitute $\bs\psi \rarrow \bs u$ in the resulting expression to obtain $\bs\alpha^{\bs u}$. 
We will use this rule of thumb from now on. 
\medskip

Let us illustrate it by discussing the case of the dressing of twisted tensorial forms. For this to be well defined, one needs only to assume that the 1-cocycle \eqref{cocycle}, $C: \Phi \times \Diff(M) \rarrow G$, controlling the equivariance \eqref{C-eq-form} of these forms has a functional expression s.t. is can be meaningfully extended to $C: \Phi \times \Diff(N, M) \rarrow G$. If so, the map 
\begin{align}
\label{twisted-dressing}
C(\bs u): \Phi &\rarrow G,  \notag\\
		\phi &\mapsto [C(\bs u)](\phi)\defeq C\big(\phi; \bs u(\phi) \big),
\end{align}
is well-defined and  is a twisted equivariant 0-form:
\begin{align}
\label{eq-twisted-dressing}
\big[ R^\star_\psi C(\bs u) \big](\phi) &= \big[ C(\bs u) \circ R_\psi \big] (\phi) \defeq C\big(\phi^\psi; \bs u(\phi^\psi) \big) = C\big(\phi^\psi;  \psi\- \circ \bs u(\phi) \big)
						       = C\big(\phi^\psi;  \psi\-  \big) \cdot C\big(\phi;  \bs u(\phi) \big), \notag \\
						     &= C\big(\phi; \psi \big)\-\! \cdot C\big(\phi; \bs u (\phi)\big), \notag \\
						     & \rdefeq \big[C\big( \ \,;  \psi\big)\-\!\cdot C(\bs u) \big] (\phi),
\end{align}
where the cocycle property \eqref{cocycle} is used. 
The linearisation is  $\bs L_{X^v} C(\bs u) = \iota_{X^v} \bs d C(\bs u) = -a(X; \ \,) \cdot C(\bs u)$. 
In other words, it is a dressing field for twisted forms. It follows that, as a special case of \eqref{GT-tensorial}, its $\bs\Diff_v(\Phi)\simeq C^\infty\big(\Phi, \Diff(M) \big)$ transformation is,
\begin{align}
\label{Vert-trsf-twisted-dressing}
C(\bs u)^{\bs \psi} = C( \bs\psi )\-\! \cdot C(\bs u). 
\end{align}

 Applying e.g. the rule of thumb to $\bs\alpha \in \Omega^\bullet_\text{tens}(\Phi, C)$, by \eqref{GT-tensorial} we know that its $\bs\Diff_v(\Phi)$-transformation  is $\bs\alpha^{\bs\psi} = C(\bs\psi)\- \bs\alpha$, from which we read its dressing to be 
\begin{align}
\bs\alpha^{\bs u} = C(\bs u)\- \bs\alpha\  \in \Omega^\bullet_\text{basic}(\Phi). 
\end{align}
It is by construction invariant under vertical transformations $\bs\Diff_v(\Phi)$, thus also under gauge transformations $\bs\Aut_v(\Phi)$: which in this case is easily checked using \eqref{Vert-trsf-twisted-dressing}, 
\begin{align}
(\bs\alpha^{\bs u})^{\bs \psi} = \big( C(\bs u)^{\bs \psi} \big)\- \bs\alpha^{\bs \psi}  =  \big( C(\bs \psi)\-\! \cdot C(\bs u) \big)\-\! \cdot C(\bs \psi)\- \bs\alpha =   C(\bs u) \- \bs\alpha = \bs\alpha^{\bs u}. 
\end{align}

Of course this specialises to the case of $\bs\alpha \in \Omega^\bullet_\text{tens}(\Phi, \rho)$, so $\bs\alpha^{\bs u} =\rho(\bs u)\-\bs\alpha$. Taking the case of an Ehresmann connection, in view of  \eqref{Vert-trsf-connection} our rule of thumb ensures that $\bs\omega^{\bs u} = \bs u\-_* \bs \omega \circ \bs u + \bs u\-_* \bs{du} \in \Omega^1_\text{basic}(\Phi)$. This allows us to write an analogue of the lemma \eqref{Formula-general-rep}-\eqref{Formula-pullback-rep}, also occasionally useful: 
For $\bs\alpha, \bs{D\alpha} \in \Omega^\bullet_\text{tens}(\Phi, \rho)$, 
we have on the one hand $\bs d\, F_{\bs u}^\star \b{\bs\alpha} = \bs d\big( \rho(\bs u)\- \bs\alpha \big)$. On the other hand, since $F_{\bs u}^\star \b{\bs D} \b{\bs\alpha} = \bs d \bs\alpha^{\bs u}+ \rho_*(\bs\omega^{\bs u})\bs\alpha^{\bs u} = \rho(\bs u)\- \bs{D\alpha}$ (by our rule of thumb again), we have that,
\begin{align*}
F_{\bs u}^\star \bs d \b{\bs\alpha} &= \rho(\bs u)\- \bs{D\alpha}  -  \rho_*(\bs\omega^{\bs u})\bs\alpha^{\bs u}
				   =  \rho(\bs u)\- \bs{d\alpha} +  \rho(\bs u)\-\,  \rho_*(\bs\omega) \bs\alpha  -  \rho_*(\bs\omega^{\bs u})\bs\alpha^{\bs u}
				   = \rho(\bs u)\- \bs{d\alpha}  -  \rho_*(\bs u\-_* \bs d\bs u) \rho(\bs u)\-\bs\alpha,\\
				 &= \rho(\bs u)\- \left(  \bs{d\alpha} -  \rho_*( \bs d\bs u \circ \bs u\-) \bs\alpha \right). 
\end{align*}
By naturality of the pullback $[F_{\bs u}^\star , \bs d]=0$ (understood that the exterior derivatives belong to different spaces here) we  obtain:
\begin{align}
\label{Formula-general-rep-dress}
\bs d\big( \rho(\bs u)\- \bs\alpha \big) = \rho(\bs u)\- \left(  \bs{d\alpha} -  \rho_*( \bs d\bs u \circ \bs u\-)\, \bs\alpha \right)
\end{align}
In particular, for the pullback representation $\rho(u)\-\!= u^*$, this is:
  \begin{align}
\label{Formula-pullback-rep-dress}
\bs d\big( \bs u^* \bs\alpha \big) = \bs u^* \left(  \bs{d\alpha} + \mathfrak L_{\bs d\bs u \circ \bs u\-} \bs\alpha \right). 
\end{align}
The latter identity appears in the covariant phase space and edge mode literature, e.g.  \cite{DonnellyFreidel2016, Speranza2018, Kabel-Wieland2022}. 
It must not be conflated  with with \eqref{Dressing-basis-1-form} as the two results have distinct geometric origins and meaning.

\subsubsection{Dressing field and flat  connections}  
\label{Dressing field and flat  connections}  

Given a field-dependent dressing field $\bs u$, the object $\bs\omega_{\text{\tiny$0$}} \defeq -\bs{du} \circ\bs u\- \in \Omega^1(\Phi)$ is a \emph{flat Ehresmann connection}. Indeed, on the one hand, by \eqref{Vert-trsf-dressing} we have, 
\begin{align}
\label{Dress-connection-vert-prop}
{\omegaf}_{|\phi}(X^v_{|\phi})=  -\bs{du}_{|\phi}(X^v_{|\phi}) \circ \bs u(\phi)\- = -\bs L_{X^v} \bs u \circ  \bs u(\phi)\- = X \circ \bs u(\phi) \circ  \bs u(\phi)\- = X \ \in\diff(M). 
\end{align}
On the other hand, by definition \eqref{Field-dep-dressing} and using the naturality of $\bs d$ (commuting with pullbacks), 
\begin{align}
\label{Dress-connection-equiv-prop}
R^\star_\psi \omegaf &=  -R^\star_\psi \bs{du}  \circ (R^\star_\psi \bs u)\- =  - \bs d (R^\star_\psi \bs u) \circ  (R^\star_\psi \bs u)\- = - \bs d (\psi\- \circ \bs u) \circ (\psi\- \circ \bs u)\-
				= \psi\-_* \bs{du} \circ \bs u\- \circ \psi ,  \notag\\
				&=  \psi\-_*\,\omegaf\!\circ \psi. 
\end{align}
These are indeed the defining properties \eqref{Variational-connection} of an Ehresmann connection. It is immediate to see that 
\begin{align}
\bs d\omegaf + \tfrac{1}{2} [\omegaf, \omegaf]_{\text{{\tiny $\diff(M)$}}} \equiv0,
\end{align}
i.e. that a dressing field supplies a \emph{flat} connection on $\Phi$.
An immediate corollary is that, 
\begin{align}
\label{Vert-trsf-flat-connection}
\omegaf^{\bs \psi}= \bs\psi\-_*\, \omegaf \circ \bs\psi + \bs\psi\-_*\bs{d\psi}, \qquad \text{ and } \qquad
\bs L_{\bs X^v} \omegaf =\bs{dX}+ [\omegaf, \bs X]_{\text{{\tiny $\diff(M)$}}\,},
\end{align}
as a special case of \eqref{Vert-trsf-connection} and  \eqref{Inf-Vert-trsf-connection}. This may be compared to eq.(3.19) in \cite{DonnellyFreidel2016}. 

Thus, \eqref{Dressing-basis-1-form} can also be written as
\begin{align}
\label{Dressing-basis-1-form-bis}
\bs d\phi^{\bs u}=  \bs u^* \big( \bs d\phi  -  \mathfrak L_{  \omegaf} \phi \big) \quad \in \ \Omega^1_{\text{basic}}(\Phi).
\end{align}
Being basic by construction, we know it is $\bs\Diff_v(\Phi)$-invariant (and also gauge-invariant, under $\bs\Aut_v(\Phi)$), but let us check it as an exercise: Using   \eqref{GT-basis-1-form}, \eqref{Vert-trsf-dressing} and \eqref{Vert-trsf-flat-connection}, we have
\begin{align*}
(\bs d \phi^{\bs u})^{\bs \psi} &=   (\bs u^{\bs \psi})^* \big( \bs d\phi^{\bs \psi}  -  \mathfrak L_{  \omegaf^{\bs \psi}}\, \phi^{\bs \psi} \big), \\
					   &= (\bs \psi\- \circ \bs u)^* \left(  \bs\psi^* \big( \bs d\phi   +   \mathfrak L_{ [\bs d \bs\psi  \circ \bs\psi\-]} \phi \big)  -  \mathfrak L_{  \bs\psi\-_*\, \omegaf \circ \bs\psi + \bs\psi\-_*\bs{d\psi} }\, \bs \psi^*\phi \right), \\
					   &=  \bs u^* {\bs\psi\-}^{_{\text{\scriptsize	$*$} }}\,  \bs\psi^*   ( \bs d\phi   +   \mathfrak L_{ [\bs d \bs\psi  \circ \bs\psi\-]} \phi ) 
					       \ -\  \bs u^* {\bs\psi\-}^{_{\text{\scriptsize	$*$} }} \mathfrak L_{  \bs\psi\-_*\, \omegaf \circ \bs\psi }\,  \bs\psi^* \phi
					       \ - \ \bs u^* {\bs\psi\-}^{_{\text{\scriptsize	$*$} }} \mathfrak L_{  \bs\psi\-_*\bs{d\psi}  }\, \bs\psi^* \phi, \\
					    &=    \bs u^* \bs d\phi  +  \bs u^* \mathfrak L_{ [\bs d \bs\psi  \circ \bs\psi\-]}\,\phi 
					       \ - \   \bs u^*\mathfrak L_{\omegaf} \,\phi     \ - \   \bs u^*\mathfrak L_{ [\bs d \bs\psi  \circ \bs\psi\-]} \,\phi, \\
					    &=  \bs u^* \big( \bs d\phi  -  \mathfrak L_{  \omegaf} \phi \big) \rdefeq \bs d\phi^{\bs u}. 
\end{align*}
Where the lemma \eqref{lemma1} is used in the 3rd to 4th line. 
Now, given \eqref{Dressing-basis-1-form-bis}, for $\bs \alpha=\alpha(\bs d \phi; \phi)\, \in \Omega^1_\text{inv}(\Phi)$,  \eqref{Dressing-form-bis} may be written,
\begin{equation}
\label{dress-inv-1form}
\begin{aligned}
\bs\alpha^{\bs u} =\alpha\big( \bs u^*( \bs d\phi  -  \mathfrak L_{  \omegaf}) ; \phi^{\bs u} \big) = \alpha\big(  \bs d\phi  -  \mathfrak L_{  \omegaf}; \phi \big) 
			  &=\bs\alpha - \alpha\big(   \mathfrak L_{  \omegaf}; \phi \big),  \\
			  &=\bs\alpha -\iota_{[\omegaf(\ )]^v} \bs\alpha \quad \in \Omega^1_\text{basic}(\Phi).
\end{aligned}
\end{equation}

The formulae \eqref{Dressing-basis-1-form-bis} and \eqref{dress-inv-1form} may be compared to \eqref{Horiz-basis-1-form} and \eqref{Horiz-1-form}/\eqref{Horiz-forms-explicit}: It shows that using the DFM or Ehresmann connections to build a basic counterparts of  invariant 1-forms,  $\bs \alpha \in \Omega^1_\text{inv}(\Phi)$, gives formally very analogous results.\footnote{An observation first hinted at in \cite{Gomes-et-al2018} in the context of the study of the covariant phase space approach to field theory (over bounded regions) with internal gauge symmetries: there $\bs\alpha=\bs\theta \in \Omega^1_\text{inv}(\Phi)$ is the presymplectic potential of the theory. See also \cite{Francois-et-al2021} for a detailed discussion of this point for this class of theories.} More on this in section \ref{Residual symmetries of the second kind} below. 

This also tells us something noteworthy: The existence of a flat connection, hence of a global dressing field, is a strong topological constraint on a bundle.\footnote{In the finite dimensional case of a bundle $P$ over a simply connected manifold, it implies that $P$ is trivial.} In some cases, a dressing field  indeed gives  a global trivialisation of the bundle $\Phi$, yet in general the field space may not be trivial and Gribov-like obstructions may exclude the existence  of  dressing fields other than local. A generic, non-flat, connection $\bs\omega$ does not entail such constraints. 
\medskip

In the same way, a dressing field may supply a \emph{flat twisted connection} $\bs\varpi_{\text{\tiny$0$}} \defeq -\bs d C(\bs{u}) \cdot C(\bs u)\-$, satisfying the defining properties \eqref{Variational-twisted-connection} -- as can be checked using \eqref{eq-twisted-dressing} and $\bs L_{X^v} C(\bs u) = -a(X; \ \,) \cdot C(\bs u)$. 
 From which follows that, as a special case of \eqref{Vert-trsf-twisted-connection}-\eqref{Inf-Vert-trsf-twisted-connection}:
\begin{align}
\label{Vert-trsf-flat-connection}
\bs\varpi_{\text{\tiny$0$}}^{\bs\psi } 
=\Ad\big( C(\bs\psi)\- \big) \, \bs\varpi_{\text{\tiny$0$}} + C(\bs\psi)\-\bs{d} C(\bs\psi), 
 \qquad \text{ and } \qquad
\bs L_{\bs X^v}\bs\varpi_{\text{\tiny$0$}} =\bs{d} a(\bs X)+ [\bs\varpi_{\text{\tiny$0$}}, a(\bs X)]_{\text{{\tiny $\LieG$}}}. 
\end{align}

Reprising the example of $Z=\exp{iS} \in \Omega^0_{eq}(\Omega, C)$, the flat twisted connection  is $\bs\varpi_{\text{\tiny$0$}} \defeq -\bs d C(\bs{u}) \cdot C(\bs u)\-= i \int \bs d c(\ \,; \bs u)$. Then we may notice that the associated twisted covariant derivative is $\b{\bs D}_{\text{\tiny$0$}}Z \defeq \bs dZ + \bs\varpi_{\text{\tiny$0$}} Z = (i \bs d S + \bs\varpi_{\text{\tiny$0$}}  ) Z$, which implies, as special case of \eqref{Generalised-WZ-trick},  
\begin{align}
\label{Generalised-WZ-trick-flat}
i \bs d S + \bs\varpi_{\text{\tiny$0$}} = i\bs d \int L + c(\ \,;\bs u)\quad \in\ \Omega^1_\text{basic} (\Phi). 
\end{align}
The object $L'\defeq L + c(\ \,;\bs u)$ is what one may call a Wess-Zumino improved Lagrangian,  $c(\ \,;\bs u)$ playing the role of the \emph{Wess-Zumino-Witten counterterm} restoring gauge-invariance -- see \cite{Bertlmann} chap.4, \cite{GockSchuck} chap.12, or \cite{Bonora2023} chap.15.
Given the definition of the 1-cocycle $c(\ \,;\psi) \defeq \psi^*L - L$, it is clear that   $L'=  \bs u^* L \defeq L^{\bs u}    \in \Omega^0_\text{basic} (\Phi)$, i.e. it is but the dressed Lagrangian one would obtain by applying directly the DFM (and its rule of thumb) to a Lagrangian $L  \in \Omega^0_\text{eq} (\Phi)$ with equivariance $R^\star_\psi L=\psi^*L$ and therefore gauge transformation $L^{\bs \psi} =\bs\psi^* L$. Such  dressed Lagrangians are key objects of our subsequent discussions. 

\subsection{Residual symmetries}  
\label{Residual symmetries}  

In the context of the DFM, two kinds of \emph{residual} symmetries are to be discussed: The (genuine) residual symmetry coming from the elimination of only part of the original symmetry, and a ``new" symmetry that may arise (in replacement of the eliminated one) due to possible ambiguities in the choice or contruction of a dressing field. Of course, both might arise simultaneously. Let us discuss them in turn. 

\subsubsection{Residual symmetries of the first kind}  
\label{Residual symmetries of the first kind}  

Consider  $\Diff_{\text{\tiny$0$}}(M) \subset \Diff(M)$ a subgroup of diffeomorphisms satisfying some condition (boundary conditions, compact support...). 
And suppose the dressing field $ \ \bs u : \Phi \ \rarrow\ \D r[N, M]\ $ has defining equivariance given by
\begin{align}
\label{equiv-partial-dressing}
 R^\star_\vphi \bs u = \vphi\- \circ \bs u, \quad \text{for }\  \vphi \in \Diff_{\text{\tiny$0$}}(M),
\end{align}   
 i.e. $\bs u(\vphi^*\phi)= \vphi\- \circ \bs u (\phi)$.
Application of the DFM then amounts to building a bundle projection $\Phi \rarrow \Phi/ \Diff_{\text{\tiny$0$}}(M)$. For this to be a bundle map, i.e. for $\Phi/ \Diff_{\text{\tiny$0$}}(M)\rdefeq \Phi^{\bs u}$ to be principal (sub)bundle in its own right, the quotient  $\Diff(M)/ \Diff_{\text{\tiny$0$}}(M)$ must be a group, thus $\Diff_{\text{\tiny$0$}}(M)$ needs to be a \emph{normal} subgroup of $\Diff(M)$: $\Diff_{\text{\tiny$0$}}(M) \triangleleft \Diff(M)$. 
Let us assume it is so,\footnote{Such is e.g. the group $\Diff_{\text{\tiny$0$}}(M)= \Diff_{\text{\tiny c}}(M)$ of compactly supported diffeomorphisms.} and denote $\Diff_{\text{\scriptsize r}}(M) \defeq \Diff(M)/ \Diff_{\text{\tiny$0$}}(M)$ the (residual) structure group of $\Phi^{\bs u}$. 

Any dressed object $\bs\alpha^{\bs u}$ will then be $\Diff_{\text{\tiny$0$}}(M)$-basic on $\Phi$, i.e. invariant by construction under the action of $C^\infty\big(\Phi, \Diff_{\text{\tiny 0 }}(M)\big)\subset C^\infty\big(\Phi, \Diff(M)\big) \simeq \bs\Diff_v(\Phi)$ -- and $\bs\Diff_{\text{\tiny$0$}}(M)\subset \bs\Diff(M)$ in particular. 
But it is expected to exhibit \emph{residual transformations} under $C^\infty\big(\Phi, \Diff_{\text{\scriptsize r}}(M)\big)\subset C^\infty\big(\Phi, \Diff(M)\big) \simeq \bs\Diff_v(\Phi)$ -- and $\bs\Diff_{\text{\scriptsize r}}(M)\subset \bs\Diff(M)$ in particular. 
The transformation of $\bs\alpha$ under $C^\infty\big(\Phi, \Diff_{\text{\scriptsize r}}(M)\big)$ being known by assumption, that of $\bs\alpha^{\bs u}$ boils down to determining the residual $C^\infty\big(\Phi, \Diff_{\text{\scriptsize r}}(M)\big)$-transformation of the dressing field $\bs u$, itself controlled by its equivariance: $R^\star_\psi \bs u = ?$  for $\psi \in \Diff_{\text{\scriptsize r}}(M)$.  

In that regard, there are two noteworthy cases that we can treat in a systematic way. Let us e.g. consider $\bs\alpha \in \Omega^\bullet_{\text{tens}}\big(\Phi, \rho \big)$ and $\bs \omega \in \C$, whose dressing by $\bs u$ as in \eqref{equiv-partial-dressing} above are $\bs\alpha^{\bs u}$ and $\bs\omega^{\bs u}$, 
thus $C^\infty\big(\Phi, \Diff_{\text{\tiny 0 }}(M)\big)$-invariant.

\begin{prop}
\label{prop1}
If $ \ \bs u : \Phi \ \rarrow\ \D r[M, M]\ $ is s.t. 
\begin{align}
\label{Resid-eq1}
R^\star_\psi \bs u = \psi\-\! \circ \bs u \circ \psi, \quad \text{ for } \ \psi \in \Diff_{\text{\scriptsize r}}(M),
\end{align}
then $\bs\alpha^{\bs u} \in \Omega_{\text{tens}}\big(\Phi, \rho \big)$ and $\bs \omega^{\bs u} \in \C$. 
Therefore, their residual $C^\infty\big(\Phi; \Diff_{\text{\scriptsize r}}(M)\big)$-transformations are:
\begin{align}
\label{Residual-GT1}
(\bs\alpha^{\bs u})^{\bs\psi} =\rho(\bs\psi\-) \bs\alpha^{\bs u}, \qquad \text{ and } \qquad (\bs\omega^{\bs u})^{\bs\psi} =\bs\psi\-_* \, \bs\omega^{\bs u}  \circ \bs\psi + \bs\psi\-_*\bs{d\psi}. 
\end{align}
\end{prop}
Indeed, firstly: It is easy to see that if $\bs\alpha$ is horizontal, so is $\bs\alpha^{\bs u}=\rho(\bs u)\-\bs\alpha$, and $R^\star_\psi\bs\alpha^{\bs u} = \rho\big(R^\star_\psi \bs u \big)\- R^\star_\psi \bs\alpha = \rho\big(\psi\-\! \circ \bs u \circ \psi \big) \- \rho(\psi)\-\bs\alpha = \rho(\psi\-) \bs\alpha^{\bs u}$. 
Secondly, on the one hand, given the linear version of \eqref{Resid-eq1} is 
\begin{align}
\label{Resid-eq1-linear}
\bs L_{X^v} \bs u = \iota_{X^v}\bs{du}= \tfrac{d}{d\tau}\, R^\star_{\psi_\tau} \bs u\,\big|_{\tau=0} = -X \circ \bs u + \bs u_* X, \quad \text{ for } \ X\defeq \tfrac{d}{d\tau} \psi_\tau\,\big|_{\tau=0} \in \diff_{\text{\scriptsize r}}(M),
\end{align}
 one has $\bs\omega^{\bs u}(X^v)= \bs u\-_* \bs\omega(X^v) \circ \bs u + \bs u\-_*\bs d \bs u (X^v)= \bs u\-_* X \circ \bs u +  \bs u\-_*\big(-X \circ \bs u + \bs u_* X \big) =X$.
Then, on the other hand, given \eqref{Resid-eq1} and  \eqref{Variational-connection}, it is easy to show that: 
$R^\star_\psi \bs\omega^{\bs u} = \psi\-_*\, \bs\omega^{\bs u} \circ \psi$, for $\psi \in \Diff_{\text{\scriptsize r}}(M)$. 
So indeed, $\bs\omega^{\bs u}$ satisfies the defining properties \eqref{Variational-connection} of a $\Diff_{\text{\scriptsize r}}(M)$-principal connection. 
The transformations \eqref{Residual-GT1} thus follow as a special case of \eqref{GT-tensorial} and \eqref{Vert-trsf-connection}. 
From both follows in particular that $\bs\Omega^{\bs u} \in \Omega^2_{\text{tens}}\big(\Phi, \diff_{\text{\scriptsize r}}(M) \big)$, so $(\bs\Omega^{\bs u})^{\bs\psi} =\bs\psi\-_*\, \bs\Omega^{\bs u} \circ \bs\psi$. 

The other interesting case is:
\begin{prop}
\label{prop2}
If  $\, \bs u : \Phi \ \rarrow\ \D r[N, M]\, $ is s.t. 
\begin{align}
\label{Resid-eq2}
R^\star_\psi \bs u = \psi\-\! \circ \bs u \circ C(\ \, ;\psi), \quad \text{ for } \ \psi \in \Diff_{\text{\scriptsize r}}(M), 
\end{align}
and where $C:\Phi \times \Diff_{\text{\scriptsize r}}(M) \rarrow \Diff(N)$ is a special case of  1-cocycle as defined in \eqref{cocycle}, thus satisfying the relation 
$C(\phi; \psi'\circ \psi) =C(\phi; \psi') \circ C(\phi^{\psi'}; \psi)$.

Then $\bs\alpha^{\bs u} \in \Omega_{\text{tens}}\big(\Phi, C \big)$ and $\bs \omega^{\bs u} \in \b\C$. 
Therefore, their residual $C^\infty\big(\Phi, \Diff_{\text{\scriptsize r}}(M)\big)$-transformations are:
\begin{align}
\label{Residual-GT2}
(\bs\alpha^{\bs u})^{\bs\psi} =\rho\big( C(\bs\psi)\big)\- \bs\alpha^{\bs u}, \qquad \text{ and } \qquad (\bs\omega^{\bs u})^{\bs\psi} =C(\bs\psi)\-_* \bs\omega^{\bs u}  \circ C(\bs\psi) + C(\bs\psi)\-_*\bs d C(\bs\psi). 
\end{align}
\end{prop}
As above we have, firstly, that if $\bs\alpha$ is horizontal, so is $\bs\alpha^{\bs u}=\rho(\bs u)\-\bs\alpha$, and $R^\star_\psi\bs\alpha^{\bs u} = \rho\big(R^\star_\psi \bs u \big)\- R^\star_\psi \bs\alpha = \rho\big(\psi\-\! \circ \bs u \circ C(\ \,;\psi) \big) \- \rho(\psi)\-\bs\alpha = \rho\big(C(\ \,;\psi) \big)\- \bs\alpha^{\bs u}$. That is, modulo $\rho$, $\bs\alpha^{\bs u}$ is a twisted tensorial form. Remark in particular that if $\rho$ is the pullback action, $\rho(\psi)\-=\psi^*$, then 
\begin{align}
\label{Special-twisted-tens}
R^\star_\psi \bs\alpha^{\bs u}  =C(\ \,;\psi)^* \bs\alpha^{\bs u}, \quad \text{so} \quad (\bs\alpha^{\bs u})^{\bs\psi} = C(\bs\psi)\big)^* \bs\alpha^{\bs u}.
\end{align}
Secondly, on the one hand, given the linear version of \eqref{Resid-eq2} is 
\begin{align}
\label{Resid-eq2-linear}
\bs L_{X^v} \bs u= \iota_{X^v}\bs{du}= \tfrac{d}{d\tau}\, R^\star_{\psi_\tau} \bs u\,\big|_{\tau=0} = -X \circ \bs u + \bs u_*\, \tfrac{d}{d\tau} C(\ \,;\psi_\tau)\,\big|_{\tau=0}= -X \circ \bs u + \bs u_* \, a(X;\ \,), \quad \text{ for } \ X  \in \diff_{\text{\scriptsize r}}(M),
\end{align}
 one has 
 $\bs\omega^{\bs u}(X^v)= \bs u\-_* \bs\omega(X^v) \circ \bs u + \bs u\-_*\bs d \bs u (X^v)= \bs u\-_* X \circ \bs u +  \bs u\-_*\big(-X \circ \bs u + \bs u_* \, a(X;\ \,) \big) =a(X;\ \, )\ \in \diff(N)$.
Then, on the other hand, given \eqref{Resid-eq2} and  \eqref{Variational-connection}, it is easy to show that: 
$R^\star_\psi \bs\omega^{\bs u} = C(\ \, ;\psi)\-_*\, \bs\omega^{\bs u} \circ C(\ \, ;\psi) + C(\ \, ;\psi)\-_*\bs d C(\ \, ;\psi)$, for $\psi \in \Diff_{\text{\scriptsize r}}(M)$. 
So indeed, $\bs\omega^{\bs u}$ satisfies the defining properties \eqref{Variational-twisted-connection}-\eqref{Special-eq-twisted} of a $\Diff_{\text{\scriptsize r}}(M)$-twisted connection. 
The~transformations \eqref{Residual-GT2}-\eqref{Special-twisted-tens} thus follow as a special case of \eqref{GT-tensorial} and \eqref{Vert-trsf-twisted-connection}. 
From both follows in particular that $\bs\Omega^{\bs u} \in \Omega^2_{\text{tens}}\big(\Phi, C \big)$, so $(\bs\Omega^{\bs u})^{\bs\psi} =C(\bs\psi)\-_*\, \bs\Omega^{\bs u} \circ C(\bs\psi)$. 
\medskip

We observe that Proposition \ref{prop2} reproduces (up to left/right-pushforward/pullback convention) the construction of ``relational observables"  in section 3.4.4 of \cite{Hoehn-et-al2022} -- which we already noticed to be twisted tensorial 0-forms in section \ref{Twisted connections} : compare indeed
eq.(3.182) and eq. (3.185) to \eqref{Resid-eq2}, then 
eq.(3.183) and eq.(3.190)  to \eqref{Special-twisted-tens}. In their language, the field-dependent dressing field $\bs u(\phi)$ is referred to as a ``frame field" (noted $R[\phi]$).

We may also signal that exact analogues of Proposition \ref{prop1} and \ref{prop2} are used in the (finite dimensional) context of Cartan conformal geometry to build the twistor/tractor connection and twistor and tractor fields -- in an application of the DFM for field theories with internal gauge groups \cite{Attard-Francois2016_I, Attard-Francois2016_II}, see also \cite{Attard_et_al2017}.\footnote{
Considering the conformal Cartan geometry $(P,  A)$, with $A$  one among all conformal Cartan connections $\A$,  one can build a $\A$-dependent dressing field  $\bs u : \A \rarrow \D r(K, K)$, $ A \mapsto \bs u( A)$, for $K$ the (abelian) subgroup of special conformal transformations. The residual symmetry is $\SO(1,3) \times \W$, i.e. the Lorentz group and Weyl rescalings. The dressing satisfies Proposition \ref{prop1} w.r.t. Lorentz, and Proposition \ref{prop2} w.r.t. Weyl rescalings. Using it to dress the conformal Cartan connection $ A$ itself one gets $ A^{\bs u}$, which is then a \emph{mixed} connection:  Ehresmann for $\SO(1,3)$ and twisted for $\W$. 
This mixed Cartan connection $ A^{\bs u}$ is none other than the twistor/tractor connection 1-form.
Conformal tractors and twistors are obtained by dressing sections of the associated $\RR$- and $\CC$-vector bundles $E$ and $\sf E$ respectively. See  \cite{Attard_et_al2017, Attard-Francois2016_I, Attard-Francois2016_II} }

\subsubsection{Residual symmetries of the second kind}  
\label{Residual symmetries of the second kind}  

The second kind of residual symmetry, may arise even when the original $\Diff(M)$-symmetry has been entirely neutralised via DFM. 
Indeed, if by definition of the dressing field $\, \bs u : \Phi \ \rarrow\ \D r[N, M]\, $ its equivariance $R^\star_\psi \bs u=\psi\- \circ \bs u$ means that $\Diff(M)$ acts on its target space as the map $\bs u(\phi): N \rarrow M$, there may be a priori a natural right action of $\Diff(N)$ on its source space: 
\begin{align}
\D r[N, M] \times \Diff(N) &\rarrow \D r[N, M] \notag\\
(\bs u, \vphi) &\mapsto \bs u \circ \vphi \rdefeq \bs u^\vphi.  \label{dressing-ambiguity}
\end{align} 
Another way to see it, is to remark that two candidate dressing fields $\bs u'$ and $\bs u$ can a priori be related by an element $\vphi \in \Diff(N)$ acting on their common source space, the model manifold $N$: $\bs u '=\bs u \circ \vphi$. This indeed gives rise to the right action above,  $\bs u'= \bs u^\vphi$.  

Naturally, $\Diff(N)$ does not act on $\Phi$, a fact we may note by: $\phi^\vphi=\phi$. But given that $\phi^{\bs u} = F_{\bs u}(\phi)\defeq \bs u^* \phi$, it must be the case that $\Diff(N)$ acts on the dressed variable via its action on the dressing field:
\begin{align}
\phi^{\bs u} \mapsto   \phi^{\bs u^\vphi} \defeq (\bs u^\vphi)^*\phi = ( \bs u \circ \vphi)^*\phi =  \vphi^* ({\bs u}^*\phi) =  \vphi^* \phi^{\bs u}. 
\end{align}
Seen another way, $\phi^{\bs u}$ is a tensor field on $N$, therefore $\Diff(N)$ acts on it naturally via pullback $\phi^{\bs u} \mapsto  (\phi^{\bs u})^\vphi\defeq \vphi^* \phi^{\bs u}$. Either way, this means that the space $\Phi^{\bs u}$ of dressed fields is fibered by the right action of $\Diff(N)$:
\begin{equation}
\label{dressing-ambiguity-bis}
\begin{aligned}
  \Phi^{\bs u} \times\Diff(N) &\rarrow \Phi^{\bs u},  \\
               (\phi^{\bs u}, \vphi) &\mapsto R_\vphi \phi^{\bs u}:= \vphi ^*\phi^{\bs u},
\end{aligned}
\end{equation}
it is thus a principal bundle $ \Phi^{\bs u} \xrightarrow{\pi}   \Phi^{\bs u}/\Diff(N) \rdefeq \M^{\bs u}$. From there, the bundle geometry of $\Phi^{\bs u}$ parallels that of $\Phi$ described in previous sections. Its associated SES of groups is 
 \begin{align}
 \label{SESgroup2}
\bs{\Diff}(N) \simeq \bs{\Aut}_v(\Phi^{\bs u})  \rarrow   \bs{\Aut}(\Phi^{\bs u}) \rarrow \bs{\Diff}(\M^{\bs u}),
 \end{align}
where $\bs{\Diff}(N) := \big\{ \bs\vphi : \Phi^{\bs u} \rarrow   \Diff(N) \, |\,  R^\star_\vphi \bs\vphi  =\vphi^{-1} \circ \bs\vphi \circ \vphi   \big\} $
is the gauge group of $\Phi^{\bs u}$. The latter being a subgroup of $C^\infty\big( \Phi^{\bs u}, \Diff(N)\big)  \simeq \bs\Diff_v(\Phi^{\bs u})$, acting on $\Gamma(T\Phi^{\bs u})$ and $\Omega^\bullet(\Phi^{\bs u})$ the usual way. In particular, we may immediately write that, for  $\bs\vphi \in C^\infty\big( \Phi^{\bs u}, \Diff(N)\big)$, a (dressed) form $\bs\alpha^{\bs u}=\alpha\big( \! \w^\bullet \! \bs d\phi^{\bs u}; \phi^{\bs u}\big) \in \Omega^\bullet(\Phi^{\bs u})$ transforms as:
\begin{align}
\label{GT-general-bis} 
(\bs\alpha^{\bs u})^{\bs\vphi}  &= \alpha\big(\! \w^\bullet \! (\bs d\phi^{\bs u})^{\bs \vphi};\,  (\phi^{\bs u})^{\bs\vphi} \big),  \\[.5mm]
 \text{with}\qquad ( \bs d\phi^{\bs u})^{\bs\vphi} &= \bs\vphi^*   \big(  \bs d\phi^{\bs u}   +   \mathfrak L_{ \{\bs d \bs\vphi  \circ \bs\vphi\-\}} \phi^{\bs u}  \big).     \label{GT-basis-1-form-bis}
 \end{align}
In exact replica of \eqref{GT-general} and \eqref{GT-basis-1-form} on the original field space $\Phi$. From above, or first principles, one can also write: 
$(\bs\alpha^{\bs u})^{\bs\vphi}  =R^\star_{\bs \vphi} \bs{\alpha^u} + \iota_{\{\bs d \bs\vphi  \circ \bs\vphi\-\}^v} \bs{\alpha^u}$. 
\medskip

\noindent {\bf Affine changes of flat connection on field space.}
As observed in section \ref{Dressing field and flat connections}, a dressing field $\bs u$ supplies a flat connection  on $\Phi$, $\bs\omega_{\text{\tiny$0$}} \defeq -\bs{du} \circ\bs u\- \in \C$. A second dressing field related to the first by the action of $C^\infty\big(\Phi^{\bs u}, \Diff(N)\big) \simeq\bs\Diff_v(\Phi^{\bs u})$ (or $\bs\Diff(N)\simeq \bs\Aut_v(\Phi^{\bs u})$ as a special case),  $\bs u'=\bs u^{\bs\vphi}$, induce another flat connection related to the first as: 
\begin{align}
\label{affine-flat-residual}
\bs\omega'_{\text{\tiny$0$}} \defeq -\bs{du}' \circ{\bs u'}\- =  - \bs d (\bs u \circ \bs\vphi) \circ   \bs\vphi\- \circ \bs u\-  = -\bs{du} \circ\bs u\- - \bs u_*(\bs d \bs\vphi \circ \bs\vphi\-) \circ \bs u\- \rdefeq \bs\omega_{\text{\tiny$0$}} + \bs\beta_{\text{\tiny$0$}}. 
\end{align}

One must first observe that  $\bs\beta_{\text{\tiny$0$}}\defeq - \bs u_*(\bs d \bs\vphi \circ \bs\vphi\-) \circ \bs u\- \, \in \Omega^1_{\text{ten}}(\Phi, \diff(M))$. Indeed, $\bs\vphi$ being seen as $\phi$-dependent, $\bs d \bs\vphi \circ \bs\vphi \in \Gamma(TN)$ is a vector field near the identity in $\Diff(N)$ . So $\bs u_*(\bs d \bs\vphi \circ \bs\vphi\-) \circ \bs u\- \in \Gamma(TM)$ is the $\bs u\-$-related vector field on $M$ (remembering \eqref{Psi-relatedness}) near the identity in $\Diff(M)$.
 So $\bs\beta_{\text{\tiny$0$}} \in \Omega^1\big(\Phi, \diff(M)\big)$.
 
 Now, since by definition the $\phi$-dependence of $\bs\vphi$ is via $\phi^{\bs u}$, one has that $R_\psi^\star \bs\vphi =\bs\vphi$ for $\psi \in \Diff(M)$ -- which also secures the fact that $\bs u'=\bs u^{\bs\vphi}=\bs u \circ \bs \vphi$ remains a well-defined dressing field. Which is, at the linear level, $\bs L_{X^v} \bs\vphi = \iota_{X^v} \bs{d\vphi} =0$, for $X^v \in \Gamma(V\Phi)$ and $X \in \diff(M)$. This implies that $\bs\beta_{\text{\tiny$0$}} (X^v)=0$, i.e. $\bs\beta_{\text{\tiny$0$}} \in  \Omega^1_\text{hor}\big(\Phi, \diff(M)\big)$. This also allows to work out that the equivariance is
 \begin{align}
 R_\psi^\star \bs\beta_{\text{\tiny$0$}} =  - R_\psi^\star \big( \bs u_*(\bs d \bs\vphi \circ \bs\vphi\-) \circ \bs u\- \big) 
 							  =  -  (\psi\- \circ \bs u)_*(\bs d \bs\vphi \circ \bs\vphi\-) \circ (\psi\- \circ \bs u)\-
							  = \psi\- _* \bs\beta_{\text{\tiny$0$}}  \circ \psi,
 \end{align} 
 as one would expect. 
 So indeed $\bs\beta_{\text{\tiny$0$}} \in \Omega^1_\text{tens}\big(\Phi, \diff(M)\big)$.
 Thus, as per the discussion of Ehresmann connections in section \ref{Ehresmann connections}:  
 $\bs\omega'_{\text{\tiny$0$}} =\bs\omega_{\text{\tiny$0$}} + \bs\beta_{\text{\tiny$0$}} \in \C$. 
 Said otherwise, the action of $C^\infty\big(\Phi^{\bs u}, \Diff(N)\big) \simeq\bs\Diff_v(\Phi^{\bs u})$ -- or $\bs\Diff(N)\simeq \bs\Aut_v(\Phi^{\bs u})$ as a special case -- on the space of dressing fields induces  affine changes of flat connections \eqref{affine-flat-residual}.

One may then notice that, using \eqref{lemma1}, \eqref{GT-basis-1-form-bis} can be rewritten as a relation between (basic) forms on $\Phi$:
\begin{align}
\label{GT-basis-1-form-bis-bis}
( \bs d\phi^{\bs u})^{\bs\vphi} &= \bs\vphi^*   \big(  \bs d\phi^{\bs u}   - \bs u^*   \mathfrak L_{ -\bs u_*[\bs d \bs\vphi  \circ \bs\vphi\-] \circ \bs u\-} \phi \big), \notag \\
  					     &=\bs\vphi^*   \big(  \bs d\phi^{\bs u}   - \bs u^*   \mathfrak L_{ \bs\beta_{\text{\tiny$0$}} } \phi \big). 
 \end{align}
 Therefore, for $\bs\alpha=\alpha(\bs d \phi; \phi)\, \in \Omega^1_\text{inv}(\Phi)$, one has that
 \begin{align}
( \bs \alpha^{\bs u})^{\bs\vphi} &=  \bs \alpha^{\bs u}   -  \alpha \big( \mathfrak L_{ \bs\beta_{\text{\tiny$0$}} } \phi; \phi \big), \notag \\
  					      &= \bs \alpha^{\bs u}   - \iota_{ \bs\beta_{\text{\tiny$0$}}^v } \  \bs\alpha, \quad  \in \Omega^1_\text{basic}(\Phi).
 \end{align}
 These might be compared to  \eqref{Ambiguity-hor-dphi}-\eqref{Ambiguity-basic-1-form} which reflects the ambiguity in the construction of basic forms via the horizontalization procedure performed using  Ehresmann connections, discussed in section \ref{Ehresmann connections}. 
 
 This discussion reinforced the observation made in section \ref{Dressing field and flat connections} above, showing  that, on $\bs\alpha \in \Omega^1_\text{inv}(\Phi)$, using the DFM and Ehresmann connections to build a basic counterpart $\bs\alpha^b \in \Omega^1_\text{inv}(\Phi)$ indeed give formally very analogous result, down to the ``ambiguity" arising in both approaches. A key difference being that, when using Ehresmann connections, the ambiguity is an affine shift in $\C$ that cannot be in general ascribed to the action of a group, while it is so (a priori) when using the DFM. 
 \medskip

It must be observed that $(\phi^{\bs u})^{\vphi}$ is $\Diff(M)$-invariant for all $\vphi\in\Diff(N)$, 
so any representative in the $\Diff(N)$-orbit $\O_{\Diff(N)}[\phi^{\bs u}]$ of $\phi^{\bs u}$  is as good a coordinatisation of $[\phi]\in \M$ as any other. In other words we have, a priori, $\O_{\Diff(N)}[\phi^{\bs u}] \simeq  \O_{\Diff(M)}[\phi]$, that is $\M^{\bs u} \simeq \M$. 
Which in turn implies that the ``physical symmetries" are to be found in $\bs\Diff(\M^{\bs u}) \simeq \bs\Diff(\M)$. Given the SES \eqref{SESgroup2}, it is clear that $\bs\Diff(N)\simeq \Aut_v(\Phi^{\bs u})$  is isomorphic to  the original gauge group $\bs\Diff(M)\simeq \Aut_v(\Phi)$ -- and by extension  $\bs\Diff_v(\Phi^{\bs u}) \simeq \bs\Diff_v(\Phi)$.  
The new symmetry $ \bs\Diff_v(\Phi^{\bs u}) \supset \Aut_v(\Phi^{\bs u})$ a priori does not enjoy a more direct physical interpretation than the original  it replaces. 
 This suggests that the terminology ``physical symmetry" often encountered in the literature to describe transformations like \eqref{dressing-ambiguity}, \eqref{dressing-ambiguity-bis} and \eqref{GT-basis-1-form-bis} is essentially misleading
 -- these are also often called ``surface symmetries" or ``corner symmetries" in the literature on the covariant phase space approach to field theory,  which will be discussed in sections \ref{Covariant phase space methods} and \ref{Dressed presymplectic structure}. 
 
 
 One may wonder why bother to apply the DFM if one ends up with an equivalent gauge symmetry. The point is that  a dressing field may simply be available in a theory as a matter of fact, and its existence have implications for the formal properties and interpretation of the theory that one may as well be aware of. Such would be the case e.g. if part of the symmetry is reducible, as discussed in e.g. \ref{Residual symmetries of the first kind}, and therefore \emph{artificial}: The theory is then trimmed from its superfluous formal structure, down to its most economical (``Ockamized") and essential version. 
 
 Furthermore, it may be that the constructive process by which one obtain a dressing field $\bs u(\phi)$ from existing fields of the theory is such that the ambiguity reflected in the a priori relation \eqref{dressing-ambiguity} is minimal, so that in effect $\vphi$'s form a subgroup of $\Diff(N)$, perhaps even a discrete one  (the best possible situation would be that this subgroup is the trivial group). 
 Clearly, this possibility is forfeited if a dressing field is introduced by hand in a theory -- as is often the case in the edge mode literature. 
 We further develop the discussion of these points in section \ref{Discussion}.

%
 
 \subsection{Dressed regions and integrals} 
 \label{Dressed integrals} 
  
  As observed in section \ref{Integration map}, for $U \in \bs U(M)$ and $\bs \alpha \in \Omega^\bullet \big(\Phi, \Omega^{\text{top}}(U) \big)$,
    integrals $\bs\alpha_U=\langle \bs \alpha, U \rangle =\int_U \bs \alpha$ are objects on $\Phi \times \bs U(M)$ (with values in $\Omega^\bullet(\Phi)$). Integrals that are invariant under the action of $C^\infty\big(\Phi, \Diff(M) \big)\simeq \bs\Diff_v(\Phi)$ as defined by \eqref{Vert-trsf-int-generic} (i.e. integrals of tensorial integrand) are well-defined on the associated bundle of regions $\b{\bs U}(M)\defeq \Phi \times \bs U(M)/\sim$, quotient of the product space by the action of $\Diff(M)$  \eqref{right-action-diff}: They have well-defined projection along $\b\pi : \Phi \times \bs U(M) \rarrow \b{\bs U}(M)$, $(\phi, U) \mapsto [\phi, U]=[\psi^*\phi, \psi\-(U)] $, and can be said ``basic" w.r.t. $\b \pi$. 
    
   In section \ref{Building basic forms via dressing}, we defined dressed objects as being in Im$\,\pi^\star$, i.e. basic on $\Phi$, with the projection realised via a dressing field, $F_{\bs u} \sim \pi$.
And relying on the formal similarity of the actions of $ F_{\bs u}$ and $ \Xi \in \bs\Diff_v(\Phi) \simeq C^\infty\big(\Phi, \Diff(M) \big)$, the rule of thumb to obtain the dressing $\bs\alpha^{\bs u}$ 
of a form $\bs\alpha$ is to replace the field-dependent parameter $\bs\psi$ in $\bs\alpha^{\bs\psi}\defeq \Xi^\star \bs \alpha$ 
by the dressing field $\bs u$.  
   
    In the same way, we define dressed integrals as being basic on $\Phi \times \bs U(M)$, i.e. in Im$\,\b\pi^\star$, with the projection realised as:
\begin{equation}   
\begin{split} 
    \b F_{\bs u}: \Phi \times \bs U(M) &\rarrow \b{\bs U}(M) \simeq \Phi^{\bs u} \times \bs U(N), \\
    				(\phi, U) &\mapsto    \b F_{\bs u}(\phi, U)\defeq \big(F_{\bs u}(\phi), \bs u\-(U) \big) =  \big( \phi^{\bs u}, \bs u\-(U) \big).
\end{split}    
\end{equation}    
We highlight the fact that the region $U^{\bs u} \defeq  \bs u\-(U) \in \bs U(N)$ is a map $U^{\bs u}: \Phi \times \bs U(M) \rarrow \bs U(N)$ s.t. for $\psi \in \Diff(M)$:
\begin{align}
(U^{\bs u})^\psi \defeq \b R^\star_\psi \, U^{\bs u} = (R^\star_\psi \bs u)\- \circ \psi\-(U) = (\psi\- \circ \bs u)\-\circ \psi\-(U) = \bs u\-(U) \rdefeq U^{\bs u}. 
\end{align}
The same then holds for $\bs \psi \in C^\infty\big(\Phi, \Diff(M) \big)\simeq  \bs\Diff_v(\Phi)$: $(U^{\bs u})^{\bs\psi} \defeq \b \Xi^\star  U^{\bs u} =U^{\bs u}$. 
Therefore, $U^{\bs u}$ is a $\phi$-dependent $\Diff(M)$-invariant region of \emph{true spacetime}. It is indeed a key insight of GR that, by the conjunction of the hole argument and point-coincidence argument \cite{Norton1993, Stachel2014, Giovanelli2021}, spacetime is defined in a \emph{relational} and $\Diff(M)$-invariant way by its physical field content. This fact is tacitly encoded by the $\Diff(M)$-covariance of general relativistic theories, it is made manifest by the DFM: 
 $U^{\bs u}$~are manifestly $\Diff(M)$-invariant $\phi$-\emph{relationally} defined regions of the physical spacetime -- on which the relationally defined $\Diff(M)$-invariant fields $\phi^{\bs u}\defeq \bs u^*\phi$ live, and are to be integrated over.

Let us further observe that, as a consequence, the (physical/relational) boundary $\d U^{\bs u}$ of a physical spacetime region is of necessity $\Diff(M)$-invariant. Thus, the often encountered claim that ``boundaries in general relativistic physics break $\Diff(M)$-invariance" -- and that the ``edge modes" d.o.f. sometimes introduced as a consequence are the Goldstone bosons associated to that symmetry breaking -- is deeply misguided. It is logically equivalent to the hole argument that temporarily confused Einstein about $\Diff(M)$, and overlooks the resolution stemming from the key insight of his point coincidence argument: relationality. 

So, on the space $\Omega^\bullet \big(\Phi, \Omega^{\text{top}}(U) \big) \times \bs U(M)$ we define:
\begin{equation}   
\begin{split} 
    \t F_{\bs u}: \Omega^\bullet \big(\Phi, \Omega^{\text{top}}(U) \big) \times \bs U(M) &\rarrow \Omega^\bullet_\text{\text{basic}}\big(\Phi, \Omega^{\text{top}}(N) \big) \times \bs U(N) \\
    				(\bs \alpha, U) &\mapsto    \t F_{\bs u}^\star (\bs \alpha, U)\defeq \big(\bs \alpha^{\bs u}, \bs u\-(U) \big). 
\end{split}    
\end{equation}  
With indeed $ \Omega^\bullet_\text{\text{basic}}\big(\Phi, \Omega^{\text{top}}(N) \big) \simeq \Omega^\bullet \big(\Phi^{\bs u}, \Omega^{\text{top}}(N) \big)$, as  dressed fields $\phi^{\bs u}$ live on $N$. 
Then, in formal analogy with \eqref{Vert-trsf-int-generic}, the dressing of an integral $\bs\alpha_U \defeq \langle \bs\omega, U \rangle=\int_U \bs\alpha$ is:
\begin{equation}   
\label{Dressed-integral}
\begin{split} 
({\bs\alpha_U})^{\bs u} = \langle\ ,\ \rangle \circ  \t F_{\bs u}(\bs \alpha, U) = \langle \bs\alpha^{\bs u} , \bs u\-(U)   \rangle 
= \int_{\bs u\-(U)} \hspace{-4mm} \bs\alpha^{\bs u}. 
\end{split}    
\end{equation}  
We remark that the residual symmetry $C^\infty\big(\Phi^{\bs u}, \Diff(N) \big) \simeq \bs\Diff_v(\Phi^{\bs u})$ may then potentially act on dressed integrals, analogously to \eqref{Vert-trsf-int-generic}, as:
\begin{align}
\label{residual-trsf-dressed-int}
\left(({\bs\alpha_U})^{\bs u} \right)^{\bs \vphi} = \int_{\bs \vphi \-(U^{\bs u})}  (\bs\alpha^{\bs u})^{\bs \vphi}. 
\end{align}
Now, for $\bs \alpha \in \Omega^\bullet_\text{tens} \big(\Phi, \Omega^{\text{top}}(U) \big)$ one has: 
\begin{equation}   
\begin{split} 
{\bs\alpha_U}^{\bs u} = \langle \bs\alpha^{\bs u} , \bs u\-(U)   \rangle = \langle \bs u^*\bs\alpha, \bs u\-(U)   \rangle = \langle \bs\alpha, U   \rangle = \bs\alpha_U,
\end{split}    
\end{equation}  
by the invariance property \eqref{invariance-action} of the integration pairing. 
Therefore, ${\bs d \bs\alpha}_U = \langle \bs d \bs\alpha, U \rangle = \bs d  \langle \bs u^*\bs\alpha, \bs u\-(U)   \rangle = \bs d ({\bs\alpha_U}^{\bs u})$. The latter can also be proven in a way entirely analogous to \eqref{commute-d-vert-trsf} using the lemma  \eqref{Formula-general-rep-dress}-\eqref{Formula-pullback-rep-dress} and concluding by the invariance property  \eqref{inf-invariance-action}. This calculation also supplies the analogue of  \eqref{Vert-trsf-pairing-dalpha}/\eqref{Vert-trsf-int-dalpha}: 
\begin{equation}
\label{Dressing-int-dalpha}
\begin{aligned}
\bs (\bs{d\alpha}_U)^{\bs u}	 &= \bs{d\alpha}_U + \langle \mathfrak L_{\bs{du}\circ \bs u\-} \bs\alpha,  U \rangle,  \\
 \langle  \bs{d\alpha}, U \rangle ^{\bs u} &= \langle\bs{d\alpha}, U \rangle + \langle \mathfrak L_{\bs{du}\circ \bs u\-} \bs\alpha,  U \rangle. 
\end{aligned}
\end{equation}
This is relevant to understand the interplay between DFM and the variational principle. 
\medskip

Indeed, consider a Lagrangian  $L \in \Omega^0_\text{tens}\big(\Phi, \Omega^\text{top}(U) \big)$ with action functional $\Phi \times \bs U(M)$: 
$S =\langle L, U \rangle= \int_U L$.
 We~have the dressed Lagrangian $L^{\bs u} =L(\phi^{\bs u})$, which is a manifestly \emph{relational} rewriting of the bare theory in term of the $\Diff(M)$-invariant relational variables $\phi^{\bs u}$.
 The corresponding dressed action is
 $S^{\bs u}=\langle L^{\bs u}, \bs u\-(U) = \langle L, U\rangle = S$. 
 Thus, on the one hand the variational principle is $\bs d S = 0 = \bs d (S^{\bs u})$, and on the other hand we have the identity, analogue of \eqref{Vert-trsf-dS}:  
\begin{equation}
\label{Dressing-dS}
\begin{aligned}
(\bs dS)^{\bs u} &= \bs d S + \langle  \mathfrak  L_{\bs d \bs u  \circ \bs u\-} L,  U  \rangle \\
			   &=\bs d S + \langle  d \iota_{\bs d \bs u  \circ \bs u\-} L,  U  \rangle
			   = \bs d S + \langle  \iota_{\bs d \bs u  \circ \bs u\-} L,  \d U  \rangle.
\end{aligned}
\end{equation}
The fact that $\bs dS$ and its dressing differ by a boundary term implies that the variational principles for the undressed and dressed fields are equivalent, and that the space $S^{\bs u}$ of dressed solutions  is isomorphic to the space $S$ of  solutions of the undressed theory. 

This is fundamental to the relational interpretation of the DFM: it implies that the field equations satisfied by the physical/relational variables $\phi^{\bs u}$ are functionally the same as those satisfied by the gauge variables $\phi$. This will be shown explicitly in section \ref{Dressed presymplectic structure}. Already we stress the fact that it clarifies the reason why, e.g., GR is applied successfully when confronted to physical observation even though the standard formulation uses in principle non-observable $\Diff(M)$-variables (such as the metric field $\phi_1=g_{\mu\nu}$): 
What we actually confront to observations is the dressed/relational theory $L^{\bs u} =L(\phi^{\bs u})$ -- often inaccurately called a ``gauge-fixed" version of the theory -- where coordinatisation is made relationally, w.r.t. other fields (other d.o.f. $\phi_2$ than the gravitational ones: the laboratories, earth, the solar system, the local group, etc.), which is formally identical to the bare version.    
 So, the fact that it is possible to successfully test GR without explicitly solving first the issue of its observables is precisely due to its relational character, as encoded by its $\Diff(M)$-equivariance, which technically implies that its bare and dressed/physical version are functionally identical.

\subsection{Discussion}
\label{Discussion}

Let us concisely sum up the main technical idea of this section, before offering some more conceptual comments. 
The DFM is a systematic and formal way to obtain objects, built from fields $\phi$ and their variations $\bs d \phi$, that are (partially or fully) invariant under diffeomorphisms of $M$:   \emph{basic} forms on field space $\Phi$, in the bundle theoretic terminology. 
It is a \emph{conditional} proposition: \emph{If} one can find/build a dressing field, \emph{then} it gives the algorithm to produce these invariants (and analyse possible residual symmetries).
\medskip

We must observe  that the existence of global dressing fields are not guaranteed, global being meant in two ways. 
Firstly, it is clear that $\phi$-independent dressing fields of the type  $\D r[\RR^n, M]$, i.e. $u: \RR^n \rarrow M$, are nothing but \emph{global} coordinate charts, which may not exist depending on the topology of $M$. Topological obstructions remain relevant in the $\phi$-dependent case $\bs u: \RR^n \rarrow M$, as they may imply that no field $\phi$ is globally defined, or non-vanishing, everywhere on $M$ and fit to serve as a global coordinatisation.\footnote{One may  e.g. think of the Poincaré–Hopf theorem (or its special case, the Hairy Ball theorem) stating that there is no nowhere vanishing vector fields on compact manifold with vanishing Euler characteristic. Admittedly, this particular case may be of limited (fundamental) physical importance, as spacetime is modelled by Lorentzian 4-dimensional manifolds requiring precisely the existence of such a nowhere vanishing vector field -- hence spacetime is either non-compact or has vanishing Euler characteristic. See \cite{ONeill1983} p.149.}

Secondly, as noted in section \ref{Dressing field and flat connections},  dressing fields  globally defined  \emph{on field space} $\Phi$  provide  flat Ehresmann connections, which in turn imply the global triviality of $\Phi$ as a bundle. A priori, if $\Phi$ is globally non-trivial, dressing fields would only be available locally, over open subsets $\U \subset \M$ (being compatible with the local triviality of $\Phi$). 
\medskip

Remark that the formalism extends familiar notions of standard differential geometry. Indeed, as just noted, $\phi$-independent dressings $\bs u \rarrow u \in \D r[\RR^n, U \subset M]$ are just coordinates charts of $M$ (thus non-variational objects, $\bs d u=0$). In that case, dressed variables $\phi^u$ are the usual coordinate representations of the fields $\phi$, while the formula defining dressed integrals \eqref{Dressed-integral}  reproduces the \emph{definition} of integration on $M$ -- the general case of \eqref{Dressed-integral} may thus be understood as defining \emph{integration on spacetime}. 
Furthermore,  $\Diff(N)=\Diff(\RR^n)$ is then just the group of transition functions of $M$ (i.e. coordinate changes), so that relations  \eqref{dressing-ambiguity-bis}, \eqref{GT-basis-1-form-bis}, \eqref{GT-general-bis}  and  \eqref{residual-trsf-dressed-int} reduce to standard coordinate change formulas for fields, field-dependent volume forms, and integrals.
\medskip

Explicit in the DFM, is the proposition that the dressing field is to be found among, or built from, the existing set of fields $\phi$ of a theory: it is key to the natural relational interpretation of the formalism. This is the situation of most fundamental physical significance for the analysis of a $\Diff(M)$-theory.  

It is not so for dressing fields introduced by \emph{fiat} as d.o.f. independent from the original field space $\Phi$, i.e. admitting $\bs d u \neq 0$:  It amounts to extending the original field space to $\Phi'=\Phi + u$, and thus to considering an altogether different theoretical framework  where  both the kinematics and the \emph{character} of the symmetry are altered.
 Two common situations are covered by this case of the DFM.

The first is when starting from a $\Diff(M)$-theory on $\Phi$, and one tries to reduce the symmetry by introducing such an \emph{ad hoc} dressing field. When $\Diff(M)$ is seemingly broken by a background structure, like a non-relational boundary, the move is often understood as ``restoring $\Diff(M)$-invariance". This is typically the premise of the literature on ``edge modes", see e.g. \cite{DonnellyFreidel2016, Speranza2018, Speranza2022, Ciambelli2023}. 
An immediate drawback of such a move was pointed out earlier, in section \ref{Residual symmetries of the second kind}: The original $\Diff(M)$ symmetry is replaced  by an isomorphic, and a priori  no more physical, $\Diff(N)$ symmetry -- think again of the case $N=\RR^n$ and $u \in \D r[\RR^n, U \subset M]$.
Clearly, there is little chance for $\Diff(N)$ to be a symmetry giving relevant \emph{new} information about the theory under consideration (not already given by $\Diff(M)$).
And the \emph{ad hoc} dressing field being introduced non-constructively, there is no compelling argument for ever having  $\Diff(N)$ reduced to a subgroup (small or discrete) either. Therefore, all ``issues" one had with $\Diff(M)$ one has now with $\Diff(N)$, defeating the purpose of introducing a dressing field in the first place.  
 
 The second situation is when starting from a theory devoid of $\Diff(M)$ symmetry, and one tries to enforce it via the introduction of an \emph{ad hoc} dressing field. In that context,  \emph{ad hoc} dressing fields are known as  (gravitational) Stueckelberg fields, and the move is again seen as ``restoring $\Diff(M)$-invariance". This is typically the case of the literature on massive gravity, bi-gravity, and sometimes string theory \cite{deRham2014, Green-Thorn1991, Siegel1994}. 
Such an artificially enforced $\Diff(M)$ symmetry have no physical content, and is there mainly to hide background non-dynamical structures -- often some reference metric (preserved by a subgroup of $\Diff(M)$, which is the real symmetry of the theory). 
For that reason these $\Diff(M)$ symmetries introduced via \emph{ad hoc} dressing fields, i.e. Stueckelberg fields, are called ``\emph{artificial}" in the philosophy of physics literature \cite{Pitts2005, Pitts2008, Pitts2009, Pitts2012}.\footnote{Indeed, enforcing a symmetry via a Stueckelberg trick, is an instance of ``Kretschmannisation": rewriting a theory so as to display a strictly formal symmetry without physical signature or content. The name stems from the famous ``Kretschmann objection" against general covariance as a fundamental feature of GR, according to which any theory can be made $\Diff(M)$-invariant, or generally covariant. Analysis of the objection resulted in the realisation that one must distinguish between \emph{substantive} $\Diff(M)$/general covariance as a distinctive native feature of a theory, with physical content/signature, from \emph{artificial} $\Diff(M)$/general covariance which is implemented by hand (often via the introduction of extra structures/fields, e.g. à la Stueckelberg) and thus without physical content. A comparable crucial distinction exists for (internal) gauge symmetries, in response to a ``generalised Kretschmann objection" to the gauge principle \cite{Pitts2009, Francois2018, Berghofer-et-al2023}: There, a key physical signature of substantive gauge symmetries is the trade-off between gauge invariance and locality of a theory, absent for artificial gauge symmetries.}
The class of theories displaying such artificial $\Diff(M)$ symmetries is thus ``non-general-relativistic" in a deep sense: They are only superficially and formally alike GR, 
but  betray its key physical insight, i.e. the relationality and absence of background structures/fields that a substantive $\Diff(M)$ symmetry encodes. 
%
\medskip

Field-dependent dressing fields $\bs u=\bs u(\phi)$, whose d.o.f. are extracted from the original field space $\Phi$, are much more interesting. 
%
They allow in principle a formal implementation of a \emph{relational description} of the physics of general relativistic field theory.
As already noted early in section \ref{Building basic forms via dressing}, the invariant dressed fields $\phi^{\bs u(\phi)}$ defined by  \eqref{F-map}
are readily understood as a relational description of the physical d.o.f., in that
 the  d.o.f. of the fields $\phi$ are \emph{coordinatised relative to each other}.
As a result, only $\Diff(M)$-invariant, physical relational d.o.f. are manifest.\footnote{Said otherwise, by dressing $\phi \mapsto \phi^{\bs u(\phi)}$, one reshuffles the d.o.f. in presence so as to exhibits only the relational and physical ones. }
It~has long been advocated that the (completely) relational character of general relativistic physics is indeed one of its key innovating feature, tacitly encoded by the $\Diff(M)$ covariance -- as understood by the conjunction the hole and point-coincidence arguments. 
See e.g. \cite{Norton1993, Rovelli2002, Rovelli2004, Tamborino2012, Stachel2014, Rovelli2014, Giovanelli2021}. The DFM can be understood as a way to reformulate a general relativistic field theory in a $\Diff(M)$-invariant and manifestly relational way. 
Accordingly, one may suggest to rebrand General Relativity as ``\emph{General Relationality}". We shall here tacitly adopt this nomenclature shift whenever writing ``GR". 

Relatedly, as observed in section \ref{Residual symmetries of the second kind}, only by \emph{constructively} producing a $\phi$-dependent dressing field  is there  any chance  for the ``new" gauge symmetry  $\Diff(N)$ to be reduced to a small, discrete, or trivial subgroup. In concrete situations, it may be reduced to a discrete choice: that of reference coordinatizing field among the collection $\phi$. 
\medskip

Finally, we draw attention on the fact that the DFM as developed here is the common geometric framework underpinning a number of notions appearing in recent years in the literature on gravity, or $\Diff(M)$-theories more generally:  the already mentioned massive gravity and bi-gravity theories  \cite{deRham2014}, but also the notion of ``gravitational dressings" as proposed in  \cite{Giddings-Donnelly2016, Giddings-Donnelly2016-bis, Giddings-Kinsella2018, Giddings2019, Giddings-Weinberg2020}, or yet that of ``dynamical reference frames” as proposed in  \cite{carrozza-et-al2022, Hoehn-et-al2022}. 
Notably, dressing fields underlie the  notions of ``\emph{gravitational edge modes}" as introduced by \cite{DonnellyFreidel2016} in the covariant phase space approach to the presymplectic structure of gravity over bounded regions, which spurred some amount of activity since its inception --  e.g. \cite{Geiller2017, Speranza2018, Geiller2018, Chandrasekaran_Speranza2021} or \cite{Freidel-et-al2020-1, Freidel-et-al2020-2, Freidel-et-al2020-3}, 
as well as the derived and closely related notion  of ``embedding fields" \cite{Speranza2019, Freidel:2021, Ciambelli-et-al2022, Speranza2022, Ciambelli2023}  -- see also  \cite{Kabel-Wieland2022}.
The following sections \ref{Covariant phase space methods} and \ref{Relational formulation}, will substantiate this claim. 
 In addition to the technical streamlining stemming from the DFM, the insights it offers and conceptual clarifications just discussed should be kept in mind when looking at the cited literature through the DFM lense.


\section{Covariant phase space methods and bundle geometry}  
\label{Covariant phase space methods}  

%

The covariant phase space approach was introduced to study the (pre) symplectic structure of gauge field theory. As (re-) introduced by  \cite{Zuckerman1986, CrnkovicWitten1986, Crnkovic1987}, the aim was 
to associate a symplectic structure to a gauge field theory given by a Lagrangian $L$ over some region $U \subset M$, and so doing while keeping spacetime symmetries manifest (with in mind a possible covariant canonical or geometric quantization). Nowadays, it is mainly used to derive charges as physically interesting quantities, realising the symmetries of the theory through their poisson algebra: it finds notable applications in the study of asymptotic symmetries and  gravitational waves physics, as well as in the study of bounded subsystems and their symmetries (like black holes).\footnote{In this case, Noether charges capture e.g.  the Hawking-Finkelstein entropy \cite{Wald1993}, or the Komar mass (see  \cite{Choquet-Bruhat2009} def. 4.6, eq. (4.8), p.460). } 
  Classical references are \cite{Lee-Wald1990, Ashtekar-et-al1990}, and modern introductions are \cite{Compere-Fiorucci2018, Harlow-Wu2020}  (see also \cite{Farajollahi-Luckock2002} for a compact  summary). 
  
  Its roots go further back though. The recent review \cite{Gieres2021}  gives historical context and shows the relation of covariant phase space methods to other approaches such as the multisymplectic formalism -- about which we recommend  \cite{Helein2012} -- and the variational bicomplex \cite{Anderson1992}. 

In what follows we propose our own account of the topic, articulating covariant phase space approach for $\Diff(M)$-theories with the bundle geometry of field space $\Phi$ as described in section \ref{Geometry of field space}. Our main goal in so doing, is both to streamline the derivation of key results -- such as charges for both field-independent and field-dependent gauge parameter, and their Poisson bracket -- and provide clear understanding of new ones, such as the vertical/gauge transformations of the symplectic potential and 2-form, that are necessary to apply immediately the DFM and thus give the general form of the \emph{basic presymplectic structure} of a theory (encompassing existing constructions, e.g.  ``edge mode extended phase space").

\subsection{Covariant phase space for ${\Diff(M)}$-theories}  
\label{Covariant phase space for Diff(M) theories}  

Consider a theory given by the Lagrangian $n$-form $L'=\mathcal L'\, dx^n \in \Omega^0(\Phi)$, $n=$ dim$M$,  s.t.: 

\begin{equation}
 \begin{aligned}
L'=L+d \ell : \Phi &\rarrow \Omega^n(M), \\
			\phi &\mapsto L'(\phi). 
\end{aligned}
\end{equation}
The two Lagrangians $L'$ and $L$ belong to the same ($n^\text{th}$) DeRham cohomology group of $M$. 
The exact term $\ell$ is called a boundary Lagrangian, as  over a region $U\in \bs U(M)$ with boundary $\d U$, the associated action functional 
$S'=\langle L', U\rangle= \int_U L'$ is: $S'=S + \langle \ell, \d U\rangle =S+ \int_{\d U} \ell$. 
The variational principle, finding extrema of the action $\bs d S' = \langle \bs d L', U\rangle \equiv 0$, requires  to consider the 1-form $\bs d L' \in \Omega^1(\Phi)$, which is generically s.t.:
\begin{equation}
 \begin{aligned}
 \label{Variational-principle}
\bs d L'= \bs d L+d \bs d \ell = \bs E + d\big(\bs\theta + \bs d\ell \big) = \bs E + d\bs\theta',  
\end{aligned}
\end{equation}
where $\bs E= E(\bs d\phi; \phi)$ are the field equations 1-form, and $\bs\theta^{(\prime)}= \theta^{(\prime)}(\bs d\phi; \phi)$ is the presymplectic potential current.
We~see that both $L'$ and $L$  have the same field equations but distinct presymplectic potential currents, $\bs\theta'$ and $\bs\theta$: this is what a boundary term/Lagrangian does, it contributes to the presymplectic structure by shifting the potential. 
The~presymplectic 2-form current of the theory is defined by $\bs\Theta \defeq \bs d\bs \theta'=\bs d\bs \theta\, \in \Omega^2(\Phi)$, i.e. $\Theta\big(\bs d\phi \wedge \bs d\phi; \phi \big)$. It is the same for both $L'$ and $L$, it is thus cohomologically well defined. 
Given a codimension 1 submanifold $\Sigma$ of $U\subset M$, the presymplectic potential and presymplectic 2-form are objects on $\Omega^\bullet(\Phi) \times \bs U(M)$, so on $\Phi \times \bs U(M)$:
\begin{align}
\bs\theta_\Sigma &\defeq \mathcal I(\bs\theta, \Sigma) = \langle \bs\theta, \Sigma \rangle= \int_\Sigma \bs \theta, \\
\bs\Theta_\Sigma &\defeq  \mathcal I(\bs\Theta, \Sigma) = \langle \bs\Theta, \Sigma \rangle= \int_\Sigma \bs \Theta.\\
				&\, = \bs d \mathcal I(\bs\theta, \Sigma)=  \mathcal I( \bs d \bs\theta, \Sigma) \notag
\end{align}

The configuration space is the space of solutions $\S \defeq \{\phi \in \Phi \, | \, \bs E_{|\phi}=0 \}$, a subbundle of $\Phi$ whose base $\M_\S \defeq \S/\Diff(M)$ is the reduced phase space. We remark that there is then a reduced version of the associated bundle of regions: $\b{\bs U}(M)_{|\S} \defeq \S \times_\text{{\tiny $\Diff(M)$}}  \bs U(M)=\S \times \bs U(M)/ \sim$. On-shell integrated forms are objects on $\S \times \bs U(M)$, and those that are invariant under the induced action of  $\bs\Diff_v(\Phi)\simeq C^\infty\big(\Phi, \Diff(M)\big)$ are well-defined on $\b{\bs U}(M)_{|\S} \rarrow \M_\S$ -- as discussed in section \ref{Integration map}.

 One original motivation of the formalism was to associate a symplectic space to the gauge theory  $L'$ over $U\subset M$: 
for this $\bs\Theta_\Sigma$ must be well-defined on $\b{\bs U}(M)_{|\S}$. This would happens  if $\bs\Theta$ is tensorial or basic, see again section \ref{Integration map},  or if its transformation is a boundary term and either $\d\Sigma = \emptyset$ or boundary conditions (b.c.) are given. 
The trouble is, generically none of this occurs. In particular, when $\d\Sigma \neq \emptyset$ and no reasonable b.c. can be imposed,  
 we have an occurrence of ``boundary problem". 

Another instance of ``boundary problem"  arises when, studying the symmetries of a field theory, one establishes the relation between Noether charges and  $\bs\Theta_\Sigma$: 
Indeed generically, the Noether charges fails to be Hamiltonian generator of the action of $\diff(M)$ on $\Phi$ ($X^v$ are not Hamiltonian vector fields) due to boundary contributions that cannot be ``integrated" into a redefinition of the charges (an issue typical of ``open" systems). 
Said otherwise, Noether charges fails to be a moment map for $\diff(M)$. 

\medskip
To establish such boundary problems in their most generic form,  in as clear a way as possible, in the following our goals are: 
First, to derive an expression for the Noether charges associated to the action of $\diff(M)$ and exhibit their relation to $\bs\Theta_\Sigma$. This will allow us to also address the question of the definition of their Poisson bracket. 

Second, we will derive the vertical/gauge transformations -- a.k.a. field-dependent gauge transformations -- of all objects in presence, $L^{(\prime)}$, $\bs dL^{(\prime)}$, $\bs E$, $\bs \theta^{(\prime)}$ and $\bs\Theta$: This will both highlight the general form of the first boundary problem (the ``non-basicity" of $\bs\Theta_\Sigma$), and help to give its formal solution in the form of the dressed/basic presymplectic structure obtained via the DFM (applying its simple rule of thumb). 
\medskip

For this tasks, we will need the equivariance and verticality properties of these objects: 
Since they are all are form on $\Phi$ with values in $\Omega^\bullet(M)$, their $\Diff(M)$-equivariance is  given by the pullback representation, and their  $\diff(M)$-equivariance given by the Lie derivative representation. So, for $\psi \in \Diff(M)$ and $X \in \diff(M)\simeq \Gamma(TM)$:
\begin{align}
R^\star_\psi L^\prime &= \psi^* L^\prime \qquad \text{so} \qquad \bs L_{X^v} L^\prime = \mathfrak L_X L^\prime \rdefeq \alpha(X;\phi), \label{Lagrangian} \\
R^\star_\psi \bs dL^\prime&= \psi^* \bs dL^\prime \qquad \text{so} \qquad \bs L_{X^v} \bs d L^\prime = \mathfrak L_X \bs d L^\prime= \bs d \alpha(X;\phi),  \label{d-Lagrangian}\\
R^\star_\psi \bs E &= \psi^*\bs E \qquad \text{so} \qquad \bs L_{X^v} \bs E = \mathfrak L_X \bs E, \label{E}\\
R^\star_\psi \bs\theta^\prime&= \psi^*\bs\theta^\prime \qquad \text{so} \qquad \bs L_{X^v} \bs\theta^\prime = \mathfrak L_X \bs\theta^\prime\rdefeq \bs\alpha(X), \label{potential} \\
R^\star_\psi \bs\Theta&= \psi^*\bs\Theta \qquad \text{so} \qquad \bs L_{X^v} \bs\Theta = \mathfrak L_X \bs\Theta = \bs d \bs\alpha(X), \label{non-symplectomorph}
\end{align}
The term $\alpha(X;\phi)$ can be called the Lagrangian anomaly, a 0-form on $\Phi$ linear in $X\in \diff(M)$, while $\bs\alpha(X)$ may be called the ``symplectic anomaly", a 1-form on $\Phi$ linear in $X$. Given \eqref{Variational-principle},  the two are related by:
\begin{align}
\label{Identity-anomalies}
\bs d\alpha(X;\phi) = d \bs\alpha(X) + \mathfrak L_X \bs E = d\big(  \bs\alpha(X) +  \iota_X \bs E \big),
\end{align}
where we used the fact that $d \bs E=0$ since $\bs E $ is a top form on $M$.  

We may notice already that, on account of \eqref{non-symplectomorph}, a priori $\Diff(M)$ does not act as symplectomorphisms. So we should expect obstruction from $X^v \in \Gamma(V\Phi)$ being Hamiltonian vector fields for a moment map/Noether charge.

Finally,  the $C^\infty\big( \Phi, \Diff(M) \big)\simeq \bs\Diff_v(\Phi)$-transformations of the integrals of the above objects will be obtained using \eqref{Vert-trsf-int-generic} from section \ref{Integration map}.

\subsubsection{Noether charges for field-independent gauge parameters}
\label{Noether charges for field-independent gauge parameters}  

Let us proceed anyway to find the expression of Noether charges for the action of $\diff(M)$.
As we've already established in section \ref{Differential forms and their derivations} (just below \eqref{twisted-tens-forms}), and reiterated just above, the $\diff(M)$-equivariance of the Lagrangian as a 0-form on $\Phi$ is 
\begin{align}
\label{Diff-anomaly}
\bs L_{X^v} L' = \iota_{X^v} \bs dL' &= \mathfrak L_X L' = d(\iota _X L') \rdefeq d\beta(X, \phi) = \alpha(X;\phi),   \notag\\
		      &=  \mathfrak L_X L +  \mathfrak L_X \, d \ell = d\big(\iota _X L + \mathfrak L_X \ell \big) \rdefeq d\big(  \iota _X L + \beta_\ell(X, \phi) \big). 
\end{align}
Where we used the fact that $n$-forms on $M$ are $d$-closed. 
From this and \eqref{Variational-principle} we get: 
$-\iota_{X^v}\bs E = d\big( \iota_{X^v} \bs\theta'  - \beta(X, \phi) \big)$. 
Which tells us that there is an on-shell ($\bs E=0$) $d$-closed quantity, the Noether $(n-1)$-form current:  
\begin{equation}
\begin{aligned}
\label{Noether-current}
J(X; \phi) \defeq&\,  \iota_{X^v} \bs\theta'  - \beta(X, \phi)  -d \gamma(X;\phi), \\
             =&\,   \iota_{X^v} \bs\theta  - \iota_X L   -d \gamma(X;\phi), \quad\ \in\, \Omega^{n-1}(M).
\end{aligned}
\end{equation}
By definition then, $-\iota_{X^v}\bs E = dJ(X, \phi)$.
The current is independent of the boundary Lagrangian $\ell$, so depends only on the DeRham cohomology class  $[L']=[L]$, but still defined a priori up to a $d$-exact term $d \gamma(X;\phi)$ (the sign is conventional).
Given a codimension 1 submanifold $\Sigma\subset U \in \bs U(M)$, the Noether charge is defined on $\Phi \times \bs U(M)$ as:
\begin{align}
\label{Noether-charge1}
Q_\Sigma (X; \phi)\defeq \langle J(X; \phi), \Sigma \rangle =\int_\Sigma J(X; \phi) =\int_\Sigma  \iota_{X^v} \bs\theta  - \iota_X L   -d \gamma(X;\phi).
\end{align}
Notice this can be seen as a map
\begin{equation}
\label{Moment map}
\begin{aligned}
Q_\Sigma ({\color{gray} X}; \ \,): \Phi &\rarrow  \diff(M)^*, \\
				\phi& \mapsto Q_\Sigma ({\color{gray} X} ; \phi). 
\end{aligned}
\end{equation}
where $ \diff(M)^*$ is the dual of  $\diff(M)$, as indeed $Q_\Sigma (\ \, ; \phi): \diff(M) \rarrow \RR$, $X \mapsto Q_\Sigma (X; \phi)$,  is a  linear map.

Now, in the spirit of the covariant phase space, all quantities ultimately must be expressed on-shell. So, throughout, we will endeavor to find expressions of our key objects in terms of the field equations. Case in point, using  \eqref{Identity2} derived in  \ref{A3} (from the assumptions that we are dealing with the field space of a gauge theory given by a first-order Lagrangian), we find that the Noether current and charge are:
\begin{align}
\label{Noether-current2}
J(X; \phi)&=d \big( \theta(\iota_X \phi; \phi) - \gamma(X ; \phi) \big) \, -\, E(\iota_X \phi; \phi), \\[2mm]
\label{Noether-charge2}
Q_\Sigma (X; \phi)&=\int_{\d\Sigma}   \theta(\iota_X \phi; \phi) - \gamma(X ; \phi) \,  - \int_{\Sigma} E(\iota_X \phi; \phi).
\end{align}
The charge is manifestly  a boundary term on-shell. 
This result is consistent with the expectation that,  the current being $d$-closed on-shell by construction, and assuming the region $U$ has trivial topology, it should indeed be $d$-exact. The fact is made manifest by the above forms.\footnote{\label{Myfootnote} Remark that had we postulated that there is another $d$-exact contribution to the $\Diff(M)$-anomaly \eqref{Diff-anomaly}, $\alpha_\gamma(X; \phi)=d\beta_\gamma(X; \phi)$, the current would have been $J(X; \phi)=d \big( \theta(\iota_X \phi; \phi) - \gamma(X ; \phi) \big) - \beta_\gamma(X; \phi) - E(\iota_X \phi; \phi)$. But since by definition $\iota_{X^v} \bs E = - dJ(X; \phi)$, it follows that on-shell $d\beta_\gamma(X; \phi)=0$. So, assuming trivial topology, we get $\beta_\gamma(X; \phi)=d\b \gamma(X;\phi)$. A term that can be reabsorbed in the $d$-ambiguity of the current: $\gamma \mapsto \t\gamma = \gamma + \b \gamma$. } 
Remark that from \eqref{Noether-current2} follows $\iota_{X^v}\bs E = dE(\iota_X \phi; \phi)$. 
\medskip

Then, one must assess the relation of the Noether charge $Q_\Sigma (X; \phi)$ to the presymplectic 2-form $\bs\Theta_\Sigma$, or of $J(X; \phi)$ to $\bs\Theta$. This will allow to see under which conditions  the Noether charge is ``integrable" into a Hamiltonian generator for $X^v \in \Gamma(V\Phi)$, or in other words, if it defines via \eqref{Moment map}  a moment map for the action of $\diff(M)$ on field space $\Phi$. 

The most direct way is to look at the  $\diff(M)$-equivariance of $\bs\theta'$: 
\begin{equation}
\label{Symplectic anomaly}
\begin{aligned}
\bs L_{X^v} \bs \theta'& = \mathfrak L_X \bs \theta' \rdefeq \bs \alpha(X), \\
\hookrightarrow\quad & \iota_{X^v} \bs \Theta = -\bs d \iota_{X^v} \bs\theta' + \bs\alpha(X), \\
				& \iota_{X^v} \bs \Theta = -\bs d \iota_{X^v} \bs\theta - \bs d\beta_\ell(X;\phi) + \bs\alpha(X),
\end{aligned}
\end{equation}
It is easy to work out an expression for $\bs \alpha(X)$ in terms of the field equation:
\begin{align}
 \bs\alpha(X)\defeq \mathfrak L_X \bs \theta' &= \iota_X d\bs\theta' + d \iota_X \bs \theta', \notag\\
 							     &= \iota_X (\bs dL'- \bs E) + d \iota_X \bs \theta', \notag\\
							      &= \iota_X (\bs dL- \bs E) + d \iota_X \bs \theta + \bs d \beta_\ell(X;\phi).   \label{symplect-anomaly}
\end{align}
Inserting this into \eqref{Symplectic anomaly}, and given the definition of the Noether current \eqref{Noether-current}, we finally obtain: 
\begin{equation}
\label{Moment-map-equation}
\begin{aligned}
\iota_{X^v} \bs \Theta 
				 &= -\bs d J(X;\phi) +d \left(  \iota_X \bs \theta - \bs d \gamma(X;\phi) \right)  - \iota_X \bs E, \\[2mm]
\iota_{X^v} \bs \Theta_\Sigma &= -\bs d Q_\Sigma (X;\phi) + \int_{\d\Sigma }   \iota_X \bs \theta - \bs d\gamma(X;\phi)\  \, - \int_\Sigma  \iota_X \bs E.
\end{aligned}
\end{equation}
We  observe that, even on-shell, there is a boundary obstruction preventing the Noether charge to be a moment map. 
The integrand of the boundary term in \eqref{Moment-map-equation} is often called the ``flux term" or ``symplectic flux", and is sometimes interpreted as meaning that there is physical flux leaking through the boundary $\d\Sigma$, making the system ``open". 

To ``close" the system and reestablish integrability of the Noether charge, one needs to impose boundary condition s.t. $\bs d\gamma(X;\phi) =   \iota_X \bs \theta$, or to restrict the admissible class of diffeomorphisms, considering $X \in \diff(M)$ vanishing at $\d\Sigma$. An instance of the first approach suggests itself when requiring that the variational principle for $L'$ be well-posed on a bounded region: $\bs d S' =0 \Rightarrow \bs E=0$, which necessitates $\bs\theta + \bs d\ell=0_{|\d U\supset \d\Sigma}$, in turn implying $\iota_X \bs \theta =- \bs d \iota_X\ell$. In other words we set $\gamma(X;\phi)=-\iota_X\ell$, so that on-shell:
\begin{align}
\label{BC-imposed}
Q_\Sigma (X; \phi)&=\int_{\d\Sigma}   \theta(\iota_X \phi; \phi) +\iota_X\ell \ _{|\S} \qquad \text{and}\qquad \iota_{X^v} \bs \Theta_\Sigma = -\bs d Q_\Sigma (X;\phi) \ _{|\S}.
\end{align}
The boundary Lagrangian makes its return into the expression of the current/charge. 
This may be compared to Eqs.(2.46)-(2.47)/(2.50) of \cite{Harlow-Wu2020}, where the presymplectic potential is assumed to decompose as $\bs \theta = -\bs d \ell + d \bs C\,_{|\d\Sigma}$ -- 
or to  Eq.(2.20)-(2.21) in \cite{Chandrasekaran_Speranza2021} where it is assumed $\bs \theta' =\bs \theta  - \bs d \ell + d \bs C\,_{|\d\Sigma}$ -- for some 1-form $\bs C$ in $\Phi$, so a term $\iota_{X^v} \bs C$ adds to the charge. In GR with $\Lambda = 0$, \eqref{BC-imposed} applies for $L'=L_\text{EGH}+d \ell_\text{YGH}$, with $\ell_\text{YGH}$ the York-Gibbons-Hawking boundary Lagrangian. 
\medskip

The relation \eqref{Moment-map-equation} allows to investigate the possibility of defining a Poisson bracket of Noether charges.
For $X, Y \in \diff(M)$ generating $X^v, Y^v \in \Gamma(V\phi)$, the standard prescription is:
\begin{align}
\label{Poisson-bracket-def}
\big\{Q_\Sigma (X; \phi), Q_\Sigma (Y; \phi) \big\} \defeq \bs\Theta_\Sigma \big(X^v, Y^v \big).
\end{align}
We need only to work out  the right-hand side. For this, using \eqref{Symplectic anomaly} we have, 
\begin{align}
 \bs\Theta \big(X^v, Y^v \big) =  \iota_{Y^v}  \iota_{X^v} \bs\Theta = - \iota_{Y^v} \left( \bs d \iota_{X^v} \bs\theta' +   \bs\alpha(X) \right)
 												&= -  \bs L_{Y^v} \iota_{X^v} \bs\theta' + \iota_{Y^v}  \bs\alpha(X), \notag\\
											&= \iota _{[X^v, Y^v]} \bs\theta' - \iota_{X^v}  \bs L_{Y^v} \bs\theta' + \iota_{Y^v}  \bs\alpha(X),\notag \\
											&=  \iota_{([X, Y]_{\text{{\tiny $\diff(M)$}}})^v} \bs\theta' 	- \iota_{X^v}  \bs \alpha(Y) +  \iota_{Y^v}  \bs\alpha(X).  \label{PB-step}
\end{align} 
where we used \eqref{identity-L-iota} with \eqref{FN-bracket-field-dep-diff=BT-bracket} (or \eqref{extended-bracket1}/\eqref{extended-bracket2}), as well as \eqref{demo1} from appendix \eqref{Lie algebra (anti)-isomorphisms}. 
 The last two terms can be related to the Lagrangian anomaly which, as stated below \eqref{twisted-tens-forms}, satisfies an Abelian version of the 1-cocycle relation (consistency condition) \eqref{inf-cocycle}:
\begin{align}
 \iota_{X^v} \bs d\alpha(Y;\phi) -  \iota_{Y^v}\bs d\alpha(X;\phi) = \alpha\big([X, Y]_{\text{{\tiny $\diff(M)$}}};\phi \big)
\end{align}
stemming from the commutation relation of Nijenhuis-Lie derivative \eqref{NL-derr-ext-bracket}-\eqref{FN-bracket-field-dep-diff=BT-bracket} (or \eqref{extended-bracket1}/\eqref{extended-bracket2}). Indeed, using \eqref{Identity-anomalies} we get:
\begin{align}
 &\iota_{X^v}  d\big(   \bs \alpha(Y) + \iota_Y \bs E\big) -  \iota_{Y^v} d \big(   \bs \alpha(X) + \iota_X \bs E \big)= d \beta \big([X, Y]_{\text{{\tiny $\diff(M)$}}};\phi \big), \notag \\[2mm]
 \hookrightarrow &\quad
  \iota_{X^v}  \bs \alpha(Y) -  \iota_{Y^v}   \bs \alpha(X)  + \iota_{X^v} \iota_Y \bs E  -  \iota_{Y^v}  \iota_X \bs E  -  \beta \big([X, Y]_{\text{{\tiny $\diff(M)$}}};\phi \big) \rdefeq - d\mathscr{A}\big( \lfloor X, Y\! \rfloor ;\phi \big).  \label{2-cocycle}
\end{align}
The notation $d\mathscr{A}\big(\lfloor X, Y\! \rfloor ;\phi \big)$ is meant to indicate that it is bilinear and antisymmetric in $X, Y$ (as the left-hand side shows). 
 Inserting this into \eqref{PB-step}, and using the definition \eqref{Noether-current} of the Noether current, we get 
\begin{align}
\label{pre-bracket}
 \bs\Theta \big(X^v, Y^v \big) &=  \iota_{([X, Y]_{\text{{\tiny $\diff(M)$}}})^v} \bs\theta' - \beta \big([X, Y]_{\text{{\tiny $\diff(M)$}}};\phi \big) 
 						+ d\mathscr{A}\big(\lfloor X, Y\! \rfloor ;\phi \big) \ + \iota_{X^v} \iota_Y \bs E  -  \iota_{Y^v}  \iota_X \bs E, \notag\\
						&= J \big([X, Y]_{\text{{\tiny $\diff(M)$}}};\phi \big) + d \gamma \big([X, Y]_{\text{{\tiny $\diff(M)$}}};\phi \big)
						+ d\mathscr{A}\big( \lfloor X, Y\! \rfloor ;\phi \big) \ + \iota_{X^v} \iota_Y \bs E  -  \iota_{Y^v}  \iota_X \bs E. 
 \end{align} 
Which gives the formal expression, 
\begin{align}
\label{Poisson-bracket-result}
\big\{Q_\Sigma (X; \phi), Q_\Sigma (Y; \phi) \big\} = Q_\Sigma \big([X, Y]_{\text{{\tiny $\diff(M)$}}};\phi \big) 
					+ \int_{\d\Sigma} \mathscr{A}\big( \lfloor X, Y\! \rfloor ;\phi \big)  + \gamma \big([X, Y]_{\text{{\tiny $\diff(M)$}}};\phi \big)
					+ \int_{\Sigma} \iota_{X^v} \iota_Y \bs E  -  \iota_{Y^v}  \iota_X \bs E. 
\end{align}
Which is on-shell, 
\begin{align}
\label{Poisson-bracket-on-shell}
\big\{Q_\Sigma (X; \phi), Q_\Sigma (Y; \phi) \big\} = Q_\Sigma \big([X, Y]_{\text{{\tiny $\diff(M)$}}};\phi \big)  + \mathscr C\big( \lfloor X, Y\! \rfloor ;\phi \big)_{\ | \S}\,,
\end{align}
defining the map 
\begin{equation}
\begin{aligned}
\label{2-cocycle-map}
 \mathscr C \big(\ \ ; {\color{gray}\phi}\big): \diff(M) \times \diff(M) &\rarrow C^\infty(\Phi),\\
 									(X, Y)		 &  \mapsto  \mathscr C\big( \lfloor X, Y\! \rfloor ;{\color{gray}\phi} \big) \defeq \int_{\d\Sigma} \mathscr{A}\big( \lfloor X, Y\! \rfloor ;{\color{gray}\phi} \big)  + \gamma \big([X, Y]_{\text{{\tiny $\diff(M)$}}};{\color{gray}\phi} \big),
\end{aligned}
\end{equation}
which is clearly bilinear and antisymmetric. If it can be shown that it furthermore satisfies 
\begin{align}
\label{2-cocycle-relation}
 \mathscr C \big( \lfloor X, [Y, Z]_{\text{{\tiny $\diff(M)$}}}\rfloor ; {\color{gray}\phi} \big) 
+ \mathscr C \big( \lfloor Y, [Z, X]_{\text{{\tiny $\diff(M)$}}}\rfloor ; {\color{gray}\phi} \big)  
+  \mathscr C \big( \lfloor Z, [X, Y]_{\text{{\tiny $\diff(M)$}}}\rfloor ; {\color{gray}\phi} \big)  \equiv0,
\end{align}
 then it would be a 2-cocycle on $\diff(M)$, meaning that  the Poisson bracket \eqref{Poisson-bracket-on-shell}  is a Lie bracket\footnote{ Indeed, since by definition on-shell $\mathscr C\big( \lfloor X, Y\! \rfloor ;\phi \big)=\big\{Q_\Sigma (X; \phi), Q_\Sigma (Y; \phi) \big\} - Q_\Sigma \big([X, Y]_{\text{{\tiny $\diff(M)$}}};\phi \big)$, \eqref{2-cocycle-relation} is equivalent to the Jacobi identity holding for both $[\ ,\ ]_{\text{{\tiny $\diff(M)$}}}$ and the Poisson bracket $\{\ ,\ \}$.} 
 and that the Poisson algebra of Noether charges, PAlg$[Q_\Sigma]$, is a central extension of $\diff(M)$: i.e. there is a short exact sequence of Lie algebras, 
\begin{align}
\label{PB-central-ext}
\makebox[\displaywidth]{
\hspace{-18mm}\begin{tikzcd}[column sep=large, ampersand replacement=\&]
\      \&  \big( C^\infty(\Phi); \ \cdot \ \big)   \arrow[r, "\iota"  ]          \&   \text{PAlg}[Q_\Sigma]= \big( Q_\Sigma (\ \,; \phi) ; \{\ ,\ \} \big)     \arrow[r, "\pi"]      \&      \diff(M)=\big( \Gamma(TM) ; -[\ \, , \ ]_{\text{{\tiny $\Gamma(TM)$}}} \big)    \arrow[l, bend left=20,  start anchor={[xshift=4ex]}, end anchor={[xshift=-1ex]}, "Q_\Sigma(\, \ ;\,\phi)"]         
\end{tikzcd}}  \raisetag{8.5ex}
\end{align}
with $\im \iota \subset Z\big(\text{PAlg}[Q_\Sigma] \big)$ and $Q_\Sigma(\, \ ;\phi): \diff(M) \rarrow \text{PAlg}[Q_\Sigma]$, $X \mapsto Q_\Sigma(X ;\phi)$, a linear map s.t. $\pi \circ Q_\Sigma(\, \ ;\phi) =\id_{\,\diff(M)}$ (it is a section of $\pi$, \emph{not} a Lie algebra morphism). 
Since $\pi$ is a Lie algebra morphism, and since on-shell we have $\mathscr C\big( \lfloor X, Y\! \rfloor ;\phi \big)=\big\{Q_\Sigma (X; \phi), Q_\Sigma (Y; \phi) \big\} - Q_\Sigma \big([X, Y]_{\text{{\tiny $\diff(M)$}}};\phi \big)$ by \eqref{Poisson-bracket-on-shell}, it is clear that $\mathscr C\big( \lfloor X, Y\! \rfloor ;\phi \big) \in \ker \pi = \im \iota \subset Z\big(\text{PAlg}[Q_\Sigma] \big)$.

 There is no guarantee a priori that \eqref{2-cocycle-relation} holds.
  To ascertain the fact of the matter, 
 we compute the 2-cocycle condition on $\mathscr C$ in Appendix \ref{A4}, and from \eqref{2-cocycle-condition-diff}  we obtain:
 \begin{align}
\label{Jacobiator-PB}
 \mathscr C \big( \lfloor X, [Y, Z]_{\text{{\tiny $\diff(M)$}}}\rfloor ; \phi \big) + c.p.  = \int_\Sigma \big( \bs L_{X^v} \bs\Theta \big)(Y^v, Z^v) + c.p.=  \int_\Sigma \big(\bs d \iota_{X^v} \bs\Theta \big) (Y^v, Z^v) + c. p.
\end{align}
Therefore, we see that the on-shell bracket \eqref{Poisson-bracket-on-shell} 
 is a Poisson-Lie bracket if either 
{\bf a)}  
fundamental vector fields $X^v \in \Gamma(V\Phi)$ generated by $\diff(M)$
act as symplectomorphisms -- which, as we remarked already, is not generically the case on account of \eqref{non-symplectomorph} -- or 
{\bf b)} if  $X^v \in \Gamma(V\Phi)$ are Hamiltonian, i.e. if the Noether charges are moment maps for $\diff(M)$: 
Given \eqref{Moment-map-equation}, this happens {\bf 1)} \emph{on-shell} (as is already assumed here) and {\bf 2)} given adequate \emph{boundary conditions}, as indeed \eqref{Jacobiator-PB} is written as
 \begin{align}
\label{Jacobiator-PB-bis}
 \mathscr C \big( \lfloor X, [Y, Z]_{\text{{\tiny $\diff(M)$}}}\rfloor ; \phi \big) + c.p.  &=  \bs d \left( \int_{\d\Sigma }   \iota_X \bs \theta - \bs d\gamma(X;\phi)\  \, - \cancel{\int_\Sigma  \iota_X \bs E} \ \right)  (Y^v, Z^v) + c. p. 
\end{align}
If pressed for an interpretation, we may remark that on-shell, the obstruction to \eqref{Poisson-bracket-def}-\eqref{Poisson-bracket-result} to be a Lie bracket is given by a boundary term often interpreted as the ``symplectic flux" leaking through $\d\Sigma$,\footnote{
From  \eqref{non-symplectomorph} one has obviously that \eqref{Jacobiator-PB}  is rewritten $ \mathscr C \big( \lfloor X, [Y, Z]_{\text{{\tiny $\diff(M)$}}}\rfloor ; \phi \big) + c.p. = \left(\int_{\Sigma } \mathfrak L_X\bs\Theta \right)    (Y^v, Z^v) + c. p.$ This expression of the integrand would indeed lends itself easily to the terminology ``symplectic flux".
} 
suggesting the obstruction would come from the system being ``open".

For the sake of completeness, and definiteness -- and because the result will be used later -- let us write the concrete expression of the map $\mathscr C\big( \lfloor X, Y\! \rfloor ;\phi \big)$ \eqref{2-cocycle-map}. A straightforward computation, detailed in Appendix \ref{A5}, gives
\begin{align}
\label{Cocycle-concrete}
\mathscr C\big( \lfloor X, Y\! \rfloor ;\phi \big) &= \int_{\d\Sigma} \mathscr{A}\big( \lfloor X, Y\! \rfloor ; \phi \big)  + \gamma \big([X, Y]_{\text{{\tiny $\diff(M)$}}};\phi \big), \notag\\
								    &= \int_{\d\Sigma} \mathfrak L_X\,  \theta(\iota_Y\phi; \phi) - \mathfrak L_Y\, \theta(\iota_X\phi; \phi) + \iota_X\iota_Y L 
								          +  \gamma \big([X, Y]_{\text{{\tiny $\diff(M)$}}}; \phi \big)
								            \ \ +\ \ \iota_Y E(\iota_X\phi; \phi) - \iota_X E(\iota_Y\phi; \phi). 
\end{align}
So, on-shell and with the particular b.c. such that the Noether charges are \eqref{BC-imposed} (ensuring that the system is ``closed"), it is a 2-cocycle expressed in terms of $L$, $\ell$ and $\theta$:
\begin{align}
\label{Cocycle-concrete-on-shell-b-c}
\mathscr C\big( \lfloor X, Y\! \rfloor ;\phi \big) = \int_{\d\Sigma} \mathfrak L_X\,  \theta(\iota_Y\phi; \phi) - \mathfrak L_Y\, \theta(\iota_X\phi; \phi) + \iota_X\iota_Y L 
								          -   \iota_{[X, Y]_{\text{{\tiny $\Gamma(TM)$}}} } \ell.  
\end{align}
These results will enter the concrete expression for the vertical/gauge transformation of the presymplectic 2-form. 
\bigskip

We may observe that it is ``\emph{loisible}", as French say, to modify the standard prescription \eqref{Poisson-bracket-def} for the bracket of charges, substracting the 2-cocyle $\mathscr C$ to as to define via \eqref{Poisson-bracket-on-shell} an  on-shell bracket representing $\diff(M)$:
\begin{align}
\label{Poisson-bracket-redef}
\big\{Q_\Sigma (X; \phi), Q_\Sigma (Y; \phi) \big\}_{\text{{\tiny ``new"}}} \defeq   \bs\Theta_\Sigma \big(X^v, Y^v \big)   - \mathscr C\big( \lfloor X, Y\! \rfloor ;\phi \big) = Q_\Sigma \big([X, Y]_{\text{{\tiny $\diff(M)$}}};\phi \big)_{\ | \S}\,.
\end{align}
The value of such a move of course depends on what one's aim is. One issue is that contrary to the original prescription, it is not clear how the bracket \eqref{Poisson-bracket-redef} would generate $\diff(M)$-transformations of functions on $\Phi$. Indeed, for $\bs f \in C^\infty(\Phi)$, admitting we have the associated hamiltonian vector field $V_{\bs f}$ defined via $\iota_{V_{\bs f}} \bs \Theta_\Sigma =- \bs d \bs f$, the standard bracket by a charge $Q_\Sigma (X; \phi)$ associated to $X^v \in \Gamma(V\Phi)$  is:
\begin{align}
\big\{Q_\Sigma (X; \phi), \bs f \big\} = \iota_{V_{\bs f}} \iota_{X^v} \bs \Theta_\Sigma = - \iota_{X^v}\iota_{V_{\bs f}} \bs \Theta_\Sigma = \iota_{X^v}\bs d \bs f = \bs L_{X^v} \bs f.
\end{align}
This is indeed by definition the infinitesimal equivariance of $\bs f$. 
We notice it could be computed, on-shell,  once the expression for $V_{\bs f}$ found via its definition and using  \eqref{BC-imposed}: 
 $ \bs L_{X^v} \bs f = \iota_{V_{\bs f}} \iota_{X^v} \bs \Theta_\Sigma = - \iota_{V_{\bs f}} \bs d Q_\Sigma (X;\phi) \ _{|\S} $.

 \subsection{Vertical and gauge transformations}  
\label{Vertical and gauge transformations}  
 
 We now turn to the issue of the vertical/gauge transformations -- a.k.a. \emph{field-dependent} gauge transformations -- of the objects defined above. 
 These are obtained  geometrically,  relying on the conceptual and technical resources of section~\ref{Geometry of field space}. 
 This is not an idle application: it will first allow to see explicitly that the variational principle remains well-defined under the action of field-dependent diffeomorphism, a non-trivial result 
 Then it will also allow to  extend the construction to charges and Poisson bracket for field-dependent parameters (vector fields), section \ref{Noether charges for field-dependent gauge parameters}.
 And finally, it will be the basis  from which to apply the rule of thumb of the DFM, easily obtaining the basic (relational) counterparts of all objects considered, section \ref{Dressed presymplectic structure}.  
 \medskip
 
 Let us recall that the group of general vertical diffeomorphisms of $\Phi$ is  $\bs\Diff_v(\Phi) \simeq C^\infty\big(\Phi, \Diff(M) \big)$, and  contains in particular the vertical automorphisms group/gauge group of $\Phi$, $\bs\Aut_v(\Phi)\simeq\bs\Diff(M)$: vertical transformations are given by the action of these groups via pullback. 
 Their Lie algebras are   $\bs\diff_v(\Phi) \simeq C^\infty\big(\Phi, \diff(M) \big)$ and $\bs\aut_v(\Phi)\simeq\bs\diff(M)$: these can be seen to be elements $ \Omega^0(\Phi, V\Phi) \subset \Omega^\bullet(\Phi, T\Phi)$, so infinitesimal vertical transformations are given by the action of these Lie algebras via  the Nijenhuis-Lie derivative \eqref{NL-derivative}. 
 
   To find the vertical transformations of $L'$, $\bs d L'$,$\bs E$, $\bs\theta$ and $\bs\Theta$, we need only to apply the natural geometric definition \eqref{GT-geometric} of section \ref{General vertical transformations, and gauge transformations}: Collect the $\Diff(M)$-equivariance and verticality properties, and use  \eqref{pushforward-X}
    \begin{align}
 \Xi_\star \mathfrak X = R_{\bs\psi \star} \mathfrak X + \left\{   \bs\psi\-_* \bs d \bs\psi(\mathfrak X)   \right\}^v 
 					    =R_{\bs\psi \star} \left( \mathfrak X +  \left\{  \bs d \bs\psi(\mathfrak X)  \circ \bs\psi\-  \right\}^v \right),
 \end{align}
 for $\bs\psi \in C^\infty\big(\Phi, \Diff(M) \big)$ (or $\bs\Diff(M)$) corresponding to $\Xi \in \bs\Diff_v(\Phi)$ (or $\bs\Aut_v(\Phi)$), to get  \eqref{GT-geometric} for $\bs\alpha \in \Omega^\bullet(\Phi)$,
 \begin{equation}
\begin{aligned}
\bs\alpha^{\bs\psi} (\mathfrak X, \ldots ) =  \Xi^\star \bs\alpha (\mathfrak X, \ldots)= \bs\alpha (\Xi_\star \mathfrak X, \ldots) &= \bs\alpha \left( R_{\bs\psi \star} \big( \mathfrak X +  \left\{  \bs d \bs\psi(\mathfrak X)  \circ \bs\psi\-  \right\}^v \big), \ldots\right),  \\
&=R_{\bs\psi}^\star \bs\alpha \left(  \mathfrak X +  \left\{  \bs d \bs\psi (\mathfrak X)  \circ \bs\psi\-  \right\}^v, \ldots  \right),
\end{aligned}
\end{equation}
whose linear version is given by \eqref{Inf-2-Vert-Trsf}/\eqref{Inf-GT-general}. 
Alternatively, or as a crosscheck, given the expression $\bs\alpha = \alpha\big(\! \w^\bullet \!\bs d\phi; \phi \big)$, one can rely on  \eqref{GT-basis-1-form}
\begin{align}
  \bs d\phi^{\bs\psi}   \defeq \Xi^\star \bs d \phi = \bs\psi^*   \big(  \bs d\phi   +   \mathfrak L_{ [\bs d \bs\psi  \circ \bs\psi\-]} \phi  \big).   
 \end{align}
 to determine (by multi-linearity) \eqref{GT-general}: $\bs\alpha^{\bs\psi}  = \alpha\big(\! \w^\bullet \! \bs d\phi^{\bs\psi} ;\,  \phi^{\bs\psi} \big)$. 
Then, the transformations of corresponding integrated quantities -- $S'$, $\bs d S'$, $\bs\theta_\Sigma$ and $\bs\Theta_\Sigma$ -- as objects on $\Phi \times \bs U(M)$, are obtained via \eqref{Vert-trsf-int-generic}.

  \medskip
  For $L'\in \Omega_\text{tens}^0(\Phi)$, verticality is trivial, so the vertical transformation is  controlled by its equivariance  \eqref{Lagrangian}:
  \begin{align}
  \label{GT-L}
 {L'}^{\,\bs\psi} \defeq \Xi^\star L' = R^\star_{\bs\psi} L' = \bs\psi^* L'. 
  \end{align}
  From this, or from \eqref{Inf-GT-general}-\eqref{Inf-GT-eq}, follows that the $\bs\diff_v(\Phi) \simeq C^\infty\big(\Phi, \diff(M) \big)$ transformation is simply:  
\begin{align}  
\label{NL-der-L'}
  \bs L_{\bs X^v} L' = \mathfrak L_{\bs X} L'= \alpha(\bs X; \phi)= d\beta(\bs X;\phi),
  \end{align}
  with $\bs X \in C^\infty\big(\Phi, \diff(M) \big)$ a field-dependent vector field of $M$.
  
  For $\bs d L'\in \Omega_\text{eq}^1(\Phi)$, from \eqref{Lagrangian}-\eqref{d-Lagrangian} we get $R^\star_\psi\, \bs dL'=\psi^* \bs d L'$ and $\iota_{X^v} \bs dL' = \alpha(X; \phi)$. So, 
 \begin{align}
 {\bs dL'}^{\,\bs\psi} (\mathfrak X) \defeq \Xi^\star \bs d L' (\mathfrak X) &= R^\star_{\bs\psi}\, \bs d L '  \left( \mathfrak X +  \left\{  \bs d \bs\psi(\mathfrak X)  \circ \bs\psi\-  \right\}^v \right), \notag \\
 												    &= \bs\psi^* \left( \bs d L' (\mathfrak X)+ \alpha\big(  \bs d \bs\psi(\mathfrak X)  \circ \bs\psi\- ;\  \phi \big) \right), \notag \\[2mm]
 \hookrightarrow    \quad  {\bs dL'}^{\,\bs\psi}  &=  \bs\psi^* \left( \bs d L' + \alpha\big(  \bs d \bs\psi  \circ \bs\psi\- ;\  \phi \big) \right).    \label{GT-dL}
 \end{align}
 Of course, this could have been also obtained from \eqref{GT-L}, since the \emph{naturality of the pullback} $[\Xi^\star, \bs d]=0$ implies ${\bs dL'}^{\,\bs\psi}=\bs d({L'}^{\,\bs\psi})$.
 The corresponding linear version is, from \eqref{GT-dL} or \eqref{Inf-GT-general}-\eqref{Inf-GT-eq}/\eqref{Inf-GT-diff-rep}: 
 \begin{align}  
 \label{GT-dL-fied-dep}
  \bs L_{\bs X^v} \bs dL' = \mathfrak L_{\bs X} \bs d L' +  \alpha(\bs{dX}; \phi), 
  \end{align}
which could have equally been found via \eqref{NL-der-L'}, since $[\bs L_{\bs X^v}, \bs d]=0$. 

 For $\bs E\in \Omega_\text{eq}^1(\Phi)$, we have  $R^\star_\psi\, \bs E=\psi^* \bs E$ and $\iota_{X^v} \bs E = -d J(X; \phi)= dE\big(\iota_X \phi ;\phi\big)$. So, 
 \begin{align}
 &\bs E^{\,\bs\psi} (\mathfrak X) \defeq \Xi^\star \bs E (\mathfrak X) = R^\star_{\bs\psi}\, \bs E \left( \mathfrak X +  \left\{  \bs d \bs\psi(\mathfrak X)  \circ \bs\psi\-  \right\}^v \right) 
 												    = \bs\psi^* \left( \bs E (\mathfrak X)+ d E\big(   \iota_{\ \bs d \bs\psi(\mathfrak X)  \circ \bs\psi\-}\phi ;\  \phi \big) \right), \notag \\[2mm]
 \hookrightarrow   & \quad  \bs E^{\,\bs\psi}  =  \bs\psi^* \left( \bs E + dE\big(  \iota_{\bs d \bs\psi  \circ \bs\psi\-}\phi ;\  \phi \big) \right).    \label{GT-E}
   \end{align}
 This relation tells us that the action of field-dependent diffeomorphisms $\bs\Diff_v(\Phi) \simeq C^\infty\big(\Phi, \Diff(M) \big)$ preserves the space of solutions $\S\defeq \{\phi\in \Phi\,|\, \bs E_{|\phi} =0\}$,  a  fact essential for the covariant phase space approach,
 \footnote{We may also observe -- continuing footnote \ref{Myfootnote} --  that had we postulated an additional  contribution $\alpha_\gamma (X; \phi)$ to the Lagrangian $\Diff(M)$-anomaly $\alpha(X; \phi)$, it would have entered the vertical property of $\bs E$ through the (anomalous) current: $\iota_{X^v} \bs E = -d J(X; \phi) + \alpha_\gamma (X; \phi)= dE\big(\iota_X \phi ;\phi\big) + \alpha_\gamma (X; \phi)$. So we would have \vspace{-1mm}
 \begin{align}
 \bs E^{\,\bs\psi}  =  \bs\psi^* \left( \bs E + dE\big(  \bs d \bs\psi  \circ \bs\psi\- ;\  \phi \big) + \alpha_\gamma\big(  \bs d \bs\psi  \circ \bs\psi\- ;\  \phi \big) \right),
 \end{align}
 meaning that $\bs\Diff_v(\Phi) \simeq C^\infty\big(\Phi, \Diff(M) \big)$ would generically not preserve $\S$ (unless $\alpha_\gamma (X; \phi)=d\beta_\gamma (X; \phi)$ and given adequate b.c.). However, such additional anomalies only arise in theories with background structures, i.e. theories that are not truly general relativistic, which we do not wish to consider.}
 %
 showing that $\Diff(M)$-theories naturally enjoy a larger $\bs\Diff_v(\Phi) \simeq C^\infty\big(\Phi, \Diff(M) \big)$-covariance. 
 As observed at the end of section \ref{Integration map}, this fact was stressed in the case of GR by Bergmann and Komar \cite{Bergmann-Komar1972}.
 The~linear version of the above is, 
 \begin{align}  
  \bs L_{\bs X^v} \bs E = \mathfrak L_{\bs X} \bs E +  dE(\iota_{\bs{dX}}\phi; \phi).
  \end{align}

 Now, for $\bs \theta \in \Omega_\text{eq}^1(\Phi)$ we have still $R^\star_\psi\, \bs\theta=\psi^* \bs \theta$, and by \eqref{Identity2}:  $\iota_{X^v} \bs \theta = d\theta( \iota_X\phi; \phi) + \iota_X L  -\, E(\iota_X \phi; \phi)$. So, 
  \begin{align}
 &\bs \theta^{\,\bs\psi} (\mathfrak X) \defeq \Xi^\star \bs \theta(\mathfrak X) = R^\star_{\bs\psi}\, \bs \theta \left( \mathfrak X +  \left\{  \bs d \bs\psi(\mathfrak X)  \circ \bs\psi\-  \right\}^v \right) 
 							 = \bs\psi^*\bs \theta  \left(\mathfrak X + \ldots  \phantom{\big|^v}
							              \right),  \notag \\[2mm]
 \hookrightarrow   & \quad  \bs \theta^{\,\bs\psi}  =  \bs\psi^* \left( \bs \theta +  d \theta \big( \iota_{\bs d \bs\psi \circ \bs\psi\-}\phi ;\  \phi \big) 
							      + \iota_{  \bs d \bs\psi  \circ \bs\psi\-} L  
							      -\, E(\iota_{  \bs d \bs\psi \circ \bs\psi\-}\phi; \phi) \right).    \label{GT-theta}
   \end{align}
Linearly, we thus get
 \begin{align}  
 \label{Inf-vert-trsf-theta}
  \bs L_{\bs X^v} \bs \theta = \mathfrak L_{\bs X} \bs \theta +  d \theta \big( \iota_{\bs{dX}}\phi ;\  \phi \big) 
							      + \iota_{ \bs{dX}} L  
							      -\, E(\iota_{ \bs{dX}}\phi; \phi).
  \end{align}
 as also found via by \eqref{Inf-GT-general}-\eqref{Inf-GT-eq}/\eqref{Inf-GT-diff-rep}. 
For the presymplectic potential $\bs \theta_\Sigma\defeq \int_{\,\Sigma} \bs\theta$,  
by \eqref{Vert-trsf-int-generic}  we have thus:
\begin{align}
{\bs \theta_\Sigma}^{\bs\psi}_{\phantom 0}  &   \defeq \int_{\bs\psi\-(\Sigma)} \hspace{-3mm}  \bs\theta^{\,\bs \psi} =  \int_{\bs\psi\-(\Sigma) } \bs\psi^* \left( \bs \theta +  d \theta \big(  \iota_{\bs d \bs\psi \circ \bs\psi\-}\phi ;\  \phi \big) 
							      + \iota_{  \bs d \bs\psi  \circ \bs\psi\-} L  
							      -\, E(\iota_{  \bs d \bs\psi \circ \bs\psi\-}\phi; \phi) \right), \notag \\
					  &=  \int_{\Sigma } \bs \theta +  d \theta \big( \iota_{\bs d \bs\psi \circ \bs\psi\-}\phi ;\  \phi \big) 
							      + \iota_{  \bs d \bs\psi  \circ \bs\psi\-} L  
							      -\, E(\iota_{  \bs d \bs\psi \circ \bs\psi\-}\phi; \phi).   \label{GT-theta-sigma}
\end{align}
From which one reads the linear version: $\delta_{\bs X} {\bs \theta_\Sigma} = \int_\Sigma d \theta \big(  \iota_{\bs d \bs X}\phi ;\  \phi \big) 
							      + \iota_{  \bs d \bs X } L  
							      -\, E(\iota_{  \bs d \bs X }\phi; \phi)$,
cross-checked via \eqref{linear-versions-integral},
\begin{equation}
\label{inf-vert-trsf-pot}
 \begin{aligned}
\delta_{\bs X} {\bs \theta_\Sigma} \defeq&\, \langle \bs L_{\bs X^v} \bs\theta,\,  \Sigma \rangle  + \langle \bs\theta,\, -\bs X(\Sigma)   \rangle, \\
						            =&\, \langle \mathfrak L_{\bs X} \bs \theta +  d \theta \big( \iota_{\bs{dX}}\phi ;\  \phi \big) 
							      + \iota_{ \bs{dX}} L  
							      -\, E(\iota_{ \bs{dX}}\phi; \phi),  \Sigma \rangle  + \langle \bs\theta,\, -\bs X(\Sigma)   \rangle, \\
					                     =&\, \langle   d \theta \big( \iota_{\bs{dX}}\phi ;\  \phi \big) 
							      + \iota_{ \bs{dX}} L  
							      -\, E(\iota_{ \bs{dX}}\phi; \phi),  \Sigma \rangle  
                                                      + \underbrace{ \langle \mathfrak L_{\bs X} \bs \theta ,  \Sigma \rangle + \langle \bs\theta,\, -\bs X(\Sigma)   \rangle}_{=0 \text{ by \eqref{inf-invariance-action} }  }.
\end{aligned}
\end{equation}
   From   \eqref{GT-theta}-\eqref{GT-theta-sigma}-\eqref{inf-vert-trsf-pot} we observe that, even on-shell, in general the lack of invariance of the presymplectic potential under 
   $\bs\Diff_v(\Phi) \simeq C^\infty\big(\Phi, \Diff(M) \big)$ is not only a \emph{boundary problem} -- unless $\iota_X L=0$ for a specific $L$ (like GR) -- but a bulk issue: the introduction of \emph{edge modes}, d.o.f. located  at $\d\Sigma$, cannot solve this issue (and ``restore invariance").

Let us remark that from \eqref{GT-E},  \eqref{GT-theta} and  \eqref{GT-dL}, one finds: 
\begin{equation}
\label{GT-Var-Principle}
\begin{aligned}
\langle \bs E^{\bs \psi}  + d \bs \theta^{\bs \psi}, \bs\psi\-(U) \rangle &= \langle \bs dL + \mathfrak L_{\bs{d\psi}\circ \bs\psi\-} L, U \rangle 
												      =\langle \bs dL^{\bs \psi},  \bs\psi\-(U)\rangle, \\	
		 										&=\bs d S + \langle \mathfrak L_{\bs{d\psi}\circ \bs\psi\-} L, U \rangle = (\bs dS)^{\bs\psi}, 
\end{aligned}
\end{equation}
as expected from \eqref{Vert-trsf-dS}. This highlights again the fact that the variational principle is well-defined, and the space of solutions $\S$ stable, under the action of $\bs\Diff_v(\Phi) \simeq C^\infty\big(\Phi, \Diff(M) \big)$. 
   \medskip
   
Turning finally to  $\bs \Theta \in \Omega_\text{eq}^2(\Phi)$ we have $R^\star_\psi\, \bs\Theta=\psi^* \bs \Theta$, and by \eqref{Moment-map-equation}:  
$\iota_{X^v} \bs \Theta = -\bs d J(X;\phi) +d \left(  \iota_X \bs \theta - \bs d \gamma(X;\phi) \right)  - \iota_X \bs E$. 
So, for $\mathfrak X, \mathfrak Y \in \Gamma(T\Phi)$:
\begin{equation}
 \begin{aligned}
 \bs \Theta^{\bs\psi} (\mathfrak X, \mathfrak Y) \defeq \Xi^\star \bs \Theta(\mathfrak X, \mathfrak Y) 
 	                                                           &= R^\star_{\bs\psi}\, \bs \Theta \left( \mathfrak X +  \left\{  \bs d \bs\psi(\mathfrak X)  \circ \bs\psi\-  \right\}^v\!,
												\ \mathfrak Y +  \left\{  \bs d \bs\psi(\mathfrak Y)  \circ \bs\psi\- \right\}^v \right) ,\\
								&= {\bs\psi}^* \bs \Theta \left( \mathfrak X +  \left\{  \bs d \bs\psi(\mathfrak X)  \circ \bs\psi\-  \right\}^v\!,
												\ \mathfrak Y +  \left\{  \bs d \bs\psi(\mathfrak Y)  \circ \bs\psi\- \right\}^v \right), \\
								&=  {\bs\psi}^* \left[ \, \bs \Theta \big( \mathfrak X,  \mathfrak Y\big) 
											+  \bs \Theta \left( \mathfrak X,   \left\{  \bs d \bs\psi(\mathfrak Y)  \circ \bs\psi\- \right\}^v \right)
											+  \bs \Theta \left( \left\{  \bs d \bs\psi(\mathfrak X)  \circ \bs\psi\- \right\}^v,  \mathfrak Y \right) 
 \right.  \\ &\hspace{4.5cm}                       + \left. \bs \Theta \left( \left\{  \bs d \bs\psi(\mathfrak X)  \circ \bs\psi\- \right\}^v, 
 													\left\{  \bs d \bs\psi(\mathfrak Y)  \circ \bs\psi\- \right\}^v \right)
								\, \right]
\end{aligned}
\end{equation}
The last term is related to the expression of the bracket \eqref{Poisson-bracket-def}, so by \eqref{pre-bracket} it is immediately:
\begin{equation}
\label{c1}
    \begin{aligned}
\bs \Theta \left( \left\{  \bs d \bs\psi(\mathfrak X)  \circ \bs\psi\- \right\}^v, \left\{  \bs d \bs\psi(\mathfrak Y)  \circ \bs\psi\- \right\}^v \right)  
 	&= J \big([ \bs d \bs\psi(\mathfrak X)  \circ \bs\psi\-\!,\   \bs d \bs\psi(\mathfrak Y)  \circ \bs\psi\-]_{\text{{\tiny $\diff(M)$}}};\phi \big)   \\
	&\quad + d \gamma \big([ \bs d \bs\psi(\mathfrak X)  \circ \bs\psi\-\!, \  \bs d \bs\psi(\mathfrak Y)  \circ \bs\psi\-]_{\text{{\tiny $\diff(M)$}}};\phi \big) \\
	&\quad + d\mathscr{A}\big( \lfloor  \bs d \bs\psi(\mathfrak X)  \circ \bs\psi\-\!, \  \bs d \bs\psi(\mathfrak Y)  \circ \bs\psi\-\! \rfloor ;\phi \big)  \\
	  	&\quad + \iota_{\left\{  \bs d \bs\psi(\mathfrak X)  \circ \bs\psi\- \right\}^v} \iota_{ \bs d \bs\psi(\mathfrak Y)  \circ \bs\psi\-} \bs E  
			-  \iota_{\left\{  \bs d \bs\psi(\mathfrak Y)  \circ \bs\psi\- \right\}^v}  \iota_{ \bs d \bs\psi(\mathfrak X)  \circ \bs\psi\-} \bs E. 
\end{aligned}
\end{equation}
The two middle terms are can be written at once via the ``moment map/integrability relation" \eqref{Moment-map-equation}:  
\begin{equation}
\label{c2}
 \begin{aligned}
& - \iota_{\mathfrak X} \iota_{  \left\{  \bs d \bs\psi(\mathfrak Y)  \circ \bs\psi\- \right\}^v} \bs \Theta 
 + \iota_{\mathfrak Y} \iota_{  \left\{  \bs d \bs\psi(\mathfrak X)  \circ \bs\psi\- \right\}^v} \bs \Theta  = \\
&\hspace{3cm}   - \iota_{\mathfrak X}  \left(  -\bs d \,J\big( \bs d \bs\psi(\mathfrak Y)  \circ \bs\psi\-;\ \munderline{ForestGreen}{\phi}\big) 
		+ d \left(  \iota_{ \bs d \bs\psi(\mathfrak Y)  \circ \bs\psi\-} \bs \theta 
		- \bs d \,\gamma\big( \bs d \bs\psi(\mathfrak Y)  \circ \bs\psi\-;\ \munderline{ForestGreen}{\phi}\big) \right)  
		- \iota_{ \bs d \bs\psi(\mathfrak Y)  \circ \bs\psi\-} \bs E \right) \\
&\hspace{3cm} + \iota_{\mathfrak Y}  \left(  -\bs d\, J\big( \bs d \bs\psi(\mathfrak X)  \circ \bs\psi\-;\ \munderline{ForestGreen}{\phi}\big) 
		+ d \left(  \iota_{ \bs d \bs\psi(\mathfrak X)  \circ \bs\psi\-} \bs \theta 
		- \bs d\, \gamma\big( \bs d \bs\psi(\mathfrak X)  \circ \bs\psi\-;\ \munderline{ForestGreen}{\phi}\big) \right)  
		- \iota_{ \bs d \bs\psi(\mathfrak X)  \circ \bs\psi\-} \bs E \right).
\end{aligned}
\end{equation}
In the above expressions, it should be noticed that terms like
$\bs d \bs\psi (\mathfrak Y)  \circ \bs\psi\-$ are considered $\phi$-independent so that in $\iota_{\mathfrak X}  \bs d  \,  J\left( \bs d \bs\psi(\mathfrak Y)  \circ \bs\psi\-;\  \munderline{ForestGreen}{\phi} \right) = \mathfrak X \cdot   J\left( \bs d \bs\psi(\mathfrak Y)  \circ \bs\psi\-;\  \munderline{ForestGreen}{\phi} \right)$, the derivative $\bs d$ or vector field ${\mathfrak X} $ act only on the usual (underlined) $\phi$-dependence of $J$. 

 Yet,  noticing that  
   $\bs d \big( -\bs d \bs\psi  \circ \bs\psi\- \big) + \tfrac{1}{2}[ \bs d \bs\psi  \circ \bs\psi\-\!,\   \bs d \bs\psi \circ \bs\psi\-]_{\text{{\tiny $\diff(M)$}}} =0$ 
 and using Koszul formula for $\bs d$, the current term in \eqref{c1} is:
\begin{align*}
 J \left( \big[\bs d \big( \bs d \bs\psi  \circ \bs\psi\- \big)\big] (\mathfrak X, \mathfrak Y);\phi \right) =   
    J\left(  \mathfrak X \cdot  [\bs d \bs\psi(\mathfrak Y)  \circ \bs\psi\-];\ \phi \right)  
  -  J\left(  \mathfrak Y \cdot [\bs d \bs\psi(\mathfrak X)  \circ \bs\psi\-];\ \phi \right)   
  - J\left(    [\bs d \bs\psi  \circ \bs\psi\-] ([\mathfrak X, \mathfrak Y]);\ \phi \right). 
\end{align*}
This combines with the current terms in  \eqref{c2}, so that:
\begin{align}
 &\ J\left(  \mathfrak X \cdot  [\bs d \bs\psi(\mathfrak Y)  \circ \bs\psi\-];\ \phi \right)  
  -  J\left(  \mathfrak Y \cdot [\bs d \bs\psi(\mathfrak X)  \circ \bs\psi\-];\ \phi \right)   
  - J\left(    [d \bs\psi  \circ \bs\psi\-] ([\mathfrak X, \mathfrak Y]);\ \phi \right)  \notag\\
  &+ \mathfrak X \cdot   J\left( \bs d \bs\psi(\mathfrak Y)  \circ \bs\psi\-;\ \munderline{ForestGreen}{\phi}\right)  
  - \mathfrak Y \cdot    J\left( \bs d \bs\psi(\mathfrak X)  \circ \bs\psi\-;\ \munderline{ForestGreen}{\phi} \right)  \notag \\
=&\ \mathfrak X \cdot   J\left( \bs d \bs\psi(\mathfrak Y)  \circ \bs\psi\-;\ \phi \right)  
  - \mathfrak Y \cdot    J\left( \bs d \bs\psi(\mathfrak X)  \circ \bs\psi\-;\ \phi \right) 
  - J\left(    [d \bs\psi  \circ \bs\psi\-] ([\mathfrak X, \mathfrak Y]);\ \phi \right) \notag\\
=&\!:  \bs d \big[ J\big( \bs d \bs\psi  \circ \bs\psi\- ; \phi\big) \big]  (\mathfrak X, \mathfrak Y). 
\end{align}
We may then already write a nice expression for the $\bs\Diff_v(\Phi) \simeq C^\infty\big(\Phi, \Diff(M) \big)$-transformation of $\bs\Theta$:
\begin{equation}
\label{c3}
\begin{aligned}
 \bs \Theta^{\bs\psi} =  {\bs\psi}^*  \left[ \phantom{\tfrac{1}{2}}\!\! \bs \Theta  
 				\right.& +   \bs d  J\big( \bs d \bs\psi  \circ \bs\psi\- ; \phi\big)  \\[1mm]
 				& + \tfrac{1}{2} \left(   d\mathscr{A}\big( \lfloor  \bs d \bs\psi  \circ \bs\psi\-\!, \  \bs d \bs\psi \circ \bs\psi\-\! \rfloor ;\phi \big)
				   + d \gamma \big([ \bs d \bs\psi \circ \bs\psi\-\!, \  \bs d \bs\psi \circ \bs\psi\-]_{\text{{\tiny $\diff(M)$}}};\phi \big) \right)\\[1mm]
				& + d \left(  \iota_{ \bs d \bs\psi  \circ \bs\psi\-} \bs \theta   - \bs d \,\gamma\big( \bs d \bs\psi  \circ \bs\psi\-;\ \munderline{ForestGreen}{\phi}\big) \right) \\[1mm]
				& +  \iota_{\left\{  \bs d \bs\psi \circ \bs\psi\- \right\}^v} \iota_{ \bs d \bs\psi  \circ \bs\psi\-} \bs E 
						- \iota_{ \bs d \bs\psi \circ \bs\psi\-} \bs E \left. \phantom{\tfrac{1}{2}}\!\!  \right].
\end{aligned}
\end{equation}
Using further the expression for the current \eqref{Noether-current2} we further find:
\begin{equation}
\begin{aligned}
 \bs \Theta^{\bs\psi} =  {\bs\psi}^*  \left[ \phantom{{\tfrac{1}{2}}}\!\! \bs \Theta  
 				\right.& 
		+  d\  \bs d\,  \left( \theta\big(\iota_{\bs d \bs\psi  \circ \bs\psi\-} \phi; \phi\big) - \gamma\big(\bs d \bs\psi  \circ \bs\psi\- ; \phi\big) \right)\\[1mm]
 				& + \tfrac{1}{2} \left(   d\mathscr{A}\big( \lfloor  \bs d \bs\psi  \circ \bs\psi\-\!, \  \bs d \bs\psi \circ \bs\psi\-\! \rfloor ;\phi \big)
				   + d \gamma \big([ \bs d \bs\psi \circ \bs\psi\-\!, \  \bs d \bs\psi \circ \bs\psi\-]_{\text{{\tiny $\diff(M)$}}};\phi \big) \right)\\[1mm]
				& + d \left(  \iota_{ \bs d \bs\psi  \circ \bs\psi\-} \bs \theta   - \bs d \,\gamma\big( \bs d \bs\psi  \circ \bs\psi\-;\ \munderline{ForestGreen}{\phi}\big) \right) \\[1mm]
				&   +  \iota_{\left\{  \bs d \bs\psi \circ \bs\psi\- \right\}^v} \iota_{ \bs d \bs\psi  \circ \bs\psi\-} \bs E  
							- \iota_{ \bs d \bs\psi  \circ \bs\psi\-} \bs E 
							- \bs d\, E(\iota_{\bs d \bs\psi  \circ \bs\psi\-} \phi; \phi)  \left. \phantom{\tfrac{1}{2}}\!\!  \right].
\end{aligned}
\end{equation}
So, the transformation of the presymplectic 2-form is, by \eqref{Vert-trsf-int-generic}: 
\begin{equation}
\label{GT-Theta-Sigma1}
\begin{aligned}
{\bs \Theta_\Sigma}^{\bs\psi}_{\phantom 0}  \defeq \int_{\bs\psi\-(\Sigma)} \hspace{-3mm}  \bs\Theta^{\bs \psi}=  \int_{\Sigma}  \bs \Theta  
		+  \int_{\d\Sigma}\  &\bs d\,  \left( \theta\big(\iota_{\bs d \bs\psi  \circ \bs\psi\-} \phi; \phi\big) - \gamma\big(\bs d \bs\psi  \circ \bs\psi\- ; \phi\big) \right)\\[1mm]
 				& + \tfrac{1}{2} \left(   \mathscr{A}\big( \lfloor  \bs d \bs\psi  \circ \bs\psi\-\!, \  \bs d \bs\psi \circ \bs\psi\-\! \rfloor ;\phi \big)
				   +  \gamma \big([ \bs d \bs\psi \circ \bs\psi\-\!, \  \bs d \bs\psi \circ \bs\psi\-]_{\text{{\tiny $\diff(M)$}}};\phi \big) \right)\\[1mm]
				& +  \left(  \iota_{ \bs d \bs\psi  \circ \bs\psi\-} \bs \theta   - \bs  \,\gamma\big( \bs d \bs\psi  \circ \bs\psi\-;\ \munderline{ForestGreen}{\phi}\big) \right) \\[1mm]
		+ \int_{\Sigma}   &    \iota_{\left\{  \bs d \bs\psi \circ \bs\psi\- \right\}^v} \iota_{ \bs d \bs\psi  \circ \bs\psi\-} \bs E  
							- \iota_{ \bs d \bs\psi  \circ \bs\psi\-} \bs E 
							- \bs d\, E(\iota_{\bs d \bs\psi  \circ \bs\psi\-} \phi; \phi) 
\end{aligned}
\end{equation}
This expression  makes conceptual structures manifest: the last line obviously vanishes on-shell, while the third line vanishes when imposing b.c.   necessary for integrability of charges (and the Poisson bracket to satisfy the Jacobi identity). What remains then are pure boundary terms, with a linear contribution (in the field-dependent transformation parameter $\bs d \bs\psi  \circ \bs\psi\- \in \Omega^1\big(\Phi; \diff(M)\big)$) due to the current term, and a quadratic contribution due to the 2-cocycle term $\mathscr C$ involved in the Poisson bracket \eqref{Poisson-bracket-on-shell}. 

It is of course possible to give another, more ``concrete" (if less conceptually transparent) expression. Going back to \eqref{c1}-\eqref{c2}, we must notice that the way the current terms have been worked out applies exactly the same for the $\gamma(\bs d \bs\psi  \circ \bs\psi\-; \phi)$ terms. So we find the form, 
\begin{equation}
\begin{aligned}
 \bs \Theta^{\bs\psi} =  {\bs\psi}^*  \left[ \phantom{\tfrac{1}{2}}\!\! \bs \Theta  
 				\right.& +   \bs d  \left( J\big( \bs d \bs\psi  \circ \bs\psi\- ; \phi\big) + d\,\gamma\big( \bs d \bs\psi  \circ \bs\psi\-; \phi\big)  \right)\\[1mm]
 				& + \tfrac{1}{2} \  d\mathscr{A}\big( \lfloor  \bs d \bs\psi  \circ \bs\psi\-\!, \  \bs d \bs\psi \circ \bs\psi\-\! \rfloor ;\phi \big) 
				\  +\  d \, \iota_{ \bs d \bs\psi  \circ \bs\psi\-} \bs \theta    \\[1mm]
				&   +  \iota_{\left\{  \bs d \bs\psi \circ \bs\psi\- \right\}^v} \iota_{ \bs d \bs\psi  \circ \bs\psi\-} \bs E 
						- \iota_{ \bs d \bs\psi  \circ \bs\psi\-} \bs E  \left. \phantom{\tfrac{1}{2}}\!\!  \right].
\end{aligned}
\end{equation}
Which may also have been found from \eqref{c3} using  $\bs d \big( -\bs d \bs\psi  \circ \bs\psi\- \big) + \tfrac{1}{2}[ \bs d \bs\psi  \circ \bs\psi\-\!,\   \bs d \bs\psi \circ \bs\psi\-]_{\text{{\tiny $\diff(M)$}}} =0$. 
Using once more  the expression \eqref{Noether-current2}  for the current, 
and $\iota_{X^v}\bs E = dE(\iota_X \phi; \phi)$ to rewrite the first term in the last line, we get
\begin{equation}
\begin{aligned}
 \bs \Theta^{\bs\psi} =  {\bs\psi}^*  \left[ \phantom{{\tfrac{1}{2}}}\!\! \bs \Theta  
 				\right.& 
		+  d\  \bs d\,  \theta\big(\iota_{\bs d \bs\psi  \circ \bs\psi\-} \phi; \phi\big) \\[1mm]
 				& + \tfrac{1}{2} \   d\mathscr{A}\big( \lfloor  \bs d \bs\psi  \circ \bs\psi\-\!, \  \bs d \bs\psi \circ \bs\psi\-\! \rfloor ;\phi \big) 		
				    + d \,  \iota_{ \bs d \bs\psi  \circ \bs\psi\-} \bs \theta   \\[1mm]
				&   + \iota_{ \bs d \bs\psi  \circ \bs\psi\-} d E \big(\iota_{\bs d \bs\psi \circ \bs\psi\-}\phi :\phi\big) 
							- \iota_{ \bs d \bs\psi  \circ \bs\psi\-} \bs E 
							- \bs d\, E(\iota_{\bs d \bs\psi  \circ \bs\psi\-} \phi; \phi)  \left. \phantom{\tfrac{1}{2}}\!\!  \right].
\end{aligned}
\end{equation}
From the result \eqref{dA} of appendix \ref{A5} (used to write the 2-cocycle \eqref{Cocycle-concrete}), we have immediately 
\begin{align*}
\label{c4}
\tfrac{1}{2} \   d\mathscr{A}\big( \lfloor  \bs d \bs\psi  \circ \bs\psi\-\!, \  \bs d \bs\psi \circ \bs\psi\-\! \rfloor ;\phi \big) 
 = d\left( \, \mathfrak L_{ \bs d \bs\psi  \circ \bs\psi\-} \, \theta\big(\iota_{ \bs d \bs\psi  \circ \bs\psi\-} \phi ;\phi \big) + \tfrac{1}{2}\ \iota_{ \bs d \bs\psi  \circ \bs\psi\-} \iota_{ \bs d \bs\psi  \circ \bs\psi\-} L  
					             \ \  - \ \  \iota_{ \bs d \bs\psi  \circ \bs\psi\-}   E\big(\iota_{ \bs d \bs\psi  \circ \bs\psi\-} \phi ;\phi \big)\, \right).
\end{align*}
We thus  obtain the final expression: 
\begin{equation}
\begin{aligned}
 \bs \Theta^{\bs\psi} =  {\bs\psi}^*  \left[ \phantom{{\tfrac{1}{2}}}\!\! \bs \Theta  
 				\right.& 
		+  d\  \bs d\,  \theta\big(\iota_{\bs d \bs\psi  \circ \bs\psi\-} \phi; \phi\big) \\[1mm]
 				& + d\left( \, \mathfrak L_{ \bs d \bs\psi  \circ \bs\psi\-} \, \theta\big(\iota_{ \bs d \bs\psi  \circ \bs\psi\-} \phi ;\phi \big) + \tfrac{1}{2}\ \iota_{ \bs d \bs\psi  \circ \bs\psi\-} \iota_{ \bs d \bs\psi  \circ \bs\psi\-} L  \right)	
				    + d \,  \iota_{ \bs d \bs\psi  \circ \bs\psi\-} \bs \theta   \\[1mm]
				&  	- \iota_{ \bs d \bs\psi\circ \bs\psi\-} \bs E 
					- \bs d\, E(\iota_{\bs d \bs\psi  \circ \bs\psi\-} \phi; \phi)  \left. \phantom{\tfrac{1}{2}}\!\!  \right].
\end{aligned}
\end{equation}
Which is of course also derivable from \eqref{GT-theta} via: $ \bs \Theta^{\bs\psi} =\bs d \big(  \bs \theta^{\bs\psi} \big)$ -- which flows from the naturality of the pullback, $[\Xi^\star, \bs d]=0$ for $\Xi \in \bs\Diff_v(\Phi)$. 
The final expression for  the $\bs\Diff_v(\Phi) \simeq C^\infty\big(\Phi, \Diff(M) \big)$ transformation of the  presymplectic 2-form $\bs \Theta_\Sigma$ is thus:
\begin{equation}
\label{GT-Theta-Sigma2}
\begin{aligned}
{\bs \Theta_\Sigma}^{\bs\psi}_{\phantom 0} =   \int_{\Sigma} \phantom{{\tfrac{1}{2}}}\!\! \bs \Theta  
 				& +    \int_{\d \Sigma}    \bs d\,  \theta\big(\iota_{\bs d \bs\psi  \circ \bs\psi\-} \phi; \phi\big) 
				    + \iota_{ \bs d \bs\psi  \circ \bs\psi\-} \bs \theta 
 				    + \mathfrak L_{ \bs d \bs\psi  \circ \bs\psi\-} \, \theta\big(\iota_{ \bs d \bs\psi  \circ \bs\psi\-} \phi ;\phi \big) 
				    + \tfrac{1}{2}\ \iota_{ \bs d \bs\psi  \circ \bs\psi\-} \iota_{ \bs d \bs\psi  \circ \bs\psi\-} L      \\[1mm]
				&  -    \int_{\Sigma}	 \iota_{ \bs d \bs\psi \circ \bs\psi\-} \bs E 
					+ \bs d\, E(\iota_{\bs d \bs\psi  \circ \bs\psi\-} \phi; \phi).
\end{aligned}
\end{equation}
Here again we see manifestly that, on-shell, the transformation involves only boundary terms. 
\medskip

 The results of this section are necessary to apply the DFM and to obtain the dressed versions of $L$, $\bs dL$, $\bs dS$, $\bs E$, $\bs\theta_\Sigma$ and $\bs\Theta_\Sigma$. We will use them in the final section \ref{Relational formulation}. 
 These results are also useful to generalise what has been done in section \ref{Noether charges for field-independent gauge parameters} to the case of $\phi$-dependent gauge parameters, i.e. to derive the Noether current and charge associated to the action of $C^\infty\big(\Phi, \diff(M) \big) \simeq \bs\diff_v(\Phi)$. 
 It is not our main goal to do so, but as it appears relevant to the covariant phase space literature, we still do so in the next section.

\subsubsection{Noether charges for  field-dependent gauge parameters}  
\label{Noether charges for  field-dependent gauge parameters}  

We emulate the analysis of section \ref{Noether charges for field-independent gauge parameters} replacing $X \in \diff(M)\ \rarrow \ \bs X \in C^\infty\big(\Phi, \diff(M) \big)$.
From \eqref{NL-der-L'} we have:
\begin{equation}
\label{Lagrangian-anomaly-field-dep}
\begin{alignedat}{2}
  \bs L_{\bs X^v} L' &= \mathfrak L_{\bs X} L'= \alpha(\bs X; \phi) &&= d\beta(\bs X;\phi), \\
  		              & \phantom{\mathfrak L_{\bs X} L'= \alpha(\bs X; \phi)}	 &&=  d\big(\iota _{\bs X} L + \mathfrak L_{\bs X} \ell \big) \rdefeq d\big(  \iota _{\bs X} L + \beta_\ell(\bs X, \phi) \big).\\
			     &= \iota_{\bs X^v} \bs dL = 	\iota_{\bs X^v}	\big( \bs E + &&d\bs \theta' \big).
\end{alignedat}
\end{equation}
Which imply that the Noether current,
\begin{equation}
\begin{aligned}
\label{Noether-current-phi-dep}
J(\bs X; \phi) \defeq&\,  \iota_{\bs X^v} \bs\theta'  - \beta(\bs X, \phi)  -d \gamma(\bs X;\phi), \\
             =&\,   \iota_{\bs X^v} \bs\theta  - \iota_{\bs X} L   -d \gamma(\bs X;\phi),
\end{aligned}
\end{equation}
satisfies $-\iota_{\bs X^v} \bs E = dJ(\bs X; \phi)$, and is thus $d$-closed on-shell, as \eqref{Noether-current}.
For a codimension 1 submanifold $\Sigma \in \bs U(M)$, the associated Noether charge is defined on $\Phi \times \bs U(M)$ as:
\begin{align}
\label{Noether-charge1-phi-dep}
Q_\Sigma (\bs X; \phi)\defeq \langle J(\bs X; \phi), \Sigma \rangle =\int_\Sigma J(\bs X; \phi) =\int_\Sigma  \iota_{ \bs X^v} \bs\theta  - \iota_{\bs X} L   -d \gamma(\bs X;\phi).
\end{align}
in exact analogy to \eqref{Noether-charge1}. 
It is a map,
\begin{equation}
\label{Moment map-phi-dep}
\begin{aligned}
Q_\Sigma ({\color{gray} \bs X}; \ \,): \Phi &\rarrow C^\infty\big(\Phi, \diff(M) \big)^*\simeq \bs \diff_v(\Phi)^*, \\
				\phi& \mapsto Q_\Sigma ({\color{gray} \bs X} ; \phi). 
\end{aligned}
\end{equation}
with $C^\infty\big(\Phi, \diff(M) \big)^*$ the dual of $C^\infty\big(\Phi, \diff(M) \big)$, as indeed $Q_\Sigma (\ \, ; \phi): C^\infty\big(\Phi, \diff(M) \big) \rarrow \RR$, $\bs X \mapsto Q_\Sigma (\bs X; \phi)$,  is a  linear map. Now, from  \eqref{Identity2} derived in  \ref{A3}, we get the results in terms of the field equations, similar to \eqref{Noether-current2}-\eqref{Noether-charge2}:
\begin{align}
\label{Noether-current2-phi-dep}
J(\bs X; \phi)&=d \big( \theta(\iota_{\bs X} \phi; \phi) - \gamma(\bs X ; \phi) \big) \, -\, E(\iota_{\bs X} \phi; \phi), \\[2mm]
\label{Noether-charge2-phi-dep}
Q_\Sigma (\bs X; \phi)&=\int_{\d\Sigma}   \theta(\iota_{\bs X} \phi; \phi) - \gamma(\bs X ; \phi) \,  - \int_{\Sigma} E(\iota_{\bs X} \phi; \phi).
\end{align}
From which follows that $\iota_{\bs X^v}\bs E = dE(\iota_{\bs X} \phi; \phi)$. 
As expected, the charge is a boundary/corner term on-shell. 
Compare this e.g. to eq.(2.11)-(2.12) of \cite{Freidel-et-al2021}.

We now need to relate this current to the symplectic form current $\bs \Theta =\bs{d\theta}$. 
Naturally, one should start with the relation \eqref{Inf-vert-trsf-theta}, special case of \eqref{Inf-GT-diff-rep}: 
\begin{align}
\bs L_{\bs X^v} \bs \theta' & = \mathfrak L_{{\bs X}} \bs\theta' + \iota_{\{\bs{dX}\}^v} \bs\theta',  \notag\\
\hookrightarrow\quad & \iota_{\bs X^v} \bs \Theta = -\bs d \iota_{\bs X^v} \bs\theta' + \iota_{\{\bs{dX}\}^v} \bs\theta' +  \mathfrak L_{{\bs X}} \bs\theta', \notag \\
				& \iota_{\bs X^v} \bs \Theta = -\bs d \iota_{\bline{\bs X}^v} \bs\theta' +  \mathfrak L_{{\bs X}} \bs\theta',   \label{sympl-anom-phi-dep}
\end{align}
The notation $\bline{\bs X}$ here is meant to indicate that the object is considered $\phi$-constant (from the field space viewpoint), and we have by linearity of the inner product: $-\bs d \iota_{\bs X^v} \bs\theta' = -\bs d \iota_{\bline{\bs X}^v} \bs\theta -  \iota_{\{\bs{dX}\}^v} \bs\theta'$. 
The result  \eqref{sympl-anom-phi-dep} is fornally the same as in the $\phi$-independent case \eqref{Symplectic anomaly}, 
so in an analogous  way we get:
\begin{equation}
\label{Moment-map-equation-phi-dep}
\begin{aligned}
\iota_{\bs X^v} \bs \Theta 
				 &= -\bs d J(\bline{\bs X};\phi) +d \left(  \iota_{{\bs X}} \bs \theta^{\phantom '} - \bs d \gamma(\bline{\bs X};\phi) \right)  - \iota_{{\bs X}} \bs E, \\
				 & = -\bs d J({\bs X};\phi) +  J(\bs{dX};\phi)) +d \left(  \iota_{{\bs X}} \bs \theta^{\phantom '} - \bs d \gamma({\bs X};\phi)  + \gamma({\bs{dX}}; \phi) \right) - \iota_{{\bs X}} \bs E, 
 \end{aligned}
\end{equation}
by linearity of $J({\color{gray} \bs X}; \phi)$ and $\gamma({\color{gray} \bs X}; \phi)$ in their first argument. 
So, we get the relation of the Noether charge to the symplectic 2-form:
\begin{equation}
\label{Moment-map-equation-phi-dep-bis}
\begin{aligned}
\iota_{\bs X^v} \bs \Theta_\Sigma &= -\bs d Q_\Sigma (\bline{\bs X};\phi)  + \int_{\d\Sigma }   \iota_{{\bs X}} \bs \theta - \bs d\gamma(\bline{\bs X}; \phi)\  \, - \int_\Sigma  \iota_{{\bs X}} \bs E,\\
\iota_{\bs X^v} \bs \Theta_\Sigma &= -\bs d Q_\Sigma (\bs X;\phi) + Q_\Sigma (\bs{dX} ;\phi) + \int_{\d\Sigma }   \iota_{{\bs X}} \bs \theta - \bs d\gamma({\bs X}; \phi) + \gamma({\bs{dX}}; \phi)\  \, - \int_\Sigma  \iota_{{\bs X}} \bs E,
\end{aligned}
\end{equation}
which manifestly reproduces \eqref{Moment-map-equation}  in the limit $\bs X \rarrow X$. 
This result can be compared to eq.(2.18), (2.19) and (3.31) in \cite{Freidel-et-al2021} 
(where the  last three terms on the right-hand side, noted $\mathcal F_{\bs X}$, are referred to as  the ``Noetherian flux"). 
The~discussion on restoring charge integrability, around \eqref{BC-imposed}, applies here too. In particular, upon imposing the b.c.  $\bs\theta + \bs d\ell=0_{\,|\d U\supset \d\Sigma}$ and $\bs{dX}=0_{\, |\d U\supset \d\Sigma}$, one  can set $\gamma(\bs X;\phi)=-\iota_{\bs X} \ell$ so that,
\begin{align}
\label{BC-imposed-phi-dep}
Q_\Sigma ({\bs X}; \phi)&=\int_{\d\Sigma}   \theta(\iota_{ \bs X} \phi; \phi) +\iota_{{\bs X}} \ell \ _{|\S} \qquad \text{and}\qquad \iota_{{\bs X}^v} \bs \Theta_\Sigma = -\bs d Q_\Sigma ({\bs X};\phi) \ _{|\S},
\end{align}
which applies e.g. to GR with $\Lambda = 0$, with the York-Gibbons-Hawking boundary Lagrangian $\ell=\ell_\text{YGH}$. 
\medskip

Now, it remains to address the question of the bracket of the charges \eqref{Noether-charge2-phi-dep} induced by $\bs\Theta_\Sigma$ via the usual prescription $\big\{Q_\Sigma (\bs X; \phi), Q_\Sigma (\bs Y; \phi) \big\} \defeq \bs\Theta_\Sigma \big({\bs X}^v, {\bs Y}^v \big)$. 
As is natural, we start back from \eqref{sympl-anom-phi-dep}:
\begin{align}
\bs L_{\bs X^v} \bs \theta' & = \mathfrak L_{{\bs X}} \bs\theta' + \iota_{\{\bs{dX}\}^v} \bs\theta' = \bs\alpha({\bs X}) + \iota_{\{\bs{dX}\}^v} \bs\theta' \rdefeq \t{\bs\alpha}(\bs X),  \notag\\[1mm]
\hookrightarrow\quad & \iota_{\bs X^v} \bs \Theta = -\bs d \iota_{\bs X^v} \bs\theta' + \t{\bs\alpha}(\bs X), \notag \\
\text{so } \quad &\iota_{\bs Y^v} \iota_{\bs X^v} \bs \Theta = -\iota_{\bs Y^v} \bs d \iota_{\bs X^v} \bs\theta' + \iota_{\bs Y^v} \t{\bs\alpha}(\bs X).
\end{align}
And by using the definition and properties  \eqref{Relations-L-iota}-\eqref{FN-NL-der} of the Nijenhuis-Lie derivative, 
we find:
\begin{align}
\iota_{\bs Y^v} \iota_{\bs X^v} \bs \Theta &= -\bs L_{\bs Y^v} \iota_{\bs X^v} \bs\theta' + \iota_{\bs Y^v} \t{\bs\alpha}(\bs X)\notag\\
				&=  -\iota_{\bs X^v} \bs L_{\bs Y^v}  \bs\theta' + \iota_{[\bs X^v, \bs Y^v]_{\text{{\tiny FN}}} } \bs\theta' + 
				     \iota_{\bs Y^v} \t{\bs\alpha}(\bs X), \notag\\
				&=   \iota_{\{\bs X, \bs Y\}^v} \bs\theta' -\iota_{\bs X^v}\t{\bs\alpha}(\bs Y)  + \iota_{\bs Y^v} \t{\bs\alpha} ({{\bs X}}).
				  \label{PB-step-phi-dep}
\end{align}
Remark that, using the definition of $\t{\bs\alpha}({\color{gray}{\bs X}})$ and the properties   \eqref{NR-bracket-id}-\eqref{NR-bracket-field-dep-diff} of the Nijenhuis-Richardson bracket, we also find that the above could be written as: 
$\iota_{\bs Y^v} \iota_{\bs X^v} \bs \Theta  = \iota_{([\bs X, \bs Y]_{\text{{\tiny $\diff(M)$}}})^v} \bs\theta' 	- \iota_{{\bs X}^v}  \bs \alpha({{\bs Y}}) +  \iota_{{\bs Y}^v}  \bs\alpha({{\bs X}})$. A result formally identical the one \eqref{PB-step} for $\phi$-independent parameters.

Our goal is to reconstruct the current \eqref{Noether-current2-phi-dep}, which contains the exact Lagrangian anomaly \eqref{Lagrangian-anomaly-field-dep}.
The last two terms on the r.h.s. of  \eqref{PB-step-phi-dep} can indeed be related to 
the Lagrangian anomaly $\alpha(\bs X; \phi)$, which satisfies the abelian version of the 1-cocycle relation \eqref{BB}:
\begin{equation}
\label{phi-dep-anomaly}
\begin{aligned}
& \iota_{\bs X^v} \bs d \big( \alpha(\bs Y; \phi) \big) -   \iota_{\bs Y^v}\bs d \big( \alpha(\bs X; \phi) \big)    - \alpha(\{\bs X, \bs Y\}; \phi)   =0. 
\end{aligned}
\end{equation}
To work out this relation, generalising \eqref{Identity-anomalies}, we observe that $[\bs L_{X^v}, \bs d] L'= 0$ and that on the one hand $\bs d \bs L_{X^v} L' =\bs d \alpha(\bs X; \phi)$. While on the other hand, by \eqref{GT-dL-fied-dep} (or \eqref{Inf-GT-diff-rep}): 
\begin{equation}
\begin{aligned}
\bs L_{X^v} \bs d L'&= \mathfrak L_{\bs X}  \bs d L' + \alpha(\bs{dX}; \phi), \\
			     &= \mathfrak L_{\bs X} \big(\bs E + d\bs\theta' \big) + \alpha(\bs{dX}; \phi), \\
			     &= \mathfrak L_{\bs X} \bs E + d \bs \alpha({\bs X}) + \alpha(\bs{dX}; \phi), \\
			     &= d\iota_{\bs X} \bs E + d\big(\t{\bs\alpha}({{\bs X}}) - \iota_{\bs{dX}^v} \bs\theta'  \big)  + d\beta(\bs{dX}; \phi). 
\end{aligned}
\end{equation}
This gives us the relation: 
\begin{equation}
\label{Identity-anomalies-phi-dep}
\begin{aligned}
\bs d\alpha(\bs X; \phi) &= d\left(\t{\bs\alpha}({{\bs X}}) -   \iota_{\bs{dX}^v} \bs\theta' + \beta(\bs{dX}; \phi) + \iota_{\bs X} \bs E^{\phantom '}\! \right), \\
				   &= d\left(\t{\bs\alpha}({{\bs X}}) -   J(\bs{dX}; \phi) - d\gamma(\bs{dX}; \phi) + \iota_{\bs X} \bs E^{\phantom '}\! \right). 
\end{aligned}
\end{equation}
It is clear that that the first line reduces to  \eqref{Identity-anomalies} when $\bs X \rarrow X$. We used the definition \eqref{Noether-current-phi-dep} of the current to obtain the second line. Combining \eqref{Identity-anomalies-phi-dep} and \eqref{phi-dep-anomaly} we get: 
\begin{equation}
\begin{aligned}
\iota_{\bs X^v}\t{\bs\alpha}({{\bs Y}}) - \iota_{\bs Y^v}\t{\bs\alpha}({{\bs X}}) 
&\ -\  \beta\big( \{ \bs X, \bs Y\}; \phi\big) 
 \ +\  \iota_{\bs X^v} \iota_{\bs Y} \bs E -  \iota_{\bs Y^v} \iota_{\bs X} \bs E \\
& \ -\   J\!\left(\bs{X}^v(\bs Y)- \bs{Y}^v(\bs X) ; \phi \right) 
 - d\gamma\!\left(\bs{X}^v(\bs Y)- \bs{Y}^v(\bs X) ; \phi \right) 
\rdefeq - d\mathscr{A}\big( \lfloor \bs X, \bs Y\! \rfloor ;\phi \big).  \label{2-cocycle-phi-dep}
\end{aligned}
\end{equation}
This expression reduces to  \eqref{2-cocycle} when $\bs X \rarrow X$ (it can actually be rewritten in exactly the same way). 
From this,  \eqref{PB-step-phi-dep} is:
\begin{align}
\iota_{\bs Y^v} \iota_{\bs X^v} \bs \Theta  = \iota_{\{\bs X, \bs Y\}^v} \bs\theta' - \beta\big( \{ \bs X, \bs Y\}; \phi\big) 
															+ d\mathscr{A}\big( \lfloor \bs X, \bs Y\! \rfloor ;\phi \big)
									\ &- \  J\!\left(\bs{X}^v(\bs Y)- \bs{Y}^v(\bs X) ; \phi \right) 
									   - d\gamma\!\left(\bs{X}^v(\bs Y)- \bs{Y}^v(\bs X) ; \phi \right) \notag \\
									\ &+\  \iota_{\bs X^v} \iota_{\bs Y} \bs E   -  \iota_{\bs Y^v} \iota_{\bs X} \bs E, \notag\\[1mm]
						= J\big( \{\bs X, \bs Y\} ;\phi \big) + d \gamma\big( \{\bs X, \bs Y\} ;\phi \big)
									+ d\mathscr{A}\big( \lfloor \bs X, \bs Y\! &\rfloor ;\phi \big)
									\ - \  J\!\left(\bs{X}^v(\bs Y)- \bs{Y}^v(\bs X) ; \phi \right) 
									   - d\gamma\!\left(\bs{X}^v(\bs Y)- \bs{Y}^v(\bs X) ; \phi \right) \notag \\
						\ &\hspace{7.5mm} +\  \iota_{\bs X^v} \iota_{\bs Y} \bs E   -  \iota_{\bs Y^v} \iota_{\bs X} \bs E, \notag\\[1mm]
					        = J\big( \{\bs X, \bs Y\} ;\phi \big) +  d\mathscr{A}\big( \lfloor \bs X, \bs Y\! \rfloor ;\phi \big) 
					        							  + d \gamma\big( [\bs X, \bs Y&]_{\text{{\tiny $\diff(M)$}}} ;\phi \big)
												  \ - \  J\!\left(\bs{X}^v(\bs Y)- \bs{Y}^v(\bs X) ; \phi \right)  \\
								      \  &\hspace{13.5mm}+\  \iota_{\bs X^v} \iota_{\bs Y} \bs E   -  \iota_{\bs Y^v} \iota_{\bs X} \bs E. \notag
\end{align}
We have again used  the definition \eqref{Noether-current-phi-dep} of the Noether current. We have therefore the bracket of charges:
\begin{equation}
\begin{aligned}
\left\{ Q_\Sigma(\bs X; \phi); Q_\Sigma(\bs Y; \phi) \right\} = Q_\Sigma\big( \{\bs X, \bs Y\} ; \phi \big) + \b{\mathscr C}\big( \lfloor \bs X, \bs Y\! \rfloor ;\phi \big)
																			+ \int_\Sigma \iota_{\bs X^v} \iota_{\bs Y} \bs E   -  \iota_{\bs Y^v} \iota_{\bs X} \bs E
\end{aligned}
\end{equation}
where we define the map
\begin{equation}
\begin{aligned}
\label{2-cocycle-map-phi-dep}
\b{ \mathscr C} \big(\ \ ; {\color{gray}\phi}\big): C^\infty\big( \Phi, \diff(M)\big) \times C^\infty\big( \Phi, \diff(M)\big) &\rarrow C^\infty(\Phi),\\
 									(\bs X, \bs Y)		 &  \mapsto  \b{\mathscr C}\big( \lfloor \bs X, \bs Y\! \rfloor ;{\color{gray}\phi} \big) \defeq \int_{\d\Sigma} \mathscr{A}\big( \lfloor \bs X, \bs Y\! \rfloor ;{\color{gray}\phi} \big)  + \gamma \big([\bs X, \bs Y]_{\text{{\tiny $\diff(M)$}}};{\color{gray}\phi} \big) \\
							                                                   &\hspace{4cm}		-\int_\Sigma  J\!\left(\bs{X}^v(\bs Y)- \bs{Y}^v(\bs X) ; \phi \right),
\end{aligned}
\end{equation}
which is clearly antisymmetric and bilinear. It is manifest that it reproduces the map $ \mathscr C$ \eqref{2-cocycle-map} in the limit $\bs X\rarrow X$. 
Given the expressions of the current and charge \eqref{Noether-current2-phi-dep}-\eqref{Noether-charge2-phi-dep}, we can write it as:
\begin{equation}
\begin{aligned}
\label{Relation-2-cocycles}
 \b{\mathscr C}\big( \lfloor \bs X, \bs Y\! \rfloor ; \phi \big)& =  \mathscr C\big( \lfloor \bs X, \bs Y\! \rfloor ; \phi \big) + Q_\Sigma\big(\bs{X}^v(\bs Y)- \bs{Y}^v(\bs X); \phi \big), \\[1mm]
	 &= \int_{\d\Sigma} \mathscr{A}\big( \lfloor \bs X, \bs Y\! \rfloor ; \phi \big)  + \gamma \big(\{X, \bs Y\}; \phi \big)
	 					- \theta\big(\iota_{[\bs X^v(\bs Y) - \bs Y^v(\bs X)]} \phi; \phi \big) 
						\ + \ \int_\Sigma E\big(\iota_{[\bs X^v(\bs Y) - \bs Y^v(\bs X)]} \phi; \phi \big). 
\end{aligned}
\end{equation}
Using the result \eqref{Cocycle-concrete} for  $\mathscr C$ in  the $\phi$-independent case, we have the concrete expression:
\begin{equation}
\begin{aligned}
\label{Cocycle-concrete-phi-dep}
 \b{\mathscr C}\big( \lfloor \bs X, \bs Y\! \rfloor ; \phi \big) &= \int_{\d\Sigma} \mathfrak L_{\bs X}\,  \theta(\iota_{\bs Y}\phi; \phi) - \mathfrak L_{\bs Y}\, \theta(\iota_{\bs X}\phi; \phi) + \iota_{\bs X}\iota_{\bs Y} L - \theta\big(\iota_{[\bs X^v(\bs Y) - \bs Y^v(\bs X)]} \phi; \phi \big)
								         + \gamma \big(\{\bs X, \bs Y\}; \phi \big)\\[1mm]
					 &\hspace{2cm}	         + \int_{\d\Sigma}  \iota_{\bs Y} E(\iota_{\bs X}\phi; \phi) - \iota_{\bs X} E(\iota_{\bs Y}\phi; \phi)
								         + \int_\Sigma E\big(\iota_{[\bs X^v(\bs Y) - \bs Y^v(\bs X)]} \phi; \phi \big). 
\end{aligned}
\end{equation}

On-shell, the bracket is:
\begin{equation}
\label{Final-on-shell-bracket}
\begin{aligned}
\left\{ Q_\Sigma(\bs X; \phi); Q_\Sigma(\bs Y; \phi) \right\} = Q_\Sigma\big( \{\bs X, \bs Y\} ; \phi \big) + \b{\mathscr C}\big( \lfloor \bs X, \bs Y\! \rfloor ;\phi \big)_{\, |\S}
\end{aligned}
\end{equation}
and by \eqref{Relation-2-cocycles}-\eqref{Cocycle-concrete-phi-dep} the map $\b{\mathscr C}$  is manifestly a  boundary term. 
We can further ascertain the conditions under which it is a 2-cocycle (as we did in Appendix \ref{A4}). Writing that on-shell:
\begin{equation}
\label{C-on-shell}
\begin{aligned}
\b{\mathscr C}\big( \lfloor \bs X, \bs Y\! \rfloor ;\phi \big) \defeq&\,  \big\{Q_\Sigma (\bs X; \phi), Q_\Sigma (\bs Y; \phi) \big\} - Q_\Sigma (\{\bs X, \bs Y\} ; \phi), \\
									=&\ \ \bs\Theta_\Sigma \big( \bs X^v, \bs Y^v \big) - Q_\Sigma (\{\bs X, \bs Y\}; \phi),
\end{aligned}
\end{equation}
we have that,
\begin{equation}
\label{2-cocycle-condition-phi-dep}
\begin{aligned}
 \b{\mathscr C} \big( \lfloor \bs X, \{\bs Y, \bs Z\}\rfloor ; \phi \big) + c.p.  &= \bs\Theta_\Sigma \big( \bs X^v, ( \{\bs X, \bs Y\})^v \big) - Q_\Sigma (\big\{\bs X, \{\bs Y, \bs Z\} \big\}; \phi) + c.p.  \\
 											&= \bs\Theta_\Sigma \big( \bs X^v, [\bs Y^v, \bs Z^v] \big) - Q_\Sigma (\big\{\bs X, \{\bs Y, \bs Z\} \big\}; \phi)  + c.p. \\
											&= \int_\Sigma \big( \bs L_{\bs X^v} \bs\Theta \big)(\bs Y^v, \bs Z^v) + c.p. \\
											&=  \int_\Sigma \big( \bs d\iota_{\bs X^v} \bs\Theta \big)(\bs Y^v, \bs Z^v) + c.p.
\end{aligned}
\end{equation}
where we used  the property of the  Frölicher-Nijenhuis (extended) bracket \eqref{extended-bracket1}/\eqref{FN-bracket-field-dep-diff=BT-bracket}, which satisfies Jacobi, and the identity \eqref{Cool-identity}. 
In view of  \eqref{Moment-map-equation-phi-dep},  we  see that $ \b{\mathscr C}$ satisfies a 2-cocycle condition, and thus \eqref{Final-on-shell-bracket} is a central extension of $ C^\infty\big( \Phi, \diff(M)\big) \simeq \bs\diff_v(\Phi)$, when the charges are moment maps: i.e. when we are {\bf 1)} \emph{on-shell}, as is already assumed, and {\bf 2)} given adequate boundary conditions, as indeed \eqref{2-cocycle-condition-phi-dep} is written as
 \begin{equation}
 \begin{aligned}
\label{flux-2-cocyle-phi-dep}
 \b{\mathscr C}\big( \lfloor \bs X, \{\bs Y,\bs  Z\} \rfloor ; \phi \big) + c.p.  = 
  \bs d \left( \int_{\d\Sigma}  \right.& \theta(\iota_{\bs{dX}} \phi; \phi) - \gamma(\bs{dX} ; \phi) \,  - \cancel{\int_{\Sigma} E(\iota_{\bs{dX}} \phi; \phi)}  \\
                   &+\  \left.\int_{\d\Sigma }   \iota_X \bs \theta - \bs d\gamma(X;\phi)+  \gamma(\bs{dX};\phi)\  \, - \cancel{\int_\Sigma  \iota_X \bs E} \ \right)  (Y^v, Z^v) + c. p. 
\end{aligned}
\end{equation}
So is the case, e.g. imposing the b.c.  $\bs\theta + \bs d\ell=0_{\,|\d U\supset \d\Sigma}$ and $\bs{dX}=0_{\, |\d U\supset \d\Sigma}$, and setting  $\gamma(\bs X;\phi)=-\iota_{\bs X} \ell$, as in \eqref{BC-imposed-phi-dep}.
However {\bf 2)} is obtained, the algebra of charges is on-shell the central extension:
\begin{align}
\label{PB-central-ext-phi-dep}
\makebox[\displaywidth]{
\hspace{-18mm}\begin{tikzcd}[column sep=large, ampersand replacement=\&]
\      \&  \big( C^\infty(\Phi); \ \cdot \ \big)   \arrow[r, "\iota"  ]          \&   \text{PAlg}[Q_\Sigma]= \big( Q_\Sigma (\ \,; \phi) ; \{\ ,\ \} \big)     \arrow[r, "\pi"]      \&      \bs\diff_v(\Phi)\simeq\big(  C^\infty\big( \Phi, \diff(M)\big); \{\ \, , \ \} \big)    \arrow[l, bend left=20,  start anchor={[xshift=4ex]}, end anchor={[xshift=-1ex]}, "Q_\Sigma(\, \ ;\,\phi)"]         
\end{tikzcd}}  \raisetag{8.5ex}
\end{align}
with $\im \iota \subset Z\big(\text{PAlg}[Q_\Sigma] \big)$ and $Q_\Sigma(\, \ ;\phi): C^\infty\big( \Phi, \diff(M)\big) \rarrow \text{PAlg}[Q_\Sigma]$, $\bs X \mapsto Q_\Sigma(\bs X ;\phi)$, is a linear map s.t. \mbox{$\pi \circ Q_\Sigma(\, \ ;\phi) =\id$} (i.e. it is a section of $\pi$, not a Lie algebra morphism). 
Since $\pi$ is a Lie algebra morphism, and since on-shell we have $\mathscr C\big( \lfloor \bs X, \bs Y\! \rfloor ;\phi \big)=\big\{Q_\Sigma (\bs X; \phi), Q_\Sigma (\bs Y; \phi) \big\} - Q_\Sigma \big(\{\bs X, \bs Y\};\phi \big)$ by \eqref{C-on-shell}, we have $\mathscr C\big( \lfloor \bs X, \bs Y\! \rfloor ;\phi \big) \in \ker \pi = \im \iota \subset Z\big(\text{PAlg}[Q_\Sigma] \big)$.
\medskip

Depending on one's goal -- see e.g. \cite{Freidel-et-al2021} -- one may modify the standard prescription for the bracket of charges, substracting the 2-cocyle $\b{\mathscr C}$ to as to define via \eqref{Final-on-shell-bracket} an  on-shell bracket representing $\bs\diff_v(\Phi)\simeq C^\infty\big( \Phi, \diff(M)\big)$:
\begin{align}
\label{Poisson-bracket-redef-phi-dep}
\big\{Q_\Sigma (\bs X; \phi), Q_\Sigma (\bs Y; \phi) \big\}_{\text{{\tiny ``new"}}} \defeq   \bs\Theta_\Sigma \big(\bs X^v, \bs Y^v \big)   - \b{\mathscr C}\big( \lfloor \bs X, \bs Y\! \rfloor ;\phi \big) = Q_\Sigma \big(\{\bs X, \bs Y\};\phi \big)_{\ | \S}\,.
\end{align}
An issue is that, contrary to the original prescription, it is not clear that the bracket \eqref{Poisson-bracket-redef-phi-dep}  generates $\bs\diff_v(\Phi)$-transformations of functions on $\Phi$. Indeed, for $\bs f \in C^\infty(\Phi)$, with  associated hamiltonian vector field $V_{\bs f}$ defined via $\iota_{V_{\bs f}} \bs \Theta_\Sigma =- \bs d \bs f$, the standard bracket by a charge $Q_\Sigma (\bs X; \phi)$ associated to $\bs X^v \in \bs\diff_v(\Phi)$  is:
\begin{align}
\big\{Q_\Sigma (\bs X; \phi), \bs f \big\} = \iota_{V_{\bs f}} \iota_{\bs X^v} \bs \Theta_\Sigma = - \iota_{\bs X^v}\iota_{V_{\bs f}} \bs \Theta_\Sigma = \iota_{\bs X^v}\bs d \bs f = \bs L_{\bs X^v} \bs f.
\end{align}
This is indeed by definition the infinitesimal vertical $\bs\diff_v(\Phi) \simeq C^\infty\big( \Phi, \Diff(M)\big)$-transformation of $\bs f$. 

\bigskip

\section{Relational formulation}  
\label{Relational formulation}  

Exploiting the work done in previous sections, we are now in a position to showcase in a very concise way how to apply the DFM to obtain an invariant and relational reformulation of a $\Diff(M)$-theory. Let us remind (section~\ref{The dressing field method}) that technically, the DFM is a tool to formally and systematically produce dressed/\emph{basic} objects on field space $\Phi$ depending on the dressed fields $\phi^{\bs u}$, and that in favorable circumstances the latter can be readily understood as invariant (Dirac) relational variables -- i.e. physical d.o.f.  --  so that the former can be seen as relationally defined quantities. 

Furthermore, dressed regions $U^{\bs u}=\bs u\-(U)$ or $\Sigma^{\bs u}=\bs u\-(\Sigma)$, on which $\phi^{\bs u}$ live, are relationally defined $\Diff(M)$-invariant regions of the physical spacetime. The DFM then supplies dressed integrals as defining integration on spacetime. 
It should be noticed that relational, physical, boundaries $\d U^{\bs u}=\bs u\-(\d U)$ or $\d \Sigma^{\bs u}=\bs u\-(\d\Sigma)$, are obviously $\Diff(M)$-invariant. This dispels the assertion (logically equivalent to the ``hole argument'') that boundaries in general relativistic theories can ``break $\Diff(M)$-invariance''. 

\subsection{Basic presymplectic structure}  
\label{Dressed presymplectic structure}  

In what follows, we will exploit the structural results of the DFM, i.e. its rule of thumb: To obtain the dressed/basic /relational version of an object on $\Phi$, take its $\bs\Diff_v(\Phi) \simeq C^\infty \big(\Phi, \Diff(M) \big)$-transformation, and replace the gauge parameter by the dressing field, $\bs\psi \rarrow \bs u$. 

Starting from a theory specified by a Lagrangian $L=L(\phi)$, and given a dressing field $\bs u=\bs u(\phi)$, the $\Diff(M)$-invariant manifestly relational theory
is, from \eqref{GT-L}:
  \begin{align}
  \label{Dressed-L}
 {L}^{\,\bs u}  = \bs u^* L=L(\phi^{\bs u}). 
  \end{align}
  We remind that, as a dressing \emph{is not a gauge-fixing}, \eqref{Dressed-L} \emph{is not} a gauge-fixed version of $L$. 
  The dressed field equations are then, from \eqref{GT-E}:
 \begin{align}
   \label{Dressed-E}
  \bs E^{\,\bs u}  =  \bs u^* \left( \bs E^{\phantom{'}}\! + dE\big(  \iota_{\bs d \bs u  \circ \bs u\-}\phi ;\  \phi \big) \right).  
   \end{align}
   Let us show explicitly that $\bs E^{\,\bs u}$ is invariant under  $\bs\Diff_v(\Phi) \simeq C^\infty\big( \Phi, \Diff(M)\big)$, i.e. field-dependent diffeomorphisms.
   We do so by checking its basicity. First, it is easily checked that it is horizontal, by $\iota_{X^v}\bs E = dE(\iota_X \phi; \phi)$ and \eqref{Dress-connection-vert-prop}:
\begin{equation}
\label{Horiz-E-u}
\begin{aligned}
 \bs E^{\,\bs u}(X^v) &= \bs u^* \left( \bs E (X^v)+ dE\big(  \iota_{\bs d \bs u  \circ \bs u\-(X^v)}\phi ;\  \phi \big) \right), \\
 			       &=  \bs u^* \left( dE\big(  \iota_X\phi ;\  \phi \big)+ dE\big(  \iota_{(-X)}\phi ;\  \phi \big) \right) \equiv 0. 
\end{aligned}
\end{equation}
   Then, one shows the triviality of equivariance (i.e. invariance). Using $R^\star_\psi\,  \bs{du}\! \circ\! \bs u\- = \psi\-_*  (\bs{du}\! \circ\! \bs u\- )\circ \psi$ \eqref{Dress-connection-equiv-prop}, we get:
\begin{equation}
\label{Inv-E-u}
\begin{aligned}
R^\star_\psi \,\bs E^{\,\bs u} &= (R^\star_\psi\, \bs u)^* \left(R^\star_\psi \bs E + R^\star_\psi dE\big(  \iota_{\bs d \bs u  \circ \bs u\-}\phi ;\  \phi \big) \right), \\
 			       &=  (\psi\- \circ \bs u)^* \left(\psi^*\bs E + dE\big(  \iota_{\psi\-_* ( \bs d \bs u  \circ \bs u\-) \circ \psi}\, \psi^*\phi ;\ \psi^* \phi \big) \right),\\
			       &= \bs u^*{\psi\-}^* \left(\psi^*\bs E + dE\big( \psi^*\, \iota_{ \bs d \bs u  \circ \bs u\-} \phi ;\ \psi^* \phi \big) \right),\\
			       &= \bs u^*{\psi\-}^*\, \psi^*\left(\bs E + dE\big( \iota_{ \bs d \bs u  \circ \bs u\-} \phi ;\ \phi \big) \right) =  \bs E^{\,\bs u},
\end{aligned}  
\end{equation}  
   where a property analogous to lemma \eqref{lemma1} is used. We remark that these two computations rely on the properties \eqref{Dress-connection-vert-prop}-\eqref{Dress-connection-equiv-prop} of $\bs{du}\! \circ\! \bs u\-$ as an Ehresmann connection on $\Phi$. By \eqref{Horiz-E-u}-\eqref{Inv-E-u} is established that $\bs E^{\,\bs u}  \in \Omega^1_{\text{basic}}(\Phi)$, thus that it is indeed  $\bs\Diff_v(\Phi) \simeq C^\infty\big( \Phi, \Diff(M)\big)$-invariant, as expected. 

Now, the dressed presymplectic potential current is, from \eqref{GT-theta}:
  \begin{align}
   \label{Dressed-theta}
 \bs \theta^{\,\bs u}  =  \bs u^* \left( \bs \theta^{\phantom{'}}\!  +  d \theta \big( \iota_{\bs d \bs u \circ \bs u\-}\phi ;\  \phi \big) 
							      + \iota_{  \bs d \bs u  \circ \bs u\-} L  
							      -\, E(\iota_{  \bs d \bs u \circ \bs u\-}\phi; \phi) \right).   
   \end{align}
    So, by the general prescription \eqref{Dressed-integral} to dress integrals or from \eqref{GT-theta-sigma}, the dressed presymplectic potential is:
\begin{align}
\label{Dressed-theta-sigma}
{\bs \theta_\Sigma}^{\bs u}_{\phantom 0}     \defeq&\, \int_{\Sigma^{\bs u}}   \bs\theta^{\,\bs u} =  \int_{\bs u\-(\Sigma) } \bs u^* \left( \bs \theta^{\phantom{'}}\!  +  d \theta \big(  \iota_{\bs d \bs u \circ \bs u\-}\phi ;\  \phi \big) 
							      + \iota_{  \bs d \bs u  \circ \bs u\-} L  
							      -\, E(\iota_{  \bs d \bs u \circ \bs u\-}\phi; \phi) \right), \notag \\[1mm]
					  =&\,  \int_{\Sigma } \bs \theta +  d \theta \big( \iota_{\bs d \bs u \circ \bs u\-}\phi ;\  \phi \big) 
							      + \iota_{  \bs d \bs u  \circ \bs u\-} L  
							      -\, E(\iota_{  \bs d \bs u \circ \bs u\-}\phi; \phi).   
\end{align}
This result, we notice, immediately reproduces and generalises the so-called ``extended potential" derived in the edge mode literature: compare e.g. to eq.(3.22) in \cite{DonnellyFreidel2016}, eq.(4.6)-(4.9) in \cite{Speranza2018}, or eq.(2.13) in \cite{Geiller2018}. 
What is done in this literature may then be understood, reinterpreted, in light of the DFM conceptual framework -- in particular, edge modes (a misnomer, given the bulk contribution in \eqref{Dressed-theta-sigma}) are special cases of \emph{ad hoc} dressing fields, see section \ref{Discussion}. 

  The dressing of the  variational 1-form $\bs dL$ associated to $L$ is, from \eqref{GT-dL}: 
 \begin{equation}
  \label{Dressed-dL}
  \begin{aligned}
  {\bs dL}^{\,\bs u}  &=  \bs u^* \left( \bs d L + \alpha\big(  \bs d \bs u  \circ \bs u\- ;\  \phi \big) \right),   \\
                                &=  \bs u^* \left( \bs d L + \mathfrak L_{\bs d \bs u  \circ \bs u\-} L \right),
 \end{aligned}
 \end{equation}
consistent with the lemma \eqref{Formula-pullback-rep-dress}.
We then observed that:
\begin{equation}
\label{Dressed-Var-Principle} 
\begin{aligned}
\langle \bs E^{\bs u}  + d \bs \theta^{\bs u}, \bs u\-(U) \rangle &= \langle \bs dL + \mathfrak L_{\bs{du}\circ \bs u\-} L, U \rangle 
												      =\langle \bs dL^{\bs u},  \bs u\-(U)\rangle, \\	
		 										&=\bs d S + \langle \mathfrak L_{\bs{du}\circ \bs u\-} L, U \rangle = (\bs dS)^{\bs u}, 
\end{aligned}
\end{equation}
which is also found  from \eqref{GT-Var-Principle}. 
This is in accord with the general result \eqref{Dressing-int-dalpha}, giving for the dressed variational 1-form associated to the action $S=\langle L', U\rangle$  \eqref{Dressing-dS}:
\begin{equation}
\begin{aligned}
(\bs dS)^{\bs u} &= \bs d S + \langle  \mathfrak  L_{\bs d \bs u  \circ \bs u\-} L,  U  \rangle \\
			   &=\bs d S + \langle  d \iota_{\bs d \bs u  \circ \bs u\-} L,  U  \rangle
			   = \bs d S + \langle  \iota_{\bs d \bs u  \circ \bs u\-} L,  \d U  \rangle.
\end{aligned}
\end{equation}
 As previewed at the end of section \ref{Dressed integrals}, this  indicates, first, that $\bs E^{\bs u}$ are indeed the field equations associated to $ {L}^{\,\bs u}$, and second, that these equations for the relational variables $\phi^{\bs u}$ are functionally the same as those, $\bs E$, for the gauge variables $\phi$ of the bare theory $L'$: i.e. the variational principle is stable under dressing, i.e. under relational reformulation (as one would expect). 
  As is manifest from \eqref{Dressed-E}, the space of dressed/relational solutions is isomorphic to the space of bare solutions, $\S^{\bs u} \simeq \S$. 
 Formula \eqref{Dressed-E} also shows that, when one can neglect the contribution from the variation of the dressing field (i.e. the coordinatizing reference frame, or d.o.f.), the physical/relational field equations are, up to $\bs u^*$, the bare field equations. 
 
 These facts, we argue, explain why it is possible to successfully test general relativistic theories before solving the issue of their observables.
 In~the~case e.g. of GR, some authors (notably Rovelli \cite{Rovelli1991, Rovelli2002, Rovelli2002b}) have argued that what is compared to experiments is a ``gauge-fixed" version of the theory, tacitly presuming some privileged physical reference frame: We suggest that a more accurate statement would be that it is tacitly a dressed/relational version of GR that is compared to experiments, in the specified regime of neglecting variation of the chosen dressing field. 
 It is worth pondering what happens when such variation cannot be neglected, in a ``strongly relational regime". We will do so in a follow-up paper.

\medskip
Turning our attention to the symplectic 2-form current, we find its dressing to be, from \eqref{c3}: 
\begin{equation}
\label{Dressed-Theta}
\begin{aligned}
 \bs \Theta^{\bs u} =  {\bs u}^*  \left[ \phantom{\tfrac{1}{2}}\!\! \bs \Theta  
 				\right.& +   \bs d  J\big( \bs d \bs u  \circ \bs u\- ; \phi\big)  \\[1mm]
 				& + \tfrac{1}{2} \left(   d\mathscr{A}\big( \lfloor  \bs d \bs u  \circ \bs u\-\!, \  \bs d \bs u \circ \bs u\-\! \rfloor ;\phi \big)
				   + d \gamma \big([ \bs d \bs u \circ \bs u\-\!, \  \bs d \bs u \circ \bs u\-]_{\text{{\tiny $\diff(M)$}}};\phi \big) \right)\\[1mm]
				& + d \left(  \iota_{ \bs d \bs u  \circ \bs u\-} \bs \theta   - \bs d \,\gamma\big( \bs d \bs u  \circ \bs u\-;\ \munderline{ForestGreen}{\phi}\big) \right) \\[1mm]
				& +  \iota_{\left\{  \bs d \bs u \circ \bs u\- \right\}^v} \iota_{ \bs d \bs u  \circ \bs u\-} \bs E 
						- \iota_{ \bs d \bs u \circ \bs u\-} \bs E \left. \phantom{\tfrac{1}{2}}\!\!  \right].
\end{aligned}
\end{equation}
From  \eqref{GT-Theta-Sigma1}, or using the expression for the current \eqref{Noether-current2}-\eqref{Noether-current2-phi-dep}, we then find the dressed symplectic 2-form to be:
\begin{equation}
\begin{aligned}
{\bs \Theta_\Sigma}^{\bs u}_{\phantom 0}  \defeq \int_{\Sigma^{\bs u}}  \bs\Theta^{\bs u}=  \int_{\Sigma}  \bs \Theta  
		+  \int_{\d\Sigma}\  &\bs d\,  \left( \theta\big(\iota_{\bs d \bs u  \circ \bs u\-} \phi; \phi\big) - \gamma\big(\bs d \bs u  \circ \bs u\- ; \phi\big) \right)\\[1mm]
 				& + \tfrac{1}{2} \left(   \mathscr{A}\big( \lfloor  \bs d \bs u  \circ \bs u\-\!, \  \bs d \bs u \circ \bs u\-\! \rfloor ;\phi \big)
				   +  \gamma \big([ \bs d \bs u \circ \bs u\-\!, \  \bs d \bs u \circ \bs u\-]_{\text{{\tiny $\diff(M)$}}};\phi \big) \right)\\[1mm]
				& +  \left(  \iota_{ \bs d \bs u  \circ \bs u\-} \bs \theta   - \bs  \,\gamma\big( \bs d \bs u  \circ \bs u\-;\ \munderline{ForestGreen}{\phi}\big) \right) \\[1mm]
		+ \int_{\Sigma}   &    \iota_{\left\{  \bs d \bs u \circ \bs u\- \right\}^v} \iota_{ \bs d \bs u  \circ \bs u\-} \bs E  
							- \iota_{ \bs d \bs u  \circ \bs u\-} \bs E 
							- \bs d\, E(\iota_{\bs d \bs u  \circ \bs u\-} \phi; \phi) 
\end{aligned}
\end{equation}
A more explicit expression is, from \eqref{GT-Theta-Sigma2}:
\begin{equation}
\label{Dressed-Theta-Sigma2}
\begin{aligned}
{\bs \Theta_\Sigma}^{\bs u}_{\phantom 0}  \defeq \int_{\Sigma^{\bs u}} \bs\Theta^{\bs u} =   \int_{\Sigma} \phantom{{\tfrac{1}{2}}}\!\! \bs \Theta  
 				& +    \int_{\d \Sigma}    \bs d\,  \theta\big(\iota_{\bs d \bs u  \circ \bs u\-} \phi; \phi\big) 
				    + \iota_{ \bs d \bs u  \circ \bs u\-} \bs \theta 
 				    + \mathfrak L_{ \bs d \bs u  \circ \bs u\-} \, \theta\big(\iota_{ \bs d \bs u  \circ \bs u\-} \phi ;\phi \big) 
				    + \tfrac{1}{2}\ \iota_{ \bs d \bs u  \circ \bs u\-} \iota_{ \bs d \bs u  \circ \bs u\-} L      \\[1mm]
				&  -    \int_{\Sigma}	 \iota_{ \bs d \bs u  \circ \bs u\-} \bs E 
					+ \bs d\, E(\iota_{\bs d \bs u  \circ \bs u\-} \phi; \phi).
\end{aligned}
\end{equation}
This result immediately reproduces and generalises the ``extended symplectic form" computed in the edge mode literature: compare e.g. eq.(4.12) in \cite{Speranza2018}, or eq.(2.24) in \cite{Geiller2018}. See also eq.(3.132) in \cite{Hoehn-et-al2022}, where $\bs u$ are called ``dynamical reference frames" and an explicit relational interpretation is advocated as well.  

The results \eqref{Dressed-theta-sigma}-\eqref{Dressed-Theta-Sigma2} constitute the dressed, or relational, presymplectic structure of the theory $L$. As these are basic by construction, so horizontal in particular, there is no Noether charges nor bracket associated to $\diff(M)$. 
But there may be charges associated to possible residual symmetries, as detailed in section  \ref{Residual symmetries}.

\subsection{Charges  for residual symmetries and their bracket}  
\label{Charges  for residual symmetries and their bracket}  

As discussed in section \ref{Residual symmetries of the first kind}, and illustrated in Proposition \ref{prop1} and \ref{prop2},
if the chosen dressing field only eliminates a normal subgroup  $\Diff_{\text{\tiny$0$}}(M)$ of $\Diff(M)$, the above dressed objects would display residual transformations under the remaining subgroup   $\Diff_{\text{\tiny r}}(M) \subset \Diff(M)$.
In that case, the analysis and results of section \ref{Covariant phase space methods} applies mutatis mutandis: in particular Noether charges associated to 
$\diff_{\text{\tiny r}}(M)$ are expressed via \eqref{Noether-charge2} in terms of ${\bs \theta_\Sigma}^{\bs u}$ and $\bs E^{\bs u}$, while their (on-shell Poisson) bracket is induced by ${\bs \Theta_\Sigma}^{\bs u}$. 
\medskip

As detailed in section \ref{Residual symmetries of the second kind},
if the dressing field eliminates $\Diff(M)$ completely, a priori there may  be  residual transformations of the second kind stemming from  ambiguities  \eqref{dressing-ambiguity} in choosing the dressing field: Transformations  under $\Diff(N)$, the structure group of the space of dressed fields $\Phi^{\bs u}$, and under $\bs\Diff_v(\Phi^{\bs u}) \simeq C^\infty\big( \Phi^{\bs u}, \Diff(N) \big)$, its group of vertical diffeomorphisms. 
In this case, results of section \ref{Noether charges for field-independent gauge parameters} again applies mutatis mutandis.

 Noether charges associated to $\mathsf X \in \diff(N)$ are immediately found to be, as in \eqref{Noether-charge1}-\eqref{Noether-charge2}:
\begin{equation}
\label{Noether-charges-residual}
\begin{aligned}
Q_{\Sigma^{\bs u}} (\mathsf X; \phi^{\bs u})\defeq \langle J(\mathsf X; \phi^{\bs u}), \Sigma^{\bs u} \rangle =\int_{\Sigma^{\bs u}} J(\mathsf X; \phi^{\bs u}) &=\int_{\Sigma^{\bs u}}  \iota_{{\mathsf X}^v} \bs\theta^{\bs u}  - \iota_{\mathsf X} L^{\bs u}   -d \gamma(\mathsf X;\phi^{\bs u}),\\[1mm]
&=\int_{\d\Sigma^{\bs u}}   \theta(\iota_{\mathsf X} \phi^{\bs u}; \phi^{\bs u}) - \gamma(\mathsf X ; \phi^{\bs u}) \,  - \int_{\Sigma^{\bs u}} E(\iota_{\mathsf X} \phi^{\bs u}; \phi^{\bs u}),
\end{aligned}
\end{equation}
and to satisfy, as in \eqref{Moment-map-equation}, 
\begin{equation}
\label{Moment-map-equation-residual}
\begin{aligned}
\iota_{{\mathsf X}^v} {\bs \Theta_\Sigma}^{\bs u} = -\bs d Q_{\Sigma^{\bs u}} (\mathsf X;\phi^{\bs u}) + \int_{\d\Sigma^{\bs u} }   \iota_{\mathsf X} \bs \theta^{\bs u} - \bs d\gamma(\mathsf X;\phi^{\bs u})\  \, - \int_{\Sigma^{\bs u}}  \iota_{\mathsf X} \bs E^{\bs u},
\end{aligned}
\end{equation}
so that their bracket is induced by ${\bs \Theta_\Sigma}^{\bs u}$ as in \eqref{Poisson-bracket-result}-\eqref{2-cocycle-map} :
\begin{equation}
\label{Poisson-bracket-residual}
\begin{aligned}
\big\{Q_{\Sigma^{\bs u}} (\mathsf X; \phi^{\bs u}), Q_{\Sigma^{\bs u}} (\mathsf Y; \phi^{\bs u}) \big\} \defeq {\bs\Theta_\Sigma}^{\bs u} \big(\mathsf X^v, \mathsf Y^v \big)
				         = Q_{\Sigma^{\bs u}} \big([\mathsf X, \mathsf Y]_{\text{{\tiny $\diff(N)$}}};\phi^{\bs u} \big) 
					&+ \int_{\d\Sigma^{\bs u}} \mathscr{A}\big( \lfloor \mathsf X, \mathsf Y\! \rfloor ;\phi^{\bs u} \big)  + \gamma \big([\mathsf X, \mathsf Y]_{\text{{\tiny $\diff(M)$}}};\phi^{\bs u} \big) \\
					&+ \int_{\Sigma^{\bs u}} \iota_{\mathsf X^v} \iota_{\mathsf Y} \bs E^{\bs u}  -  \iota_{\mathsf Y^v}  \iota_{\mathsf X} \bs E^{\bs u},
\end{aligned}
\end{equation}
with $\mathsf X, \mathsf Y \in \diff(N)$ and $\mathsf X^v, \mathsf Y^v\! \in \Gamma(V\Phi^{\bs u})$ the corresponding fundamental vertical vector fields on the space of dressed fields. 
On-shell this is, 
\begin{align}
\label{Poisson-bracket-on-shell-residual}
\big\{Q_{\Sigma^{\bs u}} (\mathsf X; \phi^{\bs u}), Q_{\Sigma^{\bs u}} (\mathsf Y; \phi^{\bs u}) \big\} = Q_\Sigma \big([\mathsf X, \mathsf Y]_{\text{{\tiny $\diff(N)$}}};\phi^{\bs u} \big)  + \mathscr C\big( \lfloor \mathsf X, \mathsf Y\! \rfloor ;\phi^{\bs u} \big)_{\,| \S}\,,
\end{align}
defining the map 
\begin{equation}
\begin{aligned}
\label{2-cocycle-map-residual}
 \mathscr C \big(\ \ ; {\color{gray}\phi}\big): \diff(N) \times \diff(N) &\rarrow C^\infty(\Phi^{\bs u}),\\[-2mm]
 									(X, Y)		 &  \mapsto  \mathscr C\big( \lfloor \mathsf X, \mathsf Y\! \rfloor ;{\color{gray}\phi} \big) \defeq \int_{\d\Sigma} \mathscr{A}\big( \lfloor \mathsf X, \mathsf Y\! \rfloor ;{\color{gray}\phi^{\bs u}} \big)  + \gamma \big([\mathsf X, \mathsf Y]_{\text{{\tiny $\diff(N)$}}};{\color{gray}\phi^{\bs u}} \big),
\end{aligned}
\end{equation}
with concrete expression as in \eqref{Cocycle-concrete}-\eqref{Cocycle-concrete-on-shell-b-c}. The latter is a 2-cocycle on-shell and under adequate b.c. so that \eqref{Poisson-bracket-on-shell-residual} is a central extension of $\diff(N)$.
\medskip

Obviously the results for charges $Q_{\Sigma^{\bs u}} (\bs{\mathsf X}; \phi^{\bs u})$ associated to $\phi^{\bs u}$-dependent parameters $\bs{\mathsf X} \in C^\infty\big( \Phi^{\bs u}, \diff(N)\big)\simeq \bs\diff_v(\Phi^{\bs u})$ and their bracket are just copy/pasted from  section \eqref{Noether charges for  field-dependent gauge parameters}. 

Also, vertical transformations of all relevant dressed objects under $\bs\Diff_v(\Phi^{\bs u}) \simeq C^\infty\big( \Phi^{\bs u}, \Diff(N) \big)$, by \eqref{GT-general-bis},  are the same as those of their bare counterpart under $\bs\Diff_v(\Phi) \simeq C^\infty\big( \Phi, \Diff(M) \big)$ as found in section \ref{Vertical and gauge transformations}. 
\medskip

As argued in section \ref{Discussion}, the existence of residual symmetries of the second kind due to the ambiguity   in choosing a dressing field, thus the existence of the above corresponding formal algebra of charges, is a systematic feature of cases where   \emph{ad hoc} dressing fields $\bs u=u$ (a.k.a. gravitational Stueckelberg fields) are introduced: Then the residual symmetry $\Diff(N)$ is exactly isomorphic to the original $\Diff(M)$ symmetry, and neither contains more information nor enjoys  more immediate physical interpretation. 

Such is the case in the edge modes literature, where edge modes are indeed  examples of \emph{ad hoc} dressing fields, and the notion of ``surface symmetries" or ``corner symmetries" 
are readily understood in terms of $\Diff(N)$ and  $\bs\Diff_v(\Phi^{\bs u}) \simeq C^\infty\big( \Phi^{\bs u}, \Diff(N) \big)$ as described in section \ref{Residual symmetries of the second kind}. 
To see this, one may compare \eqref{dressing-ambiguity}  and the above results to e.g. 
 eq.(3.44) and section 3.3 in \cite{DonnellyFreidel2016},
 eq.(6.3) and section 6 in \cite{Speranza2018},
or eq.(2.42) and section 2.3 in \cite{Geiller2018}.

The above, or rather more generally section \ref{Residual symmetries of the second kind}, may also be compared to sections 3.3.6 to 3.3.9 of \cite{Hoehn-et-al2022}, where notions such as ``frame reorientations", ``relational atlases", or ``dynamical frame covariance", clearly can be understood in terms of $\Diff(N)$ and  $\bs\Diff_v(\Phi^{\bs u}) \simeq C^\infty\big( \Phi^{\bs u}, \Diff(N) \big)$.

\subsection{Relational interpretation of the DFM}  
\label{Best case}  

In case of a genuinely constructively built $\phi$-dependent dressing field $\bs u= \bs u (\phi)$, the ambiguity parametrised by $\Diff(N)$ may be greatly reduced, possibly (perhaps likely) even to a physically motivated discrete choice of reference field, or d.o.f., among the collection $\phi$. 
This is the situation where the relational interpretation of the formalism is at its strongest: The dressed theory $L^{\bs u}$, and derived basic objects 
$\bs E^{\bs u}$, ${\bs \theta_\Sigma}^{\bs u}$ and ${\bs \Theta_\Sigma}^{\bs u}$,  are  manifestly relational quantities with a clear physical interpretation. 
Such is the case in all manners of ``scalar coordinatisations" of GR, as proposed e.g. in \cite{Rovelli1991, Brown-Kuchar1995, Rovelli2002b, Pons-et-al2009, Pons-et-al2009b}. 
There, the scalar fields, clock fields, or dust fields, are exactly dressing fields $\bs u$, and the dressed metric $g^{\bs u}=\bs u^*g$ (and its derived quantities $\bs\alpha^{\bs u}=\alpha(\bs dg^{\bs u}; g^{\bs u} )$) acquire a clear physical  meaning as a relational observable -- a ``complete observable" in the terminology coined by Rovelli.

When this applies, as one may expect in modelling realistic situations, if conserved charges cannot be derived  via a Noether analysis (all objects being $\Diff(M)$-invariant), they may be derived from the field equations $\bs E^{\bs u}$.

%


\section{Conclusion}  
\label{Conclusion}  

The purpose of this paper was to give a detailed presentation of the dressing field method for $\Diff(M)$-theories, extending the original framework designed to deal with gauge theories with internal gauge symmetry groups -- of either Yang-Mills type, or gauge gravity type based on Cartan geometry.
The proper mathematics required for the formulation of this extension of the DFM is the bundle geometry of field space $\Phi$. By carefully exposing it, we clarified a number of notions often encountered in the literature. Let us mention three especially noteworthy. 

First, we have stressed the proper geometric understanding of field-independent and field-dependent diffeomorphisms: The former, $\Diff(M)$, is the structure group of $\Phi$ as a bundle.  Its action by pullback on forms of $\Phi$ defines their equivariance.  
The latter, $C^\infty\big(\Phi, \Diff(M) \big)\simeq \bs\Diff_v(\Phi)$, is the group of vertical diffeomorphisms of $\Phi$.
It contains as a subgroup $\bs\Aut_v(\Phi)$, the group of vertical \emph{automorphisms} of $\Phi$, isomorphic to its gauge group $\bs\Diff(M)$. Their actions by pullback on forms of $\Phi$ geometrically define general vertical, and gauge transformations respectively.

Secondly, we stressed that the ``extended bracket" \eqref{extended-bracket2} for field-dependent vector fields of $M$, often used in the covariant phase space, edge mode, and BMS literature -- whose introduction traces back  to \cite{Bergmann-Komar1972} and \cite{Salisbury-Sundermeyer1983}, before its resurfacing in \cite{Barnich-Troessaert2009} -- not only can be understood as a bracket of action Lie algebroid  \cite{Barnich2010}, 
but is also the degree 0 case of the Frölicher-Nijenhuis bracket for (vertical) vector-valued forms  on $\Phi$, and is tied to the properties of the Nijenhuis-Lie derivative  \cite{Kolar-Michor-Slovak} on a principal bundle. The latter expressing, on $\Phi$ and in degree 0, the action of $\bs\diff_v(\Phi) \simeq C^\infty\big(\Phi, \diff(M) \big)$. This extended bracket reduces to the standard $\diff(M)$ bracket for $ \bs\diff(M) \simeq \bs\aut_v(\Phi)$, as is standard in bundle geometry. 
Further details on this may be found in a companion  technical note \cite{Francois2023-b}, treating the finite dimensional case.

Thirdly, we have stressed the non-standard notion of twisted equivariant/tensorial forms and of the corresponding twisted connection (necessary for their covariant derivation): these are forms whose equivariance is controlled by $\Diff(M)$ 1-cocycles, rather than $\Diff(M)$ representations. These objects are key to understanding the geometry of classical and quantum  $\diff(M)$-anomalies. The Wess-Zumino consistency condition for such anomalies being e.g. encoded in the horizontality property of the twisted curvature. 
\medskip

As a natural part of the bundle geometry of $\Phi$, we gave an account of integration on $M$ as an invariant pairing on what we called the ``associated bundle of regions", a bundle associated to $\Phi$ via the defining representation of $\Diff(M)$ -- the field of open sets of $M$. This allowed to properly understand the action of $C^\infty\big(\Phi, \Diff(M) \big)$ in integrals, a technicality relevant to the variational principle. 

Exposing the bundle structure of $\Phi$ also shows the proper geometric understanding of several heuristic computations done in various work on covariant phase space literature. In so doing, we thus set the stage to give our own streamlined account of the covariance phase space methods for $\Diff(M)$-theories. 
Starting from a given Lagrangian, we derived concrete and conceptually transparent expressions for the Noether charge associated to both $\diff(M)$ and 
$\bs\diff_v(\Phi)\simeq C^\infty\big(\Phi, \diff(M) \big)$, and for their brackets induced by the symplectic 2-form $\bs\Theta_\Sigma$. These allowed for an easy reading of the conditions for the brackets to be Lie, and the algebra of charges to be  central extension of $\diff(M)$ and $\bs\diff_v(\Phi)\simeq C^\infty\big(\Phi, \diff(M) \big)$: which are, unsurprisingly {\bf 1)} being on-shell, {\bf 2)} being under b.c. killing the symplectic flux (closing the system). The expression we derive for the 2-cocycles reproduces or extends results from the covariant phase space literature.

It is reasonable to try and redefine the bracket so as to relax condition {\bf 2)}: This would be interpretable as  a Poisson bracket well-defined for open systems, i.e. where there is ``flux" through the boundary of the region. 
Such systems are  relevant when analysing electromagnetic and especially gravitational radiations. The latter subject in particular is closely related to BMS asymptotic symmetries (and its generalisations, like $\Lambda$-BMS). 
 In that context, such modified prescriptions have been proposed: Notably the Barnich-Troessaert (BT) bracket \cite{Barnich_Troessaert2011}, followed by attempts to rederive it or by other modifications, e.g. \cite{Harlow-Wu2020, Freidel-et-al2021, Freidel-et-al2021bis, Speranza2022, Kabel-Wieland2022} and references therein. 
It was not our aim to make explicit contact with those attempts. But our first principles derivation  may help to make it  conceptually clearer how one may proceed in a geometrically natural way. We shall revisit this topic in future works.

We derived geometrically the $C^\infty\big(\Phi, \Diff(M) \big)\simeq \bs\Diff_v(\Phi)$-transformations of the various key objects discussed, Lagrangian, field equation, presymplectic potential and 2-forms. This showed in particular that the variational principle, thus the space of solutions, is stable under the action of field-dependent diffeomorphisms. Which proves that $\Diff(M)$-theories have a much larger ``covariance group", $C^\infty\big(\Phi, \Diff(M) \big)\simeq \bs\Diff_v(\Phi)$, a fact hinted at first (to my knowledge) by  Bergmann and Komar in the early 70s \cite{Bergmann-Komar1972}. 

The derivation of these  $C^\infty\big(\Phi, \Diff(M) \big)\simeq \bs\Diff_v(\Phi)$-transformations was also necessary to apply the rule of thumb of the DFM, thereby obtaining the dressed, or basic, counterparts of our key objects. These reproduce and generalise results from the edge mode literature, in particular the formulae for the ``extended" presymplectic potential and 2-form. The DFM also immediately allows to derive the formulas for Noether charges associated to residual symmetries arising from possible ambiguities in the choice of dressing field. These reproduce results of the edge mode literature pertaining to the notion of ``surface/corner symmetries". We stressed that when dressing fields are \emph{ad hoc}, $\bs u=u$, residual symmetries are mere duplicates of $\Diff(M)$, encoding no more information nor susceptible to more immediate physical interpretation. Edge modes being \emph{ad hoc} dressing fields, this applies to surface/corner symmetries, contra the prevalent opinion. 

Finally we stressed that constructive dressing fields $\bs u=\bs u(\phi)$ allows the most interesting application of the DFM. It turns out to be an explicit relational framework for general relativistic theories, i.e. it allows to rewrite such theories in a $\Diff(M)$-invariant and manifestly relational way. The DFM can be seen as a systematic, if formal, implementation  in field theory of Einstein's point coincidence argument: his answer to the hole argument \cite{Norton1993, Stachel2014, Giovanelli2021}, clarifying the physical meaning of $\Diff(M)$-invariance. To register this key physical insight, we tacitly adopt a rebranding of GR as ``General Relationality". 

The DFM also happens to make clear why ``bare" versions of GR theories could be tested before solving the issue of observables: The relational version (often inaccurately referred to as a ``gauged-fixed" version) is formally identical to the bare one, and the practical deployment of the theory tacitly presume the use of its relational version (as material reference frames = dressing fields are used).
In a follow-up work, we will show how the DFM reproduces naturally several works relating to the production of observables in GR, notably the so-called ``scalar coordinatisation" \cite{Rovelli1991, Brown-Kuchar1995, Rovelli2002b, Pons-et-al2009, Pons-et-al2009b}. We will also show how to apply the formalism to cosmological perturbation theory (encompassing the notion of Bardeen variables). 

The definition of ``dressed integrals" is part of the DFM framework, and there appears naturally  the notion of ``dressed regions": i.e. $\Diff(M)$-invariant relationally defined region of \emph{spacetime} -- again in line with the insight of the point-coincidence argument. This in particular dispels the often repeated misconception that ``boundaries break $\Diff(M)$-invariance", which is logically equivalent to the hole argument: Gauge symmetries are never broken, a relationally defined, physical, boundary is obviously $\Diff(M)$-invariant. We will investigate the implications for the definition of local subsystems, black hole physics, and to asymptotic symmetries in a follow-up paper. 
\medskip

The  framework developed here extends naturally (with little adjustments) to relativistic \emph{gauge} field theories, replacing $\Diff(M)$ with $\Aut(P)$ as the structure group of $\Phi$. 
This extension, developed in the forthcoming work~\cite{JTF-Ravera2024}, gives a  relational account of ``Einstein-Yang-Mills" theories. Indeed, given $\H\simeq \Aut_v(P)$ the gauge group of a principal $H$-bundle $P$, on account of the SES of groups \mbox{$\ \H \rarrow \Aut(P) \rarrow \Diff(M)\ $}  canonically associated to $P$,\footnote{The corresponding SES of Lie algebras defines the Atiyah Lie algebroid canonically associated to $P$. See \eqref{SESgroup}-\eqref{Atiyah-Algebroid} in our case $P=\Phi$.} 
the group $\Aut(P)$ locally splits as $\Diff(M) \ltimes \H_{\text{{\tiny loc}}}$, with $ \H_{\text{{\tiny loc}}}$  the (local) gauge group of a gauge field theory.

 We will illustrate this extension of the DFM with the case of pure GR (with $\Diff(M)$ and $ \H_{\text{{\tiny loc}}}=\SO(1, 3)$), of GR + scalar EM (with $\Diff(M)$ and $ \H_{\text{{\tiny loc}}}=\U(1)$), and of conformal gravity in the Cartan geometric framework (with $\Diff(M)$ and $ \H_{\text{{\tiny loc}}}=\SO(2, 4)$). 

This  framework  has obvious implications for quantum gravity that we will also pursue.


\section*{Acknowledgment}  

This work was funded by the OP J.A.C MSCA grant, number CZ.02.01.01/00/22\_010/0003229, co-funded by the Czech government Ministry of Education, Youth \& Sports and the EU.
This research was also funded in part by the Austrian Science Fund (FWF), [P 36542].
Support from the service \emph{Physics of the Universe, Fields and Gravitation} at UMONS (BE) is also  acknowledged.


The author wishes to thank L. Ravera for sustained and stimulating discussions on relationality in general relativitic physics, and for steering him into emphasazing it as  a key feature of the extended DFM developed here. She is to be credited for the rebranding of GR as ``General Relationality", adopted in this paper. 
\appendix

\section{Appendix}
\label{Appendix}

\subsection{Lie algebra (anti)-isomorphisms}
\label{Lie algebra (anti)-isomorphisms}

We prove here that the "verticality" map $|^v :\diff(M) \rarrow \Gamma(V\Phi)$, $X \mapsto X^v$, is a morphism of Lie algebra. 
Let us write the flow through $\phi \in \Phi$ of $X^v \in \Gamma(V\Phi)$ as $\tilde \psi_\tau(\phi) := R_{\psi_\tau} \phi := \psi_\tau^* \phi$, with of course $X =\tfrac{d}{d\tau} \psi_\tau\ \big|_{\tau=0} \in \Gamma(TM)$. So that,
\begin{align}
X^v_{|\phi}=\tfrac{d}{d\tau} \tilde \psi_\tau (\phi)\, \big|_{\tau=0} \quad \left(\  =X(\phi)^v \tfrac{\delta}{\delta \phi}, \ \text{written as a derivation of $C^\infty(\Phi)$} \right).
\end{align}
Thus we can write the bracket of two vertical vector fields is, 
\begin{align}
\label{demo1}
[X^v, Y^v]_{|\phi} = {\bs L_{X^v} Y^v}_{|\phi} :=&\, \tfrac{d}{d\tau} (\tilde\psi_\tau\-)_\star Y^v_{|\tilde\psi_\tau(\phi)}\, \big|_{\tau=0}, \notag\\
								   :=&\, \tfrac{d}{d\tau} \tfrac{d}{ds} \left(  \tilde \psi_\tau\- \circ \tilde\eta_s \circ \tilde\psi_\tau \right) (\phi)\, \big|_{s=0}	   \, \big|_{\tau=0}, \notag\\
								   :=&\, \tfrac{d}{d\tau} \tfrac{d}{ds}\,  R_{\psi\-_\tau} \circ R_{\eta_s} \circ R_{\psi_\tau} \phi\, \big|_{s=0}	   \, \big|_{\tau=0},  \notag\\
								    :=&\, \tfrac{d}{d\tau} \tfrac{d}{ds}\,  R_{(\psi_\tau \circ \eta_s \circ \psi\-_\tau)} \phi\, \big|_{s=0}	   \, \big|_{\tau=0},  \notag\\
								    :=&\, \tfrac{d}{d\tau} \tfrac{d}{ds}\,  (\underbrace{\psi_\tau \circ \eta_s \circ \psi\-_\tau}_{\text{flow of $-[X,Y]$}})^*\phi\, \big|_{s=0}	   \, \big|_{\tau=0}, \notag\\
								     =&\!:\, \mathfrak L_{-[X, Y]} \, \phi =: (-[X, Y]_{\text{{\tiny $\Gamma(TM)$}}})^v_{|\phi}= ([X, Y]_{\text{{\tiny $\mathfrak{diff}(M)$}}})^v_{|\phi}.
\end{align}
In the $1^{\text{st}}$ to the $2^{\text{nd}}$, and $5^{\text{th}}$ to $6^{\text{th}}$,  lines we used the definition of $\tilde\psi_\tau$ and $\psi_\tau$-relatedness. In the first and last lines, we used the definitions of the Lie derivative of a vector field on $\Phi$ and of a  field on $M$. 
\medskip

We now prove that the map $|^v : \bs{\mathfrak{Diff}}(M) \rarrow \Gamma_{\text{\!\tiny{inv}}}(V\Phi)$, $\bs X \mapsto \bs X^v$, is a Lie algebra \emph{anti}-morphism. 
The flow through $\phi \in \Phi$ of $\bs X^v \in \Gamma_{\text{\!\tiny{inv}}}(V\Phi)$ as $\tilde{\bs\psi}_\tau(\phi) :=(R_{\bs\psi_\tau})(\phi)= R_{\bs\psi_\tau(\phi)} \phi := (\bs\psi_\tau(\phi))^* \phi$, with of course $\bs X =\tfrac{d}{d\tau} \bs\psi_\tau\ \big|_{\tau=0} \in \Gamma(TM)$. So,
\begin{align}
[\bs X^v, \bs Y^v]_{|\phi} = {\bs L_{\bs X^v} \bs Y^v}_{|\phi} :=\, \tfrac{d}{d\tau} (\tilde{\bs\psi}_\tau\-)_\star \bs Y^v_{|\tilde{\bs\psi}_\tau(\phi)}\, \big|_{\tau=0} 
				     = \tfrac{d}{d\tau} \tfrac{d}{ds} \left(  \tilde{\bs\psi}_\tau\- \circ \tilde{\bs\eta}_s \circ \tilde{\bs\psi}_\tau \right) (\phi)\, \big|_{s=0} \big|_{\tau=0} .
\end{align}
We have used the equivariance property of the elements of the gauge group. 
Now, we have, 
\begin{align}
\tilde{\bs\eta}_s \circ \tilde{\bs\psi}_\tau  (\phi) &= R_{\bs\eta_s(\tilde{\bs\psi}_\tau (\phi))}\,  \tilde{\bs\psi}_\tau (\phi)
								        = R_{\bs\eta_s\big (R_{\bs\psi_\tau (\phi)} \phi  \big)}\,  R_{\bs\psi_\tau (\phi)} \phi,  \notag\\ 
								      &= R_{\bs\psi_\tau (\phi)\- \circ \bs\eta_s(\phi) \circ \bs\psi_\tau (\phi) }\,  R_{\bs\psi_\tau (\phi)} \phi
								        = R_{\bs\eta_s(\phi) \circ \bs\psi_\tau (\phi) }\, \phi, \notag \\
								        &= \left( R_{\bs\eta_s \circ \bs\psi_\tau }\right) \phi.
\end{align}
So, 
\begin{align}
 \left(  \tilde{\bs\psi}_\tau\- \circ \tilde{\bs\eta}_s \circ \tilde{\bs\psi}_\tau \right) (\phi) =  \tilde{\bs\psi}_\tau\- \big( \tilde{\bs\eta}_s \circ \tilde{\bs\psi}_\tau  (\phi)\big) 
 				&= R_{\bs\psi_\tau \-\big( R_{\bs\eta_s(\phi) \circ \bs\psi_\tau (\phi) }\, \phi  \big)   } \, R_{\bs\eta_s(\phi) \circ \bs\psi_\tau (\phi) }\, \phi ,\notag \\
				&=R_{[\bs\eta_s(\phi) \circ \bs\psi_\tau (\phi)]\- \circ  \bs\psi_\tau (\phi)\- \circ [\bs\eta_s(\phi) \circ \bs\psi_\tau (\phi)]}\, R_{\bs\eta_s(\phi) \circ \bs\psi_\tau (\phi)}\, \phi,  \notag\\
				&= R_{ \bs\psi_\tau (\phi)\- \circ \bs\eta_s(\phi) \circ \bs\psi_\tau (\phi)}\, \phi. 
\end{align}
Finally, 
\begin{align}
\label{demo2}
[\bs X^v, \bs Y^v]_{|\phi} &=\tfrac{d}{d\tau} \tfrac{d}{ds} \,  R_{ \bs\psi_\tau (\phi)\- \circ \bs\eta_s(\phi) \circ \bs\psi_\tau (\phi)}\, \phi  \, \big|_{s=0}\big|_{\tau=0} , \notag\\
				      &= \tfrac{d}{d\tau} \tfrac{d}{ds}\,  (\underbrace{\bs\psi_\tau (\phi)\- \circ \bs\eta_s(\phi) \circ \bs\psi_\tau (\phi) }_{\text{flow of $[\bs X,\bs Y]$}})^*\phi\, \big|_{s=0}	   \, \big|_{\tau=0}, \notag\\
					&=:\, \mathfrak L_{[\bs X, \bs Y]} \, \phi =: ([\bs X, \bs Y]_{\text{{\tiny $\Gamma(TM)$}}})^v_{|\phi}= (-[\bs X, \bs Y]_{\text{{\tiny $\mathfrak{diff}(M)$}}})^v_{|\phi}.
\end{align}
Which prove the assertion. We notice that all these computations would work the same, \emph{mutadis mutandis}, for gauge theories with an internal group of symmetry.

\subsection{Pushforward by a vertical diffeomorphism of field space}
\label{Pushforward by a vertical diffeomorphism}

Consider a generic vector field $\mathfrak X \in \Gamma(T\Phi)$ with flow $\vphi_\tau$, and a vertical diffeomorphism $\Xi \in \bs\Diff_v(\phi)$ to which corresponds $\bs \psi \in C^\infty\big(\Phi, \Diff(M)\big)$. 
The pushforward of $\mathfrak X_{|\phi} \in T_\phi\Phi$ by $\bs\psi$ is
 \begin{align}
 \Xi_\star \mathfrak X_{|\phi} = \tfrac{d}{d\tau} \,  \Xi \big( \vphi_\tau(\phi) \big)  \,\big|_{\tau=0} = \tfrac{d}{d\tau} \,  R_{\bs\psi \left( \vphi_\tau(\phi) \right)}  \vphi_\tau(\phi)   \, |_{\tau=0}  
                                              &=  \tfrac{d}{d\tau} \,   R_{\bs\psi \left( \vphi_\tau(\phi) \right)}  \phi   \,\big|_{\tau=0} + \tfrac{d}{d\tau} \,  R_{\bs\psi (\phi)}  \vphi_\tau(\phi)   \, |_{\tau=0}  \notag\\
                                              &=  \tfrac{d}{d\tau} \,   R_{\bs\psi \left( \vphi_\tau(\phi) \right)}  \phi   \,\big|_{\tau=0} + R_{\bs\psi (\phi)\star}  \mathfrak X.
 \end{align}
In the last equality we used the definition of the pushforward by the right action of $\bs\psi (\phi) \in \Diff(M)$. The remaining term is clearly a vertical vector field, the question is by what element of $\Gamma(TM) = \mathfrak{Diff}(M)$ it is generated, knowing furthermore that it must be anchored at $\Xi(\phi)=R_{\bs\psi(\phi)} \phi=\bs\psi(\phi)^*\phi$. 
 \begin{align}
 \tfrac{d}{d\tau} \,   R_{\bs\psi \left( \vphi_\tau(\phi) \right)}  \phi   \,\big|_{\tau=0}  = 
 \tfrac{d}{d\tau} \,   R_{{\color{blue} \bs\psi (\phi) \circ \bs\psi(\phi)^{-1}} \circ \bs\psi \left( \vphi_\tau(\phi) \right)}  \phi   \,\big|_{\tau=0} = 
  \tfrac{d}{d\tau} \,   R_{\bs\psi(\phi)^{-1} \circ \bs\psi \left( \vphi_\tau(\phi) \right) }  \, R_{ \bs\psi (\phi)} \phi   \,\big|_{\tau=0}.   \notag
 \end{align}
So, $\bs\psi(\phi)^{-1} \circ \bs\psi \left( \vphi_\tau(\phi)\right)$ is the flow of the sought after vector field of $M$. Now, we observe that on the one hand:
 \begin{align}
  \tfrac{d}{d\tau} \,  \bs\psi\big( \vphi_\tau(\phi)  \big)  \,\big|_{\tau=0}  = \bs d \bs\psi_{|\phi} \big( \mathfrak X_{|\phi} \big) = \bs\psi_\star \mathfrak X_{|\phi}  \in T_{\bs\psi(\phi)}\Diff(M),\notag
  \end{align}
  since $\bs\psi : \Phi \rarrow \Diff(M)$. 
  On the other hand, the Maurer-Cartan form on $\Diff(M)$ is:
  \begin{align}
   L_{\psi^{-1} \star} : T_\psi \Diff(M) \rarrow&\  T_{\id}\Diff(M)=\mathfrak{Diff}(M)\simeq \Gamma(TM),  \notag\\
   			\mathcal X_{|\psi}  \mapsto &\   L_{\psi^{-1} \star} \mathcal X_{|\psi} \defeq (\psi^{-1})_* \mathcal X_{|\psi} 
  \end{align}
  So we have that, 
  \begin{align}
     L_{\bs\psi(\phi)^{-1} \star} : T_{\bs\psi(\phi)} \Diff(M) \rarrow&\  \mathfrak{Diff}(M)\simeq \Gamma(TM),  \notag\\
	       \big[  \bs d \bs\psi_{|\phi} \big( \mathfrak X_{|\phi} \big) \big]_{|\bs\psi(\phi)}  \mapsto& \   L_{\bs\psi(\phi)^{-1}\star} \big[  \bs d \bs\psi_{|\phi} \big( \mathfrak X_{|\phi} \big) \big]_{|\bs\psi(\phi)} = (\bs\psi(\phi)^{-1})_* \big[  \bs d \bs\psi_{|\phi} \big( \mathfrak X_{|\phi} \big) \big]_{|\bs\psi(\phi)} \notag\\
	       																	&\ =L_{\bs\psi(\phi)^{-1} \star}  \tfrac{d}{d\tau} \,  \bs\psi\big( \vphi_\tau(\phi)  \big)  \,\big|_{\tau=0}  
																		  =   \tfrac{d}{d\tau} \,  \bs\psi(\phi)^{-1}\circ \bs\psi\big( \vphi_\tau(\phi)  \big)  \,\big|_{\tau=0} 	
  \end{align}
  Thus, $(\bs\psi(\phi)^{-1})_* \big[  \bs d \bs\psi_{|\phi} \big( \mathfrak X_{|\phi} \big) \big]$ is the generating vector field of $M$ we looked for. Therefore, we have:
   \begin{align}
 \tfrac{d}{d\tau} \,   R_{\bs\psi \left( \vphi_\tau(\phi) \right)}  \phi   \,\big|_{\tau=0}  = 
  \tfrac{d}{d\tau} \,   R_{\bs\psi(\phi)^{-1} \circ \bs\psi \left( \vphi_\tau(\phi) \right) }  \, R_{ \bs\psi (\phi)} \phi   \,\big|_{\tau=0} =
  \left\{    (\bs\psi(\phi)^{-1})_* \big[  \bs d \bs\psi_{|\phi} \big( \mathfrak X_{|\phi} \big) \big]  \right\}^v_{|R_{\bs\psi(\phi)} \phi}.
 \end{align}
So that finally we get the result:
 \begin{align}
 \Xi_\star \mathfrak X_{|\phi} &= R_{\bs\psi(\phi) \star} \mathfrak X_{|\phi} + \left\{   \bs\psi(\phi)\-_* \bs d \bs\psi_{|\phi}(\mathfrak X_{|\phi})   \right\}^v_{|\Xi(\phi)},  \notag\\
 					    &=R_{\bs\psi(\phi) \star} \left( \mathfrak X_{|\phi} +  \left\{  \bs d \bs\psi_{|\phi}(\mathfrak X_{|\phi})  \circ \bs\psi(\phi)\-  \right\}^v_{|\phi} \right). 
\end{align}
Where in the second line we used the property \eqref{Pushforward-fund-vect} of the fundamental vertical vector fields under pushforward by the right action of the structure group. 	 
As we show in the main text, this result is essential to compute geometrically the vertical (and gauge) transformations of variational forms on $\Phi$.

\subsection{Assumptions on the set of fields and the Lagrangian functional}
\label{A3}


First, we assume that $L'/L$ is a gauge theory, based on a $H$-bundle $P$, so that the elementary fields under considerations are Ehresmann and/or Cartan connections 1-forms $A$, depending if we are doing Yang-Mills or gravity gauge theories (or both), and tensorial 0-forms $\upvarphi$ (sections of associated bundles) representing matter fields. Thus $\phi=\{A, \upvarphi\}$. The curvature of $A$ is the $\Ad$-tensorial 2-form $F=DA$, and the covariant derivative of a  $\upvarphi$ is  $\rho$-tensorial 1-form $D\upvarphi=d\upvarphi+\rho_*(A)\upvarphi$. 
The covariant derivative of a $\rho$-tensorial form $b$ is $Db=db+\rho_*(A)b$, and we have that $DF=0$ (Bianchi identity) and $DD\upvarphi=\rho_*(F)\upvarphi$. We write $D\phi=\{F, D\upvarphi \}$. Since $D^{2p} \upvarphi= \rho_*(F^p)\upvarphi$ and $D^{2p+1}\upvarphi = \rho_*(F^p)D\upvarphi$, $\{\phi, D\phi\}$ is an algebraically closed set of variables under the action of $D$. Defining the ``covariant Lie derivative" $\mathcal L^D_X \defeq [\iota_X, D]$, it is easy to probe that $\mathcal L_Xb= \mathcal L^D_X b -\rho_*(\iota_XA)b$.
On the other hand, it also easy seen that $\mathcal L_X A = D(\iota_XA) + \iota_X F$. So we can write the generic expression
\begin{align}
\label{identity1}
\iota_{X^v} \bs d \phi =\mathcal L_X \phi= \iota_X D\phi + D(\iota_X \phi) - \rho_*(\iota_X A) \phi. 
\end{align}

Now, we assume the Lagrangian is $H$-invariant, and of the form 
\begin{align}
L(\phi)=\t L(\phi; D\phi), 
\end{align}
where $\t L$ is the functional expression in terms of $\phi$ and $D\phi$. In other words, we assume that actually $L$ is defined on $J^1\Phi$, the $1^\text{st}$ jet bundle of field space. Then we have, 
\begin{align}
\bs d L_{|\phi}= \t L_0(\bs d\phi; \{\phi\}) + \t L_1(\bs d D\phi; \{\phi\}), 
\end{align}
 where $\{\phi\}$ means the collection of remaining $\phi$ and $D\phi$ in the respective functional expression $\t L_{0/1}$. 
 Using $\bs d D\phi = D(\bs d \phi) + \rho_*(\bs d A)\phi$,
 and integrating by parts, we get the formal expression of the field equation and presymplectic potential:
 \begin{equation}
 \begin{aligned}
\bs d L_{|\phi}&= \t L_0\big(\bs d\phi; \{\phi\}\big) + \t L_1\big( D(\bs d \phi) + \rho_*(\bs d A)\phi; \{\phi\}\big), \notag \\
		    &= d \t L_1\big( \bs d \phi; \{\phi\}\big) -(-)^r \t L_1\big(\bs d \phi; D\{\phi\}\big) + \t L_1\big(\rho_*(\bs dA); \{\phi\}\big) + \t L_0\big(\bs d\phi; \{\phi\}\big), \notag\\
		    &\rdefeq d\theta (\bs d\phi; \phi) + E(\bs d\phi; \phi), \\
		    & = d\bs \theta_{|\phi} + \bs E_{|\phi}. 
\end{aligned}
 \end{equation}
 Where $r=0,1$ is the $\Omega^\bullet(M)$ form degree of  $\bs d\phi$  over which the integration by part is made, and we used the $H$-invariance of $L/\t L_{(0/1)}$ to write $ \t L_1\big( \rho_*(A), \bs d \phi; \{\phi\}) + (-)^r  \t L_1\big( \bs d \phi; \rho_*(A)\{\phi\})=0$. From this, we find on the one hand the expression of the evaluation of $\bs\theta$ on a vertical vector field $X^v \in \Gamma(V\Phi)$:
 \begin{align}
 \label{Vertic-theta-formal}
 \iota_{X^v} \bs\theta = \theta ( \iota_{X^v} \bs d\phi; \phi) =\t L_1 ( \iota_{X^v} \bs d\phi; \phi)
 \end{align}
 On the other, using \eqref{identity1}, hand we also get the expression of the evaluation of the Lagrangian $n$-form on a vector field $X\in \Gamma(TM)\simeq \diff(M)$, acting as a derivation (like $\bs d$):
  \begin{align*}
 \iota_{X} L(\phi) =  \iota_{X} \t L(\phi; D\phi) &= \t L_0\big( \iota_{X} \phi; \{\phi\} \big) + \t L_1\big( \iota_{X} D\phi; \{\phi\}\big), \notag\\
 							            &= \t L_0\big( \iota_{X} \phi; \{\phi\}\big) + \t L_1\big( \iota_{X^v} \bs d \phi -  D(\iota_X \phi) + \rho_*(\iota_X A) \phi; \{\phi\} \big), \notag\\
							            &=\t L_1( \iota_{X^v} \bs d \phi ; \{\phi\}) - d\t L_1(  \iota_{X} \phi ; \{\phi\}) + \t L_1( \iota_{X} \phi ; D\{\phi\}
							            		+ \t L_1\big( \rho_*(\iota_X A) \phi; \{\phi\} \big) + \t L_0\big(\iota_X \phi; \{\phi\}\big),\notag \\
								&= \iota_{X^v} \bs \theta - d\theta( \iota_X\phi; \phi) + E( \iota_X\phi; \phi). 
 \end{align*}
Which give us the important identity, 
 \begin{align}
 \label{Identity2}
\iota_{X^v} \bs \theta - \iota_X L = d\theta( \iota_X\phi; \phi) - E( \iota_X\phi; \phi).
 \end{align}
It is used in section \ref{Noether charges for field-independent gauge parameters} to expressed the Noether current and charge,  so as to make manifest that they are $d$-exact on-shell.

\bigskip

 \subsection{Condition for the bracket \eqref{Poisson-bracket-on-shell} to be Lie }
\label{A4}

For any form $\omega \in \Omega^1(N)$ and $X,Y,Z \in \Gamma(N)$, we have by Koszul formula:
\begin{align}
d\omega(X,Y) = X \cdot \omega(Y) - Y \cdot \omega(X) - \omega([X,Y])
		      &= \iota_X d[ \omega(Y)] - \iota_Y d[ \omega(X)] - \iota_{[X, Y]} \omega,\notag \\
		      &= L_X \iota_Y \omega - L_Y \iota_X \omega -  \iota_{[X, Y]} \omega, \label{A} \\
		      &= \iota_Y L_X \omega - L_Y \iota_X \omega, \notag \\
		      &= \iota_Y L_X \omega - \iota_X L_Y \omega + \iota_{[X, Y]} \omega.  \label{B'}
\end{align}
The two numbered equalities are useful ways to express the exterior derivative of a 1-form.


Now, for a $d$-exact 2-form $\Omega=d\omega \in \Omega^2(N)$, we have by \eqref{B'}:
 \begin{align}
 \Omega\big(X, [Y,Z] \big) + c.p. &=  d\omega\big(X, [Y,Z] \big) + c.p. \notag \\
 					       &= -\iota_X  L_{[Y, Z]} \omega +   \iota_{[Y, Z]} L_X  \omega -  \iota_{\big[X, [Y, Z]\big]} \omega \ + \ c.p. \notag \\
					       &=  -\iota_X  [L_Y, L_Z] \omega +   \iota_{[Y, Z]} L_X  \omega \ + \ c.p.  \notag\\
					       &= -\iota_X  L_Y L_Z \omega +  \iota_X  L_Z L_Y \omega +  \iota_{[Y, Z]} L_X  \omega  \ + \ c.p.  \notag\\
					       &= \left( -L_Y \iota_X + \iota_{[Y,X]} \right)L_Z\omega  +  \left( L_Z \iota_X - \iota_{[Z,X]} \right)L_Y\omega + \iota_{[Y, Z]} L_X  \omega  \ + \ c.p. \notag\\[3mm]
 &=  \munderline{ForestGreen}{\left( -L_Y \iota_X + \iota_{[Y,X]} \right)L_Z\omega}  +  \left(  \munderline{red}{L_Z \iota_X} - \iota_{[Z,X]} \right)L_Y\omega + \iota_{[Y, Z]} L_X  \omega \notag \\[1mm]
   & \ \ +  \munderline{blue}{\left( -L_Z \iota_Y + \iota_{[Z,Y]} \right)L_X\omega}  +  \left(  \munderline{ForestGreen}{L_X \iota_Y} - \iota_{[X,Y]} \right)L_Z\omega + \iota_{[Z, X]} L_Y  \omega \notag \\[1mm]
   &\ \ +  \munderline{red}{\left( -L_X \iota_Z + \iota_{[X,Z]} \right)L_Y\omega}  +  \left(  \munderline{blue}{L_Y \iota_Z} - \iota_{[Y,Z]} \right)L_X\omega + \iota_{[X, Y]} L_Z  \omega, \notag\\[3mm]
   &=   \munderline{blue}{\big\{ d(L_X \omega) \big\}(Y,Z)} +   \munderline{red}{\big\{ d(L_Y \omega) \big\}(Z,X)} +   \munderline{ForestGreen}{\big\{ d(L_Z \omega) \big\}(X, Y)},\notag \\
   &= \big\{L_X \Omega\big\}  (Y, Z) + c.p. \label{Cool-identity}
\end{align}
By \eqref{A} to get the second before last line, and by $[L_X, d]=0$ to the last. Which is finally, since $d\Omega=0$:
 \begin{align}
\Omega\big(X, [Y,Z] \big) + c.p. = \big\{d \big(\iota_X \Omega \big)\big\}  (Y, Z) + c.p.
\end{align}

Applying this to the presymplectic 2-form current $\bs\Theta \in \Omega^2(\Phi)$ and $X^v, Y^v, Z^v \in \Gamma(V\Phi)$: 
\begin{align}
\label{Jacobi-type}
\bs\Theta \big(X^v, [Y^v, Z^v] \big) + c.p. = \big( \bs L_{X^v} \bs\Theta \big)(Y^v, Z^v) + c.p.= \big( \bs d\iota_{X^v} \bs\Theta \big)(Y^v, Z^v) + c.p.
\end{align}
Now, we defined a formal  bracket \eqref{Poisson-bracket-result} by $\big\{Q_\Sigma (X; \phi), Q_\Sigma (Y; \phi) \big\} \defeq \bs\Theta_\Sigma \big(X^v, Y^v \big)$,
and  from its on-shell form \eqref{Poisson-bracket-on-shell} one has:
\begin{equation}
\begin{aligned}
\mathscr C\big( \lfloor X, Y\! \rfloor ;\phi \big) \defeq&\,  \big\{Q_\Sigma (X; \phi), Q_\Sigma (Y; \phi) \big\} - Q_\Sigma ([X,Y]_{\text{{\tiny $\diff(M)$}}}; \phi), \\
									=&\ \ \bs\Theta_\Sigma \big( X^v, Y^v \big) - Q_\Sigma ([X,Y]_{\text{{\tiny $\diff(M)$}}}; \phi).
\end{aligned}
\end{equation}
So, 
\begin{equation}
\label{2-cocycle-condition-diff}
\begin{aligned}
 \mathscr C \big( \lfloor X, [Y, Z]_{\text{{\tiny $\diff(M)$}}}\rfloor ; \phi \big) + c.p.  &= \bs\Theta_\Sigma \big( X^v, ( \big[Y, Z ]_{\text{{\tiny $\diff(M)$}}})^v \big) - Q_\Sigma (\big[X, [Y, Z]_{\text{{\tiny $\diff(M)$}}}\big]_{\text{{\tiny $\diff(M)$}}}; \phi) + c.p. , \\
 											&= \bs\Theta_\Sigma \big( X^v, [Y^v, Z^v] \big) - Q_\Sigma (\big[X, [Y, Z]_{\text{{\tiny $\diff(M)$}}}\big]_{\text{{\tiny $\diff(M)$}}}; \phi)  + c.p. \\
											&=  \big( \bs d\iota_{X^v} \bs\Theta_\Sigma \big)(Y^v, Z^v) + c.p.
\end{aligned}
\end{equation}
Where we used \eqref{demo1},  the fact that $Q_\Sigma({\color{gray}X}; \phi)$ is a linear map in its first argument together with Jacobi in $\diff(M)$, and \eqref{Jacobi-type}. 
Thus, \eqref{2-cocycle-condition-diff} tells us that the 2-cocycle condition on $ \mathscr C $ holds, and the bracket \eqref{Poisson-bracket-on-shell} is Lie, if  the vertical vector fields are either symplectic or Hamiltonian for the charges, showing that the Poisson algebra of these charges is a central extension \eqref{PB-central-ext} of $\diff(M)$, as discussed in section \ref{Noether charges for field-independent gauge parameters}.

 \subsection{Concrete expression of the map \eqref{2-cocycle-map}}
\label{A5}

Here we derive the result necessary to write the concrete expression \eqref{Cocycle-concrete} for the $\mathscr C$-map \eqref{2-cocycle-map}. We have
\begin{align}
\mathscr C\big( \lfloor X, Y\! \rfloor ;\phi \big) \defeq \int_{\d\Sigma} \mathscr{A}\big( \lfloor X, Y\! \rfloor ; \phi \big)  + \gamma \big([X, Y]_{\text{{\tiny $\diff(M)$}}};\phi \big), 
\end{align}
with the definition \eqref{2-cocycle} for the first term in the integrand: 
\begin{align}
- d\mathscr{A}\big( \lfloor X, Y\! \rfloor ;\phi \big) = \iota_{X^v} \bs \alpha(Y) -  \iota_{Y^v}   \bs \alpha(X)  + \iota_{X^v} \iota_Y \bs E  -  \iota_{Y^v}  \iota_X \bs E  -  \beta \big([X, Y]_{\text{{\tiny $\diff(M)$}}};\phi \big) 
\end{align}
To computing it explicitly, using  \eqref{symplect-anomaly}, we start with the terms involving the symplectic anomaly:
\begin{align}
 \iota_{X^v} \bs \alpha(Y)  &= \iota_{X^v} \bigg(  \iota_Y (\bs dL- \bs E) + d \iota_Y \bs \theta + \bs d \beta_\ell(Y;\phi) \bigg), \notag\\
 				        &= \iota_Y \mathfrak L_X L -  \iota_{X^v} \iota_Y \bs E +  d \iota_Y ( \iota_{X^v}\bs \theta)  +  \iota_{X^v} \bs d \mathfrak L_Y \ell, \notag\\
				        &= \iota_Y d\iota_X L -  \iota_{X^v} \iota_Y \bs E + d \iota_Y \bigg( \iota_X L + d\theta(\iota_X \phi ;\phi) -  E(\iota_X \phi ;\phi) \bigg) 
				        +  \mathfrak L_Y\mathfrak L_X \ell \notag \\
				        &= \iota_Y d\iota_X L + d\iota_Y \iota_X L +  \mathfrak L_Y\mathfrak L_X \ell -  \iota_{X^v}  \iota_Y \bs E 
				        +   	d\bigg( \iota_Yd\theta(\iota_X \phi ;\phi)  -  \iota_Y   E(\iota_X \phi ;\phi) \bigg). 
\end{align}
where $dL=0$ and \eqref{Identity2} are used to get the third line. Then, using
\begin{align*}
[\mathfrak L_X, \iota_Y] L &= \iota_{[X,Y]_{\text{{\tiny $\Gamma(TM)$}}} } L \\
					  &=  \iota_X d \iota_Y L + d \iota_X \iota_Y L - \iota_Y\iota_X \cancel{dL} - \iota_Y d \iota_X L, 
\end{align*}
we have that:
\begin{align*}
 \iota_{X^v} \bs \alpha(Y)  - \iota_{Y^v} \bs \alpha(x) &= - \iota_{[X,Y]_{\text{{\tiny $\Gamma(TM)$}}} }L   - \mathfrak L_{[X, Y]_{\text{{\tiny $\Gamma(TM)$}}} }  \ell   
 											-  \iota_{X^v}  \iota_Y \bs E +   \iota_{Y^v}  \iota_X \bs E  \notag\\
 								               &\quad+ d\bigg( \iota_Yd\theta(\iota_X \phi ;\phi) - \iota_Xd\theta(\iota_Y \phi ;\phi) +  \iota_Y \iota_X L \bigg) 
										        + d\bigg( -  \iota_Y   E(\iota_X \phi ;\phi)  +   \iota_X   E(\iota_Y \phi ;\phi)  \bigg), \notag \\
										&= - \beta\big([X,Y]_{\text{{\tiny $\Gamma(TM)$}}}  ;\phi\big)  
										      -  \iota_{X^v}  \iota_Y \bs E +   \iota_{Y^v}  \iota_X \bs E  \notag\\
					&\quad  + d\bigg( \mathfrak L_Y \theta(\iota_X \phi ;\phi) -  \mathfrak L_X \theta(\iota_Y \phi ;\phi) +  \iota_Y \iota_X L \bigg) 
					              + d\bigg( -  \iota_Y   E(\iota_X \phi ;\phi)  +   \iota_X   E(\iota_Y \phi ;\phi)  \bigg),
\end{align*}
It follows indeed that the first term in the integrand of $\mathscr C$ is
\begin{align}
\label{dA}
d\mathscr{A}\big( \lfloor X, Y\! \rfloor ;\phi \big)  = d\bigg( \mathfrak L_X \theta(\iota_Y \phi ;\phi) -  \mathfrak L_Y \theta(\iota_X \phi ;\phi) +  \iota_X \iota_Y L  
					             \ \  + \ \  \iota_Y   E(\iota_X \phi ;\phi)  -   \iota_X   E(\iota_Y \phi ;\phi)   \bigg)
\end{align}
So that finally, 
\begin{align}
\mathscr C\big( \lfloor X, Y\! \rfloor ;\phi \big) = \int_{\d\Sigma}  \mathfrak L_X \theta(\iota_Y \phi ;\phi) -  \mathfrak L_Y \theta(\iota_X \phi ;\phi) +  \iota_X \iota_Y L  + \gamma \big([X, Y]_{\text{{\tiny $\diff(M)$}}};\phi \big) \ \  + \ \  \iota_Y   E(\iota_X \phi ;\phi)  -   \iota_X   E(\iota_Y \phi ;\phi). 
\end{align}
The result \eqref{Cocycle-concrete} displayed in the main text. It is a 2-cocycle on-shell and for adequate b.c.


{
\normalsize 
 \bibliography{Biblio8.5}
}

\end{document}